\documentclass{article}
\usepackage{maple2e}
\usepackage{amsmath}

\usepackage[english]{babel}
\usepackage{t1enc}
\usepackage[latin1]{inputenc}
\usepackage{graphicx}

\def\bse{\begin{subequations}}
\def\ese{\end{subequations}}
\def\comment#1{}
\def\emptyline{\vspace{12pt}}

\begin{document}
\setcounter{page}{1}
\title{Mathematical ultrashort-pulse laser physics}

\author{V.L. Kalashnikov \thanks{Author is Lise Meitner Fellow at TU Vienna and
would like to acknowledge the support of Austrian Science Fund's
grant \#M611. I am grateful to Dr. I.T. Sorokina and Dr. E.
Sorokin for their hospitality at the Photonics Institute (TU
Vienna), for stimulating discussions and experimental support of
Part IX. Also, I thank D.O. Krimer for his help in programming of
Part VI}}

\maketitle
\emptyline
\begin{center}
\textit{ Institut f\"ur Photonik, TU Wien, Gusshausstrasse 27/387,
A-1040 Vienna, Austria}
\end{center}

\begin{center}
\docLink{mailto: vladimir.kalashnikov@tuwien.ac.at}{vladimir.kalashnikov@tuwien.ac.at},
\docLink{http://www.geocities.com/optomaplev}{http://www.geocities.com/optomaplev}
\end{center}

\emptyline

\textit{Abstract:} The analytical and numerical approaches to the
analysis of the ultrashort-pulse solid-state lasers are presented.
The unique self-consistent method of the laser dynamics analysis
is based on the symbolical, numerical, programming, and graphical
capacities of Maple 6. The algorithmization of the basic
conceptions as well as sophisticated research methods is of
interest to both students and experts in the laser physics and the
nonlinear dynamics.

\textit{Application Areas/Subjects}: Optics, Laser Physics,
Nonlinear Physics, Differential Equations, and Programming

\textit{Keywords:} soliton, ultrashort pulse, mode locking,
Q-switching, solid-state laser, nonlinear Schr\"odinger equation,
nonlinear Landau-Ginzburg equation, nonlinear Klein-Gordon
equation, harmonic oscillator, nonlinear oscillator, self-phase
modulation, group-delay dispersion, self-induced transparency,
stimulated Raman scattering

\newpage

\tableofcontents

\newpage

\section{Introduction}

\emptyline

The ultrashort laser pulses, i.e. the pulses with the durations
$\sim$10$^{-10}$--10$^{-15}$ sec, have a lot of the applications,
which range ultrafast spectroscopy,  tracing chemical reactions,
precision processing of materials, optical networks and computing,
nuclear fusion and X - ray lasing, ophthalmology and surgery (for
review see \cite{Brabec}). The mechanisms of the ultrashort pulse
generation are active or passive loss switching (so-called
Q-switching, Part II) and locking of the longitudinal laser modes
(Part III) due to the active (Part V) or passive ultrafast
modulation resulting in the laser quasi-soliton formation. Such
quasi-soliton is very similar to the well-known Schr\"odinger
soliton, which runs in the optical networks (Part VI). As a matter
of fact, the model describing active mode locking is based on the
usual equation of the harmonic oscillator or its nonlinear
modifications, while for the passive mode locking description we
have primordially nonlinear Landau-Ginzburg equation (Part VII).
This equation is the dissipative analog of the nonlinear
Schr\"odinger equation and, as a result of the nonlinear
dissipation, there exist a lot of nonstationary regimes of the
ultrashort pulse generation (Part VIII). This requires the
generalization of the model, which leads to the numerical
simulations on the basis of FORTRAN (or C) codes generated by
Maple (Part IX). Simultaneously, the obtained numerical results
are supported by the analytical modelling in the framework of the
computer algebra approach. The last takes into account the main
features of the nonlinear dissipation in the mode-locked laser,
viz. the power- (Part VII) or energy-dependent (Part X) response
of the loss to the generation field. In the latter case, there is
the possibility of the so-called self-induced transparency
formation, which is described by the nonlinear Klein-Gordon
equation (Part XI).

Our considerations are based on the analytical or semi-analytical
search of the steady-state soliton-like solutions of the laser
dynamical equations and on the investigation of their stability.
Also, the breezer-like solutions are considered using the
aberrationless approximation. The computer algebra analysis is
supported by the numerical simulations on the basis of the Maple
generated FORTRAN-code. We present the analysis of these topics by
means of the powerful capacities of Maple 6 in the analytical and
numerical computations. This worksheet contains some numerical
blocks, which can take about of 12 Mb and 18 min of computation (1
GHz Athlon).

\emptyline

\section{Nonstationary lasing: passive Q-switching}\label{1}

\emptyline

\subsection{Continuous-wave oscillation}
\noindent The basic principle of Q-switching is rather simple, but
in the beginning let's consider the steady-state oscillation of
laser. The near-steady-state laser containing an active medium and
pumped by an external source of the energy (lamp, other laser or
diode, for example) obeys the following coupled equations:

$>$restart:\\
\indent $>$with(linalg):\\
\indent \indent
 $>$eq1 := diff(Phi(t),t) = (alpha(t) - rho)*Phi(t);\# field
 evolution\\
 \indent \indent \indent
  $>$eq2 := diff(alpha(t),t) = sigma[p]*(a[m]-alpha(t))*P/(h*nu[p])
  -\\
alpha(t)*sigma[g]*Phi(t)/(h*nu[g]) - alpha(t)/T[r];\# gain
evolution for quasi-two-level active medium

\[
\mathit{eq1} := {\frac {\partial }{\partial t}}\,\Phi (t)=(\alpha
 (t) - \rho )\,\Phi (t)
\]

\[
\mathit{eq2} := {\frac {\partial }{\partial t}}\,\alpha (t)=
{\displaystyle \frac {{\sigma _{p}}\,({a_{m}} - \alpha (t))\,P}{h
\,{\nu _{p}}}}  - {\displaystyle \frac {\alpha (t)\,{\sigma _{g}}
\,\Phi (t)}{h\,{\nu _{g}}}}  - {\displaystyle \frac {\alpha (t)}{
{T_{r}}}}
\]

\emptyline \noindent Here $\Phi (t)$ is the time-dependent field
intensity, $\alpha (t)$ is the dimensionless gain coefficient,
\textit{P} is the time-independent (for simplicity sake) pump
intensity, $\nu _{p}$ and $\nu _{p}$ are the frequencies of the
pump and generation fields, respectively, $\sigma _{p}$ and
$\sigma _{g}$ are the absorption and generation cross-sections,
respectively, $T_{r}$ is the gain relaxation time, $\rho $ is the
linear loss inclusive the output loss of the laser cavity, and, at
last, $a_{m}$ is the gain coefficient for the full population
inversion in the active medium. The pump increases the gain
coefficient (first term in \textit{eq2}), that results in the
laser field growth (first term in \textit{eq1}). But the latter
causes the gain saturation (second term), which can result in the
steady-state operation (so-called continuous-wave, or simply cw,
oscillation):

\emptyline $>$rhs( subs(\{alpha(t)=alpha,Phi(t)=Phi\},eq1) ) =
0;\\
\indent \indent
 $>$rhs( subs(\{alpha(t)=alpha,Phi(t)=Phi\},eq2) ) = 0;\\
\indent \indent \indent
  $>$sol := solve(\{\%,\%\%\},\{Phi,alpha\});

  \emptyline

\[
(\alpha  - \rho )\,\Phi =0
\]

\[
{\displaystyle \frac {{\sigma _{p}}\,({a_{m}} - \alpha )\,P}{h\,{
\nu _{p}}}}  - {\displaystyle \frac {\alpha \,{\sigma _{g}}\,\Phi
 }{h\,{\nu _{g}}}}  - {\displaystyle \frac {\alpha }{{T_{r}}}} =0
\]

\begin{gather}\nonumber
\mathit{sol} := \{\alpha ={\displaystyle \frac {{\sigma _{p}}\,P
\,{T_{r}}\,{a_{m}}}{{\sigma _{p}}\,P\,{T_{r}} + h\,{\nu _{p}}}} ,
\,\Phi =0\}, \,\\ \nonumber
\{\Phi = - {\displaystyle \frac {{\nu
_{g}}\,(
 - {\sigma _{p}}\,P\,{T_{r}}\,{a_{m}} + {\sigma _{p}}\,P\,{T_{r}}
\,\rho  + \rho \,h\,{\nu _{p}})}{\rho \,{\sigma _{g}}\,{\nu _{p}}
\,{T_{r}}}} , \,\alpha =\rho \}
\end{gather}

\emptyline \noindent The second solution defines the cw intensity,
which is the linear function of pump intensity:

\emptyline $>$expand( subs( sol[2],Phi ) ):\\
\indent
 $>$Phi = collect(\%,P);\\
 \indent \indent
  $>$plot( subs( \{lambda[g]=8e-5, lambda[p]=5.6e-5\},\\
  subs( \{h=6.62e-34,
sigma[g]= 3e-19, sigma[p]= 1e-19, rho=0.1, nu[g]=3e10/ lambda[g],
nu[p]=3e10/lambda[p], T[r]=3e-6, a[m]=2.5\},rhs(rho*\%) ) ), P =
4.9e4..1e5, axes=boxed, title=`output intensity vs. pump,
[W/cm$^2$]` );\# lambda is the wavelength in [cm]

\[
\Phi =({\displaystyle \frac {{\nu _{g}}\,{\sigma _{p}}\,{a_{m}}}{
\rho \,{\sigma _{g}}\,{\nu _{p}}}}  - {\displaystyle \frac {{\nu
_{g}}\,{\sigma _{p}}}{{\sigma _{g}}\,{\nu _{p}}}} )\,P -
{\displaystyle \frac {{\nu _{g}}\,h}{{\sigma _{g}}\,{T_{r}}}}
\]

\emptyline
\begin{center}
\mapleplot{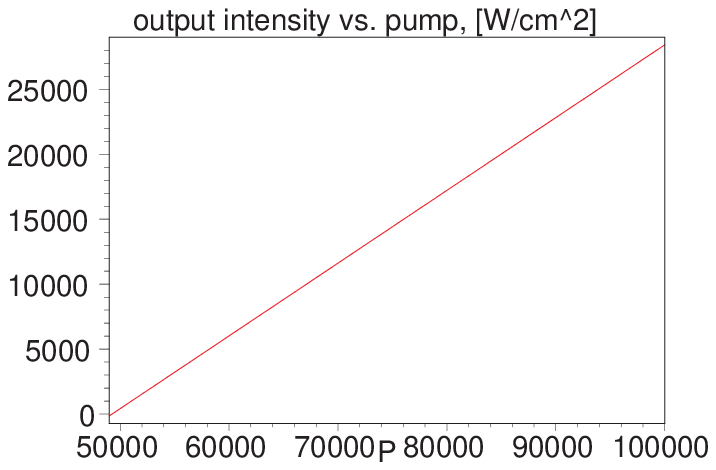}
\end{center}
\emptyline

\emptyline The pump corresponding to $\Phi $=0 defines so-called
generation threshold. Now let's consider the character of the
steady-state points \textit{sol} of our system \{\textit{eq1},
\textit{eq2}\}. The Jacobian of the system \{\textit{eq1},
\textit{eq2}\} is:

\emptyline $>$eq3 := subs(
\{Phi(t)=x,alpha(t)=y\},rhs(eq1) ):\# x = Phi(t), y = alpha(t)\\
\indent \indent $>$eq4 := subs( \{Phi(t)=x,alpha(t)=y\},rhs(eq2)
):\\
\indent \indent \indent
  $>$A := vector( [eq3, eq4] );\# vector made from the right-hand side of
system \{eq1, eq2\}\\
\indent \indent \indent \indent
   $>$B := jacobian(A, [x,y]);

\[
A :=  \left[  \! (y - \rho )\,x, \,{\displaystyle \frac {{\sigma
_{p}}\,({a_{m}} - y)\,P}{h\,{\nu _{p}}}}  - {\displaystyle \frac
{y\,{\sigma _{g}}\,x}{h\,{\nu _{g}}}}  - {\displaystyle \frac
{y}{{T_{r}}}}  \!  \right]
\]

\[
B :=  \left[ {\begin{array}{cc}
y - \rho  & x \\
 - {\displaystyle \frac {y\,{\sigma _{g}}}{h\,{\nu _{g}}}}  &  -
{\displaystyle \frac {{\sigma _{p}}\,P}{h\,{\nu _{p}}}}  -
{\displaystyle \frac {{\sigma _{g}}\,x}{h\,{\nu _{g}}}}  -
{\displaystyle \frac {1}{{T_{r}}}}
\end{array}}
 \right]
\]

\emptyline For the cw-solution the eigenvalues of the
perturbations are:

\emptyline $>$BB := eigenvalues(B):\\
\indent
 $>$BBB := subs( \{x=subs( sol[2],Phi ),y=subs( sol[2],alpha )\},BB[1]
 ):\\
 \indent \indent
  $>$BBBB := subs( \{x=subs( sol[2],Phi ),y=subs( sol[2],alpha )\},BB[2]
):\\
\indent \indent \indent
 $>$plot( [subs( \\
\{lambda[g]=8e-5, lambda[p]=5.6e-5\}\\
,subs( \\
\{h=6.62e-34, sigma[g]= 3e-19, sigma[p]= 1e-19, rho=0.1,
nu[g]=3e10/lambda[g], nu[p]=3e10/lambda[p], T[r]=3e-6, a[m]=2.5\}\\
,BBB ) ),\\
subs( \\
\{lambda[g]=8e-5, lambda[p]=5.6e-5\}\\
,subs( \\
\{h=6.62e-34, sigma[g]= 3e-19, sigma[p]= 1e-19, rho=0.1,
nu[g]=3e10/lambda[g], nu[p]=3e10/lambda[p], T[r]=3e-6, a[m]=2.5\}\\
,BBBB/1e7 ) )], P=4.9e4..1e5, axes=boxed, title=`perturbation
eigenvalues vs. pump` );\# second eigenvalue is divided by 1e7

\emptyline
\begin{center}
\mapleplot{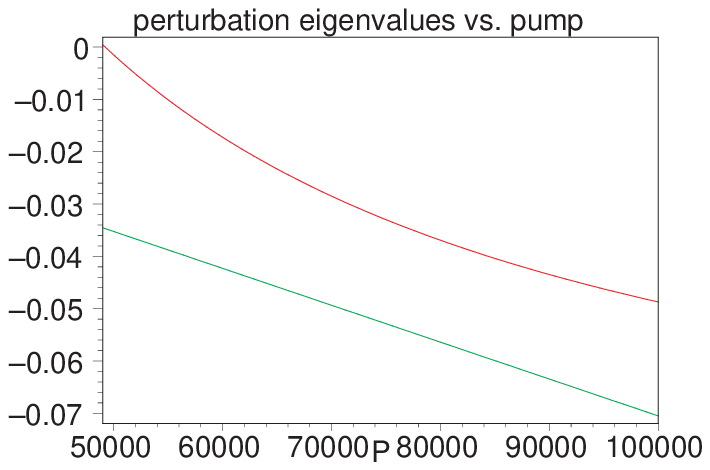}
\end{center}
\emptyline

So, cw oscillations is stable in our simple case because the
eigenvalues are negative. For the zero field solution the
perturbation eigenvalues are:

\emptyline $>$BB := eigenvalues(B):\\
\indent
 $>$BBB := subs( \{x=subs( sol[1],Phi ),y=subs( sol[1],alpha )\},BB[1]
 ):\\
\indent \indent
  $>$BBBB := subs( \{x=subs( sol[1],Phi ),y=subs( sol[1],alpha )\},BB[2]
):\\
\indent \indent
 $>$plot( [subs( \\
\{lambda[g]=8e-5, lambda[p]=5.6e-5\}\\
,subs( \\
\{h=6.62e-34, sigma[g]= 3e-19, sigma[p]= 1e-19, rho=0.1,
nu[g]=3e10/lambda[g], nu[p]=3e10/lambda[p], T[r]=3e-6, a[m]=2.5\}\\
,BBB ) ),\\
subs( \\
\{lambda[g]=8e-5, lambda[p]=5.6e-5\}\\
,subs( \\
\{h=6.62e-34, sigma[g]= 3e-19, sigma[p]= 1e-19, rho=0.1,
nu[g]=3e10/lambda[g], nu[p]=3e10/lambda[p], T[r]=3e-6, a[m]=2.5\}\\
,BBBB/1e7 ) )], P=4.9e4..1e5, axes=boxed, title=`perturbation
eigenvalues vs. pump` );\# second eigenvalue is divided by 1e7

\emptyline
\begin{center}
\mapleplot{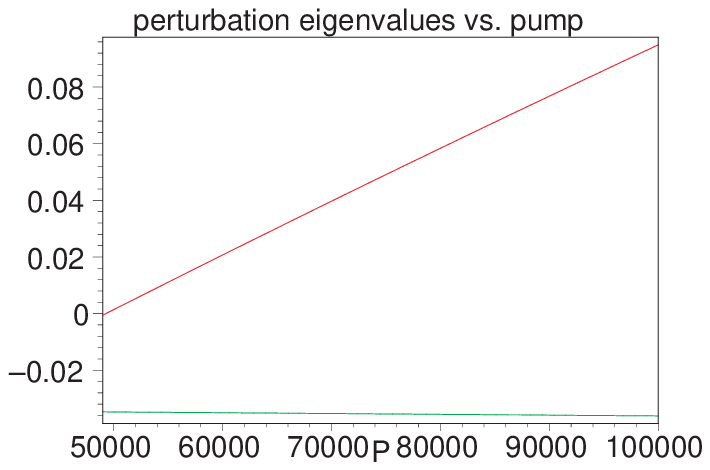}
\end{center}
\emptyline

The existence of the positive eigenvalue suggests the instability
of the zero-field steady-state solution. Hence there is the
spontaneous generation of the cw oscillation above threshold in
the model under consideration.

\subsection{Q-switching}

\emptyline \noindent The situation changes radically due to
insertion of the saturable absorber into laser cavity. In this
case, in addition to the gain saturation, the loss saturation
appears. This breaks the steady-state operation and produces the
short laser pulses.

\fbox{\parbox{.8\linewidth} {As a result of the additional
absorption, Q-factor of laser is comparatively low (high
threshold). This suppresses the generation. When $\Phi $ is small,
the gain increases in the absence of the gain saturation (see
\textit{eq2 } from the previous subsection). This causes the field
growth. The last saturates the absorption and abruptly increases
Q-factor. The absorption "switching off" leads to the explosive
generation, when the most part of the energy, which is stored in
the active medium during pumping process, converts into laser
field. The increased field saturates the gain and this finishes
the generation.}}

As the reference for the model in question see, for example,
\cite{Degnan}. To formulate the quantitative model of the laser
pulse formation let's use the next approximations: 1) the pulse
width is much larger than the cavity period, and 2) is less than
the relaxation time, 3) the pump action during the stage of the
pulse generation is negligible. We shall use the quasi-two level
schemes for the gain and loss media (the relaxation from the
intermediate levels is fast). Also, the excited-state absorption
in absorber will be taken into account.

The system of equation describing the evolution of the photon
density $\phi(t)$ is

\emptyline $>$restart:\\
\indent
 $>$with(plots):\\
\\
$>$print(`System of basic equations:`);\\
\indent $>$e1 := Diff(n[1](t),t) =\\
      -sigma[s]*c*phi(t)*n[1](t);\# EVOLUTION OF ABSORPTION. The ground
level population n[1](t) defines the absorption (quasi-two level
scheme). The relaxation is slow in the comparison with the pulse
duration, phi(t) is the photon density\\
\indent
  $>$e2 := Diff(phi(t),t) =\\
     (phi(t)/t\_cav)*\\
     (2*sigma[g]*x(t)*l - log(1/R) - L -
2*sigma[s]*n[1](t)*l[s]-2*sigma[esa]*(n[0]-n[1](t))*l[s]);\#
EVOLUTION OF PHOTON DENSITY. The field variation over cavity
round-trip is small, sigma[g] is the gain cross-section, x(t) is
the inversion in amplifier defining the gain coefficient, t\_cav
is the cavity period, l is the active medium length, R is the
output coupler refractivity, L is the linear loss, l[s] is the
absorber
length\\
\indent
         $>$e3 := Diff(x(t),t) =\\
              -gamma*sigma[g]*c*phi(t)*x(t);\# EVOLUTION OF GAIN. c is
the light velocity, gamma is the parameter of the inversion
reduction (2 for the pure three-level scheme and 1 for the pure
four-level scheme)

\[
\mathit{System\ of\ basic\ equations:}
\]

\[
\mathit{e1} := {\frac {\partial }{\partial t}}\,{n_{1}}(t)= - {
\sigma _{s}}\,c\,\phi (t)\,{n_{1}}(t)
\]

\[
\mathit{e2} := {\frac {\partial }{\partial t}}\,\phi (t)=
{\displaystyle \frac {\phi (t)\,(2\,{\sigma _{g}}\,\mathrm{x}(t)
\,l - \mathrm{ln}({\displaystyle \frac {1}{R}} ) - L - 2\,{\sigma
 _{s}}\,{n_{1}}(t)\,{l_{s}} - 2\,{\sigma _{\mathit{esa}}}\,({n_{0
}} - {n_{1}}(t))\,{l_{s}})}{\mathit{t\_cav}}}
\]

\[
\mathit{e3} := {\frac {\partial }{\partial t}}\,\mathrm{x}(t)= -
\gamma \,{\sigma _{g}}\,c\,\phi (t)\,\mathrm{x}(t)
\]

\emptyline Now we shall search the ground state population in the
absorber as a function of the initial population inversion in
amplifier:

\emptyline $>$Diff(n1(n),n) =\\
subs(\{n[1](t)=n[1](n),x(t)=x\},rhs(e1)/rhs(e3));\# devision of e1
by e3\\
\indent
 $>$Int(1/y,y=n[0]..n[1]) = Int(zeta/z,z=x\_i..x);\# zeta =
sigma[s]/gamma/sigma[g], x\_i is the initial inversion in
amplifier, n[0] is the concentration of active ions in absorber\\
\indent
   $>$solve(value(\%),n[1]):\\
   \indent \indent
    $>$n[1] = expand(\%);\\
    \indent \indent \indent
     $>$print(`Ground state population in absorber:`);\\
     \indent \indent \indent \indent
      $>$sol\_1 := n[1] = n[0]*(x/x\_i)$^{\zeta}$;

\[
{\frac {\partial }{\partial n}}\,\mathrm{n1}(n)={\displaystyle
\frac {{\sigma _{s}}\,{n_{1}}(n)}{\gamma \,{\sigma _{g}}\,x}}
\]

\[
{\displaystyle \int _{{n_{0}}}^{{n_{1}}}} {\displaystyle \frac {1
}{y}} \,dy={\displaystyle \int _{\mathit{x\_i}}^{x}}
{\displaystyle \frac {\zeta }{z}} \,dz
\]

\[
{n_{1}}={\displaystyle \frac {x^{\zeta }\,{n_{0}}}{\mathit{x\_i}
^{\zeta }}}
\]

\[
\mathit{Ground\ state\ population\ in\ absorber:}
\]

\[
\mathit{sol\_1} := {n_{1}}={n_{0}}\,({\displaystyle \frac {x}{
\mathit{x\_i}}} )^{\zeta }
\]

\emptyline The similar manipulation allows to find the photon
density as a function of inversion in amplifier:

\emptyline
$>$A := Diff(phi(x),x) =\\
     simplify( subs( n[1](t)=rhs(sol\_1),subs(
\{x(t)=x\},rhs(e2)/rhs(e3) ) ) );\# division of e2 by e3\\
\indent
  $>$B := numer( rhs(A) );\\
  \indent \indent
   $>$C := denom( rhs(A) );\\
   \indent \indent \indent
    $>$BB := subs(\\
\{op(4,B)=(x/x\_i)$^{\zeta}$*log(1/T[0]$^2$),\\
2*sigma[esa]*l[s]*n[0]=ln(1/T[s
]$^2$),\\op(6,B)=-ln(1/T[s]$^2$)*(x/x\_i)$^{\zeta}$\}, B );\# T[0]
is the initial transmission of the absorber (log(1/T[0]$^2$)=
2*sigma[s]*l[s]*n[0], l[s] is the absorber length)\\
\indent
      $>$CC := 2*l\_cav*gamma*sigma[g]*x;\# l\_cav=t\_cav*c/2 is the cavity
length, t\_cav is the cavity period

\emptyline \begin{gather}\nonumber A := {\frac {\partial
}{\partial x}}\,\phi (x)=\\ \nonumber {\displaystyle \frac { -
2\,{\sigma _{g}}\,x\,l + \mathrm{ln}({\displaystyle \frac {1}{R}}
) + L + 2\,{\sigma _{s}}\,{n_{0}}\,({\displaystyle \frac
{x}{\mathit{x\_i}}} )^{\zeta }\,{l_{s}} + 2\,{\sigma _{
\mathit{esa}}}\,{l_{s}}\,{n_{0}} - 2\,{\sigma _{\mathit{esa}}}\,{
l_{s}}\,{n_{0}}\,({\displaystyle \frac {x}{\mathit{x\_i}}} )^{
\zeta }}{\mathit{t\_cav}\,\gamma \,{\sigma _{g}}\,c\,x}}
\end{gather}

\[
B :=  - 2\,{\sigma _{g}}\,x\,l + \mathrm{ln}({\displaystyle \frac
{1}{R}} ) + L + 2\,{\sigma _{s}}\,{n_{0}}\,({\displaystyle \frac
{x}{\mathit{x\_i}}} )^{\zeta }\,{l_{s}} + 2\,{\sigma _{
\mathit{esa}}}\,{l_{s}}\,{n_{0}} - 2\,{\sigma _{\mathit{esa}}}\,{
l_{s}}\,{n_{0}}\,({\displaystyle \frac {x}{\mathit{x\_i}}} )^{
\zeta }
\]

\[
C := \mathit{t\_cav}\,\gamma \,{\sigma _{g}}\,c\,x
\]

\[
\mathit{BB} :=  - 2\,{\sigma _{g}}\,x\,l + \mathrm{ln}(
{\displaystyle \frac {1}{R}} ) + L + ({\displaystyle \frac {x}{
\mathit{x\_i}}} )^{\zeta }\,\mathrm{ln}({\displaystyle \frac {1}{
{T_{0}}^{2}}} ) + \mathrm{ln}({\displaystyle \frac {1}{{T_{s}}^{2
}}} ) - \mathrm{ln}({\displaystyle \frac {1}{{T_{s}}^{2}}} )\,(
{\displaystyle \frac {x}{\mathit{x\_i}}} )^{\zeta }
\]

\[
\mathit{CC} := 2\,\mathit{l\_cav}\,\gamma \,{\sigma _{g}}\,x
\]

\emptyline $>$print(`Evolution of the photon density:`);\\
\indent
       $>$e4 := diff(phi(x),x) =\\
subs(ln(1/(T[s]$^2$))=delta*ln(1/(T[0]$^2$)),BB)/CC;\#
delta=sigma[esa]/sigma[s] = ln(T[s])/ln(T[0]) is the parameter
defining the contribution of an excited-state absorption with
cross-section sigma[esa], T[s] is the fully saturated transmission
of absorber

\emptyline
\[
\mathit{Evolution\ of\ the\ photon\ density:}
\]

\begin{gather}\nonumber
\mathit{e4} := {\frac {\partial }{\partial x}}\,\phi (x)=\\
\nonumber {\displaystyle \frac {1}{2}} \,{\displaystyle \frac { -
2\,{ \sigma _{g}}\,x\,l + \mathrm{ln}({\displaystyle \frac {1}{R}}
)
 + L + ({\displaystyle \frac {x}{\mathit{x\_i}}} )^{\zeta }\,
\mathrm{ln}({\displaystyle \frac {1}{{T_{0}}^{2}}} ) + \delta \,
\mathrm{ln}({\displaystyle \frac {1}{{T_{0}}^{2}}} ) - \delta \,
\mathrm{ln}({\displaystyle \frac {1}{{T_{0}}^{2}}} )\,(
{\displaystyle \frac {x}{\mathit{x\_i}}} )^{\zeta }}{\mathit{
l\_cav}\,\gamma \,{\sigma _{g}}\,x}}
\end{gather}

\emptyline Hence the photon density is:

\emptyline $>$dsolve(\{e4,phi(x\_0)=0\},phi(x)):\\
\indent
 $>$simplify( subs(x\_0=x\_i,\%) ):\\
 \indent \indent
  $>$print(`This is the basic dependence for photon density:`);\\
  \indent \indent \indent
   $>$sol\_2 :=
collect(combine(\%,ln),\{log(1/T[0]$^2$),sigma[g],zeta\});

\emptyline
\[
\mathit{This\ is\ the\ basic\ dependence\ for\ photon\ density:}
\]

\maplemultiline{ \mathit{sol\_2} := \phi (x)= - {\displaystyle
\frac {1}{2}} \, {\displaystyle \frac {2\,x\,l -
2\,l\,\mathit{x\_i}}{\mathit{
l\_cav}\,\gamma }} + \\
\mbox{}{\displaystyle \frac {{\displaystyle \frac {1}{2}} \,
{\displaystyle \frac {\delta \,(\mathrm{ln}(x) - \mathrm{ln}(
\mathit{x\_i}))\,\mathrm{ln}({\displaystyle \frac {1}{{T_{0}}^{2}
}} )}{\mathit{l\_cav}\,\gamma }}  - {\displaystyle \frac {1}{2}}
\,{\displaystyle \frac { - \mathrm{ln}({\displaystyle \frac {1}{R
}} )\,(\mathrm{ln}(x) - \mathrm{ln}(\mathit{x\_i})) - L\,(
\mathrm{ln}(x) - \mathrm{ln}(\mathit{x\_i}))}{\mathit{l\_cav}\,
\gamma }} }{{\sigma _{g}}}}  \\
\mbox{} - {\displaystyle \frac {1}{2}} \,{\displaystyle \frac {(
 - ({\displaystyle \frac {x}{\mathit{x\_i}}} )^{\zeta } + \delta
\,({\displaystyle \frac {x}{\mathit{x\_i}}} )^{\zeta } - \delta
 + 1)\,\mathrm{ln}({\displaystyle \frac {1}{{T_{0}}^{2}}} )}{
\mathit{l\_cav}\,\gamma \,{\sigma _{g}}\,\zeta }}  }

\emptyline So, we have:

\[ \boxed{
\frac {l\, \left(  \! {x_{i}} - x - \frac {\mathrm{ln}(\frac {{x
_{i}}}{x})\,(\mathrm{ln}(\frac {1}{R}) + L + \delta \,\mathrm{ln}
(\frac {1}{{T_{0}}^{2}}))}{2\,{\sigma _{g}}\,l} - \frac {(1 - (
\frac {x}{{x_{i}}})^{\zeta })\,\mathrm{ln}(\frac {1}{{T_{0}}^{2}}
)\,(1 - \delta )}{2\,{\sigma _{g}}\,l\,\zeta } \!  \right) }{{l_{
\mathit{cav}}}\,\gamma }}
\] (\textbf{Eq. 1})

\emptyline Now let's define the key Q-switching parameters:

\emptyline $>$subs( n[1](t)=n[0],rhs(e2)*t\_cav/phi(t)) = 0;\#
Q-switching start, n[1](0) = n[0]\\
\indent
 $>$print(`Solution for the initial inversion:`);\\
\indent \indent
  sol\_3 := x\_i = subs( 2*sigma[s]*n[0]*l[s]=log(1/T[0]$^2$),
solve(\%,x(t)) );

\emptyline
\[
2\,{\sigma _{g}}\,\mathrm{x}(t)\,l - \mathrm{ln}({\displaystyle
\frac {1}{R}} ) - L - 2\,{\sigma _{s}}\,{n_{0}}\,{l_{s}}=0
\]

\[
\mathit{Solution\ for\ the\ initial\ inversion:}
\]

\[
\mathit{sol\_3} := \mathit{x\_i}={\displaystyle \frac {1}{2}} \,
{\displaystyle \frac {\mathrm{ln}({\displaystyle \frac {1}{R}} )
 + L + \mathrm{ln}({\displaystyle \frac {1}{{T_{0}}^{2}}} )}{{
\sigma _{g}}\,l}}
\]

\emptyline So, the initial inversion defining the gain at
Q-switching start is

\[ \boxed {x_{i}=\frac {\mathrm{ln}(\frac {1}{R}) + \mathrm{ln}(\frac {1
}{{T_{0}}^{2}}) + L}{2\,{\sigma _{g}}\,l}}
 \](\textbf{Eq. 2})

\emptyline $>$e5 := numer( rhs(e4) ) = 0;\# definition of the
pulse maximum\\
\indent
 $>$print(`The inversion at the pulse maximum:`);\\
\indent \indent
  $>$sol\_4 := x\_t = solve(e5,x);

\emptyline
\[
\mathit{e5} :=  - 2\,{\sigma _{g}}\,x\,l + \mathrm{ln}(
{\displaystyle \frac {1}{R}} ) + L + ({\displaystyle \frac {x}{
\mathit{x\_i}}} )^{\zeta }\,\mathrm{ln}({\displaystyle \frac {1}{
{T_{0}}^{2}}} ) + \delta \,\mathrm{ln}({\displaystyle \frac {1}{{
T_{0}}^{2}}} ) - \delta \,\mathrm{ln}({\displaystyle \frac {1}{{T
_{0}}^{2}}} )\,({\displaystyle \frac {x}{\mathit{x\_i}}} )^{\zeta
 }=0
\]

\[
\mathit{The\ inversion\ at\ the\ pulse\ maximum:}
\]

\emptyline
\begin{gather}\nonumber
\mathit{sol\_4} := \mathit{x\_t}= \\
\nonumber e^{\mathrm{RootOf}(2\,e^{
\mathit{\_Z}}\,l\,\mathit{x\_i}\,{\sigma _{g}} - \mathrm{ln}(
\frac {1}{R}) - L - e^{(\mathit{\_Z}\,\zeta )}\,\mathrm{ln}( \frac
{1}{{T_{0}}^{2}}) - \delta \,\mathrm{ln}(\frac {1}{{T_{0}} ^{2}})
+ \delta \,\mathrm{ln}(\frac {1}{{T_{0}}^{2}})\,e^{(
\mathit{\_Z}\,\zeta )})}\,\mathit{x\_i}
\end{gather}
\emptyline

$>$e6 := numer( simplify( rhs(sol\_2) ) ) = 0;\# definition of the
Q-switching finish\\
\indent
 $>$print(`The inversion at Q-switching finish`);\\
\indent \indent
  $>$sol\_5 := x\_f = solve(e6,x);

\maplemultiline{ \mathit{e6} :=  - 2\,l\,x\,{\sigma _{g}}\,\zeta +
2\,l\,\mathit{ x\_i}\,{\sigma _{g}}\,\zeta  +
\mathrm{ln}(x)\,\delta \,\mathrm{ ln}({\displaystyle \frac
{1}{{T_{0}}^{2}}} )\,\zeta  + \mathrm{ln
}(x)\,\mathrm{ln}({\displaystyle \frac {1}{R}} )\,\zeta  +
\mathrm{ln}(x)\,L\,\zeta  \\
\mbox{} - \mathrm{ln}(\mathit{x\_i})\,\delta \,\mathrm{ln}(
{\displaystyle \frac {1}{{T_{0}}^{2}}} )\,\zeta  - \mathrm{ln}(
\mathit{x\_i})\,\mathrm{ln}({\displaystyle \frac {1}{R}} )\,\zeta
  - \mathrm{ln}(\mathit{x\_i})\,L\,\zeta  + ({\displaystyle
\frac {x}{\mathit{x\_i}}} )^{\zeta }\,\mathrm{ln}({\displaystyle
\frac {1}{{T_{0}}^{2}}} ) \\
\mbox{} - \delta \,\mathrm{ln}({\displaystyle \frac {1}{{T_{0}}^{
2}}} )\,({\displaystyle \frac {x}{\mathit{x\_i}}} )^{\zeta } +
\delta \,\mathrm{ln}({\displaystyle \frac {1}{{T_{0}}^{2}}} ) -
\mathrm{ln}({\displaystyle \frac {1}{{T_{0}}^{2}}} )=0 }

\[
\mathit{The\ inversion\ at\ Q-switching\ finish}
\]

\maplemultiline{ \mathit{sol\_5} := \mathit{x\_f}=\\
e^{\mathrm{RootOf}(2\,e^{ \mathit{\_Z}}\,l\,\mathit{x\_i}\,{\sigma
_{g}}\,\zeta  - 2\,l\, \mathit{x\_i}\,{\sigma _{g}}\,\zeta  -
\delta \,\mathrm{ln}( \frac {1}{{T_{0}}^{2}})\,\zeta
\,\mathit{\_Z} - \mathrm{ln}( \frac {1}{R})\,\zeta \,\mathit{\_Z}
- L\,\zeta \,\mathit{\_Z} - e ^{(\mathit{\_Z}\,\zeta
)}\,\mathrm{ln}(\frac {1}{{T_{0}}^{2}})}
 \\
^{\mbox{} + \delta \,\mathrm{ln}(\frac {1}{{T_{0}}^{2}})\,e^{(
\mathit{\_Z}\,\zeta )} - \delta \,\mathrm{ln}(\frac {1}{{T_{0}}^{
2}}) + \mathrm{ln}(\frac {1}{{T_{0}}^{2}}))}\mathit{x\_i} }

\emptyline Additionally, we define the inversion at the pulse
maximum, when $\zeta $ tends to infinity:

\emptyline $>$subs( (x/x\_i)$^{\zeta}$=0,e5 );\# by virtue of x\_i
$>$ x and zeta $\rightarrow infty$\\
\indent
 $>$print(`Inversion at the pulse maximum for large zeta`);\\
\indent \indent
  $>$sol\_6 := x\_t0 = solve(\%,x);

\[
 - 2\,{\sigma _{g}}\,x\,l + \mathrm{ln}({\displaystyle \frac {1}{
R}} ) + L + \delta \,\mathrm{ln}({\displaystyle \frac {1}{{T_{0}}
^{2}}} )=0
\]

\[
\mathit{Inversion\ at\ the\ pulse\ maximum\ for\ large\ zeta}
\]

\[
\mathit{sol\_6} := \mathit{x\_t0}={\displaystyle \frac {1}{2}} \,
{\displaystyle \frac {\mathrm{ln}({\displaystyle \frac {1}{R}} )
 + L + \delta \,\mathrm{ln}({\displaystyle \frac {1}{{T_{0}}^{2}}
} )}{{\sigma _{g}}\,l}}
\]

\emptyline So, we have the expressions for ${x_{i}}$ (initial
inversion, \textit{sol\_3}), ${x_{t}}$ (the inversion at pulse
maximum, \textit{sol\_4} and \textit{e5}), ${x_{t, \,0}}$ (the
inversion at pulse maximum when $\zeta$ $\rightarrow$ $\infty$,
\textit{sol\_6}), ${x_{f}}$ (the final inversion, \textit{sol\_5}
and \textit{e6}) and the photon density $\phi $ as function of
inversion \textit{x} (\textit{sol\_2}).

As an example, we consider the real situation of Yb/Er-glass laser
with the crystalline Co:MALO saturable absorber. The obtained
expressions allow to plot the typical dependencies for the pulse
parameters:

\emptyline $>$print(`Pulse energy:`);\\
\indent
 $>$fun\_1 :=
 (h*nu*S)/(2*sigma[g]*gamma)*log(1/R)*log(x\_i/x\_f);\\
\indent \indent
  $>$print(`Pulse power:`);\\
  \indent \indent \indent
   $>$fun\_2 := (h*nu*S*l)/(t\_cav*gamma)*log(1/R)*(x\_i - x\_t -
x\_t0* log(x\_i/x\_t) -
(x\_i-x\_t0)*(1-(x\_t/x\_i)$^{\alpha}$)/$\alpha$);\\
\indent \indent \indent \indent
     $>$print(`Pulse width:`);\\
     \indent \indent \indent \indent \indent
$>$fun\_3 := simplify(fun\_1/fun\_2);

\emptyline
\[
\mathit{Pulse\ energy:}
\]

\[
\mathit{fun\_1} := {\displaystyle \frac {1}{2}} \,{\displaystyle
\frac {h\,\nu \,S\,\mathrm{ln}({\displaystyle \frac {1}{R}} )\,
\mathrm{ln}({\displaystyle \frac {\mathit{x\_i}}{\mathit{x\_f}}}
)}{{\sigma _{g}}\,\gamma }}
\]

\[
\mathit{Pulse\ power:}
\]

\[
\mathit{fun\_2} := {\displaystyle \frac {h\,\nu \,S\,l\,\mathrm{
ln}({\displaystyle \frac {1}{R}} )\, \left(  \! \mathit{x\_i} -
\mathit{x\_t} - \mathit{x\_t0}\,\mathrm{ln}({\displaystyle \frac
{\mathit{x\_i}}{\mathit{x\_t}}} ) - {\displaystyle \frac {(
\mathit{x\_i} - \mathit{x\_t0})\,(1 - ({\displaystyle \frac {
\mathit{x\_t}}{\mathit{x\_i}}} )^{\alpha })}{\alpha }}  \!
 \right) }{\mathit{t\_cav}\,\gamma }}
\]

\[
\mathit{Pulse\ width:}
\]

\begin{gather} \nonumber
\mathit{fun\_3} := -\\ \nonumber {\displaystyle \frac {1}{2}} \,
{\displaystyle \frac {\mathrm{ln}({\displaystyle \frac {\mathit{
x\_i}}{\mathit{x\_f}}} )\,\mathit{t\_cav}\,\alpha }{{\sigma _{g}}
\,l\,( - \mathit{x\_i}\,\alpha  + \mathit{x\_t}\,\alpha  +
\mathit{x\_t0}\,\mathrm{ln}({\displaystyle \frac {\mathit{x\_i}}{
\mathit{x\_t}}} )\,\alpha  + \mathit{x\_i} - \mathit{x\_i}\,(
{\displaystyle \frac {\mathit{x\_t}}{\mathit{x\_i}}} )^{\alpha }
 - \mathit{x\_t0} + \mathit{x\_t0}\,({\displaystyle \frac {
\mathit{x\_t}}{\mathit{x\_i}}} )^{\alpha })}}
\end{gather}

\emptyline This numerical procedure plots the dependence of the
output pulse energy on the reflectivity of the output mirror:

\emptyline $>$En := proc(gam,L,T\_0,l,w0,i) \# definition of
system's parameters\\
\indent
 $>$delta := 0.028:\\
 \indent \indent
 $>$sigma[g] := 7e-21:\# the gain cross-section in [cm$^2$]\\
 \indent \indent \indent
  $>$sigma[s] := 3.5e-19:\# the absorption cross-section in
  Co:MALO\\
  \indent \indent \indent \indent
   $>$alpha := sigma[s]/(sigma[g]*gam):\\
   \indent \indent \indent
    $>$h := 6.63e-34:\# J*s\\
    \indent \indent
     $>$nu := evalf(3e8/1.535e-6):\# lasing frequency for 1.54
     micrometers\\
     \indent
      $>$S := Pi*w0$^2$/2:\# area of Gaussian beam in amplifier, w0 is the
beam radius\\
     $>$R := 0.5+0.5*i/100:\\

\indent $>$x\_i := 1/2*(ln(1/R) + L +
ln(1/(T\_0$^2$)))/(l*sigma[g]):\\
\indent \indent
 $>$eq :=\\
ln(x)*delta*ln(1/(T\_0$^2$))*alpha -
ln(x\_i)*delta*ln(1/(T\_0$^2$))*alpha - 2*l* x*sigma[g]* alpha +
2*l*x\_i*sigma[g]*alpha + ln(x)*L* alpha + ln(x)*ln(1/R)*alpha -
ln(x\_i)*L*alpha - ln(x\_i)*ln(1/R)*alpha +
(x/x\_i)$^{\alpha}$*ln(1/(T\_0$^2$)) - delta*
ln(1/(T\_0$^2$))*(x/x\_i)$^{\alpha}$ - ln(1/(T\_0$^2$)) +
delta*ln(1/(T\_0$^2$)) = 0:\# e6\\

\indent $>$sol\_f := fsolve(eq, x, avoid=\{x=0\}):\\
\indent \indent
 $>$sol\_En := evalf( 1/2*h*nu*S*ln(1/R)*ln(x\_i/sol\_f)/(sigma[g]*gam) *
1e3 ):\# [mJ]\\

\indent end:\\

$>$print(`The parameters are:`);\\
$>$print(`1) inversion reduction factor gamma`);\\
$>$print(`2) linear loss L`);\\
$>$print(`3) initial transmission of absorber T[0]`);\\
$>$print(`4) gain medium length l in cm`);\\
$>$print(`5) beam radius in amplifier w0 in cm`);

\emptyline
\[
\mathit{The\ parameters\ are:}
\]

\[
\mathit{1)\ inversion\ reduction\ factor\ gamma}
\]

\[
\mathit{2)\ linear\ loss\ L}
\]

\[
\mathit{3)\ initial\ transmission\ of\ absorber\ T[0]}
\]

\[
\mathit{4)\ gain\ medium\ length\ l\ in\ cm}
\]

\[
\mathit{5)\ beam\ radius\ in\ amplifier\ w0\ in\ cm}
\]

\emptyline For the comparison we use the experimental data
(crosses in Figure):

\emptyline $>$points :=\\
\{seq([0.5 +
0.5*k/100,En(1.9,0.04,0.886,4.9,0.065,k)],k=1..100)\}:\\
\indent
 $>$points\_exp := \{[0.793,10.5],[0.88,9],[0.916,5.5]\}:\\
 \indent \indent
 $>$plot(points,x=0.5..1, style=point, axes=boxed, symbol=circle, \\
 color=black, title=`Pulse energy vs. R, [mJ]`):\\

\indent
$>$plot(points\_exp,x=0.5..1, style=point, symbol=cross,\\
 axes=boxed, color=red):\\
$>$display(\{\%,\%\%\});

\emptyline
\begin{center}
\mapleplot{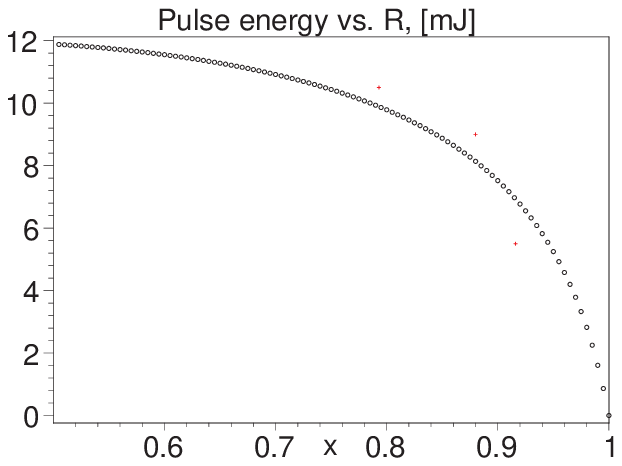}
\end{center}

\emptyline Similarly, for the output power we have:

\emptyline $>$Pow := proc(gam,L,T\_0,l,l\_cav,w0,i) \# definition
of system's parameters\\
\indent
 $>$delta := 0.028:\\
 \indent \indent
 $>$sigma[g] := 7e-21:\# in [cm$^2$]\\
 \indent \indent \indent
  $>$sigma[s] := 3.5e-19:\# Co:MALO\\
  \indent \indent \indent \indent
   $>$alpha := sigma[s]/(sigma[g]*gam):\\
   \indent \indent \indent
    $>$h := 6.63e-34:\# J*s\\
    \indent \indent
     $>$nu := evalf(3e8/1.354e-6):\# frequency for 1.54
     micrometers\\
     \indent
      $>$S := Pi*w0$^2$/2:\# area for Gaussian beam\\
     $>$R := 0.5 + 0.5*i/100:\\
     \indent
    $>$c := 3e10:\\
\indent \indent
   $>$refractivity := 1.6:\# refractivity coefficient for the active
medium\\
\indent \indent \indent
  $>$t\_cav := 2*(l\_cav - l)/c + 2*(l*refractivity)/c:\\

$>$x\_i := 1/2*(ln(1/R) + L + ln(1/(T\_0$^2$)))/(l*sigma[g]):\\
\indent  $>$x\_t0 :=\\
1/2*(ln(1/R) + L + delta*ln(1/(T\_0$^2$)))/(sigma[g]*l):\\
\indent
 $>$eq :=\\
-2*sigma[g]*x*l + ln(1/R) + L +
(x/x\_i)$^{\alpha}$*ln(1/(T\_0$^2$)) + delta*ln(1/(T\_0$^2$)) -
delta*ln(1/(T\_0$^2$))*(x/x\_i)$^{\alpha}$ = 0:\# e5

\indent $>$sol\_t[i] := fsolve(eq, x, avoid=\{x=0\}):\\
\indent \indent
 $>$sol\_Pow[i] := h*nu*S*l*ln(1/R)*\\
(x\_i-sol\_t[i] - x\_t0*ln(x\_i/sol\_t[i]) -
(x\_i-x\_t0)*(1-(sol\_t[i]/x\_i)$^{\alpha}$
)/alpha)/ (t\_cav*gam)/1e3:\# [kW]\\

$>$end:

$>$print(`The parameters are:`);\\
$>$print(`1) inversion reduction factor gamma`);\\
$>$print(`2) linear loss L`);\\
$>$print(`3) initial transmission of absorber T[0]`);\\
 $>$print(`4) gain medium length l in cm`);\\
 $>$print(`5) cavity length l\_cav in cm`);\\
 $>$print(`6) beam radius in amplifier w0 in cm`);

\emptyline
\[
\mathit{The\ parameters\ are:}
\]

\[
\mathit{1)\ inversion\ reduction\ factor\ gamma}
\]

\[
\mathit{2)\ linear\ loss\ L}
\]

\[
\mathit{3)\ initial\ transmission\ of\ absorber\ T[0]}
\]

\[
\mathit{4)\ gain\ medium\ length\ l\ in\ cm}
\]

\[
\mathit{5)\ cavity\ length\ l\_cav\ in\ cm}
\]

\[
\mathit{6)\ beam\ radius\ in\ amplifier\ w0\ in\ cm}
\]

\emptyline $>$points :=\\
\{seq([0.5+0.5*k/100,Pow(1.6,0.01,0.886,4.9,35,0.06,k)],k=1..100)\}:\\
 $>$plot(points,x=0.5..1, style=point, symbol=circle, color=black, axes=boxed
, title=`Pulse power vs. R, [kW]`);

\begin{center}
\mapleplot{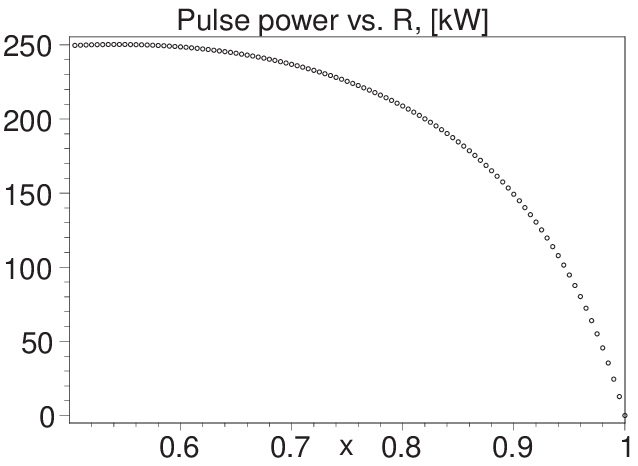}
\end{center}

\emptyline And, at last, the pulse durations are:

\emptyline $>$Width := proc(gam,L,T\_0,l,l\_cav,i) \# definition
of system's parameters\\
\indent $>$delta := 0.028:\\
\indent \indent
 $>$sigma[g] := 7e-21:\# in [cm$^2$]\\
 \indent \indent \indent
  $>$sigma[s] := 3.5e-19:\# Co:MALO\\
  \indent \indent \indent \indent
   $>$alpha := sigma[s]/(sigma[g]*gam):\\
   \indent \indent \indent \indent \indent
    $>$h := 6.63e-34:\# J*s\\
    \indent \indent \indent \indent
     $>$nu := evalf(3e8/1.354e-6):\# frequency for 1.54
     micrometers\\
     \indent \indent \indent
      $>$S := Pi*w0$^2$/2:\# area for Gaussian beam\\
      \indent \indent
     $>$R := 0.5 + 0.5*i/100:\\
     \indent
    $>$c := 3e10:\\
        $>$refractivity := 1.5:\# active medium\\
        \indent
   $>$t\_cav := 2*(l\_cav-l)/c + 2*(l*refractivity)/c:\\

$>$x\_i := 1/2*(ln(1/R) + L + ln(1/(T\_0$^2$)))/(l*sigma[g]):\\
\indent $>$x\_t0 :=\\
1/2*(ln(1/R) + L + delta*ln(1/(T\_0$^2$)))/(sigma[g]*l):\\
\indent \indent $>$eq1 :=\\
-2*sigma[g]*x*l + ln(1/R) + L +
(x/x\_i)$^{\alpha}$*ln(1/(T\_0$^2$)) + delta*ln(1/(T\_0$^2$)) -
delta*ln(1/(T\_0$^2$))*(x/x\_i)$^{\alpha}$ = 0:\# e5\\
\indent $>$eq2 :=\\
ln(x)*delta*ln(1/(T\_0$^2$))*alpha -
ln(x\_i)*delta*ln(1/(T\_0$^2$))*alpha - 2*l*x* sigma[g]*alpha +
2*l*x\_i*sigma[g]*alpha + ln(x)*L* alpha + ln(x)*ln(1/R)* alpha -
ln(x\_i)*L*alpha - ln(x\_i)*ln(1/R)*alpha +
(x/x\_i)$^{\alpha}$*ln(1/(T\_0$^2$)) - delta*
ln(1/(T\_0$^2$))*(x/x\_i)$^{\alpha}$ - ln(1/(T\_0$^2$)) +
delta*ln(1/(T\_0$^2$)) = 0:\# e6

\indent $>$sol\_f := fsolve(eq2, x, avoid=\{x=0\}):\\
\indent \indent $>$sol\_t := fsolve(eq1, x, avoid=\{x=0\}):\\
\indent \indent \indent
 $>$sol\_Width :=\\
(-1/2*ln(x\_i/sol\_f)*t\_cav*alpha/(sigma[g]*l*
(-x\_i*alpha+sol\_t*alpha+x\_t0* ln(x\_i/sol\_t)*alpha + x\_i -
x\_i*(sol\_t/x\_i)$^{\alpha}$ - x\_t0 +
x\_t0* (sol\_t/x\_i)$^{\alpha}$)))* 1e9:\# [ns]\\

$>$end:

$>$print(`The parameters are:`);\\
$>$print(`1) reducing parameter gamma`);\\
$>$print(`2) linear loss L`);\\
$>$print(`3) initial transmission of absorber T[0]`);\\
$>$print(`4) gain medium length l in cm`);\\
$>$print(`5) cavity length l\_cav in cm`);

\[
\mathit{The\ parameters\ are:}
\]

\[
\mathit{1)\ reducing\ parameter\ gamma}
\]

\[
\mathit{2)\ linear\ loss\ L}
\]

\[
\mathit{3)\ initial\ transmission\ of\ absorber\ T[0]}
\]

\[
\mathit{4)\ gain\ medium\ length\ l\ in\ cm}
\]

\[
\mathit{5)\ cavity\ length\ l\_cav\ in\ cm}
\]

\emptyline $>$points :=\\
\{seq([0.5+0.5*k/100,Width(1.9,0.04,0.89,4.9,35,k)],k=1..100)\}:\\
\indent
 $>$points\_exp := \{[0.793,70],[0.88,70],[0.916,75]\}:\\

\indent
 $>$plot(points,x=0.5..1,style=point,axes=boxed,symbol=circle,
 title=`Pulse width vs. R, [ns]`):\\
\indent
 $>$plot(points\_exp,x=0.5..1,style=point,axes=boxed,color=red):\\
 \indent \indent
 $>$display(\{\%,\%\%\},view=50..100);

\begin{center}
\mapleplot{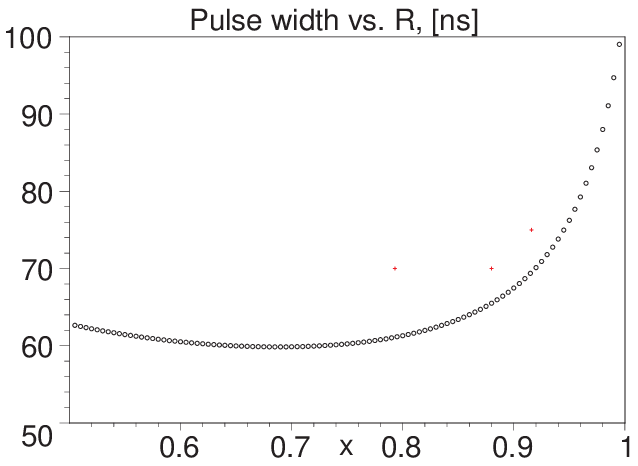}
\end{center}

\emptyline The worse agreement with the experimental data for the
pulse durations is caused by the deviation of the pulse shape from
the Gaussian profile, which was used for the analytical
estimations. More precise consideration is presented in
\cite{Liu}.

The main advantage of the analytical model in question is the
potential of the Q-switched laser optimization without any
cumbersome numerical simulations. Let's slightly transform the
above obtained expressions:

\emptyline $>$e7 := subs( x=x\_t,expand( e5/(2*sigma[g]*x\_i*l) )
);\\
\indent
 $>$e8 := subs( x=x\_f,expand( rhs(sol\_2)*l\_cav*gamma/l ) ) =
 0;\\
 \indent \indent
  $>$sol\_6;\\
  \indent \indent \indent
   $>$sol\_3;

\maplemultiline{ \mathit{e7} :=  - {\displaystyle \frac
{\mathit{x\_t}}{\mathit{ x\_i}}}  + {\displaystyle \frac
{{\displaystyle \frac {1}{2}} \, \mathrm{ln}({\displaystyle \frac
{1}{R}} )}{{\sigma _{g}}\, \mathit{x\_i}\,l}}  + {\displaystyle
\frac {{\displaystyle \frac {1}{2}} \,L}{{\sigma
_{g}}\,\mathit{x\_i}\,l}}  + {\displaystyle \frac {{\displaystyle
\frac {1}{2}} \,( {\displaystyle \frac
{\mathit{x\_t}}{\mathit{x\_i}}} )^{\zeta }\,
\mathrm{ln}({\displaystyle \frac {1}{{T_{0}}^{2}}} )}{{\sigma _{g
}}\,\mathit{x\_i}\,l}}  + {\displaystyle \frac {{\displaystyle
\frac {1}{2}} \,\delta \,\mathrm{ln}({\displaystyle \frac {1}{{T
_{0}}^{2}}} )}{{\sigma _{g}}\,\mathit{x\_i}\,l}}  -
\\
{\displaystyle \frac {1}{2}} \,{\displaystyle \frac {\delta \,
\mathrm{ln}({\displaystyle \frac {1}{{T_{0}}^{2}}} )\,(
{\displaystyle \frac {\mathit{x\_t}}{\mathit{x\_i}}} )^{\zeta }}{
{\sigma _{g}}\,\mathit{x\_i}\,l}} = 0 }

\maplemultiline{ \mathit{e8} :=  - \mathit{x\_f} + \mathit{x\_i} +
{\displaystyle \frac {{\displaystyle \frac {1}{2}} \,\delta
\,\mathrm{ln}( {\displaystyle \frac {1}{{T_{0}}^{2}}}
)\,\mathrm{ln}(\mathit{ x\_f})}{l\,{\sigma _{g}}}}  -
{\displaystyle \frac {1}{2}} \, {\displaystyle \frac {\delta
\,\mathrm{ln}({\displaystyle \frac { 1}{{T_{0}}^{2}}}
)\,\mathrm{ln}(\mathit{x\_i})}{l\,{\sigma _{g}}} }  +
{\displaystyle \frac {{\displaystyle \frac {1}{2}} \,
\mathrm{ln}({\displaystyle \frac {1}{R}} )\,\mathrm{ln}(\mathit{
x\_f})}{l\,{\sigma _{g}}}}  \\
\mbox{} - {\displaystyle \frac {1}{2}} \,{\displaystyle \frac {
\mathrm{ln}({\displaystyle \frac {1}{R}} )\,\mathrm{ln}(\mathit{
x\_i})}{l\,{\sigma _{g}}}}  + {\displaystyle \frac {
{\displaystyle \frac {1}{2}} \,L\,\mathrm{ln}(\mathit{x\_f})}{l\,
{\sigma _{g}}}}  - {\displaystyle \frac {1}{2}} \,{\displaystyle
\frac {L\,\mathrm{ln}(\mathit{x\_i})}{l\,{\sigma _{g}}}}  +
{\displaystyle \frac {{\displaystyle \frac {1}{2}} \,\mathrm{ln}(
{\displaystyle \frac {1}{{T_{0}}^{2}}} )\,({\displaystyle \frac {
\mathit{x\_f}}{\mathit{x\_i}}} )^{\zeta }}{l\,{\sigma _{g}}\,
\zeta }}  \\
\mbox{} - {\displaystyle \frac {1}{2}} \,{\displaystyle \frac {
\mathrm{ln}({\displaystyle \frac {1}{{T_{0}}^{2}}} )\,\delta \,(
{\displaystyle \frac {\mathit{x\_f}}{\mathit{x\_i}}} )^{\zeta }}{
l\,{\sigma _{g}}\,\zeta }}  + {\displaystyle \frac {
{\displaystyle \frac {1}{2}} \,\mathrm{ln}({\displaystyle \frac {
1}{{T_{0}}^{2}}} )\,\delta }{l\,{\sigma _{g}}\,\zeta }}  -
{\displaystyle \frac {1}{2}} \,{\displaystyle \frac {\mathrm{ln}(
{\displaystyle \frac {1}{{T_{0}}^{2}}} )}{l\,{\sigma _{g}}\,\zeta
 }} =0 }

\[
\mathit{x\_t0}={\displaystyle \frac {1}{2}} \,{\displaystyle \frac
{\mathrm{ln}({\displaystyle \frac {1}{R}} ) + L + \delta \,
\mathrm{ln}({\displaystyle \frac {1}{{T_{0}}^{2}}} )}{{\sigma _{g
}}\,l}}
\]

\[
\mathit{x\_i}={\displaystyle \frac {1}{2}} \,{\displaystyle \frac
{\mathrm{ln}({\displaystyle \frac {1}{R}} ) + L + \mathrm{
ln}({\displaystyle \frac {1}{{T_{0}}^{2}}} )}{{\sigma _{g}}\,l}}
\]

\emptyline $>$\# Hence\\
\indent
 $>$e9 := x\_t/x\_i = (ln(1/R) + L + delta*ln(1/(T[0]$^2$)))/ (2*l*x\_i*sigma[g]) +
(x\_t/x\_i)$^{\zeta}$*ln(1/T[0]$^2$)*(1-delta)/ (2*l*x\_i*sigma[g]);\\
\indent \indent
  $>$e10 := x\_f - x\_i = ln(x\_f/x\_i)*(ln(1/R) + L +
delta*ln(1/(T[0]$^2$)))/ (2*l*sigma[g]) -
ln(1/T[0]$^2$)*(1-delta)*(1-(x\_f/x\_i)$^{\zeta}$)/ (2*l*sigma[g]*alpha);\\
\indent \indent
   $>$simplify( sol\_6 - sol\_3 );\\
   \indent
    $>$e11 := subs(\\
ln(1/T[0]$^2$)=(x\_i-x\_t0)*2*l*sigma[g]/(1-delta), subs(ln(1/R) +
L + delta*\\ ln(1/(T[0]$^2$)) = x\_t0*2*l*sigma[g],e9));\\
\indent
     $>$e12 := subs(\\
ln(1/T[0]$^2$) = (x\_i-x\_t0)*2*l*sigma[g]/ (1-delta),subs(ln(1/R)
+ L + delta* ln(1/(T[0]$^2$)) = x\_t0*2*l*sigma[g],e10));

\[
\mathit{e9} := {\displaystyle \frac {\mathit{x\_t}}{\mathit{x\_i}
}} ={\displaystyle \frac {1}{2}} \,{\displaystyle \frac {\mathrm{
ln}({\displaystyle \frac {1}{R}} ) + L + \delta \,\mathrm{ln}(
{\displaystyle \frac {1}{{T_{0}}^{2}}} )}{{\sigma _{g}}\,\mathit{
x\_i}\,l}}  + {\displaystyle \frac {{\displaystyle \frac {1}{2}}
\,({\displaystyle \frac {\mathit{x\_t}}{\mathit{x\_i}}} )^{\zeta
}\,\mathrm{ln}({\displaystyle \frac {1}{{T_{0}}^{2}}} )\,(1 -
\delta )}{{\sigma _{g}}\,\mathit{x\_i}\,l}}
\]

\begin{gather} \nonumber
\mathit{e10} := \mathit{x\_f} - \mathit{x\_i}=\\
{\displaystyle \frac {1}{2}} \,{\displaystyle \frac
{\mathrm{ln}({\displaystyle \frac {\mathit{x\_f}}{\mathit{x\_i}}}
)\,(\mathrm{ln}( {\displaystyle \frac {1}{R}} ) + L + \delta
\,\mathrm{ln}( {\displaystyle \frac {1}{{T_{0}}^{2}}}
))}{l\,{\sigma _{g}}}}  - {\displaystyle \frac {1}{2}}
\,{\displaystyle \frac {\mathrm{ln}( {\displaystyle \frac
{1}{{T_{0}}^{2}}} )\,(1 - \delta )\,(1 - ( {\displaystyle \frac
{\mathit{x\_f}}{\mathit{x\_i}}} )^{\zeta }) }{l\,{\sigma
_{g}}\,\alpha }} \nonumber
\end{gather}

\[
\mathit{x\_t0} - \mathit{x\_i}={\displaystyle \frac {1}{2}} \,
{\displaystyle \frac {\mathrm{ln}({\displaystyle \frac {1}{{T_{0}
}^{2}}} )\,( - 1 + \delta )}{l\,{\sigma _{g}}}}
\]

\[
\mathit{e11} := {\displaystyle \frac {\mathit{x\_t}}{\mathit{x\_i
}}} ={\displaystyle \frac {\mathit{x\_t0}}{\mathit{x\_i}}}  +
{\displaystyle \frac {({\displaystyle \frac {\mathit{x\_t}}{
\mathit{x\_i}}} )^{\zeta }\,(\mathit{x\_i} - \mathit{x\_t0})}{
\mathit{x\_i}}}
\]

\[
\mathit{e12} := \mathit{x\_f} - \mathit{x\_i}=\mathrm{ln}(
{\displaystyle \frac {\mathit{x\_f}}{\mathit{x\_i}}} )\,\mathit{
x\_t0} - {\displaystyle \frac {(\mathit{x\_i} - \mathit{x\_t0})\,
(1 - ({\displaystyle \frac {\mathit{x\_f}}{\mathit{x\_i}}} )^{
\zeta })}{\alpha }}
\]

\emptyline Now, let's plot some typical dependence for
Q-switching.

\emptyline $>$eq :=
subs(\{x\_t0=x,x\_i=1\},subs(x\_t/x\_i=y,expand(e11)));\# here
x=x\_t0/x\_i, y=x\_t/x\_i

\[
\mathit{eq} := y=x + y^{\zeta } - y^{\zeta }\,x
\]

\emptyline $>$xt := proc(zeta,i)\\
\indent $>$sol := array(1..20):\\
\indent \indent
  $>$x := (i-1)/20:\\
  \indent \indent \indent
   $>$sol[i] := fsolve(y =
x+y$^{\zeta}$-y$^{\zeta}$*x,y,avoid=\{y=1\},0..infinity):\\
\indent \indent  $>$end:

\emptyline $>$points := \{seq([k/20,xt(1.5,k)],k=2..19)\}:\\
\indent
 $>$p1 := plot(points,x=0..1,style=point,axes=boxed):\\
 \indent \indent
  $>$points := \{seq([k/20,xt(3,k)],k=2..19)\}:\\
  \indent \indent \indent
   $>$p2 := plot(points,x=0..1,style=point,axes=boxed):\\
   \indent \indent
    $>$points := \{seq([k/20,xt(6,k)],k=2..19)\}:\\
    \indent
     $>$p3 := plot(points,x=0..1,style=point,axes=boxed):\\

  $>$display(\{p1,p2,p3\},view=0..1,title=`x\_t/x\_i vs x\_t0/x\_i for
different zeta`);

\begin{center}
\mapleplot{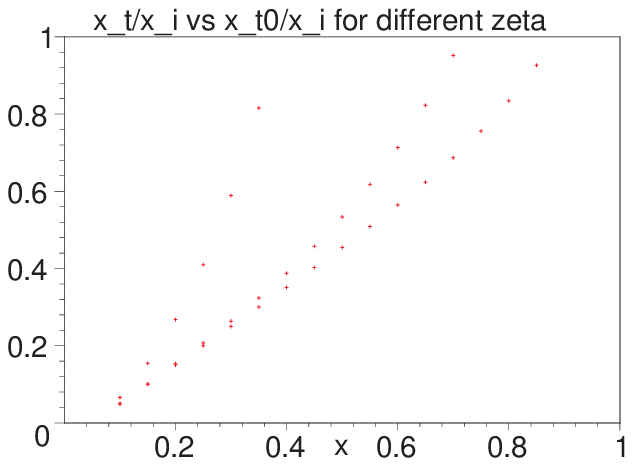}
\end{center}

\emptyline From the definition of $\mathit{x\_t0}$ this dependence
for growing $\zeta $ tends to the linear one, which correspond to
the maximal efficiency of the population inversion utilization.
For the fixed absorption and emission cross-sections, the $\zeta $
parameter increases as a result of the laser beam focusing in the
saturable absorber. Then ( $\zeta$$\rightarrow$ $\infty $), we
have:

\emptyline $>$e13 := lhs(e11) = op(1,rhs(e11));\\
\indent
 $>$e14 := expand(lhs(e12)/x\_i) = expand(op(1,rhs(e12))/x\_i);

\[
\mathit{e13} := {\displaystyle \frac {\mathit{x\_t}}{\mathit{x\_i
}}} ={\displaystyle \frac {\mathit{x\_t0}}{\mathit{x\_i}}}
\]

\[
\mathit{e14} := {\displaystyle \frac {\mathit{x\_f}}{\mathit{x\_i
}}}  - 1={\displaystyle \frac {\mathrm{ln}({\displaystyle \frac {
\mathit{x\_f}}{\mathit{x\_i}}} )\,\mathit{x\_t0}}{\mathit{x\_i}} }
\]

\emptyline and the output energy optimization can be realized by
this simple way:

\emptyline $>$\#                         Energy optimization

$>$e15 := x\_f/x\_i-1 = ln(x\_f/x\_i)*rhs(sol\_6)/x\_i;\# from e14
and expression for x\_t0\\
\indent
 $>$e16 := lhs(\%) = ln(x\_f/x\_i)*(a + b + delta*c);\# a =
ln(1/R)/\\ (2*sigma[g]*x\_i*l), b = L/(2*sigma[g]*x\_i*l), c =
ln(1/T[0]$^2$)/\\(2*sigma[g]*x\_i*l) are the relative shares of
the
output, linear and saturable loss in the net-loss\\
\indent
  $>$expand(solve(\%,a)):\\
  \indent \indent
   $>$sol\_7 := (y/log(y) - 1/log(y) - (1-delta)*b - delta)/(1-delta);\#
solution for a, y = x\_f/x\_i, we used a+b+c=1\\
\indent \indent \indent
    $>$epsilon := -a*log(y);\# dimensionless energy,\\ epsilon =
En*2*sigma[g]*gamma/(h*nu*A)/(2*sigma[g]*x\_i*l)\\ \indent \indent
    $>$simplify( subs(a=sol\_7,epsilon) );\\
    \indent
   $>$diff(\%,y) = 0;\# maximum of energy\\
   $>$print(`An optimal ratio of final and initial
   inversions:`);\\
   \indent
  $>$y\_opt := solve(\%,y);\# optimal y\\
\indent
  $>$print(`An optimal mirror:`);\\
\indent \indent
 $>$a\_opt := subs(y=y\_opt,sol\_7);\# optimal a\\
 \indent
$>$print(`A maximal dimensionless energy:`);\\
$>$epsilon\_opt := simplify( subs(\{a=a\_opt,y=y\_opt\},epsilon)
);\# optimal epsilon

\[
\mathit{e15} := {\displaystyle \frac {\mathit{x\_f}}{\mathit{x\_i
}}}  - 1={\displaystyle \frac {1}{2}} \,{\displaystyle \frac {
\mathrm{ln}({\displaystyle \frac {\mathit{x\_f}}{\mathit{x\_i}}}
)\,(\mathrm{ln}({\displaystyle \frac {1}{R}} ) + L + \delta \,
\mathrm{ln}({\displaystyle \frac {1}{{T_{0}}^{2}}} ))}{{\sigma _{
g}}\,l\,\mathit{x\_i}}}
\]

\[
\mathit{e16} := {\displaystyle \frac {\mathit{x\_f}}{\mathit{x\_i
}}}  - 1=\mathrm{ln}({\displaystyle \frac {\mathit{x\_f}}{
\mathit{x\_i}}} )\,(a + b + \delta \,c)
\]

\[
\mathit{sol\_7} := {\displaystyle \frac {{\displaystyle \frac {y
}{\mathrm{ln}(y)}}  - {\displaystyle \frac {1}{\mathrm{ln}(y)}}
 - (1 - \delta )\,b - \delta }{1 - \delta }}
\]

\[
\varepsilon  :=  - a\,\mathrm{ln}(y)
\]

\[
{\displaystyle \frac {y - 1 - b\,\mathrm{ln}(y) + b\,\mathrm{ln}(
y)\,\delta  - \delta \,\mathrm{ln}(y)}{ - 1 + \delta }}
\]

\[
{\displaystyle \frac {1 - {\displaystyle \frac {b}{y}}  +
{\displaystyle \frac {b\,\delta }{y}}  - {\displaystyle \frac {
\delta }{y}} }{ - 1 + \delta }} =0
\]

\[
\mathit{An\ optimal\ ratio\ of\ final\ and\ initial\ inversions:}
\]

\[
\mathit{y\_opt} := b - b\,\delta  + \delta
\]

\[
\mathit{An\ optimal\ mirror:}
\]

\[
\mathit{a\_opt} := {\displaystyle \frac {{\displaystyle \frac {b
 - b\,\delta  + \delta }{\mathrm{ln}(b - b\,\delta  + \delta )}}
 - {\displaystyle \frac {1}{\mathrm{ln}(b - b\,\delta  + \delta )
}}  - (1 - \delta )\,b - \delta }{1 - \delta }}
\]

\[
\mathit{A\ maximal\ dimensionless\ energy:}
\]

\begin{gather} \nonumber
\mathit{epsilon\_opt} := \\
{\displaystyle \frac {b - b\,\delta  +
\delta  - 1 - b\,\mathrm{ln}(b - b\,\delta  + \delta ) + b\,
\mathrm{ln}(b - b\,\delta  + \delta )\,\delta  - \delta \,
\mathrm{ln}(b - b\,\delta  + \delta )}{ - 1 + \delta }} \nonumber
\end{gather}

\emptyline $>$p1 := plot(\\
\{[b,subs(delta=0,a\_opt),b=0..1],[b,1-subs(delta=0,a\_opt)-b,b=0..1]\\
\},color=black,axes=boxed,title=`optimal a and c versus b`):\\
\indent
    $>$p2 :=\\
plot(\\
\{[b,subs(delta=0.05,a\_opt),b=0..1],[b,1-subs(delta=0.05,a\_opt)-b,
b=0. .1]\},\\color=red,axes=boxed,title=`optimal a and c versus
b`):\\
\indent
     $>$p3 :=\\
plot(
\{[b,subs(delta=0.1,a\_opt),b=0..1],[b,1-subs(delta=0.1,a\_opt)-b,
b=0..1 ]\},\\color=blue,axes=boxed, title=`optimal a and c versus
b`):\\

\indent \indent
     $>$display(p1,p2,p3,view=0..1);\\

     \indent
     $>$p1 := plot([b,subs(delta=0,epsilon\_opt),b=0..1],\\
axes=boxed,color=black, title=`optimal epsilon versus b`):\\
\indent
      $>$p2 := plot([b,subs(delta=0.05,epsilon\_opt),b=0..1],\\
axes=boxed,color=red, title=`optimal epsilon versus b`):\\
\indent
      $>$p3 := plot([b,subs(delta=0.1,epsilon\_opt),b=0..1],\\
axes=boxed,color=blue, title=`optimal epsilon versus b`):\\
\indent
       $>$display(p1,p2,p3);

\begin{center}
\mapleplot{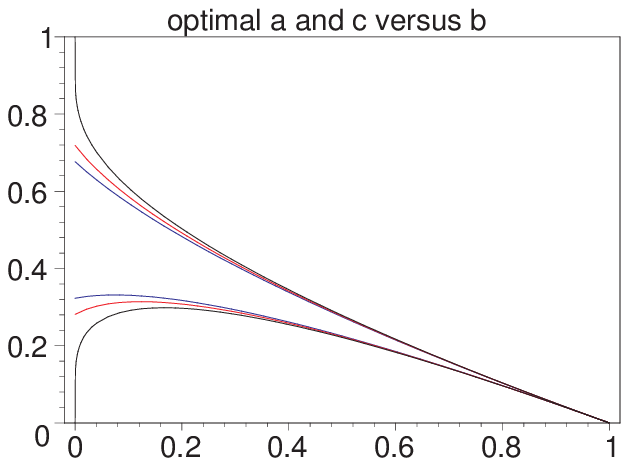}
\end{center}

\begin{center}
\mapleplot{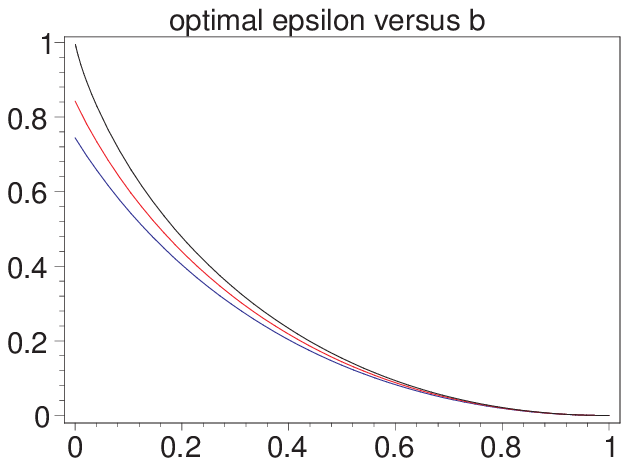}
\end{center}

\emptyline \fbox{\parbox{.8\linewidth} {From the first figure the
optimization can be performed by the simple graphical way. We have
to define the appropriate for our scheme laser's net-loss and to
determine (measure or calculate) the intracavity linear loss. This
gives the value of \textit{b}, which is the relation of the linear
loss to the net-loss. The upper group of curves gives the value of
\textit{c}, the lower curves give \textit{a} (for the different
contribution of the excited-state absorption, i. e. $\delta $).}}

\emptyline

\subsection{Two-color pulsing}

\emptyline Now let consider a more complicated situation, which
corresponds to the two-color Q-switching due to presence of the
stimulated Raman scattering in the active medium (for example,
Yb$^{3+}$:KGd(W0$_{4}$)$_{2}$, see \cite{Lagatsky}). In this case
the analytical modelling is not possible, but we can use the
numerical capacities of Maple.

Let's the gain medium length is ${l_{g}}$ and the Raman gain
coefficient is \textit{g}. We assume the exact Raman resonance and
neglect the phase and group-velocity effects. Then the evolutional
equations for the laser and scattered intensities ${I_{p}}$ and
${I_{s}}$, respectively, are:

\emptyline $>$restart:\\
\indent $>$with(DEtools):\\
\indent \indent $>$with(plots):\\

\indent $>$eq1 := diff(In[p](z),z) = -g*In[s](z)*In[p](z):\#
evolution of the laser intensity\\
\indent
 $>$eq2 := diff(In[s](z),z) = g*In[p](z)*In[s](z):\# evolution of the
Stokes component intensity\\
\indent
  $>$sys := \{eq1,eq2\};\\
\indent
   $>$IC := \{In[p](0)=In[p,0],In[s](0)=In[s,0]\};\# initial
   conditions\\
   \indent
  $>$sol := dsolve(sys union IC,\{In[s](z),In[p](z)\}):\\

\[
\mathit{sys} := \{{\frac {\partial }{\partial z}}\,{\mathit{In}_{
s}}(z)=g\,{\mathit{In}_{s}}(z)\,{\mathit{In}_{p}}(z), \,{\frac {
\partial }{\partial z}}\,{\mathit{In}_{p}}(z)= - g\,{\mathit{In}
_{s}}(z)\,{\mathit{In}_{p}}(z)\}
\]

\[
\mathit{IC} := \{{\mathit{In}_{s}}(0)={\mathit{In}_{s, \,0}}, \,{
\mathit{In}_{p}}(0)={\mathit{In}_{p, \,0}}\}
\]

\emptyline The integration produces:

\emptyline $>$subs(sol,In[s](z)):\# z is the distance in the
crystal\\
\indent
 $>$sol\_1 := In[s](z) = simplify(\%);\\
 \indent \indent
$>$subs(sol,In[p](z)):\\
\indent \indent \indent
 $>$sol\_2 := In[p](z) = simplify(\%);

\[
\mathit{sol\_1} := {\mathit{In}_{s}}(z)={\displaystyle \frac {e^{
(z\,g\,({\mathit{In}_{p, \,0}} + {\mathit{In}_{s, \,0}}))}\,{
\mathit{In}_{s, \,0}}\,({\mathit{In}_{p, \,0}} + {\mathit{In}_{s ,
\,0}})}{{\mathit{In}_{p, \,0}} + e^{(z\,g\,({\mathit{In}_{p, \,
0}} + {\mathit{In}_{s, \,0}}))}\,{\mathit{In}_{s, \,0}}}}
\]

\[
\mathit{sol\_2} := {\mathit{In}_{p}}(z)={\displaystyle \frac {({
\mathit{In}_{p, \,0}} + {\mathit{In}_{s, \,0}})\,{\mathit{In}_{p ,
\,0}}}{{\mathit{In}_{p, \,0}} + e^{(z\,g\,({\mathit{In}_{p, \,0 }}
+ {\mathit{In}_{s, \,0}}))}\,{\mathit{In}_{s, \,0}}}}
\]

\emptyline There are the amplification of the scattered field and
the depletion of the laser field, which plays a role of the pump
for the Stokes component.

Now let's transit from the intensities to the photon densities:

\emptyline $>$\#phi is the photon densities, the laser component
at 1033 nm, the Stokes component at 1139 nm

\indent $>$sol\_3 := simplify( subs(\\
\{z=2*l[g],In[p,0]=c*h*nu[p]*phi[p,0]/2,\\
In[s,0]=c*h*nu[s]*phi[s,0]/2,In[s](z)=c*h*nu[s]*phi[s](z)/2\},\\
2*sol\_1))/(c*h*nu[s]);\\

\indent $>$sol\_4 := simplify( subs(\\
\{z=2*l[g],In[p,0]=c*h*nu[p]*phi[p,0]/2,In[s,0]=c*h*nu[s]*phi[s,0]/2,\\
In[p](z)=c*h*nu[p]*phi[p](z)/2\},2*sol\_2) )/(c*h*nu[p]);

\maplemultiline{ \mathit{sol\_3} := {\phi _{s}}(z)={\displaystyle
\frac {e^{({l_{g }}\,g\,c\,h\,(\mathrm{\%1} + {\nu _{s}}\,{\phi
_{s, \,0}}))}\,{ \phi _{s, \,0}}\,(\mathrm{\%1} + {\nu
_{s}}\,{\phi _{s, \,0}})}{ \mathrm{\%1} +
e^{({l_{g}}\,g\,c\,h\,(\mathrm{\%1} + {\nu _{s}}\,
{\phi _{s, \,0}}))}\,{\nu _{s}}\,{\phi _{s, \,0}}}}  \\
\mathrm{\%1} := {\nu _{p}}\,{\phi _{p, \,0}} }

\[
\mathit{sol\_4} := {\phi _{p}}(z)={\displaystyle \frac {({\nu _{p
}}\,{\phi _{p, \,0}} + {\nu _{s}}\,{\phi _{s, \,0}})\,{\phi _{p,
\,0}}}{{\nu _{p}}\,{\phi _{p, \,0}} + e^{({l_{g}}\,g\,c\,h\,({\nu
 _{p}}\,{\phi _{p, \,0}} + {\nu _{s}}\,{\phi _{s, \,0}}))}\,{\nu
_{s}}\,{\phi _{s, \,0}}}}
\]

\emptyline Hence the photon densities evolution due to the
stimulated Raman scattering obeys:

\emptyline $>$eq3 := diff(phi[p](t),t)[raman] = ( subs(\\
\{phi[p,0]=phi[p](t),phi[s,0]=phi[s](t)\},rhs(sol\_4) ) -
phi[p](t))/t[cav];\# t[cav] is the laser cavity period\\

\indent $>$eq4 := diff(phi[s](t),t)[raman] = ( subs(\\
\{phi[p,0]=phi[p](t),phi[s,0]=phi[s](t)\},rhs(sol\_3) )- phi[s](t)
)/t[cav];

\[
\mathit{eq3} := {({\frac {\partial }{\partial t}}\,{\phi _{p}}(t)
)_{\mathit{raman}}}={\displaystyle \frac {{\displaystyle \frac {(
{\nu _{p}}\,{\phi _{p}}(t) + {\nu _{s}}\,{\phi _{s}}(t))\,{\phi
_{p}}(t)}{{\nu _{p}}\,{\phi _{p}}(t) + e^{({l_{g}}\,g\,c\,h\,({
\nu _{p}}\,{\phi _{p}}(t) + {\nu _{s}}\,{\phi _{s}}(t)))}\,{\nu
_{s}}\,{\phi _{s}}(t)}}  - {\phi _{p}}(t)}{{t_{\mathit{cav}}}}}
\]

\maplemultiline{ \mathit{eq4} := {({\frac {\partial }{\partial
t}}\,{\phi _{s}}(t) )_{\mathit{raman}}}={\displaystyle \frac
{{\displaystyle \frac {e ^{({l_{g}}\,g\,c\,h\,(\mathrm{\%1} + {\nu
_{s}}\,{\phi _{s}}(t))) }\,{\phi _{s}}(t)\,(\mathrm{\%1} + {\nu
_{s}}\,{\phi _{s}}(t))}{ \mathrm{\%1} +
e^{({l_{g}}\,g\,c\,h\,(\mathrm{\%1} + {\nu _{s}}\, {\phi
_{s}}(t)))}\,{\nu _{s}}\,{\phi _{s}}(t)}}  - {\phi _{s}}(t)
}{{t_{\mathit{cav}}}}}  \\
\mathrm{\%1} := {\nu _{p}}\,{\phi _{p}}(t) }

\emptyline The inverse and ground-state populations for the gain
medium and the absorber (Cr$^{4+}$:YAG) obey (see previous
subsection):

\emptyline $>$eq5 := diff(n(t),t) =\\
-gamma*sigma[g]*c*phi[p](t)*n(t);\\
\indent
 $>$eq6 := diff(n[0](t),t) = -rho*sigma[a]*c*phi[p](t)*n[0](t);\#
rho=S[g]/S[a] is the ratio of the beam area in the gain medium to
that in the absorber

\[
\mathit{eq5} := {\frac {\partial }{\partial t}}\,\mathrm{n}(t)=
 - \gamma \,{\sigma _{g}}\,c\,{\phi _{p}}(t)\,\mathrm{n}(t)
\]

\[
\mathit{eq6} := {\frac {\partial }{\partial t}}\,{n_{0}}(t)= -
\rho \,{\sigma _{a}}\,c\,{\phi _{p}}(t)\,{n_{0}}(t)
\]

\emptyline Contribution of gain, saturable, output and linear
intracavity loss to the field's evolution results in:

\emptyline $>$eq7 := diff(phi[p](t),t)[gain] =\\
2*(sigma[g]*n(t)*l[g] - sigma[a]*n[0](t)*l[a])*phi[p](t)/t[cav];\\
\indent
 $>$eq8 := diff(phi[p](t),t)[linear] = (-ln(1/R[p]) -
L[p])*phi[p](t)/t[cav];\\
\indent
  $>$eq9 := diff(phi[s](t),t)[linear] =\\ (-ln(1/R[s]) - L[s] -
2*kappa*sigma[a]*n[0](t)*l[a])*phi[s](t)/t[cav];

\[
\mathit{eq7} := {({\frac {\partial }{\partial t}}\,{\phi _{p}}(t)
)_{\mathit{gain}}}=2\,{\displaystyle \frac {({\sigma _{g}}\,
\mathrm{n}(t)\,{l_{g}} - {\sigma _{a}}\,{n_{0}}(t)\,{l_{a}})\,{
\phi _{p}}(t)}{{t_{\mathit{cav}}}}}
\]

\[
\mathit{eq8} := {({\frac {\partial }{\partial t}}\,{\phi _{p}}(t)
)_{\mathit{linear}}}={\displaystyle \frac {( - \mathrm{ln}(
{\displaystyle \frac {1}{{R_{p}}}} ) - {L_{p}})\,{\phi _{p}}(t)}{
{t_{\mathit{cav}}}}}
\]

\[
\mathit{eq9} := {({\frac {\partial }{\partial t}}\,{\phi _{s}}(t)
)_{\mathit{linear}}}={\displaystyle \frac {( - \mathrm{ln}(
{\displaystyle \frac {1}{{R_{s}}}} ) - {L_{s}} - 2\,\kappa \,{
\sigma _{a}}\,{n_{0}}(t)\,{l_{a}})\,{\phi _{s}}(t)}{{t_{\mathit{
cav}}}}}
\]

\emptyline \noindent where ${L_{p}}$ and ${L_{s}}$ are the linear
loss for the laser and Stokes components, respectively, ${R_{p}}$
and ${R_{s}}$ are the output mirror reflectivity at the laser and
Stokes wavelengths, respectively, $\kappa $ is the reduction
factor taking into account the decrease of the loss cross-section
at the Stokes wavelength relatively to the lasing one.

Hence, the field densities evaluate by virtue of:

\emptyline $>$eq10 := diff(phi[p](t),t) = rhs(eq3) + rhs(eq7) +
rhs(eq8);\\
\indent
 $>$eq11 := diff(phi[s](t),t) = rhs(eq4) + rhs(eq9);

\maplemultiline{ \mathit{eq10} := {\frac {\partial }{\partial
t}}\,{\phi _{p}}(t)= {\displaystyle \frac {{\displaystyle \frac
{({\nu _{p}}\,{\phi _{ p}}(t) + {\nu _{s}}\,{\phi _{s}}(t))\,{\phi
_{p}}(t)}{{\nu _{p}} \,{\phi _{p}}(t) +
e^{({l_{g}}\,g\,c\,h\,({\nu _{p}}\,{\phi _{p}} (t) + {\nu
_{s}}\,{\phi _{s}}(t)))}\,{\nu _{s}}\,{\phi _{s}}(t)}
}  - {\phi _{p}}(t)}{{t_{\mathit{cav}}}}}  \\
\mbox{} + {\displaystyle \frac {2\,({\sigma _{g}}\,\mathrm{n}(t)
\,{l_{g}} - {\sigma _{a}}\,{n_{0}}(t)\,{l_{a}})\,{\phi _{p}}(t)}{
{t_{\mathit{cav}}}}}  + {\displaystyle \frac {( - \mathrm{ln}(
{\displaystyle \frac {1}{{R_{p}}}} ) - {L_{p}})\,{\phi _{p}}(t)}{
{t_{\mathit{cav}}}}}  }

\maplemultiline{ \mathit{eq11} := {\frac {\partial }{\partial
t}}\,{\phi _{s}}(t)= {\displaystyle \frac {{\displaystyle \frac
{e^{({l_{g}}\,g\,c\,h \,(\mathrm{\%1} + {\nu _{s}}\,{\phi
_{s}}(t)))}\,{\phi _{s}}(t)\, (\mathrm{\%1} + {\nu _{s}}\,{\phi
_{s}}(t))}{\mathrm{\%1} + e^{({ l_{g}}\,g\,c\,h\,(\mathrm{\%1} +
{\nu _{s}}\,{\phi _{s}}(t)))}\,{ \nu _{s}}\,{\phi _{s}}(t)}}  -
{\phi _{s}}(t)}{{t_{\mathit{cav}}}
}}  \\
\mbox{} + {\displaystyle \frac {( - \mathrm{ln}({\displaystyle
\frac {1}{{R_{s}}}} ) - {L_{s}} - 2\,\kappa \,{\sigma _{a}}\,{n_{
0}}(t)\,{l_{a}})\,{\phi _{s}}(t)}{{t_{\mathit{cav}}}}}  \\
\mathrm{\%1} := {\nu _{p}}\,{\phi _{p}}(t) }

\emptyline In the agreement with the results of the previous
subsection we can transform the equation for the populations in
the gain and absorption media:

\emptyline $>$diff(n[0](n),n) = subs(\\
\{n(t)=n,n[0](t)=n[0](n)\},rhs(eq6)/rhs(eq5) );\\
\indent
 $>$diff(n[0](n),n) = zeta*rho*n[0](n)/n;\#
 zeta=sigma[a]/sigma[g]/gamma\\
 \indent
  $>$dsolve(\{\%\},n[0](n));

\[
{\frac {\partial }{\partial n}}\,{n_{0}}(n)={\displaystyle \frac
{\rho \,{\sigma _{a}}\,{n_{0}}(n)}{\gamma \,{\sigma _{g}}\, n}}
\]

\[
{\frac {\partial }{\partial n}}\,{n_{0}}(n)={\displaystyle \frac
{\zeta \,\rho \,{n_{0}}(n)}{n}}
\]

\[
{n_{0}}(n)=\mathit{\_C1}\,n^{(\zeta \,\rho )}
\]

\emptyline $>$eq12 := n[0] = n[0,i]*(n/n[i])$^{\zeta*\rho}$;

\[
\mathit{eq12} := {n_{0}}={n_{0, \,i}}\,({\displaystyle \frac {n}{
{n_{i}}}} )^{(\zeta \,\rho )}
\]

\emptyline $>$simplify( subs( phi[s](t)=0,subs(\\
\{n(t)=n[i],n[0](t)=n[0,i],phi[p](t)=0\},expand(\\
rhs(eq10)/phi[p](t) ) ) ) ) = 0:\# the generation start\\
\indent
 $>$numer(lhs(\%)) = 0:\\
 \indent \indent
  $>$eq13 := n[i] = subs(\\
2*sigma[a]*n[0,i]*l[a]=ln(1/T[0]$^2$),solve(\%,n[i]) );

\[
\mathit{eq13} := {n_{i}}={\displaystyle \frac {1}{2}} \,
{\displaystyle \frac {\mathrm{ln}({\displaystyle \frac {1}{{T_{0}
}^{2}}} ) + \mathrm{ln}({\displaystyle \frac {1}{{R_{p}}}} ) + {L
_{p}}}{{\sigma _{g}}\,{l_{g}}}}
\]

\emptyline Let' introduce new variables:

\emptyline $>$Phi[p](t) = 2*sigma[g]*l[g]*phi[p];\# note, that Phi
is not the dimensional intensity from the first subsection!\\
\indent
 $>$Phi[s](t) = 2*sigma[g]*l[g]*phi[s];\\
\indent \indent
  $>$tau = 2*sigma[g]*l[g]*n[i]*t/t[cav];\\
  \indent \indent \indent
   $>$Nu = nu[s]/nu[p];\\
  \indent \indent \indent \indent
    $>$G = g*l[g]*c*h*nu[p]/(2*sigma[g]*l[g]);\\
      \indent \indent \indent
     $>$y(tau) = n(t)/n[i];\\
       \indent \indent
      $>$Xi = ln(1/(T[0]$^2$))+ln(1/R[p])+L[p];\\
      \indent
       kappa = 0.38;

\[
{\Phi _{p}}(t)=2\,{\sigma _{g}}\,{l_{g}}\,{\phi _{p}}
\]

\[
{\Phi _{s}}(t)=2\,{\sigma _{g}}\,{l_{g}}\,{\phi _{s}}
\]

\[
\tau =2\,{\displaystyle \frac {{\sigma _{g}}\,{l_{g}}\,{n_{i}}\,t
}{{t_{\mathit{cav}}}}}
\]

\[
N={\displaystyle \frac {{\nu _{s}}}{{\nu _{p}}}}
\]

\[
G={\displaystyle \frac {1}{2}} \,{\displaystyle \frac {g\,c\,h\,{
\nu _{p}}}{{\sigma _{g}}}}
\]

\[
\mathrm{y}(\tau )={\displaystyle \frac {\mathrm{n}(t)}{{n_{i}}}}
\]

\[
\Xi =\mathrm{ln}({\displaystyle \frac {1}{{T_{0}}^{2}}} ) +
\mathrm{ln}({\displaystyle \frac {1}{{R_{p}}}} ) + {L_{p}}
\]

\[
\kappa =.38
\]

\emptyline Then from Eqs. (\textit{eq5, eq10, eq11, eq12}):

\emptyline $>$f[1] := diff(Phi[p](tau),tau) =
Phi[p](tau)/Xi*\\(Xi*y(tau) -
ln(1/T[0]$^2$)*y(tau)$^{\alpha*\rho}$
- (ln(1/R[p])+L[p]) +\\
((Phi[p](tau)+N*Phi[s](tau))/(Phi[p](tau) +\\
exp(G*(Phi[p](tau)+N*Phi[s](tau)))*N*Phi[s](tau)) - 1 ) );\\

\indent $>$f[2] := diff(Phi[s](tau),tau) =\\
Phi[s](tau)/Xi*(-(ln(1/R[s])+L[s]+kappa*ln(1/(T[0]$^2$))*\\
y(tau)$^{\alpha*\rho}$) +
\\(Phi[p](tau)+N*Phi[s](tau))/(Phi[p](tau)*\\exp(-G*(Phi[p](tau)+N*Phi[s](tau)))
+ N*Phi[s](tau) ) - 1 );\\

\indent $>$f[3] := diff(y(tau),tau) =
-gamma*(l[cav]/l[g])*Phi[p](tau)/Xi;

\begin{gather} \nonumber
{f_{1}} := {\frac {\partial }{\partial \tau }}\,{\Phi _{p}}(\tau
)=\\
{\displaystyle \frac {{\Phi _{p}}(\tau )\, \left(  \! \Xi \,
\mathrm{y}(\tau ) - \mathrm{ln}({\displaystyle \frac {1}{{T_{0}}
^{2}}} )\,\mathrm{y}(\tau )^{(\alpha \,\rho )} - \mathrm{ln}(
{\displaystyle \frac {1}{{R_{p}}}} ) - {L_{p}} + {\displaystyle
\frac {{\Phi _{p}}(\tau ) + N\,{\Phi _{s}}(\tau )}{{\Phi _{p}}(
\tau ) + e^{(G\,({\Phi _{p}}(\tau ) + N\,{\Phi _{s}}(\tau )))}\,N
\,{\Phi _{s}}(\tau )}}  - 1 \!  \right) }{\Xi }} \nonumber
\end{gather}

\begin{gather} \nonumber
{f_{2}} := {\frac {\partial }{\partial \tau }}\,{\Phi _{s}}(\tau
)=\\
{\displaystyle \frac {{\Phi _{s}}(\tau )\, \left(  \!  -
\mathrm{ln}({\displaystyle \frac {1}{{R_{s}}}} ) - {L_{s}} -
\kappa \,\mathrm{ln}({\displaystyle \frac {1}{{T_{0}}^{2}}} )\,
\mathrm{y}(\tau )^{(\alpha \,\rho )} + {\displaystyle \frac {{
\Phi _{p}}(\tau ) + N\,{\Phi _{s}}(\tau )}{{\Phi _{p}}(\tau )\,e
^{( - G\,({\Phi _{p}}(\tau ) + N\,{\Phi _{s}}(\tau )))} + N\,{
\Phi _{s}}(\tau )}}  - 1 \!  \right) }{\Xi }} \nonumber
\end{gather}

\[
{f_{3}} := {\frac {\partial }{\partial \tau }}\,\mathrm{y}(\tau )
= - {\displaystyle \frac {\gamma \,{l_{\mathit{cav}}}\,{\Phi _{p}
}(\tau )}{{l_{g}}\,\Xi }}
\]

\emptyline The next procedures solve the obtained system
numerically to obtain the dependencies of the normalized photon
densities at laser and Stokes wavelengths and the relative
inversion vs. cavity roundtrip, \textit{n }is the duration of the
simulation in the cavity roundtrips):

\emptyline $>$ODE\_plot1 := proc(T0,Rp,Rs,Lp,Ls,n)\\

\indent $>$gam := 1:\# gamma parameter\\

\indent $>$rho := 1:\\
\indent $>$kappa := 0.38:\\
\indent $>$alpha := evalf( 5e-18/2.8e-20 ):\#the ratio of the
cross-sections\\
\indent $>$x := 29:\# x = l[cav]/l[g]\\
\indent $>$N := evalf(1033/1139):\\
\indent $>$G := subs(\\
\{g=4.8e-9,c=3e10,h=6.62e-34,nu\_p=3e10/1.033e-4,\\
sigma\_g=2.8e-20\},g*c*h*nu\_p/(2*sigma\_g) ):\\
\indent $>$Xi := ln(1/(T0$^2$))+ln(1/Rp)+Lp:\\

\indent $>$sys := [ D(Phip)(tau) =\\
Phip(tau)*(Xi*y(tau)-ln(1/(T0$^2$))*y(tau)$^{\alpha*\rho}$\\
-ln(1/Rp)-Lp+(Phip(tau)+N*Phis(tau))/(Phip(tau)+\\
exp(G*(Phip(tau)+N*Phis(tau)))*N*Phis(tau))-1)/Xi, \\

D(Phis)(tau) =\\
Phis(tau)*(-ln(1/Rs)-Ls-kappa*ln(1/(T0$^2$))*y(tau)$^{\alpha*\rho}$\\
+(Phip(tau)+N*Phis(tau))/(Phip(tau)*\\
exp(-G*(Phip(tau)+N*Phis(tau)))+N*Phis(tau))-1)/Xi, \\

D(y)(tau) = -gam*x*Phip(tau)/Xi]:\\

\indent
$>$DEplot(sys,[Phip(tau),Phis(tau),y(tau)],tau=0..n,\\
$[[Phip(0)=1e-7,Phis(0)=1e-7,y(0)=1]]$,stepsize=0.1,\\
scene=[tau,Phip(tau)],method=classical[abmoulton],\\
axes=FRAME,linecolor=BLUE):\\ \indent $>$end:\\

\indent $>$ODE\_plot2 := proc(T0,Rp,Rs,Lp,Ls,n)\\

\indent $>$gam := 1:\\
\indent $>$rho := 1:\\
\indent $>$kappa := 0.38:\\
\indent $>$alpha := evalf( 5e-18/2.8e-20 ):\\
\indent $>$x := 29:\# x = l[cav]/l[g]\\
\indent $>$N := evalf(1033/1139):\\
\indent $>$G := subs(\\
\{g=4.8e-9,c=3e10,h=6.62e-34,nu\_p=3e10/1.033e-4,sigma\_g=2.8e-20\},\\
g*c*h*nu\_p/(2*sigma\_g)):\\
\indent $>$Xi := ln(1/(T0$^2$))+ln(1/Rp)+Lp:\\

\indent $>$sys := [ D(Phip)(tau) =\\
Phip(tau)*(Xi*y(tau)-ln(1/(T0$^2$))*y(tau)$^{\alpha*\rho}$-\\
ln(1/Rp)-Lp+(Phip(tau)+N*Phis(tau))/(Phip(tau)+\\
exp(G*(Phip(tau)+N*Phis(tau)))*N*Phis(tau))-1)/Xi, \\

D(Phis)(tau) =\\
Phis(tau)*(-ln(1/Rs)-Ls-kappa*ln(1/(T0$^2$))*y(tau)$^{\alpha*\rho}$+\\
(Phip(tau)+N*Phis(tau))/(Phip(tau)*\\
exp(-G*(Phip(tau)+N*Phis(tau)))+N*Phis(tau))-1)/Xi, \\

D(y)(tau) = -gam*x*Phip(tau)/Xi]:\\

\indent
$>$DEplot(sys,[Phip(tau),Phis(tau),y(tau)],tau=0..n,\\
$[[Phip(0)=1e-7,Phis(0)=1e-7,y(0)=1]]$,stepsize=0.1,\\
scene=[tau,Phis(tau)],method=classical[abmoulton],\\
axes=FRAME,linecolor=RED):\\ \indent $>$end:\\

\indent $>$ODE\_plot3 := proc(T0,Rp,Rs,Lp,Ls,n)\\

\indent $>$gam := 1:\\
\indent $>$rho := 1:\\
\indent $>$kappa := 0.38:\\
\indent $>$alpha := evalf( 5e-18/2.8e-20 ):\\
\indent $>$x := 29:\# x = l[cav]/l[g]\\
\indent $>$N := evalf(1033/1139):\\
\indent $>$G := subs(\\
\{g=4.8e-9,c=3e10,h=6.62e-34,nu\_p=3e10/1.033e-4,sigma\_g=2.8e-20\},\\
g*c*h*nu\_p/(2*sigma\_g)):\\
\indent $>$Xi := ln(1/(T0$^2$))+ln(1/Rp)+Lp:\\

\indent sys := [ D(Phip)(tau) =\\
Phip(tau)*(Xi*y(tau)-ln(1/(T0$^2$))*y(tau)$^{\alpha*\rho}$\\
-ln(1/Rp)-Lp+(Phip(tau)+N*Phis(tau))/(Phip(tau)+\\
exp(G*(Phip(tau)+N*Phis(tau)))*N*Phis(tau))-1)/Xi, \\

\indent D(Phis)(tau) =\\
Phis(tau)*(-ln(1/Rs)-Ls-kappa*ln(1/(T0$^2$))*y(tau)$^{\alpha*\rho}$\\
+(Phip(tau)+N*Phis(tau))/(Phip(tau)*\\
exp(-G*(Phip(tau)+N*Phis(tau)))+N*Phis(tau))-1)/Xi, \\

\indent D(y)(tau) = -gam*x*Phip(tau)/Xi]:\\

\indent
$>$DEplot(sys,[Phip(tau),Phis(tau),y(tau)],tau=0..n,\\
$[[Phip(0)=1e-7,Phis(0)=1e-7,y(0)=1]]$,stepsize=0.1,\\
scene=[tau,y(tau)],method=classical[abmoulton],\\
axes=FRAME,linecolor=BLACK):\\ \indent $>$end:

\emptyline
$>$display(ODE\_plot1(0.65,0.75,0.96,5e-2,5e-2,300),axes=boxed,\\
title=`laser photon density`);\\
\indent
$>$display(ODE\_plot2(0.65,0.75,0.96,5e-2,5e-2,300),axes=boxed,\\
title=`Stokes photon density`);\\
\indent
$>$display(ODE\_plot3(0.65,0.75,0.96,5e-2,5e-2,300),axes=boxed,\\
title=`relative inversion`);

\begin{center}
\mapleplot{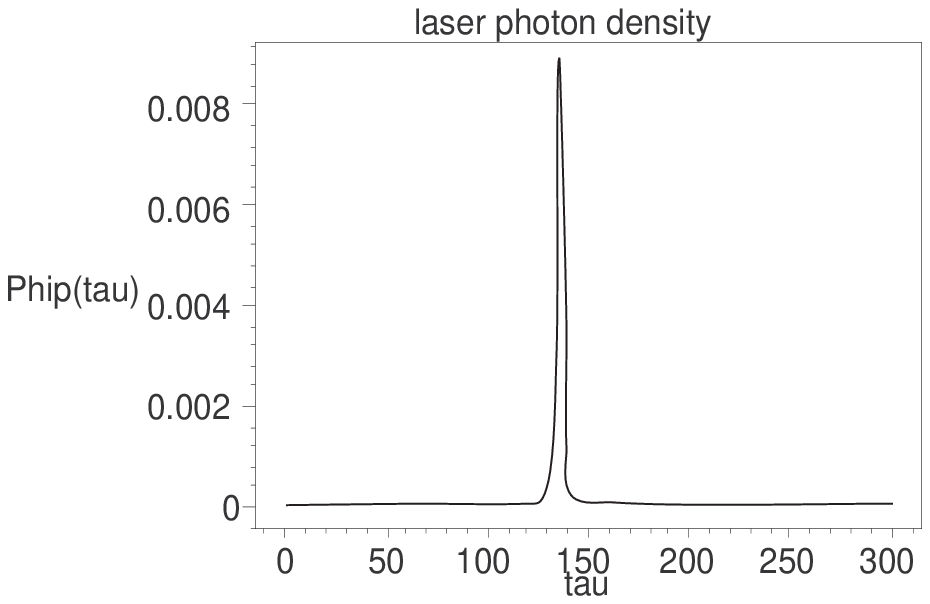}
\end{center}

\begin{center}
\mapleplot{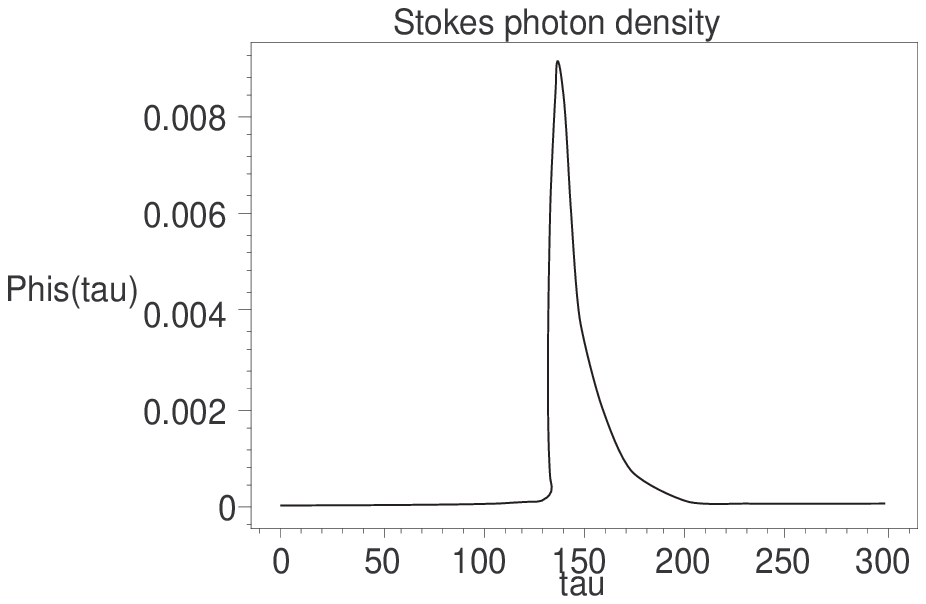}
\end{center}

\begin{center}
\mapleplot{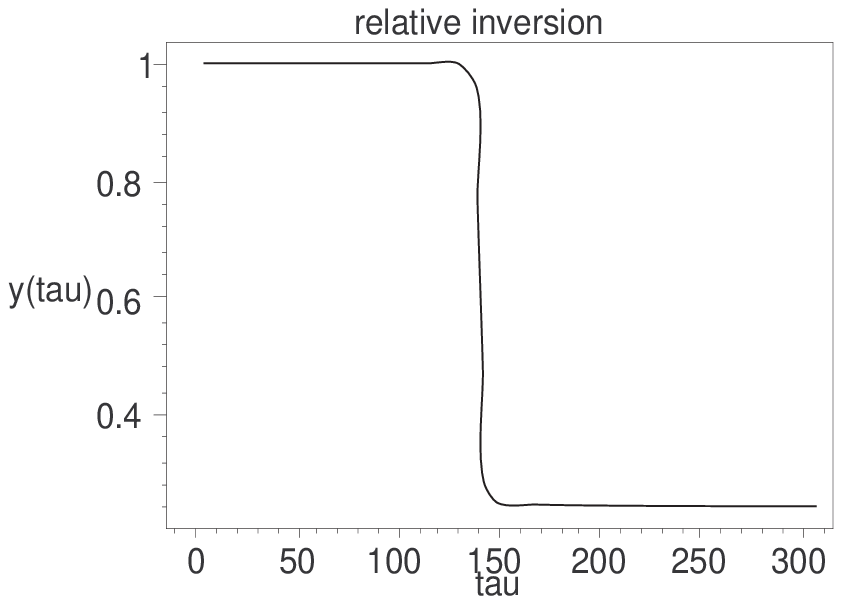}
\end{center}

\emptyline \fbox{\parbox{.8\linewidth} {So, we can obtain a quite
efficient two-color pulsing in the nanosecond time domain without
any additional wavelength conversion.}}

\emptyline The nonstationary lasing in question produces the
pulses with the durations, which are larger than the cavity
period. The cavity length shortening, i. e. the use of the
microchip lasers, decreases the pulse durations down to ten --
hundred picoseconds. But there is a method allowing the
fundamental pulse width reduction, viz. mode locking.

\section{Conception of mode locking}\label{2}

\emptyline

The laser cavity is, in fact, interferometer, which supports the
propagation of only defined light waves. Let consider a plane wave,
which is reflected from a laser mirror. The initial wave is
(\textit{cc} is the complex conjugated term):

\emptyline

$>$restart:\\
\indent \indent
with(plots):\\
\indent \indent \indent
  with(DEtools):\\
\indent \indent \indent \indent
   AI := A0*exp(I*(omega*t+k*x))+cc;

\emptyline
\[
\mathit{AI} := \mathit{A0}\,e^{(I\,(\omega \,t + k\,x))} +
\mathit{cc}
\]

\emptyline

Then the reflected wave (normal incidence and full reflectivity are
supposed) is:

\emptyline
$>$AR := A0*exp(I*(omega*t-k*x+Pi))+cc;
\emptyline

\[
\mathit{AR} := \mathit{A0}\,e^{(I\,(\omega \,t - k\,x + \pi ))}
 + \mathit{cc}
\]
\emptyline
\noindent
where $\pi $ is the phase shift due to reflection. An interference between
incident and reflected waves results in\\

\emptyline
$>$convert(AI+AR,trig):\\
\indent \indent
  expand(\%):\\
\indent \indent \indent
factor(\%);

\emptyline

\[
 - 2\,\mathit{A0}\,\mathrm{sin}(\omega \,t)\,\mathrm{sin}(k\,x)
 + 2\,I\,\mathit{A0}\,\mathrm{cos}(\omega \,t)\,\mathrm{sin}(k\,x
) + 2\,\mathit{cc}
\]

\emptyline
\noindent
This is the so-called standing wave:

\emptyline
$>$animate(sin(x)*sin(t), x=0..2*Pi,t=0..2*Pi, axes=boxed,\\
\indent
title=`Standing wave`, color=red);
\emptyline
\begin{center}
\mapleplot{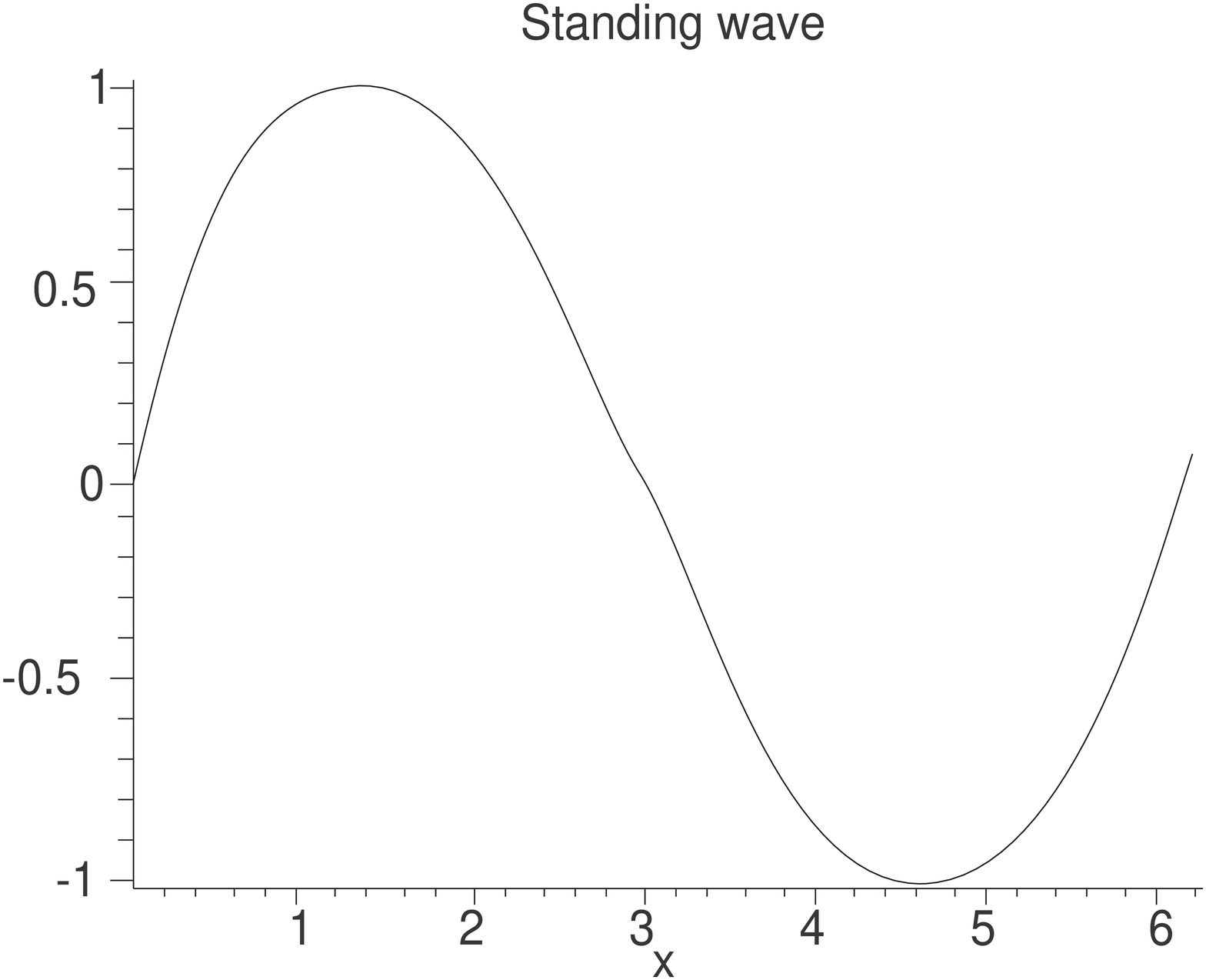}
\end{center}
\emptyline
\noindent
Note, that a wave node lies on a surface (point \textit{x}=0). The
similar situation takes place in the laser resonator. But the laser
resonator consists of two (or more) mirrors and the standing wave is
formed due to reflection from the each mirror. So, the wave in the
resonator is the standing wave with the nodes placed on the mirrors.
Such waves are called as the \textit{longitudinal laser modes}. The
laser resonator can contain a lot of modes with the different
frequencies (but its nodes have to lie on the mirrors!) and these
modes can interfere.

Let suppose, that the longitudinal modes are numbered by the index
\textit{m}. In fact, we have \textit{M} harmonic oscillators with the
phase and frequency differences \textit{dphi} and \textit{domega},
correspondingly. Let the amplitude of modes is \textit{A0}.

\emptyline
$>$mode := 1/2*A0*exp(I*(phi0+m*dphi)+\\
\indent
I*(omega0+m*domega)*t)+cc;\# amplitude of the mode numbered \\
\indent
by index m

\emptyline
\[ \boxed{
\mathit{mode} := {\displaystyle \frac {1}{2}} \,\mathit{A0}\,e^{(
I\,(\phi 0 + m\,\mathit{dphi}) + I\,(\omega 0 + m\,\mathit{domega
})\,t)} + \mathit{cc}}
\]

\emptyline
\noindent
Here
$\phi $0 and
$\omega $0 are the phase and frequency of the central mode, respectively.
The interference between these modes produces the wave packet:

\emptyline
$>$packet := sum(mode-cc,m=-(M-1)/2..(M-1)/2)+cc;\\
\indent
\# interference of longitudinal modes with constant phases

\emptyline
\begin{gather*}
\mathit{packet} := {\displaystyle \frac {1}{2}} \,{\displaystyle
\frac {\mathit{A0}\,e^{(I\,\phi 0)}\,e^{(I\,\omega 0\,t)}\,e^{(I
\,(1/2\,M + 1/2)\,\mathit{dphi})}\,e^{(I\,t\,(1/2\,M + 1/2)\,
\mathit{domega})}}{e^{(I\,\mathit{dphi})}\,e^{(I\,\mathit{domega}
\,t)} - 1}} \\
\mbox{} - {\displaystyle \frac {1}{2}} \,{\displaystyle \frac {
\mathit{A0}\,e^{(I\,\phi 0)}\,e^{(I\,\omega 0\,t)}\,e^{(I\,( - 1/
2\,M + 1/2)\,\mathit{dphi})}\,e^{(I\,t\,( - 1/2\,M + 1/2)\,
\mathit{domega})}}{e^{(I\,\mathit{dphi})}\,e^{(I\,\mathit{domega}
\,t)} - 1}}  + \mathit{cc}
\end{gather*}

\emptyline

Now, we can extract the term describing the fast oscillation on the
central ("carrier") frequency $\omega $0 from the previous expression. The obtained result is the packet's
envelope (its slowly varying amplitude):

\emptyline
$>$envelope := expand((packet-cc)/exp(I*(t*omega0+phi0)));\\
\indent
\# slowly varying envelope of the wave packet
\emptyline

\begin{gather*}
\mathit{envelope} := {\displaystyle \frac {1}{2}} \,
{\displaystyle \frac {\mathit{A0}\,e^{(1/2\,I\,\mathit{dphi}\,M)}
\,e^{(1/2\,I\,\mathit{dphi})}\,e^{(1/2\,I\,\mathit{domega}\,t\,M)
}\,e^{(1/2\,I\,\mathit{domega}\,t)}}{e^{(I\,\mathit{dphi})}\,e^{(
I\,\mathit{domega}\,t)} - 1}} \\
\mbox{} - {\displaystyle \frac {1}{2}} \,{\displaystyle \frac {
\mathit{A0}\,e^{(-1/2\,I\,\mathit{dphi}\,M)}\,e^{(1/2\,I\,
\mathit{dphi})}\,e^{(-1/2\,I\,\mathit{domega}\,t\,M)}\,e^{(1/2\,I
\,\mathit{domega}\,t)}}{e^{(I\,\mathit{dphi})}\,e^{(I\,\mathit{
domega}\,t)} - 1}}
\end{gather*}

\emptyline
\noindent
It is obviously, that this expression can be converted into following
form:

\emptyline
$>$envelope :=\\
\indent \indent
1/2*A0*sinh(1/2*I*M*(dphi+t*domega))$/$ \\
\indent \indent \indent
sinh(1/2*I*(dphi+t*domega));

\emptyline
\[ \boxed{
\mathit{envelope} := {\displaystyle \frac {1}{2}} \,
{\displaystyle \frac {\mathit{A0}\,\mathrm{sin}({\displaystyle
\frac {1}{2}} \,M\,(\mathit{dphi} + \mathit{domega}\,t))}{
\mathrm{sin}({\displaystyle \frac {1}{2}} \,\mathit{dphi} +
{\displaystyle \frac {1}{2}} \,\mathit{domega}\,t)}} }
\]

\emptyline
\noindent
The squared envelope's amplitude (i. e. a field intensity) is depicted
in the next figure for the different \textit{M}.

\emptyline
$>$plot(\{subs(\{M=5,A0=1,dphi=0.1,domega=0.1\},2*envelope$^{2}$),\\
\indent
subs(\{M=20,A0=1,dphi=0.1,domega=0.1\},2*envelope$^{2}$)\},\\
\indent
t=-80..80, axes=boxed, title=`result of modes interference`);

\begin{center}
\mapleplot{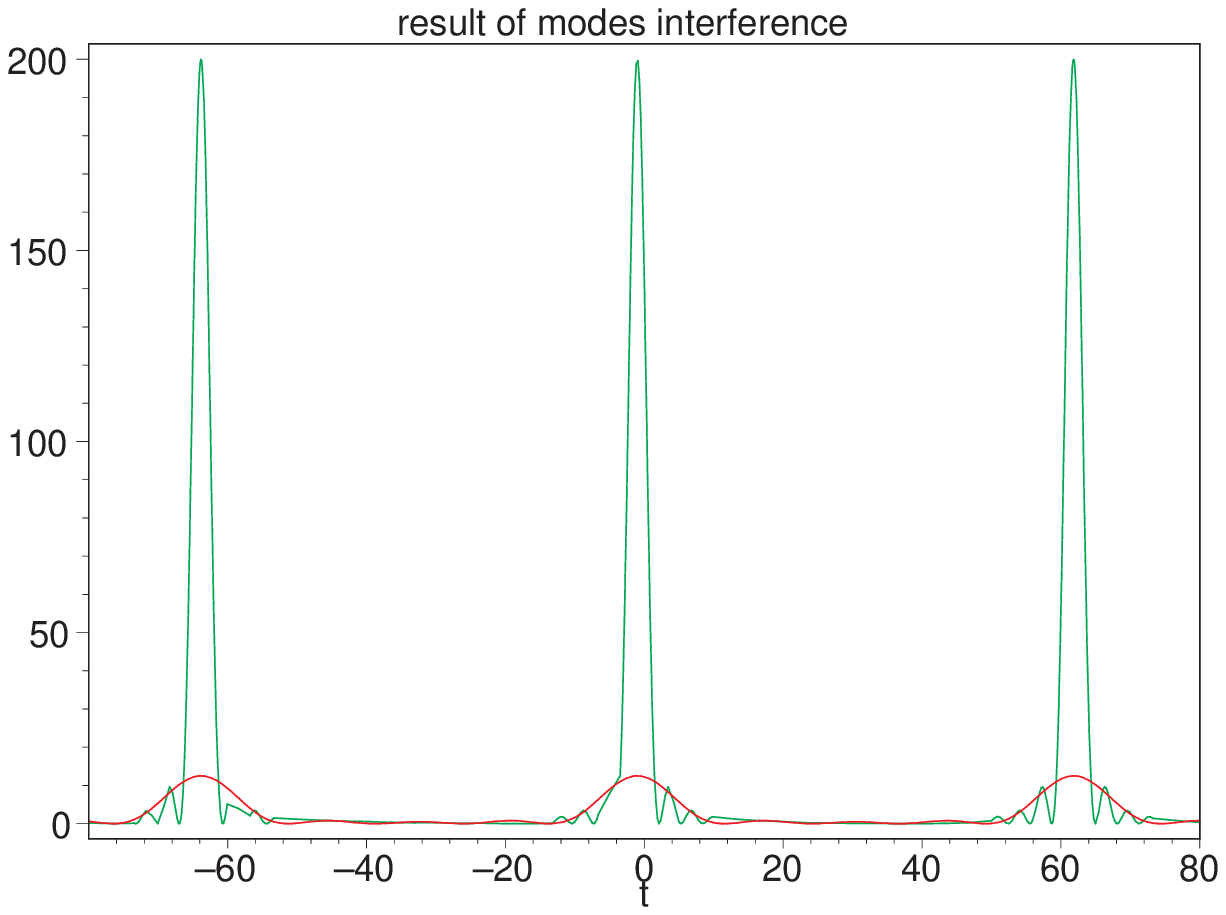}
\end{center}

\emptyline
\noindent
One can see, that the interference of modes results in the generation
of short pulses. The interval between pulses is equal to 2
$\pi $/\textit{domega}. The growth of \textit{ M} decreases the
pulse duration 2
$\pi $/(\textit{M*domega}) and to increases the pulse intensity
$M^{2}$*$\mathit{A0}^{2}$. The last is the consequence of the following relation:

\emptyline
$>$(limit(sin(M/2*x)/sin(x/2),x=0)*A0)$^{2}$;\# maximal field intensity

\emptyline
\[
M^{2}\,\mathit{A0}^{2}
\]
\emptyline
\noindent
In this example, the phase difference between neighboring modes is
constant. Such \textit{mode locking} causes the generation of short
and intense pulses. But in the reality, the laser modes are not
locked, i. e. the modes are the oscillations with the independent and
accidental phases. In this case:

\emptyline
$>$M := 20:\# 20 longitudinal modes\\
\indent \indent
 A0 := 1:\# constant dimensionless amplitude of mode\\
\indent \indent \indent
  domega := 0.1:\# constant frequency different\\
\indent \indent
   phi0 := 0:\# phase of the central mode\\
\indent
  omega0 := 1:\# dimensionless frequency of central mode\\
\indent \indent
 mode := 0:\\
\indent \indent \indent
for m from -(M-1)/2 to (M-1)/2 do:\\
\indent \indent
 die := rand(6):\\
\indent
  dphi := die():\# accidental phase difference between modes\\
\indent \indent
 mode := mode+A0*cos(phi0+m*dphi+(omega0+m*domega)*t):\\
\indent \indent \indent
od:\\
\indent \indent
plot(mode,t=-80..80, axes=boxed, title=`result of modes
interference`);

\emptyline
\begin{center}
\mapleplot{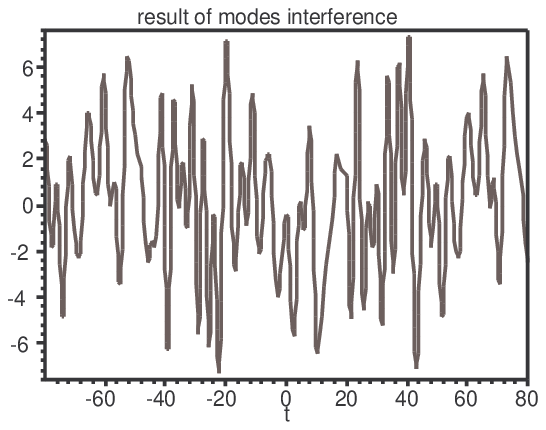}
\end{center}

\emptyline
\noindent
Thus, the interference of the unlocked modes produces the irregular
field beatings, i. e.  the noise spikes with a duration \symbol{126}
\textit{1/(M*domega)}.

What are the methods for the mode locking? Firstly, let consider the
simplest model of the harmonic oscillation in the presence of the
periodical force.

\emptyline

$>$diff\_equation := diff(diff(y(t),t),t)+omega$^{2}$*y(t)=cos(delta*t+phi);\\
\indent
\# oscillations in the presence of periodical force \\
\indent
\# (delta and phi are the frequency and\\
\indent
\#phase of the force oscillation, respectively)\\
\indent \indent
      dsolve(\{diff\_equation, y(0)=1, D(y)(0)=0\},y(t)):\\
\indent \indent \indent
            oscill1 := combine(\%);\\
\emptyline
\[
\mathit{diff\_equation} := ({\frac {\partial ^{2}}{\partial t^{2}
}}\,\mathrm{y}(t)) + \omega ^{2}\,\mathrm{y}(t)=\mathrm{cos}(
\delta \,t + \phi )
\]

\begin{gather*}
\mathit{oscill1} := \mathrm{y}(t)=(\omega \,\mathrm{cos}(\omega
\,t - \phi ) + \omega \,\mathrm{cos}(\omega \,t + \phi ) - \\
\mbox{} 2\,\mathrm{cos}(\omega \,t)\,\omega ^{3} + 2\,\omega \,\mathrm{cos}(
\omega \,t)\,\delta ^{2} - \delta \,\mathrm{cos}(\omega \,t -
\phi ) \\
\mbox{} + \delta \,\mathrm{cos}(\omega \,t + \phi ) - 2\,\mathrm{
cos}(\delta \,t + \phi )\,\omega ) \left/ {\vrule
height0.43em width0em depth0.43em} \right. \!  \! ( - 2\,\omega
^{3} + 2\,\omega \,\delta ^{2})
\end{gather*}

\emptyline
$>$animate(subs(\{omega=1,delta=0.1
\},subs(oscill1,y(t))),\\
\indent
t=0..100,phi=0..2*Pi,color=red,style=point,
axes=boxed);
\emptyline
\begin{center}
\mapleplot{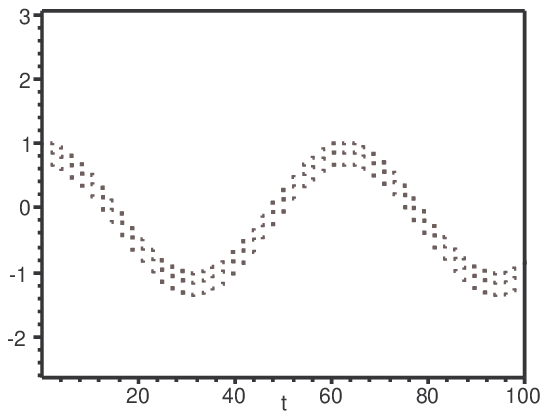}
\end{center}

\emptyline
\noindent
We can see, that the external force causes the oscillations with the
additional frequencies:
$\omega $+
$\delta $ and
$\omega $-
$\delta $. If
$\delta $ is equal to the intermode interval, the additional oscillation of
mode plays a role of the resonance external force for the neighboring
modes. Let consider such resonant oscillation in the presence of the
resonant external force:

\emptyline

$>$diff\_equation := diff(diff(y(t),t),t)+omega$^{2}$*y(t)=\\
\indent
cos(omega*t+phi);\# resonant external force\\
\indent \indent
      dsolve(\{diff\_equation,y(0)=1,D(y)(0)=0\},y(t)):\\
\indent \indent \indent
            oscill2 := combine(\%);

\emptyline
\[
\mathit{diff\_equation} := ({\frac {\partial ^{2}}{\partial t^{2}
}}\,\mathrm{y}(t)) + \omega ^{2}\,\mathrm{y}(t)=\mathrm{cos}(
\omega \,t + \phi )
\]

\begin{gather*}
\mathit{oscill2} := \mathrm{y}(t)=\\
\mbox{} {\displaystyle \frac {1}{4}} \,
{\displaystyle \frac { - \mathrm{cos}(\omega \,t - \phi ) +
\mathrm{cos}(\omega \,t + \phi ) + 4\,\mathrm{cos}(\omega \,t)\,
\omega ^{2} + 2\,t\,\mathrm{sin}(\omega \,t + \phi )\,\omega }{
\omega ^{2}}}
\end{gather*}

\emptyline
\noindent
The term, which is proportional to \textit{t} ("secular" term),
equalizes the phase of the oscillations to the phase of the external
force. It is the simplest model of a so-called \textit{active mode
locking}. Here the role of the external force can be played by the
external amplitude or phase modulator. Main condition for this
modulator is the equality of the modulation frequency to the intermode
interval that causes the resonant interaction between modes and, as
consequence, the mode locking (part \textit{4}).
\noindent
The different mechanism, a \textit{passive mode locking}, is produced
by the nonlinear interaction of modes with an optical medium. Such
nonlinearity can be caused by saturable absorption, self-focusing etc.
(see further parts of article). Now we shall consider the simplest
model of the passive mode locking. Let suppose, that there are two
modes, which oscillate with different phases and frequencies in the
cubic nonlinear medium:

\emptyline
$>$omega := 1:\\
\indent
 delta := 0.1:\\
\indent \indent
  sys := [diff(diff(y(t),t),t)+omega$^{2}$*y(t)=-(y(t)$^{3}$+z(t)$^{3}$),\\
\indent \indent
   diff(diff(z(t),t),t)+(omega+delta)$^{2}$*z(t)=-(y(t)$^{3}$+z(t)$^{3}$)];\\
\indent \indent
\#two oscillating modes with nonlinear coupling\\
\indent
 DEplot3d(sys,[y(t),z(t)],t=0..100,[[y(0)=-1,z(0)=1,D(y)(0)=0,\\
\indent
D(z)(0)=0]],stepsize=1,scene=[y(t),z(t),t],\\
\indent
axes=boxed,linecolor=BLACK,orientation=[-60,70], \\
\indent
title=`mode locking`);
\emptyline
\begin{gather*}
\mathit{sys} :=  \\
\mbox{} [({\frac {\partial ^{2}}{\partial t^{2}}}\,
\mathrm{y}(t)) + \mathrm{y}(t)= - \mathrm{y}(t)^{3} - \mathrm{z}(
t)^{3}, \,({\frac {\partial ^{2}}{\partial t^{2}}}\,\mathrm{z}(t)
) + 1.21\,\mathrm{z}(t)= - \mathrm{y}(t)^{3} - \mathrm{z}(t)^{3}]
\end{gather*}
\emptyline

\begin{center}
\mapleplot{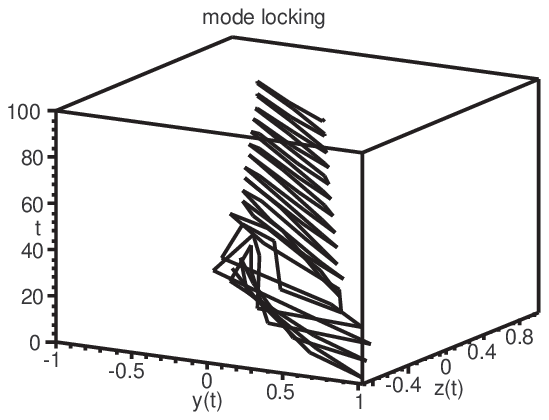}
\end{center}

\emptyline
\noindent
We can see, that the initial oscillations with the different phases
are locked due to nonlinear interaction that produces the synchronous
oscillations.

\emptyline
\noindent
\begin{center}
\fbox{\parbox{.8\linewidth}
{As conclusion, we note that the mode locking is resulted from the
interference between standing waves with constant and locked phases.
Such interference forms a train of the ultrashort pulses. The
mechanisms of the mode locking are the external modulation with
frequency, which is equal to intermode frequency interval, or the
nonlinear response of the optical medium}}
\end{center}
\emptyline
\noindent
Later on we shall consider
both methods. But firstly we have to obtain the more realistic
equations describing the ultrashort pulse generation.

\emptyline

\section{Basic model}

\emptyline
\noindent
The models describing the laser field evolution are based usually on
the so-called semi-classical approximation. In the framework of this
approximation the field obeys the classical Maxwell equation and the
medium evolution has the quantum-mechanical character. Here we shall
consider the wave equation without concretization of the medium
evolution.

The Maxwell equation for the light wave can be written as:

\emptyline
$>$restart:\\
\indent \indent
 with(PDEtools,dchange):\\
\indent \indent \indent
maxwell\_eq :=\\
\indent \indent
diff(E(z,t),z,z)-diff(E(z,t),t,t)/c$^{2}$=4*Pi*diff(P(t),t,t)/c$^{2}$;

\emptyline
\[ \boxed{
\mathit{maxwell\_eq} := ({\frac {\partial ^{2}}{\partial z^{2}}}
\,\mathrm{E}(z, \,t)) - {\displaystyle \frac {{\frac {\partial ^{
2}}{\partial t^{2}}}\,\mathrm{E}(z, \,t)}{c^{2}}} =4\,
{\displaystyle \frac {\pi \,({\frac {\partial ^{2}}{\partial t^{2
}}}\,\mathrm{P}(t))}{c^{2}}}}
\]

\emptyline
\noindent
where \textit{E(z,t)} is the field strength, \textit{P(t)} is the medium polarization,
\textit{z} is the longitudinal coordinate,
\textit{t} is the time, \textit{c} is the light velocity. The change
of the variables  \textit{z --\TEXTsymbol{>} z*}, \textit{t - z/c
--\TEXTsymbol{>} t*} produces

\emptyline

$>$macro(zs=`z*`,ts=`t*`):\\
\indent \indent
tr := \{z = zs, t = ts + zs/c\};\\
\indent
maxwell\_m := dchange(tr,maxwell\_eq,[zs,ts],simplify);
\emptyline

\[
\mathit{tr} := \{t=\mathit{t*} + {\displaystyle \frac {\mathit{z*
}}{c}} , \,z=\mathit{z*}\}
\]

\begin{gather*}
\mathit{maxwell\_m} := {\displaystyle \frac {({\frac {\partial ^{
2}}{\partial \mathit{z*}^{2}}}\,\mathrm{E}(\mathit{z*}, \,
\mathit{t*}, \,c))\,c - 2\,({\frac {\partial ^{2}}{\partial
\mathit{z*}\,\partial \mathit{t*}}}\,\mathrm{E}(\mathit{z*}, \,
\mathit{t*}, \,c))}{c}} =  \\
\mbox{} 4\,{\displaystyle \frac {\pi \,({\frac {
\partial ^{2}}{\partial \mathit{t*}^{2}}}\,\mathrm{P}(\mathit{z*}
, \,\mathit{t*}, \,c))}{c^{2}}}
\end{gather*}

\emptyline
\noindent
We does not take into account the effects connected with a wave
propagation in thin medium layer, that allows to eliminate the
second-order derivation on \textit{z*}. Then

\emptyline
$>$int(op(2,expand(lhs(maxwell\_m))),ts)-int(rhs(maxwell\_m),ts):\\
\indent
\# integration of both sides of wave equation\\
\indent \indent
    numer(\%):\\
\indent \indent \indent
         wave\_1 := expand(\%/(-2));
\emptyline
\[
\mathit{wave\_1} := ({\frac {\partial }{\partial \mathit{z*}}}\,
\mathrm{E}(\mathit{z*}, \,\mathit{t*}, \,c))\,c + 2\,\pi \,(
{\frac {\partial }{\partial \mathit{t*}}}\,\mathrm{P}(\mathit{z*}
, \,\mathit{t*}, \,c))
\]

\emptyline
\noindent
So, we reduced the order of wave equation. The inverse transformation
of the coordinates leads to the so-called shortened wave equation:

\emptyline
$>$tr := \{zs = z, ts = t - z/c\};\\
\indent \indent
wave\_2 := dchange(tr,wave\_1,[z,t],simplify);
\emptyline
\[
\mathit{tr} := \{\mathit{z*}=z, \,\mathit{t*}=t - {\displaystyle
\frac {z}{c}} \}
\]

\[
\mathit{wave\_2} := ({\frac {\partial }{\partial z}}\,\mathrm{E}(
z, \,t, \,c))\,c + ({\frac {\partial }{\partial t}}\,\mathrm{E}(z
, \,t, \,c)) + 2\,\pi \,({\frac {\partial }{\partial t}}\,
\mathrm{P}(z, \,t, \,c))
\]

\emptyline
\noindent
Next step is the transition to the slowly-varying amplitude
approximation. We shall consider field envelope
$\rho $(z,t) and polarization P(t), which are filled by the fast oscillation
with frequency
$\omega $ (\textit{k} is the wave number, \textit{N }is the concentration of
the atoms, \textit{d} is the medium length):

\emptyline
$>$Theta=omega*t-k*z;\# phase\\
\indent \indent
E := rho(z,t)*cos(Theta);\# field\\
\indent
P := N*d*(a(t)*cos(Theta)-b(t)*sin(Theta));\# polarization (a and b are
the quadrature components)

\emptyline
\[
\Theta =\omega \,t - k\,z
\]

\[
E := \rho (z, \,t)\,\mathrm{cos}(\Theta )
\]

\[
P := N\,d\,(\mathrm{a}(t)\,\mathrm{cos}(\Theta ) - \mathrm{b}(t)
\,\mathrm{sin}(\Theta ))
\]

\emptyline
\noindent
Then the wave equation can be transformed as:

\emptyline
$>$Theta:=omega*t-k*z:\\
\indent
  diff(E,z)+diff(E,t)/c+2*Pi*diff(P,t)/c:\\
\indent \indent
    combine(\%,trig):\\
\indent \indent \indent
      collect(\%,cos(Theta)):\\
\indent \indent
    wave\_3 := collect(\%,sin(Theta)):\\
\indent
  eq\_field :=\\
\indent \indent
op(1,expand(coeff(wave\_3,cos(Theta))))+op(3,expand(\\
\indent \indent \indent
coeff(wave\_3,cos(Theta))))=\\
\indent \indent
-op(2,expand(coeff(wave\_3,cos(Theta))))-\\
\indent
op(4,expand(coeff(wave\_3,cos(Theta))));\# field equation\\
\indent \indent
disp\_conditions := Int(coeff(wave\_3,sin(Theta)),t)=0;\# dispersion condition

\emptyline
\[\boxed{
\mathit{eq\_field} := ({\frac {\partial }{\partial z}}\,\rho (z,
\,t)) + {\displaystyle \frac {{\frac {\partial }{\partial t}}\,
\rho (z, \,t)}{c}} = - 2\,{\displaystyle \frac {\pi \,N\,d\,(
{\frac {\partial }{\partial t}}\,\mathrm{a}(t))}{c}}  +
{\displaystyle \frac {2\,\pi \,N\,d\,\mathrm{b}(t)\,\omega }{c}}}
\]

\begin{gather*}
\mathit{disp\_conditions} := \\
{\displaystyle \int }
{\displaystyle \frac {\rho (z, \,t)\,k\,c - 2\,\pi \,N\,d\,(
{\frac {\partial }{\partial t}}\,\mathrm{b}(t)) - \rho (z, \,t)\,
\omega  - 2\,\pi \,N\,d\,\mathrm{a}(t)\,\omega }{c}} \,dt=0
\end{gather*}

\emptyline
\begin{center}
\fbox{ \parbox{.8\linewidth}
{The obtained result is the system of the shortened wave equation and
the dispersion condition. The right-hand side of the field equation
(material part) will be different for the different applications (see
below)}}
\end{center}

\emptyline

\section{Active mode locking: harmonic oscillator}

\emptyline

\subsection{Amplitude modulation}

\emptyline
\noindent
We start our consideration with  a relatively simple technique named
as active mode locking due to amplitude modulation. The modulator,
which is governed by external signal and changes the intracavity loss
periodically, plays the role of the external "force" (see part 2). Let
consider the situation, when the modulation period is equal and the
pulse duration is much less than the cavity period. If the modulation
curve is close to cosine, then the master equation for the ultrashort
pulse evolution can be written as \cite{Kuizenga}:

\begin{center}
$\frac {\partial }{\partial z}\,\rho (z, \,t)$ = \textit{g}
$\rho (z, \,t)$ +
${t_{f}}^{2}$
${\frac {\partial ^{2}}{\partial t^{2}}}\,\rho (z, \,t)$
$ - M\,t^{2}$
$\rho (z, \,t)$,
\end{center}

\noindent
where \textit{g} is the net-gain in the laser accounting for the
saturated gain
$\alpha $ and linear loss \textit{l} (including output loss),
${t_{f}}$  is the inverse bandwidth of the spectral filter, \textit{M} is the
characteristic of the modulation strength taking into account
curvature of the modulation curve at the point of maximal loss.

Let try to solve this equation in the case of the steady-state
propagation of ultrashort pulse (when
${\frac {\partial }{\partial z}}\,\rho (z, \,t)$=0) and in the absence of detuning of modulation period from cavity
period. If the time is normalized to
${t_{f}}$ , then the obtained equation is a well-known equation of harmonic
oscillator:

\emptyline
$>$restart:\\
\indent \indent
 with(plots):\\
\indent \indent \indent
  with('linalg'):\\
\indent \indent
   ode[1] := diff(rho(t),`\$`(t,2)) + g*rho(t) - M*t$^{2}$*rho(t);
\emptyline

\[
{\mathit{ode}_{1}} := ({\frac {\partial ^{2}}{\partial t^{2}}}\,
\rho (t)) + g\,\rho (t) - M\,t^{2}\,\rho (t)
\]

\emptyline
$>$sol := dsolve(ode[1]=0,rho(t));

\emptyline

\begin{eqnarray*}
\mathit{sol} := \rho (t)={\displaystyle \frac {\mathit{\_C1}\,
\mathrm{WhittakerW}({\displaystyle \frac {1}{4}} \,
{\displaystyle \frac {g}{\sqrt{M}}} , \,{\displaystyle \frac {1}{
4}} , \,\sqrt{M}\,t^{2})}{\sqrt{t}}}  + \\
{\displaystyle \frac {
\mathit{\_C2}\,\mathrm{WhittakerM}({\displaystyle \frac {1}{4}}
\,{\displaystyle \frac {g}{\sqrt{M}}} , \,{\displaystyle \frac {1
}{4}} , \,\sqrt{M}\,t^{2})}{\sqrt{t}}}
\end{eqnarray*}

\emptyline

The next step is suggested by the asymptotic behavior of the
prospective solution:
$\lim _{t\rightarrow \infty }\,\rho (t)$ = 0. But previously it is convenient to transit from Whittaker
functions to \textit{hypergeometric} and \textit{Kummer} functions:

\begin{center}
WhittakerM(
$\mu $,
$\nu $,z)=
$e^{(\frac { - x}{2})}$
$x^{(\frac {1}{2} + \nu )}$ hypergeom(
$\frac {1}{2} + \nu  - \mu $, 1+2
$\nu $, \textit{x}),
\end{center}

\begin{center}
 WhittakerW(
$\mu $,
$\nu $,z)=
$e^{(\frac { - x}{2})}$
$x^{(\frac {1}{2} + \nu )}$ KummerU(
$\frac {1}{2} + \nu  - \mu $, 1+2
$\nu $, \textit{x}).
\end{center}

\emptyline
As result, we have (
$\mu $=
$\frac {g}{4\,\sqrt{M}}$,
$\nu $=
$\frac {1}{4}$, \textit{x}=
$\sqrt{M}\,t^{2}$):

\emptyline
\begin{center}
$\rho (t)=$
${C_{1}} e^{( - \frac {\sqrt{M}\,t^{2}}{2})}$
\textit{t} $\sqrt{\sqrt{M}}$
hypergeom(
$\frac {3}{4} - \frac {g}{4\,\sqrt{M}}$,
$\frac {3}{2}$,
$\sqrt{M}\,t^{2}$) +\\
${C_{2}} e^{( - \frac {\sqrt{M}\,t^{2}}{2})} \sqrt{\sqrt{M}}$ KummerU(
$\frac {3}{4} - \frac {g}{4\,\sqrt{M}}$,
$\frac {3}{2}$,
$\sqrt{M}\,t^{2}$)=\\
${C_{1}} e^{( - \frac {\sqrt{M}\,t^{2}}{2})} \sqrt{\sqrt{M}}$ hypergeom((
$\frac {1}{4} - \frac {g}{4\,\sqrt{M}}$)+
$\frac {1}{2}$,
$\frac {3}{2}$,
$\sqrt{M}\,t^{2}$) +\\
${C_{2}}e^{( - \frac {\sqrt{M}\,t^{2}}{2})}\sqrt{\sqrt{M}}$ hypergeom(
$\frac {1}{4} - \frac {g}{4\,\sqrt{M}}$,
$\frac {1}{2}$,
$\sqrt{M}\,t^{2}$).
\end{center}

\emptyline
\noindent
Now, the asymptotic condition
$\lim _{t\rightarrow \infty }\,\rho (t)$ = 0, which is similar to condition for quantum states in harmonic
oscillator, results in (see, for example,\cite{Flugge})

\begin{center}
$\rho _{1}$(t) =\textit{ C
${\mathit{HermiteH}_{2\,n}}$}(\textit{t
$\sqrt{\sqrt{M}}$})
$e^{( - \frac {t^{2}\,\sqrt{M}}{2})}$ \textit{ }for \textit{n= - }(
$\frac {1}{4} - \frac {g}{4\,\sqrt{M}}$) \&\& \textit{n }is integer,

\emptyline

$\rho _{2}$(t)= \textit{ C
${\mathit{HermiteH}_{2\,n + 1}}$}(\textit{t
$\sqrt{\sqrt{M}}$})
$e^{( - \frac {t^{2}\,\sqrt{M}}{2})}$ \textit{ }for \textit{n= - }(
$\frac {3}{4} - \frac {g}{4\,\sqrt{M}}$) \&\& \textit{n }is integer,
\end{center}
\emptyline

\noindent
where \textit{
${\mathit{HermiteH}_{2\,n}}$} and
${\mathit{HermiteH}_{2\,n + 1}}$ are \textit{Hermite} polynomials. Value of constant \textit{C} can be obtained
from the energy balance condition, which results from the equation of
gain saturation:

\begin{center}
$\alpha=\frac {{\alpha _{0}}}{1 + \chi \,\int _{ - \infty }^{\infty }
\rho (t)^{2}\,dt}$ .
\end{center}

\noindent
Here
$\chi $ is the inverse flux of the gain saturation energy,
${\alpha _{0}}$ is the gain for small signal defined by gain medium properties and
pump intensity (note, that \textit{g} =
$\alpha  - l$).
Now let investigate the parameters of the steady-state solution of the
master equation.
With that end in view, we have to search the generation field energy:\\

\emptyline
$>$C$^{2}$*int( HermiteH(0,t*surd(M, 4))$^{2}$ * exp( -t$^{2}$*sqrt(M) ),\\
\indent
t=-infinity..infinity ); \# energy of the first solution\\
\indent
("ground state"), where alpha=l+sqrt(M) (see above)

\emptyline
\[
{\displaystyle \frac {C^{2}\,\sqrt{\pi }}{M^{(1/4)}}}
\]

\emptyline
\noindent
The use of normalization of intensity
$\rho ^{2}$ to
$(\chi \,{t_{f}})^{(-1)}$ and energy balance condition (see above) results in

\emptyline
$>$eq := l+sqrt(M) - alpha[0]/(1+C$^{2}$*sqrt(Pi)/surd(M, 4)) = 0:\\
\indent \indent
 sol := solve(eq, C$^{2}$);\# pulse peak intensity

\emptyline
\[
\mathit{sol} := {\displaystyle \frac {\mathrm{surd}(M, \,4)\,( -
l - \sqrt{M} + {\alpha _{0}})}{\sqrt{\pi }\,(l + \sqrt{M})}}
\]

\emptyline
\noindent
Hence the ultrashort pulse obtained as result of active mode locking
can be represented as

\emptyline
$>$pulse := sol*HermiteH(0,t*surd(M, 4))$^{2}$ * exp( -t$^{2}$*sqrt(M) );\\
\indent
\# this is first solution of master equation,\\
\indent
\# which represents a Gaussian pulse profile\\
\indent \indent
 plot( subs(\{l=0.1,M=0.05,alpha[0]=1.2\},pulse),\\
\indent \indent
t=-6..6, axes=BOXED, title=`field intensity vs time` );

\[
\mathit{pulse} := {\displaystyle \frac {\mathrm{surd}(M, \,4)\,(
 - l - \sqrt{M} + {\alpha _{0}})\,e^{( - \sqrt{M}\,t^{2})}}{
\sqrt{\pi }\,(l + \sqrt{M})}}
\]

\emptyline
\begin{center}
\mapleplot{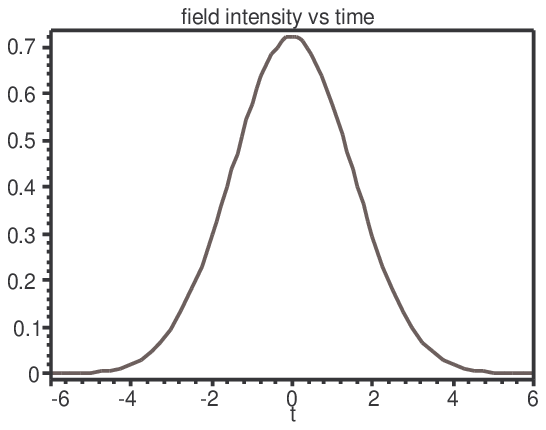}
\end{center}

\emptyline

Now we calculate the pulse duration measured on half-level of maximal
pulse intensity:

\emptyline
$>$eq := exp( -sqrt(M)*t$^{2}$ ) = 1/2:\\
\indent \indent
 sol := solve(eq, t):\\
\indent \indent \indent
  pulse\_width := simplify( sol[1] - sol[2] );

\emptyline
\[
\mathit{pulse\_width} := 2\,{\displaystyle \frac {\sqrt{M^{(3/2)}
\,\mathrm{ln}(2)}}{M}}
\]

\emptyline
$>$plot( pulse\_width,M=0.3..0.01, axes=BOXED, \\
\indent
title=`pulse width vs M`);

\emptyline
\begin{center}
\mapleplot{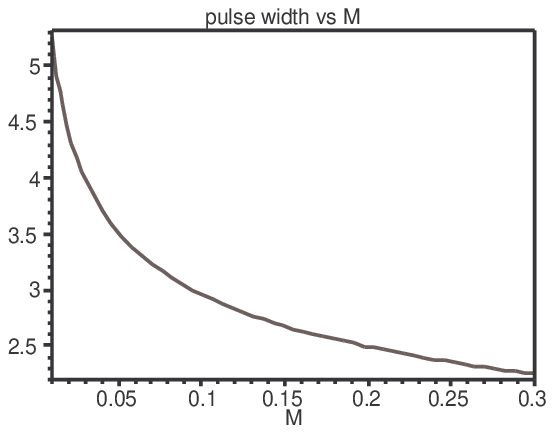}
\end{center}

\emptyline
\noindent
One can see, that the increase of modulation parameter decreases the
pulse width, but this decrease is slow (\symbol{126}1/
$\sqrt{\sqrt{M}}$).
\noindent
Next step is the taking into account the detuning of cavity and
modulation periods. The corresponding normalized parameter
$\delta $ can be introduced in the following form (see corresponding Doppler
transformation: \textit{t--\TEXTsymbol{>}t - z
$\delta $} and, as consequence,
${\frac {\partial }{\partial z}}\,\mathrm{}$\textit{--\TEXTsymbol{>}-
$\delta $
${\frac {\partial }{\partial t}}\,\mathrm{}$} for steady-state pulse, here
$\delta $ is, in fact, inverse relative velocity with dimension [time/cavity
transit], i. e. time delay on the cavity round-trip):

\emptyline
$>$ode[2] := diff(rho(t),`\$`(t,2)) + delta*diff(rho(t),t) + \\
\indent
g*rho(t) -M*t$^{2}$*rho(t);

\emptyline
\[ \boxed{
{\mathit{ode}_{2}} := ({\frac {\partial ^{2}}{\partial t^{2}}}\,
\rho (t)) + \delta \,({\frac {\partial }{\partial t}}\,\rho (t))
 + g\,\rho (t) - M\,t^{2}\,\rho (t)}
\]

\emptyline
$>$sol := dsolve(ode[2]=0,rho(t));

\emptyline
\begin{eqnarray*}
\mathit{sol} := \rho (t)={\displaystyle \frac {\mathit{\_C1}\,
\mathrm{WhittakerW}({\displaystyle \frac {1}{16}} \,
{\displaystyle \frac {4\,g - \delta ^{2}}{\sqrt{M}}} , \,
{\displaystyle \frac {1}{4}} , \,\sqrt{M}\,t^{2})\,e^{( - 1/2\,
\delta \,t)}}{\sqrt{t}}}\\
\mbox{} + {\displaystyle \frac {\mathit{\_C2}\,\mathrm{WhittakerM
}({\displaystyle \frac {1}{16}} \,{\displaystyle \frac {4\,g -
\delta ^{2}}{\sqrt{M}}} , \,{\displaystyle \frac {1}{4}} , \,
\sqrt{M}\,t^{2})\,e^{( - 1/2\,\delta \,t)}}{\sqrt{t}}}
\end{eqnarray*}

\emptyline
\noindent
The comparison with above obtained result leads to

\emptyline
\begin{center}
\[ \boxed{
\rho_{1}(t) =\mathit{ C \cdot
{\mathrm{HermiteH}_{2\,n}}}(\mathit{t
\sqrt{\sqrt{M}}})
e^{( - \frac {t\,(\delta  + \sqrt{M}\,t)}{2})}}
\]
for \[ \boxed{ \mathit{n= - }(
\frac {1}{4} + \frac {\delta ^{2} - 4\,g}{16\,\sqrt{M}})} \]
\&\& \textit{n } is integer,

\emptyline
\[ \boxed{
\rho_{2}(t) =\mathit{ C  \cdot
{\mathrm{HermiteH}_{2\,n + 1}}}(\mathit{t
\sqrt{\sqrt{M}}})
e^{( - \frac {t\,(\delta  + \sqrt{M}\,t)}{2})} }
\]
for \[ \boxed{ \mathit{n= - }(
\frac {3}{4} + \frac {\delta ^{2} - 4\,g}{16\,\sqrt{M}})} \]
\&\& \textit{n }is integer,
\end{center}
\emptyline
\noindent
and we can repeat our previous analysis

\emptyline
$>$C$^{2}$*int( HermiteH(0,t*surd(M, 4))$^{2}$ * \\
\indent
exp( -t*(delta+sqrt(M)*t) ),t=-infinity..infinity ); \\
\indent
\# energy of the first solution ("ground state"), \\
\indent
\# where alpha=l+sqrt(M)+delta$^{2/4}$

\emptyline
\[
{\displaystyle \frac {C^{2}\,e^{(1/4\,\frac {\delta ^{2}}{\sqrt{M
}})}\,\sqrt{\pi }}{M^{(1/4)}}}
\]

\emptyline
$>$eq := l+sqrt(M)+delta$^{2/4}$ - alpha[0]/(1+\%) = 0:\\
\indent
\# energy balance condition\\
\indent \indent
 sol\_0 := solve(eq, C$^{2}$);\# pulse peak intensity

\emptyline
\[
\mathit{sol\_0} :=  - {\displaystyle \frac {M^{(1/4)}\,(4\,l + 4
\,\sqrt{M} + \delta ^{2} - 4\,{\alpha _{0}})}{e^{(1/4\,\frac {
\delta ^{2}}{\sqrt{M}})}\,\sqrt{\pi }\,(4\,l + 4\,\sqrt{M} +
\delta ^{2})}}
\]

\emptyline
\noindent
Note, that there is the maximal
$ \left|  \! \,\delta \, \!  \right| $ permitting the "ground state" ultrashort pulse generation:

\emptyline
$>$-4*l-4*sqrt(M)-delta$^{2}$+4*alpha[0] > 0;\\
\indent \indent
 solve(-4*l-4*sqrt(M)-delta$^{2}$+4*alpha[0]=0, delta);

\emptyline
\[
0 <  - 4\,l - 4\,\sqrt{M} - \delta ^{2} + 4\,{\alpha _{0}}
\]
\[
2\,\sqrt{ - l - \sqrt{M} + {\alpha _{0}}}, \, - 2\,\sqrt{ - l -
\sqrt{M} + {\alpha _{0}}}
\]

\emptyline
\noindent
We see, that the upper permitted level of detuning parameter is
increased due to pump growth (rise of
${\alpha _{0}}$) and is decreased by \textit{M} growth.

It is of interest, that the growth of
$ \left|  \! \,\delta \, \!  \right| $ does not leads to generation of the "excite state" pulses because of
the corresponding limitation for these solutions is more strict:

\emptyline
$>$En := C$^{2}$*int( expand( HermiteH(1,t*surd(M, 4) ))$^{2}$ * \\
\indent
exp(-t*(delta+sqrt(M)*t) ), t=-infinity..infinity ):\\
\indent
\# energy of the second solution ("excite state")\\
\indent \indent
 eq1 := 0 = - (3/4+(delta$^{2}$-4*(alpha-l))/(16*sqrt(M))):\\
\indent \indent \indent
  solve(\%,alpha):\\
\indent \indent
   eq := \% - alpha[0]/(1+En) = 0:\# energy balance condition\\
\indent
    solve(eq, C$^{2}$);\# "excite state" pulse peak intensity\\
\indent \indent
     -12*sqrt(M)-4*l-delta$^{2}$+4*alpha[0] $>$ 0;\# condition of generation

\emptyline
\[
 - {\displaystyle \frac {M^{(5/4)}\,(12\,\sqrt{M} + 4\,l + \delta
 ^{2} - 4\,{\alpha _{0}})}{e^{(1/4\,\frac {\delta ^{2}}{\sqrt{M}}
)}\,\mathrm{surd}(M, \,4)^{2}\,\sqrt{\pi }\,(14\,\sqrt{M}\,\delta
 ^{2} + 24\,M + \delta ^{4} + 4\,l\,\delta ^{2} + 8\,l\,\sqrt{M})
}}
\]

\[
0 <  - 12\,\sqrt{M} - 4\,l - \delta ^{2} + 4\,{\alpha _{0}}
\]

\emptyline
\noindent
The dependence of pulse width on
$\delta $ has following form:

\emptyline

$>$eq := exp( -t*(delta+sqrt(M)*t) ) = 1/2:\\
\indent \indent
 sol := solve(eq, t):\\
\indent \indent \indent
  pulse\_width := simplify( sol[1] - sol[2] );\\
\indent \indent
   plot( subs(M=0.05,pulse\_width),delta=-5..5, axes=BOXED,\\
\indent
title=`pulse width vs delta` );

\emptyline
\[ \boxed{
\mathit{pulse\_width} := {\displaystyle \frac {\sqrt{M\,(\delta
^{2} + 4\,\sqrt{M}\,\mathrm{ln}(2))}}{M}}}
\]

\emptyline
\begin{center}
\mapleplot{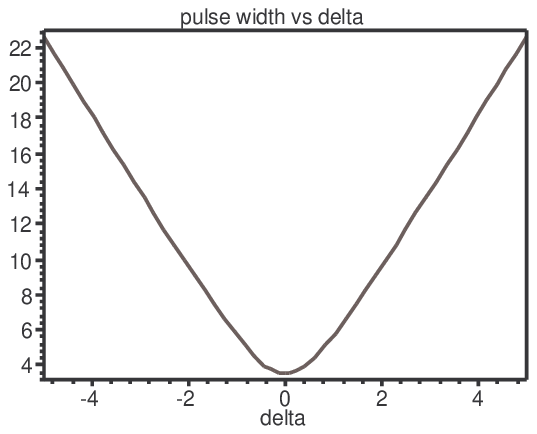}
\end{center}

\emptyline
\noindent
\begin{center}
\fbox{ \parbox{.8\linewidth}{
We can see almost linear and symmetric rise of pulse width due to
detuning increase. This is an important characteristic of active
mode-locked lasers. But in practice (see below), the dependence of the
pulse parameters on detuning has more complicated character }}
\end{center}
\emptyline
\noindent
The dependence of
the pulse maximum location on detuning parameter can
be obtained from the solution for pulse envelope's maximum:
${\frac {\partial }{\partial t}}\,e^{( - \frac {t\,(\delta  +
\sqrt{M}\,t)}{2})}$ = 0. Hence, the pulse maximum location is:

\emptyline
$>$diff(exp(-t*(delta+sqrt(M)*t)/2), t) = 0:\\
\indent \indent
 solve(\%, t);
\emptyline
\[
 - {\displaystyle \frac {1}{2}} \,{\displaystyle \frac {\delta }{
\sqrt{M}}}
\]

\emptyline
$>$plot( subs(M=0.05,\%),delta=-5..5, axes=BOXED, \\
\indent
title=`pulse shift vs delta` );

\emptyline
\begin{center}
\mapleplot{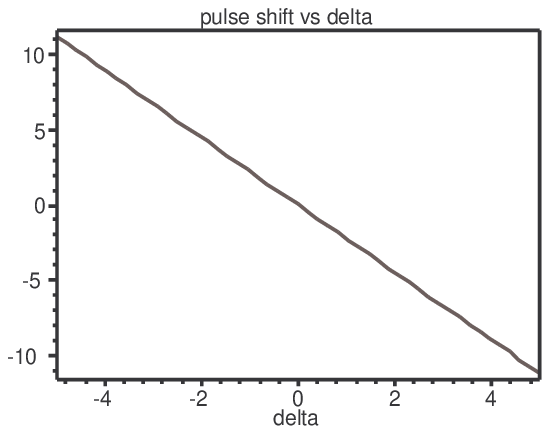}
\end{center}

\emptyline
\noindent
The increase (decrease) of the pulse round-trip frequency (
$\delta $\TEXTsymbol{<}0 and
$\delta $\TEXTsymbol{>}0, respectively) increases positive (negative) time
shift of the pulse maximum relatively modulation curve extremum.

\emptyline

\subsection{Phase modulation}

\emptyline

The external modulation can change not only field amplitude, but its
phase. In fact, this regime (phase active modulation) causes the
Doppler frequency shift of all field components with the exception of
those, which are located in the vicinity of extremum of modulation
curve. Hence, there exists the steady-state generation only for field
located in vicinity of points, where the phase is stationary.
Steady-state regime is described by equation

\emptyline
$>$ode[3] := diff(rho(t),`\$`(t,2)) + g*rho(t) + I*phi*rho(t) +\\
\indent
I*M*t$^{2}$*rho(t);
\emptyline
\[ \boxed{
{\mathit{ode}_{3}} := ({\frac {\partial ^{2}}{\partial t^{2}}}\,
\rho (t)) + g\,\rho (t) + I\,\phi \,\rho (t) + I\,M\,t^{2}\,\rho
(t)}
\]

\emptyline
\noindent
where
$\phi $ is the phase delay on the cavity round-trip.
This equation looks like previous one, but has complex character. This
suggests to search its partial solution in the form
$\rho $(\textit{t}) =
$C\,e^{((a + I\,b)\,t^{2})}$ :

\emptyline
$>$expand( subs(rho(t)=C*exp((a+I*b)*t$^{2}$), ode[3]) ):\\
\indent \indent
 expand(\%/C/exp(t$^{2}$*a)/exp(I*t$^{2}$*b)):\\
\indent \indent \indent
  eq1 := collect (coeff(\%,I), t$^{2}$ ):\\
\indent \indent \indent \indent
   eq2 := collect( coeff(\%\%,I,0), t$^{2}$ ):\\
\indent \indent \indent
    eq3 := coeff(eq1, t$^{2}$);\\
\indent \indent
     eq4 := coeff(eq1, t,0);\\
\indent
      eq5 := coeff(eq2, t$^{2}$);\\
\indent \indent
       eq6 := coeff(eq2, t,0);\\
\indent \indent \indent
        sol1 := solve(eq6=0, a);\# solution for a\\
\indent \indent \indent \indent
       sol2 := solve(subs(a=sol1,eq3=0), b);\# solution for b\\
\indent \indent \indent
      sol3 := solve(subs(b=sol2,eq4)=0, phi);\# solution for phi\\
\indent \indent \indent
\# but NB that there is eq5 defining, in fact, g:\\
\indent \indent
     solve( subs(\{a=sol1, b=sol2\},eq5)=0, g);\# solution for g

\emptyline
\[
\mathit{eq3} := 8\,a\,b + M
\]
\[
\mathit{eq4} := 2\,b + \phi
\]
\[
\mathit{eq5} := 4\,a^{2} - 4\,b^{2}
\]

\[
\mathit{eq6} := 2\,a + g
\]
\[
\mathit{sol1} :=  - {\displaystyle \frac {1}{2}} \,g
\]
\[
\mathit{sol2} := {\displaystyle \frac {1}{4}} \,{\displaystyle
\frac {M}{g}}
\]
\[
\mathit{sol3} :=  - {\displaystyle \frac {1}{2}} \,
{\displaystyle \frac {M}{g}}
\]
\[
{\displaystyle \frac {1}{2}} \,\sqrt{ - 2\,M}, \, -
{\displaystyle \frac {1}{2}} \,\sqrt{ - 2\,M}, \,{\displaystyle
\frac {1}{2}} \,\sqrt{2}\,\sqrt{M}, \, - {\displaystyle \frac {1
}{2}} \,\sqrt{2}\,\sqrt{M}
\]

\emptyline
\noindent
\begin{center}
\fbox{ \parbox{.8\linewidth}{
So, we have
$\rho $(\textit{t}) =
$C\,e^{ \left(  \!  \left(  \!  - \frac {\sqrt{\frac {M}{2}}}{2}
 + \frac {I\,\sqrt{\frac {M}{2}}}{2} \!  \right) \,t^{2} \!
 \right) }$   and
$\phi $ =
$ - \sqrt{\frac {M}{2}}$ , \textit{g=
$\sqrt{\frac {M}{2}}$.} The time-dependent parabolic phase of ultrashort pulse
is called chirp and is caused by phase modulation}}
\end{center}
\emptyline
\noindent
Pulse width for this pulse
is:

\emptyline
$>$exp(-sqrt(M/2)*t$^{2}$) = 1/2:\\
\indent \indent
 sol := solve(\%, t):\\
\indent \indent \indent
  pulse\_width := simplify( sol[1] - sol[2] );

\emptyline
\[ \boxed{
\mathit{pulse\_width} := 2\,{\displaystyle \frac {\sqrt{M^{(3/2)}
\,\sqrt{2}\,\mathrm{ln}(2)}}{M}} }
\]

\emptyline
\noindent
that is 2
$\sqrt{\mathrm{ln}(2)\,\sqrt{\frac {2}{M}}}$ . As one can see, that result is equal to one for amplitude
modulation. The main difference is the appearance of the chirp.

Let take into account the modulation detuning. In this case we may to
suppose the modification of the steady-state solution:

\emptyline
$>$ode[4] := diff(rho(t),`\$`(t,2)) + g*rho(t) + delta*diff(rho(t),t) +\\
\indent
I*phi*rho(t) + I*M*t$^{2}$*rho(t);\\
\indent \indent
    expand( subs(rho(t)=C*exp((a+I*b)*t$^{2}$ + (c+I*d)*t), ode[4]) ):\\
\indent \indent
\# c is the shift of the pulse profile from extremum \\
\indent \indent
\# of the modulation curve, d is the frequency shift of\\
\indent \indent
\# the pulse from the center of the gain band\\
\indent
    expand(\%/C/exp(t$^{2}$*a)/exp(I*t$^{2}$*b)/exp(t*c)/exp(I*t*d)):\\
\indent \indent
eq1 := collect (coeff(\%,I), t ):\\
\indent \indent \indent
 eq2 := collect( coeff(\%\%,I,0), t ):\\
\indent \indent \indent \indent
  eq3 := coeff(eq1, t$^{2}$):\\
\indent \indent \indent \indent \indent
   eq4 := coeff(eq1, t):\\
\indent \indent \indent \indent
    eq5 := coeff(eq1, t, 0):\\
\indent \indent \indent
     eq6 := coeff(eq2, t$^{2}$);\\
\indent \indent
      eq7 := coeff(eq2, t):\\
\indent
       eq8 := coeff(eq2, t, 0):\\
\indent \indent
allvalues(\\
\indent \indent
solve(subs(b=-a,\{eq3=0,eq4=0,eq5=0,eq7=0,eq8=0\}),\{a,c,d,g,phi\})
);

\emptyline

\[ \boxed{
{\mathit{ode}_{4}} := ({\frac {\partial ^{2}}{\partial t^{2}}}\,
\rho (t)) + g\,\rho (t) + \delta \,({\frac {\partial }{\partial t
}}\,\rho (t)) + I\,\phi \,\rho (t) + I\,M\,t^{2}\,\rho (t)}
\]
\[
\mathit{eq6} := 4\,a^{2} - 4\,b^{2}
\]
\begin{eqnarray}
\{c= - {\displaystyle \frac {1}{2}} \,\delta , \,\phi =
{\displaystyle \frac {1}{2}} \,\sqrt{2}\,\sqrt{M}, \,a=
{\displaystyle \frac {1}{4}} \,\sqrt{2}\,\sqrt{M}, \,g= -
{\displaystyle \frac {1}{2}} \,\sqrt{2}\,\sqrt{M} +
{\displaystyle \frac {1}{4}} \,\delta ^{2}, \,d=0\}, \nonumber \\ \nonumber
\{c= - {\displaystyle \frac {1}{2}} \,\delta , \,g=
{\displaystyle \frac {1}{2}} \,\sqrt{2}\,\sqrt{M} +
{\displaystyle \frac {1}{4}} \,\delta ^{2}, \,\phi = -
{\displaystyle \frac {1}{2}} \,\sqrt{2}\,\sqrt{M}, \,a= -
{\displaystyle \frac {1}{4}} \,\sqrt{2}\,\sqrt{M}, \,d=0\}
\end{eqnarray}

\emptyline
\noindent
We see, that there is not frequency shift of the pulse (\textit{d=}0),
but, as it was for amplitude modulation, the time delay appears (
$c= - \frac {\delta }{2}$) that changes the pulse duration as result of modulation detuning.
The rise of the detuning prevents from the pulse generation due to
saturated net-gain coefficient increase (
$g=\sqrt{\frac {M}{2}} + \frac {\delta ^{2}}{4}$).

\begin{center}
\fbox{ \parbox{.8\linewidth}{The pulse parameters behavior coincides with one for amplitude modulation}}
\end{center}

\emptyline

\subsection{Ultrashort pulse stability}

\emptyline
\noindent
Now we shall investigate the ultrashort pulse stability against low
perturbation
$\zeta $(\textit{t}). The substitution of the perturbed steady-state solution
in dynamical equation with subsequent linearization on
$\zeta $(\textit{t}) results in

\begin{center}
\[ \boxed{
{\frac {\partial }{\partial z}}\,\zeta (z, \,t) =
\alpha
\zeta (z, \,t)\mathit{ +
{\alpha _{p}}
\rho (t)  - l
\zeta (z, \,t) +
{\frac {\partial ^{2}}{\partial t^{2}}}\,\zeta (z, \,t)} -
M\,t^{2}
\zeta (z, \,t),}
\]
\end{center}
\emptyline
\noindent
where
${\alpha _{p}}$ is the perturbed saturated gain, which is obtained from the
assumption about small contribution of perturbation to gain saturation
process:

\emptyline

$>$alpha[p] = op(2,convert( series( alpha[0]/(1 + A + B), B=0,2),\\
\indent
polynom));
\emptyline
\[
{\alpha _{p}}= - {\displaystyle \frac {{\alpha _{0}}\,B}{(1 + A)
^{2}}}
\]

\emptyline
\noindent
Here \textit{A=
$\int _{ - \infty }^{\infty }a^{2}\,dt$}, \textit{B=
$\int _{ - \infty }^{\infty }a\,\zeta \,dt$} and we neglected the high-order terms relatively perturbation
amplitude. Note, that this is negative quantity.

Let the dependence of perturbation on \textit{z} has exponential form
with increment
$\lambda $. Then

\emptyline
$>$zeta(z,t) := upsilon(t)*exp(lambda*z):\# perturbation\\
\indent \indent
 ode[4] := expand( diff(zeta(z,t),z)/exp(lambda*z) ) =\\
\indent \indent
 expand( (alpha*zeta(z,t) + alpha[p]*rho(t) - l*zeta(z,t) +\\
\indent \indent
diff(zeta(z,t),`\$`(t,2)) - M*t$^{2}$*zeta(z,t))/exp(lambda*z) );

\emptyline
\[
{\mathit{ode}_{4}} := \upsilon (t)\,\lambda =\alpha \,\upsilon (t
) + {\displaystyle \frac {{\alpha _{p}}\,\rho (t)}{e^{(\lambda \,
z)}}}  - l\,\upsilon (t) + ({\frac {\partial ^{2}}{\partial t^{2}
}}\,\upsilon (t)) - M\,t^{2}\,\upsilon (t)
\]

\emptyline

Now we introduce
${\alpha _{p}}$* where \textit{B*=
$\int _{ - \infty }^{\infty }a\,\upsilon \,dt$}, that allows to eliminate the exponent from right-hand side of
${\mathit{ode}_{4}}$ (we shall eliminate the asterix below).
This is equation for eigenvalues
$\lambda $ and eigenfunctions
$\upsilon (t)$ of the perturbed laser operator. Stable generation of the ultrashort
pulse corresponds to decaying of perturbations, i.e.
$\lambda $\TEXTsymbol{<}0. For Gaussian pulse:

\emptyline
$>$ode[4] := upsilon(t)*lambda =\\
\indent
subs(rho(t)=sqrt(surd(M,4)*(alpha[0]-l-sqrt(M))/(sqrt(Pi)*(l+sqrt(M))))\\
\indent
*exp(-t$^{2}$*sqrt(M)/2),\\
\indent
alpha*upsilon(t)+alpha[p]*rho(t)-l*upsilon(t)+diff(upsilon(t),`\$`(t,2))\\
\indent
-M*t$^{2}$upsilon(t));\\
\indent \indent
sol := dsolve( ode[4],upsilon(t) );

\emptyline
\begin{eqnarray*}
\mathit{ode}_{4} := \upsilon (t)\,\lambda =\alpha \,\upsilon (t
) + {\alpha _{p}}\,\sqrt{{\displaystyle \frac {\mathrm{surd}(M,
\,4)\,( - l - \sqrt{M} + {\alpha _{0}})}{\sqrt{\pi }\,(l + \sqrt{
M})}} }\,e^{( - 1/2\,\sqrt{M}\,t^{2})}  \\
- l\,\upsilon (t) + (
{\frac {\partial ^{2}}{\partial t^{2}}}\,\upsilon (t)) - M\,t^{2}
\,\upsilon (t)
\end{eqnarray*}

\begin{eqnarray*}
\mathit{sol} := \upsilon (t)= - {\displaystyle \frac {{\alpha _{p
}}\,\sqrt{ - {\displaystyle \frac {\mathrm{surd}(M, \,4)\,(l +
\sqrt{M} - {\alpha _{0}})}{l + \sqrt{M}}} }\,e^{( - 1/2\,\sqrt{M}
\,t^{2})}}{( - \lambda  + \alpha  - l - \sqrt{M})\,\pi ^{(1/4)}}
}  \\
\mbox{} + {\displaystyle \frac {\mathit{\_C1}\,\mathrm{WhittakerM
}({\displaystyle \frac {1}{4}} \,{\displaystyle \frac { - \lambda
  + \alpha  - l}{\sqrt{M}}} , \,{\displaystyle \frac {1}{4}} , \,
\sqrt{M}\,t^{2})}{\sqrt{t}}}  \\
+ {\displaystyle \frac {\mathit{
\_C2}\,\mathrm{WhittakerW}({\displaystyle \frac {1}{4}} \,
{\displaystyle \frac { - \lambda  + \alpha  - l}{\sqrt{M}}} , \,
{\displaystyle \frac {1}{4}} , \,\sqrt{M}\,t^{2})}{\sqrt{t}}}
\end{eqnarray*}

\emptyline
\noindent
As result (see above), we have

\begin{center}
$\upsilon $(t) =\textit{
$ - \frac {{\alpha _{p}}\,\sqrt{\frac {\sqrt{\sqrt{M}}\,({\alpha
_{0}} - l - \sqrt{M})}{l + \sqrt{M}}}\,e^{( - \frac {\sqrt{M}\,t
^{2}}{2})}}{(\alpha  - l - \sqrt{M} - \lambda )\,\pi ^{(\frac {1
}{4})}}$ + C
${\mathit{HermiteH}_{2\,n}}$}(\textit{t
$\sqrt{\sqrt{M}}$})
$e^{( - \frac {t^{2}\,\sqrt{M}}{2})}$ \textit{ }for\\
\textit{n= - }(
$\frac {1}{4} - \frac {\alpha  - \lambda  - l}{4\,\sqrt{M}}$) \&\& \textit{n }is integer,

\emptyline

$\upsilon $(t) =\textit{
$ - \frac {{\alpha _{p}}\,\sqrt{\frac {\sqrt{\sqrt{M}}\,({\alpha
_{0}} - l - \sqrt{M})}{l + \sqrt{M}}}\,e^{( - \frac {\sqrt{M}\,t
^{2}}{2})}}{(\alpha  - l - \sqrt{M} - \lambda )\,\pi ^{(\frac {1
}{4})}}$ + C
${\mathit{HermiteH}_{2\,n + 1}}$}(\textit{t
$\sqrt{\sqrt{M}}$})
$e^{( - \frac {t^{2}\,\sqrt{M}}{2})}$ \textit{ }for\\
\textit{n= - }(
$\frac {3}{4} - \frac {\alpha  - \lambda  - l}{4\,\sqrt{M}}$) \&\& \textit{n }is integer,

\end{center}

\emptyline
$>$eq1 := n = - (1/4-(alpha-lambda-l)/(4*sqrt(M))):\\
\indent \indent
 eq2 := n = - (3/4-(alpha-lambda-l)/(4*sqrt(M))):\\
\indent \indent \indent
  En := subs(\\
\indent
C$^{2}$=-surd(M,4)*(l+sqrt(M)-alpha[0])/(sqrt(Pi)*(l+sqrt(M))),\\
\indent
 C$^{2}$*sqrt(Pi)/(M$^{(1/4)}$) ):\# ultrashort pulse energy (see above)\\
\indent \indent
   lambda[1,n] = solve(eq1, lambda);\# increment\\
\indent \indent \indent
    lambda[2,n] = solve(eq2, lambda);\# increment

\emptyline
\[
{\lambda _{1, \,n}}= - 4\,n\,\sqrt{M} - \sqrt{M} + \alpha  - l
\]
\[
{\lambda _{2, \,n}}= - 4\,n\,\sqrt{M} - 3\,\sqrt{M} + \alpha  - l
\]

\emptyline
\noindent
The glance on the solution is evidence of absence of solution
corresponding to
${\lambda _{1, \,0}}$. For others modes we take into account, that
$\alpha $=\textit{l}+
$\sqrt{M}$. As result, \fbox{all perturbation modes are unstable} because of

\begin{center}
\[ \boxed{
\lambda _{1,n}  =  - 4n\sqrt{M}  < 0 \  (\mathit{n=}1, 2, ...),}
\]
\[ \boxed{
\lambda _{2,n}  =  - 4n\sqrt{M} -2\sqrt{M}  < 0 \ (\mathit{n=}0, 1, ...)}
\]

\end{center}

\emptyline
\noindent
It has to note, that for the amplified pulse

\begin{center}
${\lambda _{1, \,n}}= - 4\,n\,\sqrt{M} - \sqrt{M} + \alpha  - l$,
${\lambda _{2, \,n}}= - 4\,n\,\sqrt{M} - 3\,\sqrt{M} + \alpha  -
l$,
\end{center}

\noindent
\begin{center}
\fbox{ \parbox{.8\linewidth}{
Here we see the decrease of increment as result of \textit{n} rise.
Therefore, only "ground state"  with
${\lambda _{1, \,0}}$ will be amplified predominantly. This fact provides for Gaussian
pulse generation}}
\end{center}
\emptyline
\noindent
(see, \cite{K\"artner}).

\emptyline
In the presence of detuning we have:

\emptyline

$>$eq1 := n= - (1/4+(delta$^{2}$-4*(alpha-lambda-l))/(16*sqrt(M))):\\
\indent \indent
 eq2 := n= - (3/4+(delta$^{2}$-4*(alpha-lambda-l))/(16*sqrt(M))):\\
\indent \indent \indent
  lambda[1,n] = simplify( solve(eq1, lambda) );\\
\indent \indent
   lambda[2,n] = simplify( solve(eq2, lambda) );
\emptyline
\[
{\lambda _{1, \,n}}= - 4\,n\,\sqrt{M} - \sqrt{M} -
{\displaystyle \frac {1}{4}} \,\delta ^{2} + \alpha  - l
\]
\[
{\lambda _{2, \,n}}= - 4\,n\,\sqrt{M} - 3\,\sqrt{M} -
{\displaystyle \frac {1}{4}} \,\delta ^{2} + \alpha  - l
\]

\emptyline
\begin{center}
\fbox{ \parbox{.8\linewidth}{
Surprisingly, but, as it was above, our linear analysis predicts the
pulse stability regardless of detuning
$\delta $ (because of
$\alpha $=\textit{l}+
$\sqrt{M} + \frac {\delta ^{2}}{4}$)}}
\end{center}
\emptyline
\noindent
But for the pulse the detuning growth decreases the increment that
does not favor the single pulse generation.

\emptyline

\subsection{Nonlinear processes: self-phase modulation and dynamical gain saturation}

\emptyline
Among above considered effects only gain saturation by full pulse
energy can be considered as nonlinear process, which, however, does
not affect on the pulse envelope, but governs its energy. The
time-dependent nonlinear effects, which can transform pulse profile,
are self-phase modulation (SPM) and dynamical gain saturation. First
one is the dependence of the field phase on its intensity and can play
essential role in solid-state lasers \cite{Kalashnikov1}. Second effect is
caused by the change of the gain along pulse profile and is essential
in lasers with large gain cross-sections and comparatively narrow gain
band \cite{Kalashnikov2}.

\emptyline
At first, let analyze the presence of SPM, which can be considered as
perturbation of our master equation describing active phase
modulation:

\emptyline
$>$ode[5] := diff(rho(t),`\$`(t,2)) + g*rho(t) + I*phi*rho(t) +\\
\indent
I*M*t$^{2}$*rho(t) - I*beta*C$^{2}$*exp(2*a*t$^{2}$)*rho(t);\# phase perturbation
in the form -I*beta*C$^{2}$*abs(rho(t))$^{2}$*rho(t), rho(t) has unperturbed
profile

\emptyline
\[ \boxed{
{\mathit{ode}_{5}} := ({\frac {\partial ^{2}}{\partial t^{2}}}\,
\rho (t)) + g\,\rho (t) + I\,\phi \,\rho (t) + I\,M\,t^{2}\,\rho
(t) - I\,\beta \,C^{2}\,e^{(2\,t^{2}\,a)}\,\rho (t)}
\]

\emptyline
$>$expand( subs(rho(t)=C*exp((a+I*b)*t$^{2}$),\\
\indent
ode[5])/exp(a*t$^{2}$)/exp(I*t$^{2}$*b)/C ):\\
\indent \indent
 series(\%,t=0,3):\# approximate solution as result of expansion\\
\indent
  convert(\%, polynom):\# now we shall collect the coefficients of t
and t$^{2}$ for real and imaginary parts\\
\indent
   eq1 := coeff(\%,I):\\
\indent \indent
    eq2 := coeff(\%\%,I,0):\\
\indent \indent \indent
     eq3 := coeff(eq1,t$^{2}$);\\
\indent \indent \indent \indent
      eq4 := coeff(eq1,t,0);\\
\indent \indent \indent \indent \indent
       eq5 := coeff(eq2,t$^{2}$);\\
\indent \indent \indent \indent
        eq6 := coeff(eq2,t,0);\\
\indent \indent \indent
       sol1 := solve(eq6=0, a);\# solution for a\\
\indent \indent
      sol2 := solve(subs(a=sol1,eq3=0), b);\# solution for b\\
\indent
     sol3 := solve(subs(b=sol2,eq4)=0, phi);\# solution for phi\\
\indent
\# but NB that there is eq5 defining, in fact, g:\\
\indent
    solve( subs(\{a=sol1, b=sol2\},eq5)=0, g);\# solution for g

\emptyline
\[
\mathit{eq3} := 8\,a\,b + M - 2\,a\,C^{2}\,\beta
\]
\[
\mathit{eq4} := \phi  + 2\,b - C^{2}\,\beta
\]
\[
\mathit{eq5} := 4\,a^{2} - 4\,b^{2}
\]
\[
\mathit{eq6} := 2\,a + g
\]
\[
\mathit{sol1} :=  - {\displaystyle \frac {1}{2}} \,g
\]
\[
\mathit{sol2} := {\displaystyle \frac {1}{4}} \,{\displaystyle
\frac {M + g\,C^{2}\,\beta }{g}}
\]
\[
\mathit{sol3} := {\displaystyle \frac {1}{2}} \,{\displaystyle
\frac { - M + g\,C^{2}\,\beta }{g}}
\]
\begin{eqnarray*}
\displaystyle \frac {1}{4} \,C^{2}\,\beta  + {\displaystyle
\frac {1}{4}} \,\sqrt{C^{4}\,\beta ^{2} + 8\,M}, \,
{\displaystyle \frac {1}{4}} \,C^{2}\,\beta  - {\displaystyle
\frac {1}{4}} \,\sqrt{C^{4}\,\beta ^{2} + 8\,M},  \\
 - {\displaystyle \frac {1}{4}} \,C^{2}\,\beta  + {\displaystyle
\frac {1}{4}} \,\sqrt{C^{4}\,\beta ^{2} - 8\,M},
 - {\displaystyle \frac {1}{4}} \,C^{2}\,\beta  - {\displaystyle
\frac {1}{4}} \,\sqrt{C^{4}\,\beta ^{2} - 8\,M}
\end{eqnarray*}

\emptyline
\noindent
Hence we have solution with perturbed parameters:
\begin{center}
\[ \boxed{
\rho (\mathit{t}) =
C\,e^{(( - \frac {\beta \,C^{2} + \sqrt{\beta ^{2}\,C^{4} + 8\,M
}}{4} + \frac {I\,(\beta \,C^{2} + \sqrt{\beta ^{2}\,C^{4} + 8\,M
})}{4})\,t^{2})}}\]
\end{center}
and \textit{g=
$\frac {1\,\beta \,C^{2}}{4} + \frac {1\,\sqrt{\beta ^{2}\,C^{4}
 + 8\,M}}{4}$. }
As one can see,
\begin{center}
\fbox{ \parbox{.8\linewidth}{
this solution has enlarged chirp and reduced pulse
duration due to SPM}}
\end{center}

\emptyline
\noindent
Now we shall investigate the influence of dynamical gain saturation on
the pulse characteristics in the case of active mode locking due to
amplitude modulation. Let the contribution of the dynamical gain
saturation can be considered as perturbation for Gaussian pulse (see
above). The instant energy flux of such pulse is:

\emptyline
$>$int(sol\_0*exp(-t*(delta+sqrt(M)*t)/2),t):\\
\indent \indent
 Energy := simplify( coeff( \%,\\
\indent \indent
erf(1/2*sqrt(2)*M$^{(1/4)}$*t+1/4*delta*sqrt(2)/(M$^{(1/4)}$)))*\\
\indent \indent
(1+erf(1/2*sqrt(2)*M$^{(1/4)}$*t+1/4*delta*sqrt(2)/(M$^{(1/4)}$)) ) );\\
\emptyline
\begin{gather*}
\mathit{Energy} := \\
- {\displaystyle \frac {1}{2}} \,
{\displaystyle \frac {\sqrt{2}\,(4\,l + 4\,\sqrt{M} + \delta ^{2}
 - 4\,{\alpha _{0}})\,e^{( - 1/8\,\frac {\delta ^{2}}{\sqrt{M}})}
\,(1 + \mathrm{erf}({\displaystyle \frac {1}{4}} \,
{\displaystyle \frac {\sqrt{2}\,(\delta  + 2\,\sqrt{M}\,t)}{M^{(1
/4)}}} ))}{4\,l + 4\,\sqrt{M} + \delta ^{2}}}
\end{gather*}

\emptyline
\noindent
The approximation of the small contribution of the gain saturation
allows the expansion of energy into series on \textit{t} up to second
order:

\emptyline
$>$series(Energy,t=0,2):\\
\indent \indent
 convert(\%, polynom):\\
\indent \indent \indent
  simplify(\%);
\begin{eqnarray}
 - \displaystyle \frac {1}{2 (4\,l + 4
\,\sqrt{M} + \delta ^{2})\,\sqrt{\pi }} \,(4\,l + 4
\,\sqrt{M} + \delta ^{2} - 4\,{\alpha _{0}})\,e^{( - 1/8\,\frac {
\delta ^{2}}{\sqrt{M}})}\, \nonumber \\ \nonumber
(\sqrt{2}\,\sqrt{\pi } + \sqrt{2}\,
\sqrt{\pi }\,\mathrm{erf}({\displaystyle \frac {1}{4}} \,
{\displaystyle \frac {\delta \,\sqrt{2}}{M^{(1/4)}}} ) + 2\,e^{(
 - 1/8\,\frac {\delta ^{2}}{\sqrt{M}})}\,M^{(1/4)}\,t)
\end{eqnarray}

\emptyline
\noindent
Hence we have the modified gain coefficient
$\alpha $(
$1 - {I_{0}}$\textit{t}), where
$\alpha $ is the gain coefficient at pulse peak,
${I_{0}}$ is the pulse intensity for unperturbed solution. Note, that the
additional term in brackets is resulted from the shift of pulse
maximum.
Then the master equation for perturbed solution (in our
approximation!) is

\emptyline
$>$ode[6] := diff(rho(t),`\$`(t,2)) + g*rho(t) + delta*diff(rho(t),t) -\\
\indent
(M*t$^{2}$+epsilon*t)*rho(t);\# here epsilon is alpha*I[0]

\emptyline
\[ \boxed{
{\mathit{ode}_{6}} := ({\frac {\partial ^{2}}{\partial t^{2}}}\,
\rho (t)) + g\,\rho (t) + \delta \,({\frac {\partial }{\partial t
}}\,\rho (t)) - (M\,t^{2} + \varepsilon \,t)\,\rho (t)}
\]

\emptyline
$>$dsolve(ode[6]=0, rho(t));

\emptyline
\begin{gather*}
\rho (t)=  \\
{\displaystyle \frac {\mathit{\_C1}\,\mathrm{WhittakerM}
({\displaystyle \frac {1}{16}} \,{\displaystyle \frac {4\,M\,g -
M\,\delta ^{2} + \varepsilon ^{2}}{M^{(3/2)}}} , \,
{\displaystyle \frac {1}{4}} , \,{\displaystyle \frac {1}{4}} \,
{\displaystyle \frac {(2\,M\,t + \varepsilon )^{2}}{M^{(3/2)}}} )
\,e^{( - 1/2\,\delta \,t)}}{\sqrt{{\displaystyle \frac {2\,M\,t
 + \varepsilon }{M^{(3/4)}}} }}}  \\
\mbox{} + {\displaystyle \frac {\mathit{\_C2}\,\mathrm{WhittakerW
}({\displaystyle \frac {1}{16}} \,{\displaystyle \frac {4\,M\,g
 - M\,\delta ^{2} + \varepsilon ^{2}}{M^{(3/2)}}} , \,
{\displaystyle \frac {1}{4}} , \,{\displaystyle \frac {1}{4}} \,
{\displaystyle \frac {(2\,M\,t + \varepsilon )^{2}}{M^{(3/2)}}} )
\,e^{( - 1/2\,\delta \,t)}}{\sqrt{{\displaystyle \frac {2\,M\,t
 + \varepsilon }{M^{(3/4)}}} }}}
\end{gather*}

\emptyline
\noindent
The comparison with above obtained result gives

\begin{center}
$\rho _{1}$(t) =\textit{ C
${\mathit{HermiteH}_{2\,n}}$}(\textit{t
$\sqrt{\sqrt{M}} + \frac {\varepsilon }{2\,\sqrt{M\,\sqrt{M}}}$})
$e^{( - \frac {(2\,M\,t + \varepsilon )^{2}}{8\,M\,\sqrt{M}} -
\frac {\delta \,t}{2})}$ \textit{ }for \textit{n= - }(
$\frac {1}{4} - \frac {4\,g\,M + \varepsilon ^{2} - M\,\delta ^{2
}}{16\,M\,\sqrt{M}}$) \&\& \textit{n }is integer,

\emptyline
$\rho _{2}$(t) =\textit{ C
${\mathit{HermiteH}_{2\,n + 1}}$}(\textit{t
$\sqrt{\sqrt{M}} + \frac {\varepsilon }{2\,\sqrt{M\,\sqrt{M}}}$})
$e^{( - \frac {(2\,M\,t + \varepsilon )^{2}}{8\,M\,\sqrt{M}} -
\frac {\delta \,t}{2})}$ \textit{ }for \textit{n= - }(
$\frac {3}{4} - \frac {4\,g\,M + \varepsilon ^{2} - M\,\delta ^{2
}}{16\,M\,\sqrt{M}}$) \&\& \textit{n }is integer.
\end{center}

\emptyline
\noindent
In future we shall omit
$\varepsilon ^{2}$-terms.
If we take the unperturbed intensity for calculation of perturbation
action, the perturbed pulse energy is

\emptyline
$>$C$^{2}$*HermiteH(0, t*sqrt(sqrt(M))+epsilon/(2*sqrt(M*sqrt(M))) )$^{2}$*\\
\indent
exp(-(4*M$^{2}$*t$^{2}$+4*M*t*epsilon)/(4*M*sqrt(M))-delta*t):\# field\\
\indent
\# intensity profile for "ground state"\\
\indent \indent
(1/4-(4*g*M-M*delta$^{2}$)/(16*M*sqrt(M))) = 0\\
\indent \indent \indent
  subs(epsilon=alpha*I[0],\%):\\
\indent \indent
    int(\%,t=-infinity..infinity):\\
\indent
       En := simplify(\%);\# energy of the first "ground state"

\emptyline
\[
\mathit{En} := {\displaystyle \frac {e^{ \left(  \! 1/4\,\frac {(
\alpha \,{I_{0}} + \delta \,\sqrt{M})^{2}}{M^{(3/2)}} \!
 \right) }\,C^{2}\,\sqrt{\pi }}{M^{(1/4)}}}
\]

\emptyline
$>$1/4-(4*(alpha-l)*M - M*delta$^{2}$)/(16*M*sqrt(M)):\\
\indent \indent
sol\_a := solve(\%=0,alpha):\\
\indent
  eq := sol\_a - alpha[0]/(1+subs( alpha=\%,En )) = 0:\# energy balance condition\\
\indent \indent
   sol\_I := solve(eq, C$^{2}$);\# pulse peak intensity

\emptyline
\begin{gather*}
\mathit{sol\_I} :=  \\
- {\displaystyle \frac {M^{(5/4)}\,(4\,l + 4
\,\sqrt{M} + \delta ^{2} - 4\,{\alpha _{0}})}{e^{ \left(  \! 1/64
\,\frac {(4\,{I_{0}}\,M^{(3/2)} + 4\,{I_{0}}\,M\,l + {I_{0}}\,M\,
\delta ^{2} + 4\,\delta \,M^{(3/2)})^{2}}{M^{(7/2)}} \!  \right)
}\,\sqrt{\pi }\,(4\,M^{(3/2)} + 4\,M\,l + M\,\delta ^{2})}}
\end{gather*}

\emptyline Now we plot the dependencies of the pulse intensity,
width and maximum location versus detuning parameter $\delta $.

\emptyline
$>$plot(subs( \{alpha[0]=1.2,l=0.1,M=0.05\},subs(I[0]=sol\_0,sol\_I) ),\\
\indent
delta=-2..1.5, axes=BOXED, title=`pulse intensity vs detuning`);

\begin{center}
\mapleplot{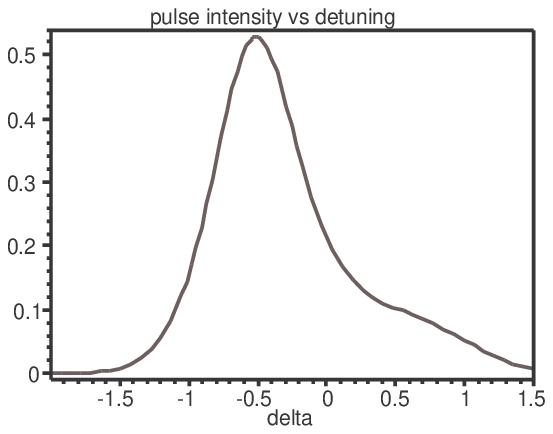}
\end{center}

\emptyline
$>$eq := exp( -(4*M$^{2}$*t$^{2}$+4*M*t*epsilon)/(4*M*sqrt(M))-delta*t ) = 1/2:\\
\indent
\# we take into account pulse profile without epsilon$^{2}$\\
\indent \indent
 sol := solve(eq, t):\\
\indent \indent \indent
  pulse\_width := simplify( subs( epsilon$^{2}$=0,sol[1] - sol[2] ) ):\\
\indent \indent
   subs(epsilon=alpha*I[0],pulse\_width):\\
\indent
    subs(\{alpha=sol\_a,I[0]=sol\_0\},\%): \# pulse width for perturbed solution

\emptyline
$>$plot(subs( \{alpha[0]=1.2,l=0.1,M=0.05\},\% ), delta=-2..1.5,\\
\indent
axes=BOXED, title=`pulse width vs detuning`);

\emptyline
\begin{center}
\mapleplot{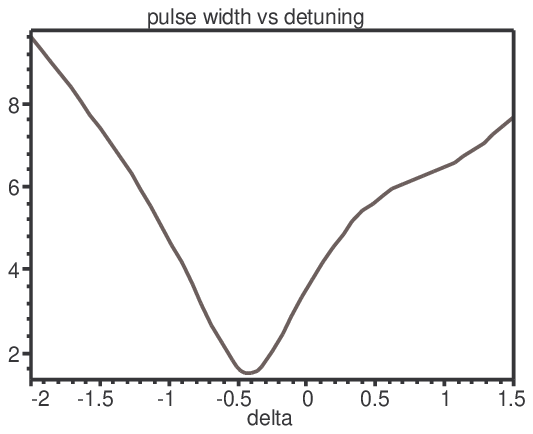}
\end{center}

\emptyline
$>$\# pulse maximum location\\
\indent
diff(exp(-(2*M*t+epsilon)$^{2}$/(8*M*sqrt(M))-delta*t/2), t) = 0:\\
\indent \indent
 solve(\%, t);

\emptyline
\[
 - {\displaystyle \frac {1}{2}} \,{\displaystyle \frac {
\varepsilon  + \delta \,\sqrt{M}}{M}}
\]

\emptyline
$>${subs(epsilon=alpha*I[0],\%):\\
\indent \indent
 subs(\{alpha=sol\_a,I[0]=sol\_0\},\%):\\
\indent
  plot(subs( \{alpha[0]=1.2,l=0.1,M=0.05\},\% ), delta=-2..1.5,\\
\indent
axes=BOXED, title=`pulse location vs detuning`);

\emptyline
\begin{center}
\mapleplot{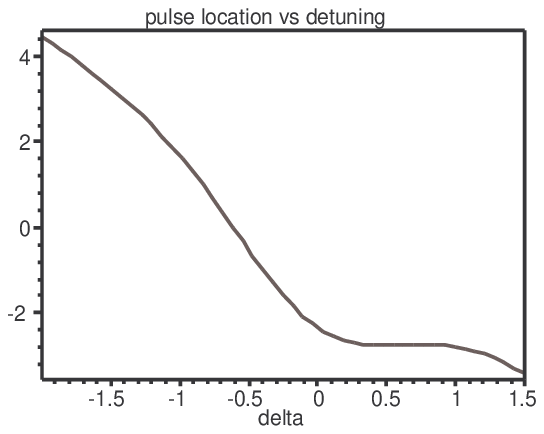}
\end{center}

\emptyline
\begin{center}
\fbox{ \parbox{.8\linewidth}{
We see, that the main peculiarity here is the asymmetric dependence of
the pulse parameters on
$\delta $. The pulse width minimum and intensity maximum don't coincide with
$\delta $=0 and the detuning characteristics have sharper behavior in negative
domain of detuning}}
\end{center}
\emptyline
\noindent
Now, as it was made in previous subsection, we estimate the condition
of the ultrashort pulse stability.
Here we take into consideration
$\varepsilon ^{2}$-term, but this does not fail our analysis because of this term
contribute only to pulse energy without shift pulse inside modulation
window.
For the sake of the simplification, we shall consider the contribution
of the destabilizing field to dynamical gain saturation, but only in
the form of the unperturbed peak intensity variation
$\zeta $ and perturbation of the saturated gain coefficient (see above). Then
the stability condition:

\emptyline
$>$eq1 := epsilon = subs(I[0]=sol\_0,sol\_a*(I[0]+zeta)):\\
\indent
\# zeta is the peak intensity variation\\
\indent \indent
-(3/4-(4*(sol\_a-lambda-l)*M+epsilon$^{2}$-M*delta$^{2}$)/(16*M*sqrt(M))):\\
\indent
 expand( solve(\%=0, lambda) ) $<$ 0;\# perturbation increment from eigenvalue problem\\
\indent \indent
  subs(epsilon=rhs(eq1),lhs(\%)):\\
\indent \indent \indent
   animate3d(subs( \{alpha[p]=0,alpha[0]=1.2,l=0.1\},\% ),\\
\indent \indent \indent
delta=-2..1.5,  M=0.005..0.05, zeta=-0.15..0.15, axes=BOXED,\\
\indent \indent \indent
title=`maximal increment of perturbation`

\emptyline
\[ \boxed{
 - 2\,\sqrt{M} + {\displaystyle \frac {{\displaystyle \frac {1}{4
}} \,\varepsilon ^{2}}{M}}  < 0}
\]
\emptyline
\begin{center}
\mapleplot{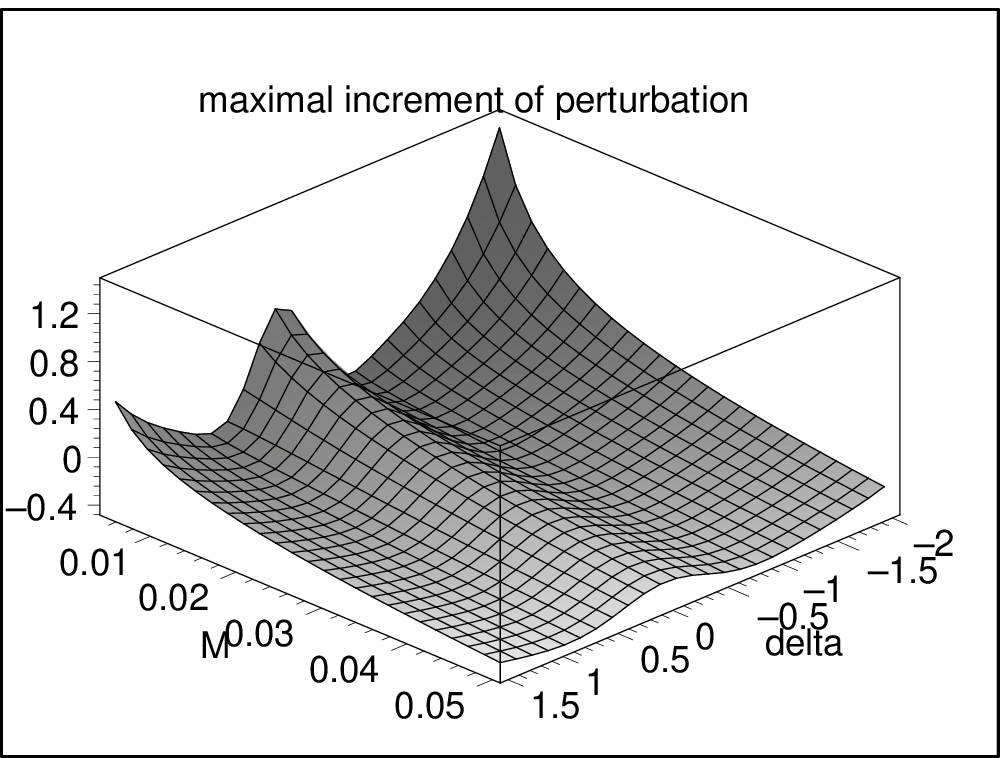}
\end{center}

\emptyline
\begin{center}
\fbox{ \parbox{.8\linewidth}{
We can see, that the perturbation growth can destabilize the pulse as
result of
$ \left|  \! \,\delta \, \!  \right| $ increase (compare with subsection \textit{"Amplitude modulation"})}}
\end{center}
\emptyline
\noindent
Also, there is the possibility of ultrashort pulse destabilization
near
$\delta $=0. So, the presence of dynamical gain saturation gives the behavior
of the ultrashort pulse parameters and stability condition, which is
close to the experimentally observed and numerically obtained (see,
for example, \cite{Dudley}.\\
Now we try to investigate the influence of dynamical gain saturation
in detail by using so-called \fbox{aberrationless approximation}
Let the pulse profile keeps its form with accuracy up to
\textit{n}-order of time-series expansion, but \textit{n} pulse
parameters are modified as result of pulse propagation. Then the
substitution of the expression for pulse profile in master equation
with subsequent expansion in \textit{t}-series produces the system of
\textit{n} ODE describing the evolution of pulse parameters.

\emptyline
$>$f1 := \\
\indent
(z,t)$->$rho0(z)*exp(-a(z)$^{2}$*t$^{2}$+b(z)*t);\# approximate pulse
profile\\
\indent \indent
     f2 := (z,tau)$->$rho0(z)*exp(-a(z)$^{2}$*tau$^{2}$+b(z)*tau):\\
\indent \indent \indent
     master := diff(rho(z,t),z) = (alpha - l)*rho(z,t) -\\
\indent \indent \indent
chi*alpha*rho(z,t)*int(rho(z,tau)$^{2}$,tau=-infinity..t) +\\
\indent \indent \indent
diff(rho(z,t),t\$2) - M*t$^{2}$*rho(z,t) + delta*diff(rho(z,t),t);\#
approximate master equation with dynamical gain saturation (parameter
chi), alpha is the gain coefficient before pulse front\\
\indent \indent
expand(lhs(subs(\{rho(z,t)=f1(z,t),rho(z,tau)=f2(z,tau)\\
\indent \indent
\},master))*exp(a(z)$^{2}$*t$^{2}$)/exp(t*b(z)) -\\
\indent \indent
rhs(subs(\{rho(z,t)=f1(z,t),rho(z,tau)=f2(z,tau)\\
\indent \indent
\},master))*exp(a(z)$^{2}$*t$^{2}$)/exp(t*b(z))):\# substitution\\
\indent
of the approximate solution\\
\indent
convert( series(\%,t=1/2*b(z)/(a(z)$^{2}$),3),polynom ):\# expansion around\\
\indent \indent
 peak at t=1/2*b(z)/(a(z)$^{2}$)\\
\indent \indent \indent
  eq1 := collect(\%,t):\\
\indent \indent \indent \indent
   eq2 :=\\
\indent \indent \indent
subs(\{diff(rho0(z),z)=u,diff(a(z),z)=v,diff(b(z),z)=w
\},coeff(eq1,t$^{2}$)):\\
\indent
    eq3 :=\\
\indent
subs(\{diff(rho0(z),z)=u,diff(a(z),z)=v,diff(b(z),z)=w
\},coeff(eq1,t)):\\
\indent \indent
     eq4 :=\\
subs(\{diff(rho0(z),z)=u,diff(a(z),z)=v,diff(b(z),z)=w
\},coeff(eq1,t,0)):\\
\indent \indent \indent
      sol := simplify (solve(\{eq2=0,eq3=0,eq4=0\},\{u,v,w\}) );

\emptyline
\[
\mathit{f1} := (z, \,t)\rightarrow \rho 0(z)\,e^{( - \mathrm{a}(z
)^{2}\,t^{2} + \mathrm{b}(z)\,t)}
\]

\begin{gather*}
\boxed{\mathit{master} := {\frac {\partial }{\partial z}}\,\rho (z, \,t)
= } \\
\boxed{(\alpha  - l)\,\rho (z, \,t) - \chi \,\alpha \,\rho (z, \,t)\,
{\displaystyle \int _{ - \infty }^{t}} \rho (z, \,\tau )^{2}\,d
\tau  +} \\
\boxed{({\frac {\partial ^{2}}{\partial t^{2}}}\,\rho (z, \,t))
 - M\,t^{2}\,\rho (z, \,t) + \delta \,({\frac {\partial }{
\partial t}}\,\rho (z, \,t))}
\end{gather*}

\begin{gather*}
\mathit{sol} := \{u=
- {\displaystyle \frac {1}{4}} \rho 0(z)(
\chi \,\alpha \,\rho 0(z)^{2}\,\sqrt{\pi }\,e^{(1/2\,\frac {
\mathrm{b}(z)^{2}}{\mathrm{a}(z)^{2}})}\,\sqrt{2}\,  \\
(
{\displaystyle \lim _{\tau \rightarrow ( - \infty )}} \, -
\mathrm{erf}({\displaystyle \frac {1}{2}} \,{\displaystyle
\frac {\sqrt{2}\,(2\,\tau \,\mathrm{a}(z)^{2} - \mathrm{b}(z))}{
\mathrm{a}(z)}} ))\,\mathrm{a}(z) \\
\mbox{} - 4\,\alpha \,\mathrm{a}(z)^{2} + 8\,\mathrm{a}(z)^{4} +
4\,l\,\mathrm{a}(z)^{2} - 2\,\rho 0(z)^{2}\,\mathrm{b}(z)\,\chi
\,\alpha \,e^{(1/2\,\frac {\mathrm{b}(z)^{2}}{\mathrm{a}(z)^{2}})
}  \\
- 4\,\mathrm{b}(z)\,\delta \,\mathrm{a}(z)^{2} - 4\,\mathrm{b}(
z)^{2}\,\mathrm{a}(z)^{2}) \left/ {\vrule height0.44em width0em depth0.44em} \right. \!
 \! \mathrm{a}(z)^{2},  \\
 \,w= - \chi \,\alpha \,\rho 0(z)^{2}\,e^{(
1/2\,\frac {\mathrm{b}(z)^{2}}{\mathrm{a}(z)^{2}})} - 2\,\delta
\,\mathrm{a}(z)^{2} - 4\,\mathrm{b}(z)\,\mathrm{a}(z)^{2},  \\
 \,v=
 - {\displaystyle \frac {1}{2}} \,{\displaystyle \frac { - M + 4
\,\mathrm{a}(z)^{4}}{\mathrm{a}(z)}} \}
\end{gather*}
\emptyline
Now try to find the steady-state points of ODE-system, which
correspond to stationary pulse parameters. For this aim let introduce
substitution
$\chi =\chi \,e^{(\frac {1\,\mathrm{b}(z)^{2}}{2\,\mathrm{a}(z)^{
2}})}$:

\emptyline
$>$sol\_main :=\\
\indent
 \{w = -chi*alpha*rho0(z)$^{2}$-2*delta*a(z)$^{2}$-4*b(z)*a(z)$^{2}$,\\
\indent
v = 1/2*(M-4*a(z)$^{4}$)/a(z), u =\\
\indent
1/4*rho0(z)*(-chi*alpha*rho0(z)$^{2}$*sqrt(Pi)*sqrt(2)*a(z)+4*alpha*a(z)$^{2}$\\
\indent
-8*a(z)$^{4}$-4*l*a(z)$^{2}$+2*rho0(z)$^{2}$*b(z)*chi*alpha+\\
\indent
4*b(z)*delta*a(z)$^{2}$+4*b(z)$^{2}$*a(z)$^{2}$)/(a(z)$^{2}$)\};

\emptyline
\begin{gather*}
\mathit{sol\_main} :=  \\
 \{u={\displaystyle \frac {1}{4}} \rho 0(z)(
 - \chi \,\alpha \,\rho 0(z)^{2}\,\sqrt{\pi }\,\sqrt{2}\,\mathrm{
a}(z) + 4\,\alpha \,\mathrm{a}(z)^{2} - 8\,\mathrm{a}(z)^{4} - 4
\,l\,\mathrm{a}(z)^{2} \\
\mbox{} + 2\,\rho 0(z)^{2}\,\mathrm{b}(z)\,\chi \,\alpha  + 4\,
\mathrm{b}(z)\,\delta \,\mathrm{a}(z)^{2} + 4\,\mathrm{b}(z)^{2}
\,\mathrm{a}(z)^{2}) \left/ {\vrule
height0.44em width0em depth0.44em} \right. \!  \! \mathrm{a}(z)^{
2},  \\
w= - \chi \,\alpha \,\rho 0(z)^{2} - 2\,\delta \,\mathrm{a}(z)^{2
} - 4\,\mathrm{b}(z)\,\mathrm{a}(z)^{2}, \,v={\displaystyle
\frac {1}{2}} \,{\displaystyle \frac {M - 4\,\mathrm{a}(z)^{4}}{
\mathrm{a}(z)}} \}
\end{gather*}

\emptyline
$>$eq1 := subs( \{rho0(z)$^{2}$=x,a(z)=sqrt(y),a(z)$^{2}$=y,a(z)$^{4}$=y$^{2}$\},\\
\indent
expand(numer( subs(sol\_main,u) )/rho0(z) )=0 );\# here we eliminated the
trivial solution\\
\indent \indent
  eq2 := subs( a(z)$^{4}$=y$^{2}$,numer( subs(sol\_main,v) )=0 );\\
\indent \indent \indent
    eq3 :=  subs( \{rho0(z)$^{2}$=x,a(z)$^{2}$=y\},subs(sol\_main,w)=0 );

\emptyline

\begin{gather*}
\mathit{eq1} := \\
  - \chi \,\alpha \,x\,\sqrt{\pi }\,\sqrt{2}\,
\sqrt{y} + 4\,\alpha \,y - 8\,y^{2} - 4\,l\,y + \\
2\,x\,\mathrm{b}(
z)\,\chi \,\alpha  + 4\,\mathrm{b}(z)\,\delta \,y + 4\,\mathrm{b}
(z)^{2}\,y=0
\end{gather*}
\[
\mathit{eq2} := M - 4\,y^{2}=0
\]
\[
\mathit{eq3} :=  - \chi \,\alpha \,x - 2\,\delta \,y - 4\,
\mathrm{b}(z)\,y=0
\]

\emptyline
$>$allvalues( solve(\{eq2,eq3\},\{y,b(z)\}) );

\emptyline
\begin{eqnarray*}
\{\mathrm{b}(z)= - {\displaystyle \frac {1}{2}} \,{\displaystyle
\frac {\chi \,\alpha \,x + \delta \,\sqrt{M}}{\sqrt{M}}} , \,y=
{\displaystyle \frac {1}{2}} \,\sqrt{M}\}, \\
\,\{\mathrm{b}(z)=
{\displaystyle \frac {1}{2}} \,{\displaystyle \frac {\chi \,
\alpha \,x - \delta \,\sqrt{M}}{\sqrt{M}}} , \,y= -
{\displaystyle \frac {1}{2}} \,\sqrt{M}\}
\end{eqnarray*}

\emptyline
\noindent
Hence
\fbox{$a^{2}=\frac {\sqrt{M}}{2}$},
$b= - \frac {\delta }{2} - \frac {\chi \,x\,\alpha }{2\,\sqrt{M}}
$ (\textit{x} is the pulse intensity). So, the shift is
\fbox{$\frac {b}{2\,a^{2}}= - \frac {\delta }{\sqrt{M}} - \frac {\chi
\,x\,\alpha }{2\,M}$}  that differs from the usual result (see above) as result of
dynamical gain saturation (last term), which shifts the pulse maximum
in negative side. This additional shift has obvious explanation. The
gain at the pulse front is greater than one at pulse tail due to gain
saturation. This shifts the pulse forward as hole.
Pulse width is:

\emptyline
$>$eq := exp( t*(-delta/2-chi*x*alpha/(2*sqrt(M))) - sqrt(M)*t$^{2}$/2 ) =
1/2:\\
\indent \indent
 sol := solve(eq, t):\\
\indent \indent \indent
  pulse\_width := simplify( sol[1] - sol[2] );

\emptyline
\[ \boxed{
\mathit{pulse\_width} := {\displaystyle \frac {\sqrt{M\,\delta ^{
2} + 2\,\delta \,\sqrt{M}\,\chi \,\alpha \,x + \chi ^{2}\,\alpha
^{2}\,x^{2} + 8\,M^{(3/2)}\,\mathrm{ln}(2)}}{M}} }
\]

\emptyline
\begin{center}
\fbox{ \parbox{.8\linewidth}{
We can see, that there is the minimum of the pulse duration in
negative domain of
$\delta $ that corresponds to result, which was obtained on the basis of
perturbation theory}}
\end{center}
\emptyline
\noindent
The pulse intensity
$\rho 0^{2}$:

\emptyline
$>$Intensity := simplify(\\
\indent
solve(\\
\indent
subs(\{y = 1/2*sqrt(M),b(z) =\\
\indent
-1/2*(delta*M+chi*sqrt(M)*x*alpha)/M\},eq1),
x)[1]);\# pulse intensity

\emptyline
\begin{gather*}
\mathit{Intensity} :=  \\
- {\displaystyle \frac {\sqrt{\pi }\,M^{(3
/4)} + \delta \,\sqrt{M} - \sqrt{M\,(\pi \,\sqrt{M} + 2\,\sqrt{
\pi }\,M^{(1/4)}\,\delta  + 4\,\alpha  - 4\,\sqrt{M} - 4\,l)}}{
\chi \,\alpha }}
\end{gather*}

\emptyline \noindent So, we have the following dependencies for
the pulse duration

\emptyline
$>$subs(x=Intensity,pulse\_width):\\
\indent \indent
animate(subs( \{alpha=1.2,l=0.1\},\% ), delta=-4..4, M=0.01..0.1,\\
\indent \indent
numpoints=200, axes=BOXED, title=`pulse width vs detuning`

\begin{center}
\mapleplot{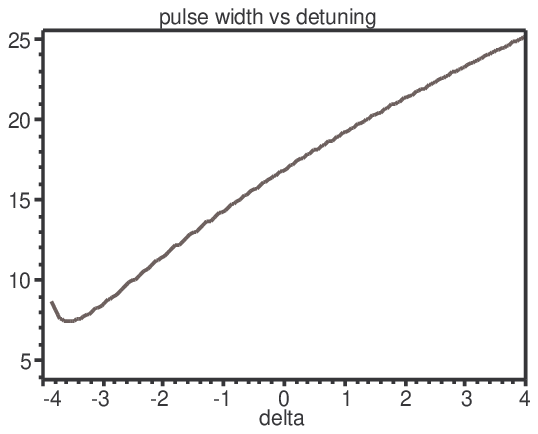}
\end{center}

\emptyline
\noindent
and pulse intensity

\emptyline
$>$animate(subs( \{alpha=1.2,l=0.1,chi=1\},Intensity ), delta=-4..4,\\
\indent
M=0.01..0.1, numpoints=200, axes=BOXED, view=0..0.45, \\
\indent
title=`pulse intensity vs detuning`

\begin{center}
\mapleplot{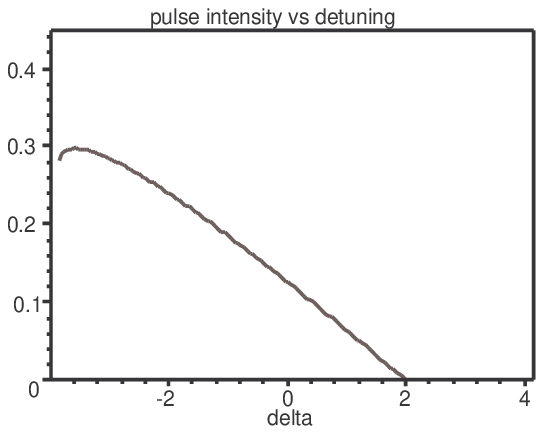}
\end{center}

\emptyline
The last step is the stability analysis. The stability of our
solutions can be estimated from the eigenvalues of Jacobian of
\textit{sol}.

\emptyline
$>$eq1 := subs(\{rho0(z)=x,a(z)=y,b(z)=z\},subs(sol\_main,u)):\\
\indent \indent
 eq2 := subs(\{rho0(z)=x,a(z)=y,b(z)=z\},subs(sol\_main,v)):\\
\indent \indent \indent
  eq3 := subs(\{rho0(z)=x,a(z)=y,b(z)=z\},subs(sol\_main,w)):\\
\indent \indent \indent \indent
m[1,1] := diff( eq1,x ):\\
\indent \indent \indent \indent \indent
 m[1,2] := diff( eq1,y ):\\
\indent \indent \indent \indent
   m[1,3] := diff( eq1,z ):\\
\indent \indent \indent
m[2,1] := diff( eq2,x ):\\
\indent \indent
 m[2,2] := diff( eq2,y ):\\
\indent
   m[2,3] := diff( eq2,z ):\\
\indent \indent
m[3,1] := diff( eq3,x ):\\
\indent \indent \indent
 m[3,2] := diff( eq3,y ):\\
\indent \indent \indent
   m[3,3] := diff( eq3,z ):\\
\indent \indent \indent \indent
A :=\\
\indent
array([[m[1,1],m[1,2],m[1,3]],[m[2,1],m[2,2],m[2,3]],[m[3,1],m[3,2],m[
3,3]]]):\# Jacobian

\emptyline
\noindent
Now we find the eigenvalues
$\lambda $ of Jacobian directly by calculation of determinant.

\emptyline
$>$evalm(A-[[lambda,0,0],[0,lambda,0],[0,0,lambda]]):\\
\indent \indent
numer( simplify(det(\%)) ):\\
\indent \indent \indent
sol := solve(\%=0,lambda):

$>$x := sqrt(Intensity):\\
\indent \indent
 y := sqrt( sqrt(M)/2 ):\\
\indent \indent \indent
  z := -delta/2-x$^{2}$*alpha/(2*sqrt(M)):\\
\indent \indent \indent
   s1 := evalf( subs(\{chi=1,l=0.1,alpha=1.2\},sol[1]) ):\\
\indent \indent
    s2 := evalf( subs(\{chi=1,l=0.1,alpha=1.2\},sol[2]) ):\\
\indent
     s3 := evalf( subs(\{chi=1,l=0.1,alpha=1.2\},sol[3]) ):\\
\indent \indent
plot3d(\{Re(s1),Re(s2),Re(s3)\}, delta=-4..2, M=0.01..0.1,\\
\indent \indent
axes=boxed,title=`Re(lambda) for initial perturbation`)

\begin{center}
\mapleplot{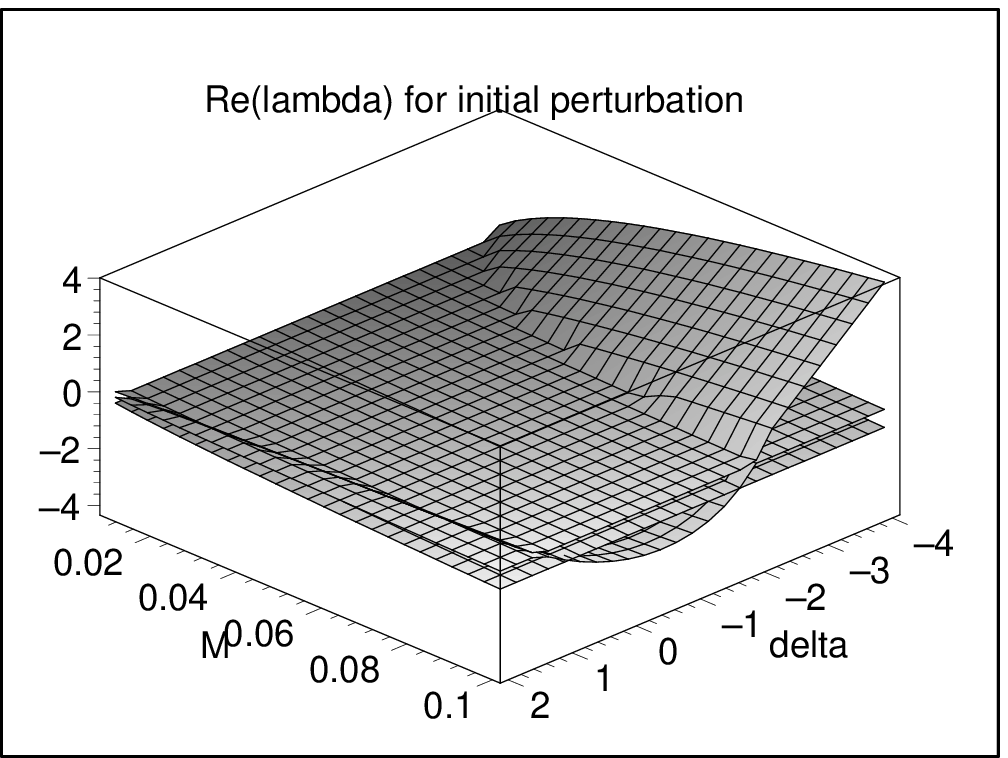}
\end{center}

\emptyline
\begin{center}
\fbox{ \parbox{.8\linewidth}{
One can see that the pulse with Gaussian-like form is stable in the
region of its existence (see two previous figures)}}
\end{center}
\emptyline
\noindent
We have to note
that the considered here perturbations belong to limited class
therefore this criterion is necessary but not sufficient condition of
pulse stability (compare with previous consideration on the basis of
perturbation theory, where we analyzed not only pulse peak variation
but also gain coefficient change).

\emptyline
\section{Nonlinear Schr\"odinger equation: construction of
the soliton solution by means of the direct Hirota's method}

\emptyline
\noindent
The Schr\"odinger equation is the well-known nonlinear equation
describing the weak nonlinear waves, in the particular, the laser
pulse propagation in fibers. In the last case, a pulse can propagate
without decaying over large distance due to  balance between two
factors: SPM and group delay dispersion (GDD). These pulses are named
as optical solitons \cite{Ablowitz}. The
ultrashort pulse evolution obeys to the next master equation:

\begin{center}
\[ \boxed{
I\,({\frac {\partial }{\partial z}}\,\rho )={k_{2}}\,({\frac {
\partial ^{2}}{\partial t^{2}}}\,\rho ) + \beta \, \left|  \! \,
\rho \, \!  \right| ^{2}\,\rho }
\]
\end{center}

\emptyline
\noindent
which is the consequence of \textit{eq\_field} from part \textit{3} in
the case of transition to local time \textit{t--\TEXTsymbol{>}t-z/c}.
The right-hand side terms describe GDD (with coefficient
${k_{2}}$) and SPM (with coefficient
$\beta $), respectively\textit{. }

\emptyline
It is very important to obtain the exact soliton solutions of
nonlinear equations. There are the inverse and direct methods to
obtain such solutions. One of the direct methods is the so-called
Hirota's method. The main steps of this method are: 1) the selection
of the suitable substitution instead of the function \textit{
$\rho $} (see the master equation), that allows to obtain the
bilinear form of the evolution equation; 2) the consideration of the
formal series of perturbation theory for this bilinear equation. In
the case of soliton solutions these series are terminated.\\

The useful substitution for the nonlinear Schr\"odinger equation is\\

\begin{center}
$\rho $\textit{(z,t) = G(z,t)/F(z,t)}.
\end{center}

Let suppose that\textit{ F} is the real function. It should be noted that we
can make any assumption about
$\rho $ to satisfy the assumptions 1) and 2). Hirota proposed to
introduce a new \textit{D-operator }in following way:

\emptyline
\begin{center}
$\mathit{Dz}^{m}\,\mathit{Dt}^{n}\,a\,b=(({\frac {\partial }{
\partial z}}\,\mathrm{}) - ({\frac {\partial }{\partial {z_{1}}
}}\,\mathrm{}))^{m}\,(({\frac {\partial }{\partial t}}\,\mathrm{
}) - ({\frac {\partial }{\partial {t_{1}}}}\,\mathrm{}))^{n}\,
\mathrm{a}(z, \,t)\,\mathrm{b}({z_{1}}, \,{t_{1}})$
\end{center}

\emptyline
\noindent
After substitution of
$\rho $\textit{ }in the terms of functions \textit{G} and \textit{F} we
obtain two bilinear differential equations with regard to the new
operator \textit{D}:

\emptyline
\begin{center}
$[I\,\mathit{Dz} + {k_{2}}\,\mathit{Dt}^{2}]\,G\,F=0$
\end{center}

\begin{center}
${k_{2}}\,\mathit{Dt}^{2}\,\mathit{FF} - \beta \,G\,\mathit{G*}=0
$        (1)
\end{center}

\emptyline
\noindent
The functions \textit{G} and \textit{F} can be expanded into the
series of the formal parameter
$\varepsilon $:

\emptyline
\begin{center}
$G=\varepsilon \,\mathit{G1} + \varepsilon ^{3}\,\mathit{G3} +
\varepsilon ^{5}\,\mathit{G5}\,;\,F=1 + \varepsilon ^{2}\,
\mathit{F2} + \varepsilon ^{4}\,\mathit{F4} + \varepsilon ^{6}\,
\mathit{F6}$
\end{center}

\emptyline
Let substitute \textit{G} and \textit{F} into Eq. (1) and treat the
terms with powers of
$\varepsilon $ as independent, to get the infinite set of the differential
equations relatively \textit{G1, G3, ...; F2, F4, ...} . These formal
series are terminated only in the case when the master equation has
exact \textit{N}-soliton solution. For instance, the set of first six
differential equations in our case is:

\emptyline
\begin{center}
$I\,({\frac {\partial }{\partial z}}\,\mathit{G1}) + {k_{2}}\,(
{\frac {\partial ^{2}}{\partial t^{2}}}\,\mathit{G1})=0\,;$\\
$\,2\,{k
_{2}}\,({\frac {\partial ^{2}}{\partial t^{2}}}\,\mathit{F2}) -
\beta \,\mathit{G1}\,\mathit{G1*}=0\,;$\\
$\,I\,({\frac {\partial }{
\partial z}}\,\mathit{G3}) + {k_{2}}\,({\frac {\partial ^{2}}{
\partial t^{2}}}\,\mathit{G3}) + [I\,\mathit{Dz} + {k_{2}}\,
\mathit{Dt}^{2}]\,\mathit{G1}\,\mathit{F2}=0\,;$\\
$\,2\,{k_{2}}\,(
{\frac {\partial ^{2}}{\partial t^{2}}}\,\mathit{F4}) + \mathit{
Dt}^{2}\,\mathit{F2F2} - \beta \,(\mathit{G3}\,\mathit{G1*} +
\mathit{G1}\,\mathit{G3*})=0\,;$\\
$\,I\,({\frac {\partial }{\partial
z}}\,\mathit{G5}) + {k_{2}}\,({\frac {\partial ^{2}}{\partial t^{
2}}}\,\mathit{G5}) + [I\,\mathit{Dz} + {k_{2}}\,\mathit{Dt}^{2}]
\,(\mathit{G3}\,\mathit{F2} + \mathit{G1}\,\mathit{F4})=0\,;$\\
$\,2\,
{k_{2}}\,({\frac {\partial ^{2}}{\partial t^{2}}}\,\mathit{F6})
 + {k_{2}}\,\mathit{Dt}^{2}\,(\mathit{F4}\,\mathit{F2} + \mathit{
F2}\,\mathit{F4}) - \beta \,(\mathit{G5}\,\mathit{G1*} + \mathit{
G3}\,\mathit{G3*} + \mathit{G1}\,\mathit{G5*})=0$
\end{center}

\emptyline
For sake of the simplification of the very cumbersome manipulations we
introduce the procedure for operator
$\mathit{Dt}^{m}\,\mathit{Dz}^{n}$, which acts on the functions \textit{a} and \textit{b}. The lasts
are the exponents (or linear combination of the exponents) in the form
$e^{\eta }$, where
$\eta (z, \,t)$ is linear function.

\emptyline
$>$restart:\\
\indent \indent
with(plots):

\subsection{Procedure
$\mathit{Dt}^{m}$
$\mathit{Dz}^{n}$ }

\emptyline
$>$Dt\_Dz := proc (a,b,m,n)\\
\indent \indent
    local Summa,k,r,result:\\
\indent \indent \indent
    Summa := 0:\\
\indent \indent \indent \indent
    if  (n$>$1) and (m$<>$0) then\\
\indent \indent \indent
        for k from 1 to n-1 do\\
\indent \indent \indent \indent
            Summa := Summa+binomial(n,k)*(-1)$^{n-k+m}$*\\
\indent \indent \indent \indent
            der( der(b,z,(n-k)),t,m)*der(a,z,k)+\\
\indent \indent \indent \indent
            binomial(n,k)*(-1)$^{n-k}$*der(b,z,(n-k))*\\
\indent \indent \indent \indent
            der( der(a,z,k),t,m)\\
\indent \indent \indent
        od:\\
\indent \indent \indent \indent
    fi:\\
\indent \indent \indent \indent

    if  (n$>$1) and (m$>$1) then\\
\indent \indent \indent \indent \indent
       for r from 1 to (m-1) do\\
\indent \indent \indent
           for k from 1 to (n-1) do\\
\indent \indent
               Summa := Summa+binomial(m,r)*\\
\indent
               binomial(n,k)*(-1)$^{n-k+m-r}$*\\
\indent
               der( der(b,z,(n-k)),t,(m-r))*\\
\indent
               der( der(a,z,k),t,r);\\
\indent
           od:\\
\indent \indent
       od:\\
\indent \indent \indent
    fi:\\
\indent \indent \indent \indent

    if  (m$>$1) and (n$<>$0) then\\
\indent \indent \indent
       for r from 1 to (m-1) do\\
\indent \indent
           Summa := Summa+binomial(m,r)*(-1)$^{m-r+n}$*\\
\indent \indent
           der( der(b,z,n),t,(m-r))*der(a,t,r)+\\
\indent \indent
           binomial(m,r)*(-1)$^{m-r}$*der(b,t,(m-r))*\\
\indent \indent
           der( der(a,z,n),t,r);\\
\indent \indent \indent
       od;\\
\indent \indent \indent \indent
    fi:\\

\indent \indent \indent \indent
    if (m$<>$0) and (n$<>$0) then\\
\indent \indent \indent
       Summa := Summa+(-1)$^{m+n}$*der(der(b,z,n),t,m)*a+\\
\indent \indent \indent \indent
       (-1)$^{m}$*der(a,z,n)*der(b,t,m)+(-1)$^{n}$*der(a,t,m)*\\
\indent \indent \indent \indent
       der(b,z,n)+der(der(a,z,n),t,m)*b; \\
\indent \indent \indent
    fi:\\

\indent \indent \indent \indent
    if (m=0) and (n$>$1) then\\
\indent \indent \indent
       Summa := Summa+(-1)$^{n}$*der(b,z,n)*a+der(a,z,n)*b;\\
\indent \indent \indent
           for k from 1 to (n-1) do\\
\indent \indent
               Summa := Summa+binomial(n,k)*(-1)$^{n-k}$*\\
\indent \indent
               der(b,z,(n-k))*der(a,z,k);\\
\indent \indent \indent
           od:\\
\indent \indent \indent \indent
    fi:\\

\indent \indent \indent \indent
    if (m=0) and (n=1) then\\
\indent \indent \indent
       Summa := der(a,z,1)*b-der(b,z,1)*a:\\
\indent \indent \indent \indent
    fi:\\

\indent \indent \indent \indent
    if (n=0) and (m$>$1) then\\
\indent \indent \indent
       Summa := Summa+(-1)$^{m}$*der(b,t,m)*a+der(a,t,m)*b;\\
\indent \indent
           for r from 1 to (m-1) do\\
\indent
               Summa := Summa+binomial(m,r)*(-1)$^{m-r}$*\\
\indent
               der(b,t,(m-r))*der(a,t,r);\\
\indent \indent
           od:\\
\indent \indent \indent
    fi:\\

\indent \indent \indent \indent
    if (n=0) and (m=1) then\\
\indent \indent \indent
       Summa := der(a,t,1)*b-der(b,t,1)*a:\\
\indent \indent
    fi:\\

\indent
    if (n=0) and (m=0) then\\
\indent \indent
       Summa := a*b\\
\indent \indent \indent
    fi:

\indent \indent \indent \indent
    result := combine(Summa,exp):\\

\indent
 end:

\emptyline
The next procedure will be used for calculation of the derivative of
$e^{\eta }$ (or combination of exponents) on \textit{t} or \textit{z}\textit{
}with further simplification of the obtained expression.

\emptyline

\subsection{Procedure der}

\emptyline
$>$der := proc (f,x,m)\\
\indent \indent
   local  difF,i,function:\\
\indent \indent \indent
   subs(eta1=eta1(x),eta1s=eta1s(x),\\
\indent \indent \indent \indent
   eta2=eta2(x),eta2s=eta2s(x),f):\\
\indent \indent \indent \indent
   difF := diff(\%,x\$m):\\

\indent \indent \indent \indent
   if (x=t) then\\
\indent \indent \indent
      function := subs(\{diff(eta1(x),x)=b1,\\
\indent \indent \indent \indent
      diff(eta2(x),x)=b2,diff(eta1s(x),x)=b1s,\\
\indent \indent \indent \indent
      diff(eta2s(x),x)=b2s\},difF)\\
\indent \indent \indent
            else\\
\indent \indent \indent \indent
      function := subs(\{diff(eta1(x),x)=a1,\\
\indent \indent \indent \indent
      diff(eta2(x),x)=a2,diff(eta1s(x),x)=a1s,\\
\indent \indent \indent \indent
      diff(eta2s(x),x)=a2s\},difF)\\
\indent \indent \indent
   fi;\\

\indent \indent \indent \indent
   subs(eta1(x)=eta1,eta1s(x)=eta1s,\\
\indent \indent \indent \indent
   eta2(x)=eta2,eta2s(x)=eta2s, function):\\

\indent
   if (m$>$1) then \\
\indent \indent
      difF := subs(\{diff(a1,x)=0,diff(a2,x)=0,\\
\indent \indent
      diff(b1,x)=0,diff(b2,x)=0,diff(a1s,x)=0,\\
\indent \indent
      diff(a2s,x)=0,diff(b1s,x)=0,diff(b2s,x)=0\},\%) \\
\indent \indent \indent \indent
            else \\
\indent \indent \indent
      combine(\%)\\
\indent \indent
   fi:\\

\indent
   collect(\%,exp):\\

\indent
end:

\emptyline
The next procedure is used to calculate an integral of
$e^{\eta }$ (or combination of exponents) on \textit{t} or \textit{z }with
further simplification of the expression.

\emptyline

\subsection{Procedure Integr}

\emptyline
$>$integr := proc (f,x,m)\\
\indent \indent
     local intF,i,g1,g1s,g2,g2s,function:\\
\indent \indent \indent
     intF := subs(eta1=g1*x,eta1s=g1s*x,eta2=g2*x,\\
\indent \indent \indent
     eta2s=g2s*x,f):\\
\indent \indent \indent

     for i from 1 to m do\\
\indent \indent \indent
         intF := int(intF,x);\\
\indent \indent
     od:\\
\indent

     if (x=t) then\\
\indent \indent
         x := t; g1 := b1; g1s := b1s; g2 := b2; g2s := b2s;\\
\indent \indent \indent
              else\\
\indent \indent \indent \indent
         x := z; g1 := a1; g1s := a1s; g2 := a2; g2s := a2s;\\
\indent \indent \indent
     fi:\\
\indent \indent

     intF;\\
\indent

     collect(\%,exp):\\
\indent \indent

     subs(b1*t=eta1,b1s*t=eta1s,b2*t=eta2,\\
\indent
     b2s*t=eta2s,a1*z=eta1,a1s*z=eta1s,a2*z=eta2,\\
\indent
     a2s*z=eta2s,\%);\\
\indent
end:

\emptyline
Now, let try to obtain \textit{a first-order soliton} for nonlinear
Schr\"odinger equation.

\emptyline
$>$macro(Gs=`G*`,G1s=`G1*`,G3s=`G3*`,G5s=`G5*`,
as=`a*`,bs=`b*`,eta0s = `eta0*`):\\
\indent \indent
G1 := exp(eta1):\# successful substitution!\\
\indent
I*der(\%,z,1)+k\_2*der(\%,t,2): \#first from the equations set\\
\indent \indent
   factor(\%);
\emptyline
\[
e^{\eta 1}\,(I\,\mathit{a1} + \mathit{k\_2}\,\mathit{b1}^{2})
\]

\emptyline
$>$\#as result we can find the parameter a1\\
\indent \indent
   a1 := I*k\_2*b1$^{2}$:\\
\indent \indent \indent
      a1s := -I*k\_2*b1s$^{2}$:

$>$G1s := exp(eta1s):\# conjugated to G1\\
\indent \indent
   G1G1s := combine(G1*G1s):\\
\indent \indent \indent
       F2 := beta/(2*k\_2)*integr(\%,t,2);\#F2 from the second equation
of the set

\emptyline
\[
\mathit{F2} := {\displaystyle \frac {1}{2}} \,{\displaystyle
\frac {\beta \,e^{(\eta 1 + \mathit{eta1s})}}{\mathit{k\_2}\,(
\mathit{b1} + \mathit{b1s})^{2}}}
\]

\emptyline
$>$\# But the next equation of the set results in\\
\indent \indent
   I\_Dz\_G1\_F2 := I*factor(Dt\_Dz(G1,F2,0,1)):\\
\indent \indent \indent
       d\_Dt2\_G1\_F2 := k\_2*factor(Dt\_Dz(G1,F2,2,0)):\\
\indent \indent \indent \indent
           factor(I\_Dz\_G1\_F2+d\_Dt2\_G1\_F2);\\
\indent \indent \indent \indent \indent
               Dt\_Dz(F2,F2,2,0);

\emptyline
\[
0
\]
\[
0
\]

\emptyline
\noindent
To obtain this result we use the trivial relationships:

\emptyline
$>$eta1=a1*z+b1*t+eta10:\\
\indent \indent
   eta2=a2*z+b2*t+eta20:\\
\indent \indent \indent
       Dz*exp(eta1)*exp(eta2)=(a1-a2)*exp(eta1+eta2);\\
\indent \indent
           Dt$^{2}$*exp(eta1)*exp(eta2)=(b1-b2)$^{2}$*exp(eta1+eta2);

\emptyline
\[
\mathit{Dz}\,e^{\eta 1}\,e^{\eta 2}=(I\,\mathit{k\_2}\,\mathit{b1
}^{2} - \mathit{a2})\,e^{(\eta 1 + \eta 2)}
\]

\[
\mathit{Dt}^{2}\,e^{\eta 1}\,e^{\eta 2}=(\mathit{b1} - \mathit{b2
})^{2}\,e^{(\eta 1 + \eta 2)}
\]

\emptyline
\noindent
As was shown above \textit{a= - i
${k_{2}}$
$b^{2}$}, hence the last term in third equation of set is equal to
\textit{0}. So, we are to choose \textit{G3} = 0 to satisfy third
equation. Furthermore
$\mathit{Dt}^{2}$\textit{F2} from fourth equation of the set is (
$\frac {\beta ^{2}}{(2(b + \mathit{bs})^{2}\,\mathit{k\_2})^{2}}$)
$\mathit{Dt}^{2}$exp(
$\eta $+
$\eta $\_s). But in concordance with above obtained relationships this
expression is equal to zero. So we can choose \textit{F4} = 0. Thus to
satisfy other equations we can keep in the expansion of functions
\textit{G} and \textit{F} only \textit{G1}, \textit{G3} and
\textit{F2}.
So, the formal series are terminated. Since
$\varepsilon $ is independent parameter we can take
$\varepsilon $=1.

\emptyline
$>$rho := G1/(1+F2);

\emptyline
\[
\rho  := {\displaystyle \frac {e^{\eta 1}}{1 + {\displaystyle
\frac {{\displaystyle \frac {1}{2}} \,\beta \,e^{(\eta 1 +
\mathit{eta1s})}}{\mathit{k\_2}\,(\mathit{b1} + \mathit{b1s})^{2}
}} }}
\]

\emptyline
$>$subs(eta1=a1*z+b1*t+eta10,eta1s=a1s*z+b1s*t+eta10s, rho):\\
\indent
soliton := expand(subs(\{b1s=b1,beta=1,k\_2=1/2\},\%));\#the choice of
k\_2 and beta is only normalization of the values in equation

\emptyline
$>$soliton :=\\
\indent \indent
exp(1/2*I*b1$^{2}$*z)*exp(b1*t)*exp(eta10)/\\
\indent \indent
(1+1/4*exp(b1*t)$^{2}$*exp(eta10)*exp(eta10s)/(b1$^{2}$));

\emptyline
\[
\mathit{soliton} := {\displaystyle \frac {e^{(1/2\,I\,\mathit{b1}
^{2}\,z)}\,e^{(\mathit{b1}\,t)}\,e^{\eta 10}}{1 + {\displaystyle
\frac {{\displaystyle \frac {1}{4}} \,(e^{(\mathit{b1}\,t)})^{2}
\,e^{\eta 10}\,e^{\mathit{eta10s}}}{\mathit{b1}^{2}}} }}
\]

\emptyline
$>$\# Now we check the obtained solution by substitution of one in
dynamical equation\\
\indent
I*der(rho,z,1)/rho+beta*exp(eta1+eta1s)/((1+1/2*beta*exp(eta1+eta1s)/\\
\indent
(k\_2*(b1+b1s)$^{2}$)))$^{2}$+k\_2*der(rho,t,2)/rho:\\
\indent \indent
simplify(\%);

\emptyline
\[
0
\]

\emptyline
\noindent
All right! This is the exact solution of the Schr\"odinger equation.
Physically \textit{b1} has a sense of the inverse pulse duration. So
it is real parameter. But what is the free parameter
$\eta $\textit{10}? Let
$\eta $\textit{10 }is real and
$e^{\eta 10}$ = b1. Then

\emptyline
$>$subs(\{exp(eta10)=2*b1, exp(eta10s)=2*b1\},soliton):\\
\indent \indent
simplify(\%);

\emptyline
\[
2\,{\displaystyle \frac {e^{(1/2\,\mathit{b1}\,(I\,z\,\mathit{b1}
 + 2\,t))}\,\mathit{b1}}{1 + e^{(2\,\mathit{b1}\,t)}}}
\]

\emptyline
\noindent
The obtained solution is the so-called \textit{first-order soliton}
with duration
$\frac {1}{\mathit{b1}}$ , amplitude \textit{b1} and phase
$\mathit{b1}^{2}$\textit{z/2}:

\begin{center}
\[ \boxed{
\rho \mathit{(z,t)=b1 \cdot sech(b1 \cdot t)exp(i \cdot
\mathit{b1}^{2}z/2)}}
\]
\end{center}

\emptyline
$>$plot(subs(\{z=0,b1=1\},\%),t=-5..5,axes=boxed, title=`first-order
soliton`);

\emptyline
\begin{center}
\mapleplot{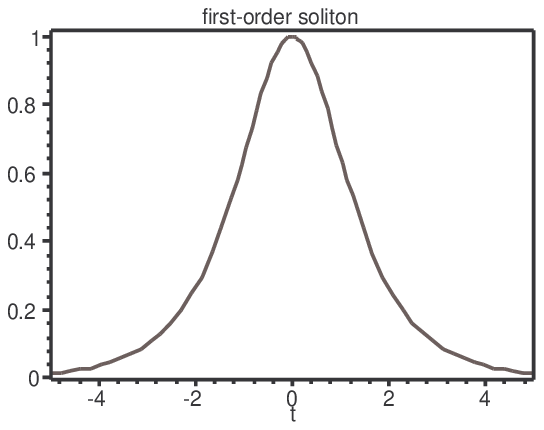}
\end{center}

\emptyline
But what is about different values of
$\eta $\textit{10}? The choice of the different values of
$\eta $\textit{10} results in the obtaining of the so-called collapsing
pulses, i. e. pulses with singularity in the dependence of their
parameters on \textit{z}.\\
\indent
The described here procedure is available for the analysis of the
higher-order solitons. For example, the substitution of \textit{G1 =
$e^{\eta 1}$+
$e^{\eta 2}$} causes the termination of the series on fifth equation (you can
prove this statement by using of the described above procedures). The
obtained solution depends on \textit{z} and is called as the
\textit{second order soliton}.


\emptyline
In the case of Schr\"odinger equation, the main feature of the described
above method is the termination of the formal series for an arbitrary
order of the solution. Such behavior results from a very rich
mathematical structure of the dynamical equation: an existence of the
infinitely many nontrivial symmetries and conservation laws, Painleve
property and integrability by means of the inverse scattering method.
The non-decaying pulse-like solutions of the integrable nonlinear
evolutional equations are called as the \fbox{solitons}.
But as we shall see later, some nonlinear equations have the
soliton-like solutions, but do not belong to integrable class. For
example, for nonlinear Landau-Ginzburg equation (see next part) there
is the soliton-like solution in the first order of the Hirota's
method. But the second-order solution does not lead to the termination
of the series. Such soliton-like solutions of the nonintegrable
dynamical equation are called as the
\fbox{quasi-solitons (or solitary waves)}.
\emptyline

\section{Nonlinear Landau-Ginzburg equation: quasi-soliton
solution}

\emptyline
\noindent
Here we shall consider a soliton-like pulse, which is generated in the
continuous-wave solid-state laser due to power-dependent saturation of
the diffraction loss in the presence of the field self-focusing in
active medium \cite{Haus1, Haus2}. The
action of the saturable loss can be described by the real cubic
nonlinear term. The energy dissipation due to spectral filtering can
be introduced by means of the real second-order derivative on
\textit{t}. Then in the absence of SPM and GDD, the dynamical equation
is a analog of the Schr\"odinger equation, but with pure real terms.

\begin{center}
\[ \boxed{
{\frac {\partial }{\partial z}}\,\rho (z, \,t) = \mathit{g
\rho (z, \,t) +
{t_{f}}^{2}
{\frac {\partial ^{2}}{\partial t^{2}}}\,\rho (z, \,t)} +
\sigma
\rho (z, \,t)^{3}}
\]
\end{center}

\emptyline
\noindent
Here \textit{g} is the net-gain in the laser taking into account the
gain and linear loss in the active medium, output loss, and
diffraction loss. For sake of simplification, we shall suppose the
normalization of time to the inverse bandwidth of the spectral filter
${t_{f}}$ (let
${t_{f}}$= 2.5 fs, that corresponds to the full generation bandwidth of Ti:
sapphire laser) and normalization of pulse intensity to the inverse
intensity of the loss saturation
$\sigma $ (the typical values of
$\sigma $ are \symbol{126}
$10^{ - 10}$
} -
$10^{ - 12}$
$\mathit{cm}^{2}$/W). We will consider the steady-state pulse propagation, then
${\frac {\partial }{\partial z}}\,\rho (z, \,t)$ = 0. So, the master equation can be transformed to the well-known
Duffing's type equation describing nonlinear oscillations without
damping:

\emptyline
$>$restart:\\
\indent \indent
  with(plots):\\
\indent \indent \indent
    with(DEtools):\\
\indent
       ode := diff(rho(t),`\$`(t,2)) + rho(t)$^{3}$ + g*rho(t);

\emptyline
\[
\mathit{ode} := ({\frac {\partial ^{2}}{\partial t^{2}}}\,\rho (t
)) + \rho (t)^{3} + g\,\rho (t)
\]

\emptyline
\noindent
Its implicit solutions are:

\emptyline
$>$sol := dsolve(ode=0,rho(t));
\emptyline
\begin{eqnarray}
\mathit{sol} := {\displaystyle \int _{\ }^{\rho (t)}} 2\,
{\displaystyle \frac {1}{\sqrt{ - 2\,\mathit{\_a}^{4} - 4\,g\,
\mathit{\_a}^{2} + 4\,\mathit{\_C1}}}} \,d\mathit{\_a} - t -
\mathit{\_C2}=0, \nonumber \\ \nonumber
{\displaystyle \int _{\ }^{\rho (t)}}  - 2\,{\displaystyle
\frac {1}{\sqrt{ - 2\,\mathit{\_a}^{4} - 4\,g\,\mathit{\_a}^{2}
 + 4\,\mathit{\_C1}}}} \,d\mathit{\_a} - t - \mathit{\_C2}=0
\end{eqnarray}

\emptyline
\noindent
The first integral of motion is:

\emptyline
$>$numer(diff(lhs(sol[1]),t)):\\
\indent \indent
   int\_motion := simplify((op(1,\%)$^{2}$-op(2,\%)$^{2}$)/2);

\emptyline
\[
\mathit{int\_motion} := 2\,({\frac {\partial }{\partial t}}\,\rho
 (t))^{2} + \rho (t)^{4} + 2\,g\,\rho (t)^{2} - 2\,\mathit{\_C1}
\]

\emptyline
\noindent
These equations describe the motion in the potential:

\emptyline
$>$pot := simplify(op(2,int\_motion)+op(3,int\_motion));

\emptyline
\[
\mathit{pot} := \rho (t)^{4} + 2\,g\,\rho (t)^{2}
\]

\emptyline
\noindent
The value of \textit{4*\_C1} plays a role of the full energy of
system. The dependence of potential on
$\rho $ for the different \textit{g} is shown in the next figure:

\emptyline
$>$plot3d(subs(rho(t)=rho,pot),g=0.05..-0.1,rho=-0.5..0.5,axes=boxed,\\
\indent \indent
title=`Potential of pendulum`);
\emptyline
\begin{center}
\mapleplot{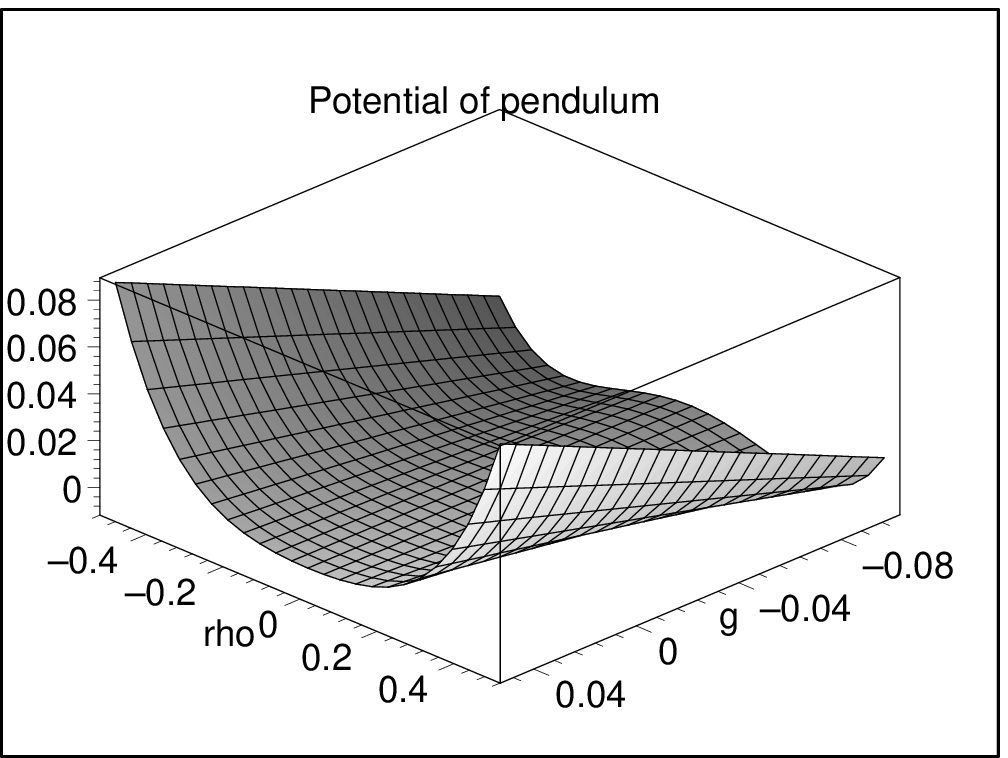}
\end{center}

\emptyline
\noindent
We can see, that for \textit{g} \TEXTsymbol{>} 0 there is one
equilibrium state of pendulum in
$\rho $ = 0 (stable state), and for \textit{g} \TEXTsymbol{<} 0 there are
three one (unstable in
$\rho $ = 0 and stable in
$\rho $ = \textit{+/-
$\sqrt{ - g}$)}. Obviously, that in this system the different oscillation regimes
is possible, that is illustrated by the phase portrait on the plane
[\textit{y,} \textit{z=d
$\rho $/dt}].

\emptyline
$>$sys := convertsys(ode = 0,[],rho(t),t,z);\\
\indent
dfieldplot([diff(rho(t),t)=z(t),diff(z(t),t)=-rho(t)$^{3}$-subs(g=-0.1,g)*
rho(t)], \\
\indent \indent
[z(t),rho(t)],t=-2..2,rho=-0.5..0.5,z=-0.1..0.1,\\
\indent \indent
arrows=LARGE,axes=boxed,title=`Nonlinear\\
\indent \indent
oscillations`,color=black);

\emptyline
\[
\mathit{sys} := [[{\mathit{YP}_{1}}={z_{2}}, \,{\mathit{YP}_{2}}=
 - {z_{1}}^{3} - g\,{z_{1}}], \,[{z_{1}}=\rho (t), \,{z_{2}}=
{\frac {\partial }{\partial t}}\,\rho (t)], \,\mathit{undefined}
, \,[]]
\]
\emptyline
\begin{center}
\mapleplot{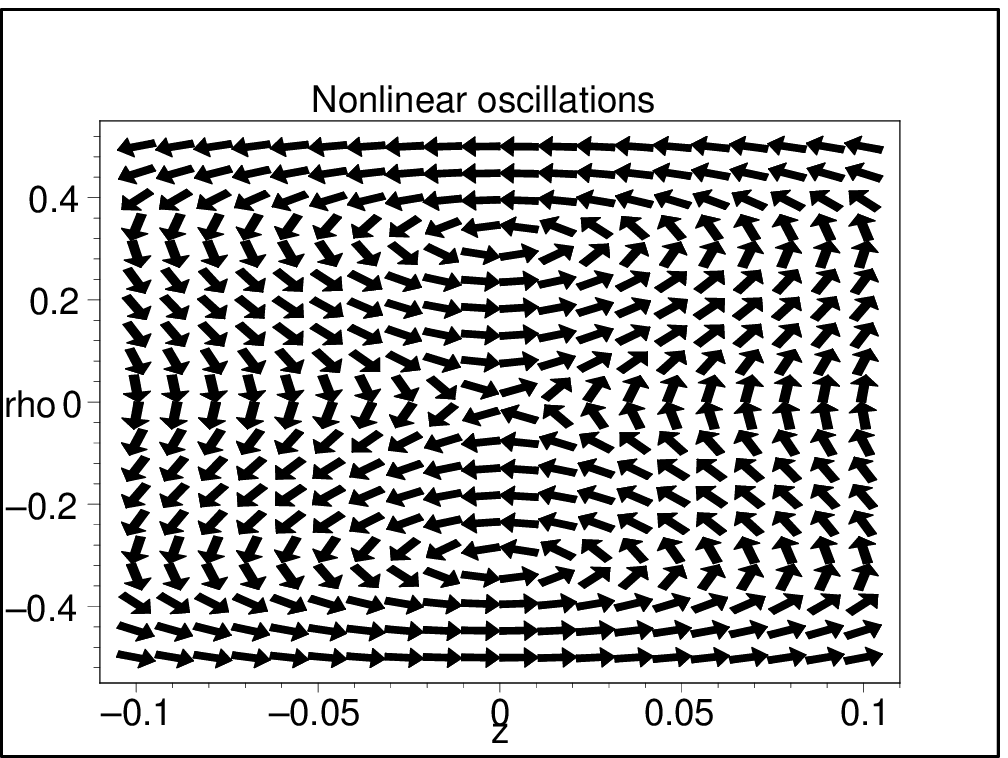}
\end{center}

\emptyline
The quasi-soliton solution of the initial equation corresponds to the
oscillation around the stable equilibrium state with infinite period.
In this case the full energy is equal to 0. Then  \textit{\_C1} = 0
and the motion begins from
$\rho $ = 0 at \textit{t} = -
$\infty $:

\emptyline
$>$plot(\{sqrt(subs(\{rho(t)=rho,g=-0.1\},-pot))/2,-sqrt(subs(\\
\indent
\{rho(t)=rho,g=-0.1\},-pot))/2\},\\
\indent
rho=0..0.45,axes=boxed,labels=[`rho(t)`,`drho(t)/dt`],color=red,\\
\indent
numpoints=200);

\begin{center}
\mapleplot{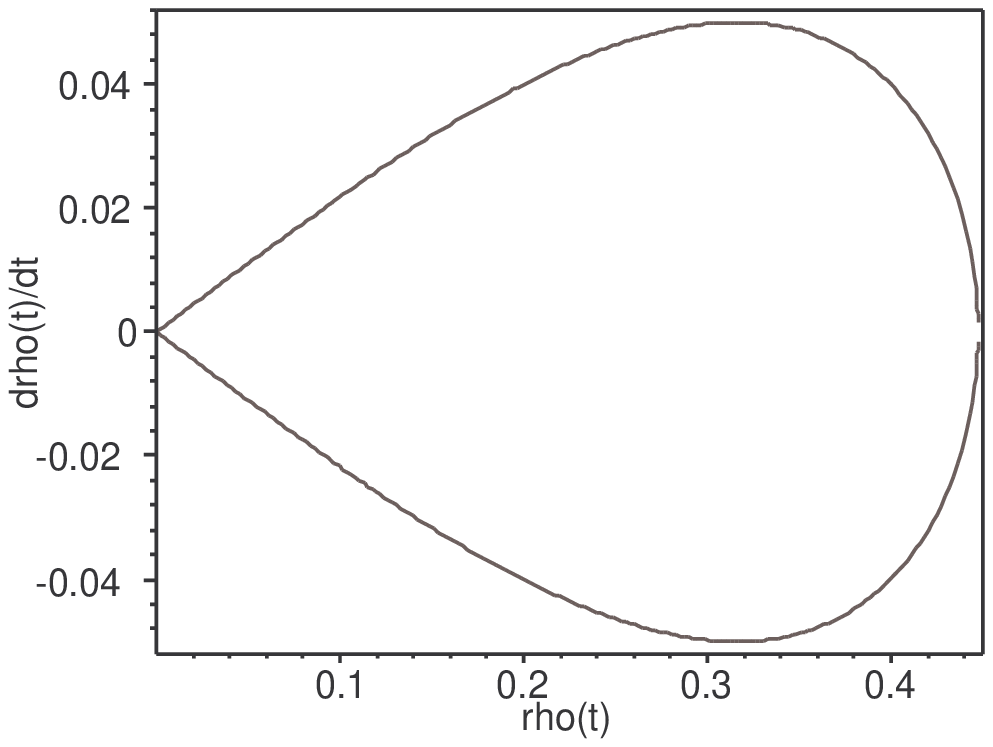}
\end{center}

\emptyline
\noindent
The amplitude of quasi-soliton can be found from the integral of
motion (i. e. pulse maximum correspond to \textit{d
$\rho $/dt }= 0 and \textit{\_C1 }= 0 in the integral of motion).

\emptyline
$>$rho0 := solve(factor(pot)/rho(t)$^{2}$=0,rho(t));

\emptyline
\[
\rho 0 := \sqrt{ - 2\,g}, \, - \sqrt{ - 2\,g}
\]

\emptyline
\noindent
The explicit integration of the solution produces:

\emptyline
$>$assume(g<0):\\
\indent \indent
  sol1\_a := numer(value(subs(\_C1=0,lhs(sol[1]))));\\
\indent \indent \indent
    sol1\_b := numer(value(subs(\_C1=0,lhs(sol[2]))));

\emptyline
\begin{eqnarray*}
\mathit{sol1\_a} := \rho (t)\,\sqrt{ - 2\,\rho (t)^{2} - 4\,
\mathit{g\symbol{126}}}\,\mathrm{arctanh}(2\,{\displaystyle
\frac {\mathit{g\symbol{126}}}{\sqrt{ - \mathit{g\symbol{126}}}\,
\sqrt{ - 2\,\rho (t)^{2} - 4\,\mathit{g\symbol{126}}}}} ) -  \\
t\,
\sqrt{ - 2\,\rho (t)^{4} - 4\,\mathit{g\symbol{126}}\,\rho (t)^{2
}}\,\sqrt{ - \mathit{g\symbol{126}}} - \mathit{\_C2}\,\sqrt{ - 2\,\rho (t)^{4} - 4\,\mathit{g
\symbol{126}}\,\rho (t)^{2}}\,\sqrt{ - \mathit{g\symbol{126}}}
\end{eqnarray*}

\begin{eqnarray*}
\mathit{sol1\_b} :=  - \rho (t)\,\sqrt{ - 2\,\rho (t)^{2} - 4\,
\mathit{g\symbol{126}}}\,\mathrm{arctanh}(2\,{\displaystyle
\frac {\mathit{g\symbol{126}}}{\sqrt{ - \mathit{g\symbol{126}}}\,
\sqrt{ - 2\,\rho (t)^{2} - 4\,\mathit{g\symbol{126}}}}} )  \\
\mbox{} - t\,\sqrt{ - 2\,\rho (t)^{4} - 4\,\mathit{g\symbol{126}}
\,\rho (t)^{2}}\,\sqrt{ - \mathit{g\symbol{126}}} - \mathit{\_C2}
\,\sqrt{ - 2\,\rho (t)^{4} - 4\,\mathit{g\symbol{126}}\,\rho (t)
^{2}}\,\sqrt{ - \mathit{g\symbol{126}}}
\end{eqnarray*}

\emptyline
\noindent
Make some transformations:

\emptyline
$>$sol2\_a :=\\
\indent
numer(simplify(expand(sol1\_a/op(1,sol1\_a)),radical,symbolic));\\
\indent
sol2\_b :=
numer(simplify(expand(sol1\_b/op(1,sol1\_b)),radical,symbolic));

\emptyline
\[
\mathit{sol2\_a} := \mathrm{arctanh}(2\,{\displaystyle \frac {
\mathit{g\symbol{126}}}{\sqrt{ - \mathit{g\symbol{126}}}\,\sqrt{
 - 2\,\rho (t)^{2} - 4\,\mathit{g\symbol{126}}}}} ) - t\,\sqrt{
 - \mathit{g\symbol{126}}} - \mathit{\_C2}\,\sqrt{ - \mathit{g
\symbol{126}}}
\]

\[
\mathit{sol2\_b} := \mathrm{arctanh}(2\,{\displaystyle \frac {
\mathit{g\symbol{126}}}{\sqrt{ - \mathit{g\symbol{126}}}\,\sqrt{
 - 2\,\rho (t)^{2} - 4\,\mathit{g\symbol{126}}}}} ) + t\,\sqrt{
 - \mathit{g\symbol{126}}} + \mathit{\_C2}\,\sqrt{ - \mathit{g
\symbol{126}}}
\]

\emptyline
\noindent
When at the pulse maximum
$\rho $\textit{(0)} =
${\rho _{\mathit{max}}}$, the value of \textit{\_C2} can be found as:

\emptyline
$>$i\_C2\_a := solve(subs(\{t=0,rho(t)=rho\_max\},sol2\_a)=0,\_C2);\\
\indent
i\_C2\_b := solve(subs(\{t=0,rho(t)=rho\_max\},sol2\_b)=0,\_C2);

\emptyline
\[
\mathit{i\_C2\_a} := {\displaystyle \frac {\mathrm{arctanh}(2\,
{\displaystyle \frac {\mathit{g\symbol{126}}}{\sqrt{ - \mathit{g
\symbol{126}}}\,\sqrt{ - 2\,\mathit{rho\_max}^{2} - 4\,\mathit{g
\symbol{126}}}}} )}{\sqrt{ - \mathit{g\symbol{126}}}}}
\]

\[
\mathit{i\_C2\_b} :=  - {\displaystyle \frac {\mathrm{arctanh}(2
\,{\displaystyle \frac {\mathit{g\symbol{126}}}{\sqrt{ - \mathit{
g\symbol{126}}}\,\sqrt{ - 2\,\mathit{rho\_max}^{2} - 4\,\mathit{g
\symbol{126}}}}} )}{\sqrt{ - \mathit{g\symbol{126}}}}}
\]

\emptyline
\noindent
Let \textit{tanh(i\_C2\_a*sqrt(-g))} = \textit{tanh(-i\_C2\_b*sqrt(-g)
)} = \textit{-/+
$\upsilon $}, then from an expression for the tangents of sum of arguments:
\textit{tanh(a + b)} = \textit{tanh(a)} + \textit{tanh(b)}/(1 +
\textit{tanh(a)tanh(b)}), the equations can be transformed as:

\emptyline
$>$sol3\_a :=\\
\indent
simplify(tanh(op(1,sol2\_a)))=(tanh(-op(2,sol2\_a))+upsilon)/\\
\indent
(1+tanh(-op(2,sol2\_a))*upsilon);\\
\indent \indent
sol3\_b :=\\
\indent
simplify(tanh(op(1,sol2\_b)))=(tanh(-op(2,sol2\_b))+upsilon)/\\
\indent
(1+tanh(-op(2,sol2\_b))*upsilon);

\emptyline
\[
\mathit{sol3\_a} := 2\,{\displaystyle \frac {\mathit{g
\symbol{126}}}{\sqrt{ - \mathit{g\symbol{126}}}\,\sqrt{ - 2\,\rho
 (t)^{2} - 4\,\mathit{g\symbol{126}}}}} ={\displaystyle \frac {
\mathrm{tanh}(t\,\sqrt{ - \mathit{g\symbol{126}}}) + \upsilon }{1
 + \mathrm{tanh}(t\,\sqrt{ - \mathit{g\symbol{126}}})\,\upsilon }
}
\]

\[
\mathit{sol3\_b} := 2\,{\displaystyle \frac {\mathit{g
\symbol{126}}}{\sqrt{ - \mathit{g\symbol{126}}}\,\sqrt{ - 2\,\rho
 (t)^{2} - 4\,\mathit{g\symbol{126}}}}} ={\displaystyle \frac {
 - \mathrm{tanh}(t\,\sqrt{ - \mathit{g\symbol{126}}}) + \upsilon
}{1 - \mathrm{tanh}(t\,\sqrt{ - \mathit{g\symbol{126}}})\,
\upsilon }}
\]

\emptyline
$>$sol4\_a := solve(sol3\_a, rho(t));\\
\indent \indent
sol4\_b := solve(sol3\_b, rho(t));

\emptyline
\begin{eqnarray*}
\mathit{sol4\_a} := 2\,{\displaystyle \frac {\sqrt{2\,\mathit{g
\symbol{126}} - 2\,\mathit{g\symbol{126}}\,\upsilon ^{2}}\,e^{(t
\,\sqrt{ - \mathit{g\symbol{126}}})}}{\mathrm{\%1} - 1 + \upsilon
 \,\mathrm{\%1} + \upsilon }} , \, - 2\,{\displaystyle \frac {
\sqrt{2\,\mathit{g\symbol{126}} - 2\,\mathit{g\symbol{126}}\,
\upsilon ^{2}}\,e^{(t\,\sqrt{ - \mathit{g\symbol{126}}})}}{
\mathrm{\%1} - 1 + \upsilon \,\mathrm{\%1} + \upsilon }} \\
\mathrm{\%1} := (e^{(t\,\sqrt{ - \mathit{g\symbol{126}}})})^{2}
\end{eqnarray*}

\begin{eqnarray*}
\mathit{sol4\_b} := 2\,{\displaystyle \frac {\sqrt{2\,\mathit{g
\symbol{126}} - 2\,\mathit{g\symbol{126}}\,\upsilon ^{2}}\,e^{(t
\,\sqrt{ - \mathit{g\symbol{126}}})}}{ - \mathrm{\%1} + 1 +
\upsilon \,\mathrm{\%1} + \upsilon }} , \, - 2\,{\displaystyle
\frac {\sqrt{2\,\mathit{g\symbol{126}} - 2\,\mathit{g\symbol{126}
}\,\upsilon ^{2}}\,e^{(t\,\sqrt{ - \mathit{g\symbol{126}}})}}{ -
\mathrm{\%1} + 1 + \upsilon \,\mathrm{\%1} + \upsilon }}  \\
\mathrm{\%1} := (e^{(t\,\sqrt{ - \mathit{g\symbol{126}}})})^{2}
\end{eqnarray*}

\emptyline
Now, we must note, that the transit
${\rho _{\mathit{max}}}$ -\TEXTsymbol{>}
$\rho $\textit{0} (see above) corresponds to the transit
$\upsilon $ --\TEXTsymbol{>}
$\infty $. Then the final solutions result from the next operations:

\emptyline
$>$sol\_fin\_1 := limit(sol4\_a[1],upsilon=infinity);\\
\indent \indent
 sol\_fin\_2 := limit(sol4\_a[2],upsilon=infinity);\\
\indent \indent \indent
   sol\_fin\_1 := limit(sol4\_b[1],upsilon=infinity);\\
\indent \indent
     sol\_fin\_2 := limit(sol4\_b[2],upsilon=infinity);

\emptyline
\[ \boxed{
\mathit{sol\_fin\_1} := 2\,{\displaystyle \frac {\sqrt{ - 2\,
\mathit{g\symbol{126}}}\,e^{(t\,\sqrt{ - \mathit{g\symbol{126}}})
}}{e^{(2\,t\,\sqrt{ - \mathit{g\symbol{126}}})} + 1}} }
\]

\[
\mathit{sol\_fin\_2} :=  - 2\,{\displaystyle \frac {\sqrt{ - 2\,
\mathit{g\symbol{126}}}\,e^{(t\,\sqrt{ - \mathit{g\symbol{126}}})
}}{e^{(2\,t\,\sqrt{ - \mathit{g\symbol{126}}})} + 1}}
\]

\[
\mathit{sol\_fin\_1} := 2\,{\displaystyle \frac {\sqrt{ - 2\,
\mathit{g\symbol{126}}}\,e^{(t\,\sqrt{ - \mathit{g\symbol{126}}})
}}{e^{(2\,t\,\sqrt{ - \mathit{g\symbol{126}}})} + 1}}
\]

\[
\mathit{sol\_fin\_2} :=  - 2\,{\displaystyle \frac {\sqrt{ - 2\,
\mathit{g\symbol{126}}}\,e^{(t\,\sqrt{ - \mathit{g\symbol{126}}})
}}{e^{(2\,t\,\sqrt{ - \mathit{g\symbol{126}}})} + 1}}
\]

\emptyline
\noindent
The positive roots satisfy to the initial condition and the result is
the quasi-soliton pulse with \textit{sech} - shape envelope, which has
the duration \fbox{\textit{1/
$\sqrt{ - g}$}} and the peak amplitude
\fbox{$\rho $\textit{0 =
$\sqrt{ - 2\,g}$}}. The pulse intensity profile is shown in the next figure (
$\sigma $ =
$10^{ - 11}$
$\mathit{cm}^{2}$/W,
$\iota $ is the time normalized to
${t_{f}}$):

\emptyline
$>$animate(evalf(subs(t=iota/(2.5e-15),sol\_fin\_1)$^{2}$*100),\\
\indent
iota=-1e-13..1e-13,g=-0.05..-0.01,frames=50,\\
\indent
axes=boxed,color=red,labels=[`time, fs`,`rho$^{2}$, GW/cm$^{2}$`],\\
\indent
title=`Pulse envelope`);

\emptyline
\begin{center}
\mapleplot{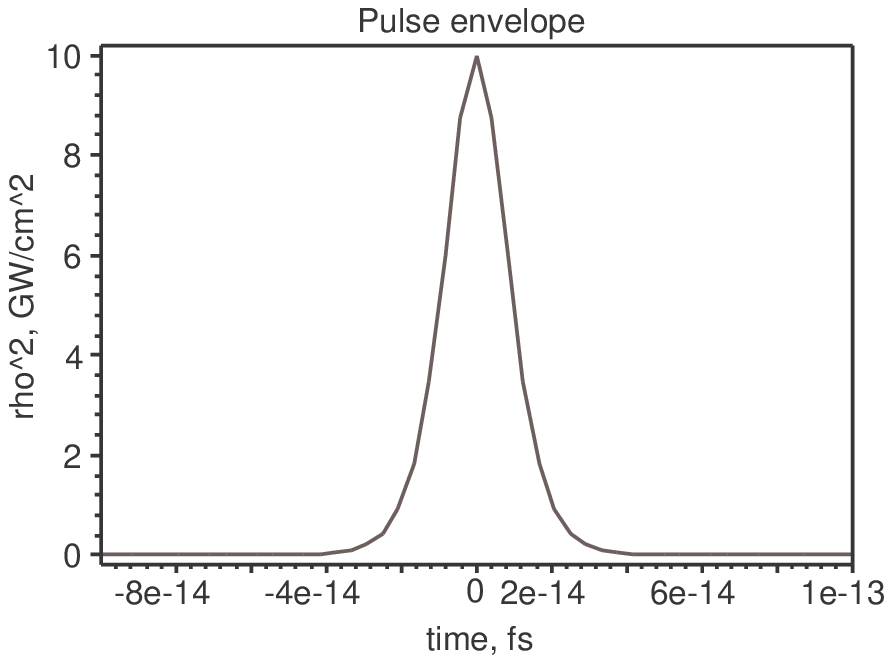}
\end{center}

\begin{center}
\fbox{ \parbox{.8\linewidth}{
So, there is the close analogy between an ultrashort pulse generation
in the laser with power-dependent loss saturation and an oscillation
of nonlinear pendulum. The considered solution has the character of
steady-state quasi-soliton}}
\end{center}
\emptyline
\noindent
In the next part we are going to
demonstrate some analogies of the high-order soliton behavior (laser
breezers).

\section{Autooscillations of quasi-solitons in the laser}

\emptyline \noindent The main efforts in the design of the modern
femtosecond lasers aim to the stabilization of the ultrashort
pulse parameters. For example, as it was shown \cite{Sergeev}, the
pulse destabilization are very important factors limiting the
operation of laser with so-called fast saturable absorber (the
bleaching of diffraction loss due to Kerr-lensing is the example
of such saturable absorber, see previous part). The
destabilization of the laser quasi-soliton can produce its
destruction or regular/nonregular autooscillations. The last can
be considered as the analog of a high-order soliton formation (see
part VI), which is reminiscent of the breezers of nonlinear
dynamical equation. We shall consider the master laser equation,
which joins the Landau - Ginzburg and Schr\"odinger equations. The
main nonlinear factors now are SPM and power-dependent loss
saturation. GDD and spectral filtering will be taken into
consideration, too.

\emptyline
$>$restart:\\
\indent \indent
  with(DEtools):\\
\indent \indent \indent
    with(plots):\\
\indent
       master\_1 := diff(rho(z,t),z) =\\
\indent
alpha*rho(z,t)-gamma*rho(z,t)+I*phi*rho(z,t)+tf$^{2}$*diff(rho(z,t),`\$`(t,\\
\indent
2))+I*k\_2*diff(rho(z,t),`\$`(t,2))+sigma*rho(z,t)$^{2}$*conjugate(rho(z,t))\\
\indent
-I*beta*rho(z,t)$^{2}$*conjugate(rho(z,t));

\begin{gather*}
\boxed{ \mathit{master\_1} := {\frac {\partial }{\partial z}}\,\rho (z,
\,t)=}  \\
\boxed{ \alpha \,\rho (z, \,t) - \gamma \,\rho (z, \,t) + I\,\phi \,
\rho (z, \,t) + \mathit{tf}^{2}\,({\frac {\partial ^{2}}{
\partial t^{2}}}\,\rho (z, \,t)) + I\,\mathit{k\_2}\,({\frac {
\partial ^{2}}{\partial t^{2}}}\,\rho (z, \,t))} \\
\boxed{ \mbox{} + \sigma \,\rho (z, \,t)^{2}\,\overline{(\rho (z, \,t))}
 - I\,\beta \,\rho (z, \,t)^{2}\,\overline{(\rho (z, \,t))}}
\end{gather*}

\emptyline
\noindent
Here
$\phi $ is the phase delay on the full cavity round trip,
$\alpha $ and
$\gamma $ are the gain and loss coefficients, respectively. The general exact
solution of this equation is not known, but there is the quasi-soliton
solution in the following form:

\emptyline
$>$f1 := (z,t)$->$rho0(z)*sech(t*tau(z))$^{1+I*psi(z)}$;

\emptyline
\[
\mathit{f1} := (z, \,t)\rightarrow \rho 0(z)\,\mathrm{sech}(t\,
\tau (z))^{(1 + I\,\psi (z))}
\]

\emptyline
\noindent
Here
$\rho $0 is the pulse amplitude,
$\tau $ is the inverse pulse width,
$\psi $ is the chirp. This solution obeys the condition of steady-state
propagation, when
${\frac {\partial }{\partial z}}\,\rho $ = 0, i.e. the pulse parameters are constant. To describe the
nonstationary pulse propagation, we shall use the aberrationless
approximation: the changes of the pulse parameters do not cause the
aberration of the pulse form. Next step is the substitution of
\textit{f1} into \textit{master\_1} with following expansion in
\textit{t} - series. The coefficients of the expansion produce the set
of ODE for the evaluating pulse parameters.

\emptyline
$>$assume(tau(z),real):\\
\indent \indent
  assume(t,real):\\
\indent

expand(lhs(subs(rho(z,t)=f1(z,t),master\_1))-rhs(subs(rho(z,t)=f1(z,t),\\
\indent \indent
master\_1))):\\
\indent \indent \indent
  eq :=\\
\indent \indent
subs(\\
\indent
\{alpha=alpha(z),diff(rho0(z),z)=a,diff(tau(z),z)=b,diff(psi(z),z)=c\},\\
\indent \indent
convert(series(\%,t=0,3),polynom)):\\
\indent \indent \indent
  assume(rho0(z),real):\\
\indent \indent \indent \indent
eq1 := evalc(coeff(eq,t$^{2}$)):\\
\indent \indent \indent
  eq2 := evalc(coeff(eq,t)):\\
\indent \indent
    eq3 := evalc(coeff(eq,t,0)):\\
\indent
       eq4 := coeff(eq1,I):\\
\indent \indent
          eq5 := coeff(eq1,I,0):\\
\indent \indent \indent
             eq6 := coeff(eq3,I):\\
\indent \indent
                eq7 := coeff(eq3,I,0):\\
\indent
solve(\{eq4=0,eq5=0,eq6=0,eq7=0\},\{a,b,c,phi\}):\\
\indent \indent
sys :=\\
\indent
diff(rho0(z),z)=subs(\%,a),diff(tau(z),z)=subs(\\
\indent
\%,b),diff(psi(z),z)=subs(\%,c):

\emptyline
\noindent
The obtained system \textit{sys} have to be supplemented with equation
for gain evolution:

\emptyline
$>$sys :=\\
\indent
\{\\
\indent
\%,diff(alpha(z),z)=alpha(z)*exp(-2*xi*rho0(z)$^{2}$/tau(z)-1/Tr-Pump)+\\
\indent
Pump*alphamx*(1-exp(-1/Tr-Pump))/(Pump+1/Tr)-alpha(z)\};

\emptyline
\begin{gather*}
\mathit{sys} :=  \left\{  \right. \!  \! {\frac {
\partial }{\partial \mathit{z\symbol{126}}}}\,\rho 0(\mathit{z
\symbol{126}})=  \\
\alpha (\mathit{z\symbol{126}})\,\rho 0(\mathit{z\symbol{126}})
 + \sigma \,\rho 0(\mathit{z\symbol{126}})^{3} - \gamma \,\rho 0(
\mathit{z\symbol{126}}) - \\
\mathit{tf}^{2}\,\rho 0(\mathit{z
\symbol{126}})\,\tau (\mathit{z\symbol{126}})^{2} + \mathit{k\_2}
\,\rho 0(\mathit{z\symbol{126}})\,\psi (\mathit{z\symbol{126}})\,
\tau (\mathit{z\symbol{126}})^{2},  \\
{\frac {\partial }{\partial \mathit{z\symbol{126}}}}\,\tau (
\mathit{z\symbol{126}})=\\
\sigma \,\rho 0(\mathit{z\symbol{126}})^{
2}\,\tau (\mathit{z\symbol{126}}) - 2\,\mathit{tf}^{2}\,\tau (
\mathit{z\symbol{126}})^{3} + 3\,\mathit{k\_2}\,\tau (\mathit{z
\symbol{126}})^{3}\,\psi (\mathit{z\symbol{126}}) + \mathit{tf}^{
2}\,\psi (\mathit{z\symbol{126}})^{2}\,\tau (\mathit{z
\symbol{126}})^{3},  \\
{\frac {\partial }{\partial \mathit{z\symbol{126}}}}\,\psi (
\mathit{z\symbol{126}})=\\
 - 2\,\mathit{tf}^{2}\,\psi (\mathit{z
\symbol{126}})\,\tau (\mathit{z\symbol{126}})^{2} - 4\,\mathit{
k\_2}\,\psi (\mathit{z\symbol{126}})^{2}\,\tau (\mathit{z
\symbol{126}})^{2} - 2\,\beta \,\rho 0(\mathit{z\symbol{126}})^{2
} - 4\,\mathit{k\_2}\,\tau (\mathit{z\symbol{126}})^{2} \\
\mbox{} - 2\,\psi (\mathit{z\symbol{126}})\,\sigma \,\rho 0(
\mathit{z\symbol{126}})^{2} - 2\,\mathit{tf}^{2}\,\psi (\mathit{z
\symbol{126}})^{3}\,\tau (\mathit{z\symbol{126}})^{2},  \\
{\frac {\partial }{\partial \mathit{z\symbol{126}}}}\,\alpha (
\mathit{z\symbol{126}})=\\
\alpha (\mathit{z\symbol{126}})\,e^{( - 2
\,\frac {\xi \,\rho 0(\mathit{z\symbol{126}})^{2}}{\tau (\mathit{
z\symbol{126}})} - \frac {1}{\mathit{Tr}} - \mathit{Pump})} +
{\displaystyle \frac {\mathit{Pump}\,\mathit{alphamx}\,(1 - e^{(
 - \frac {1}{\mathit{Tr}} - \mathit{Pump})})}{\mathit{Pump} +
{\displaystyle \frac {1}{\mathit{Tr}}} }}  - \alpha (\mathit{z
\symbol{126}}) \! \! \left.  \right\}
\end{gather*}

\emptyline \noindent Here \textit{Tr} is the gain relaxation time
normalized to the cavity period, \textit{Pump }is the
dimensionless pump (see part \textit{9}), $\xi $ is the inverse
gain saturation energy.

At first let's find the parameters of a steady-state
quasi-soliton, viz. the solution of \textit{sys }independent
on\textit{ z.}

\emptyline $>$f[1] :=\\
-2*tf$^2$*psi*Tau - 4*k\_2*psi$^2$*Tau - 4*k\_2*Tau -
2*tf$^2$*psi$^3$*Tau - 2*sigma*Phi *psi - 2*beta*Phi;\# Tau =
tau$^2$, Phi = rho$^2$\\

\indent $>$f[2] := alpha + sigma*Phi - gamma - tf$^2$*Tau +
k\_2*psi*Tau;\\

\indent $>$f[3] := 3*k\_2*Tau*psi + tf$^2$*psi$^2$*Tau -
2*tf$^2$*Tau + sigma*Phi;

\[
{f_{1}} :=  - 2\,\mathit{tf}^{2}\,\psi \,T - 4\,\mathit{k\_2}\,
\psi ^{2}\,T - 4\,\mathit{k\_2}\,T - 2\,\mathit{tf}^{2}\,\psi ^{3
}\,T - 2\,\sigma \,\Phi \,\psi  - 2\,\beta \,\Phi
\]

\[
{f_{2}} := \alpha  + \sigma \,\Phi  - \gamma  - \mathit{tf}^{2}\,
T + \mathit{k\_2}\,\psi \,T
\]

\[
{f_{3}} := 3\,\mathit{k\_2}\,\psi \,T + \mathit{tf}^{2}\,\psi ^{2
}\,T - 2\,\mathit{tf}^{2}\,T + \sigma \,\Phi
\]

\emptyline $>$sol1 := solve(\{f[2]=0,f[3]=0\},\{Tau,Phi\});\\
\indent
 $>$subs(\{Tau=subs(sol1,Tau),Phi=subs(sol1,Phi)\},f[1]):\\
 \indent \indent
  $>$simplify(\%):\\
  \indent \indent \indent
   $>$numer(\%)/2/(-alpha+gamma):\\
   \indent \indent \indent \indent
    $>$sol2 := solve(\%=0,psi);

\[
\mathit{sol1} := \{\Phi = - {\displaystyle \frac {(\alpha  -
\gamma )\,(3\,\mathit{k\_2}\,\psi  + \mathit{tf}^{2}\,\psi ^{2}
 - 2\,\mathit{tf}^{2})}{(2\,\mathit{k\_2}\,\psi  + \mathit{tf}^{2
}\,\psi ^{2} - \mathit{tf}^{2})\,\sigma }} , \,T={\displaystyle
\frac {\alpha  - \gamma }{2\,\mathit{k\_2}\,\psi  + \mathit{tf}^{
2}\,\psi ^{2} - \mathit{tf}^{2}}} \}
\]

\maplemultiline{ \mathit{sol2} := \\
{\displaystyle \frac {1}{2}}
\,{\displaystyle \frac { - 3\,\beta \,\mathit{k\_2} +
3\,\mathit{tf}^{2}\,\sigma
 + \sqrt{9\,\beta ^{2}\,\mathit{k\_2}^{2} - 2\,\beta \,\mathit{
k\_2}\,\mathit{tf}^{2}\,\sigma  + 9\,\mathit{tf}^{4}\,\sigma ^{2}
 + 8\,\beta ^{2}\,\mathit{tf}^{4} + 8\,\mathit{k\_2}^{2}\,\sigma
^{2}}}{\beta \,\mathit{tf}^{2} + \mathit{k\_2}\,\sigma }} ,  \\
{\displaystyle \frac {1}{2}} \,{\displaystyle \frac { - 3\,\beta
\,\mathit{k\_2} + 3\,\mathit{tf}^{2}\,\sigma  - \sqrt{9\,\beta ^{
2}\,\mathit{k\_2}^{2} - 2\,\beta \,\mathit{k\_2}\,\mathit{tf}^{2}
\,\sigma  + 9\,\mathit{tf}^{4}\,\sigma ^{2} + 8\,\beta ^{2}\,
\mathit{tf}^{4} + 8\,\mathit{k\_2}^{2}\,\sigma ^{2}}}{\beta \,
\mathit{tf}^{2} + \mathit{k\_2}\,\sigma }}  }

\emptyline Normalization of the time to \textit{tf} and the
intensity to $\beta $ allows simple plotting of the obtained
result:

\emptyline $>$plot(\{subs( \{beta=1,tf=1,sigma=1\},sol2[1]
),\\
subs(
\{beta=1,tf=1,sigma=1\},sol2[2])\},\\
k\_2=-100..10,axes=boxed,view=-10..1, title=`chirp vs. GDD`);

\begin{center}
\mapleplot{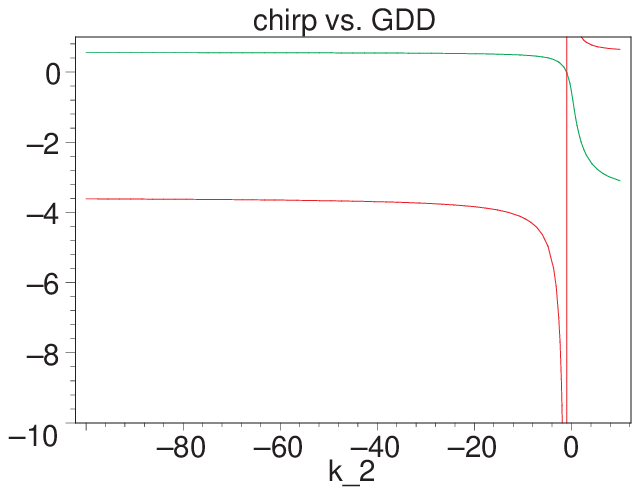}
\end{center}

\emptyline To plot the pulse duration and its intensity we have to
take into account the dependence of $\alpha $ on the pulse energy.
However for the sake of simplification we suppose
$\alpha$=const$<$ $\gamma$ (see previous section). Then

\emptyline $>$plot(subs( \{beta=1,tf=1,sigma=1,gamma=0.05,\\
alpha=0.04\},subs(psi=sol2[2],1/sqrt(subs(sol1,Tau)))),\\
k\_2=-100..10,axes=boxed, title=`pulse width vs. GDD`);

\begin{center}
\mapleplot{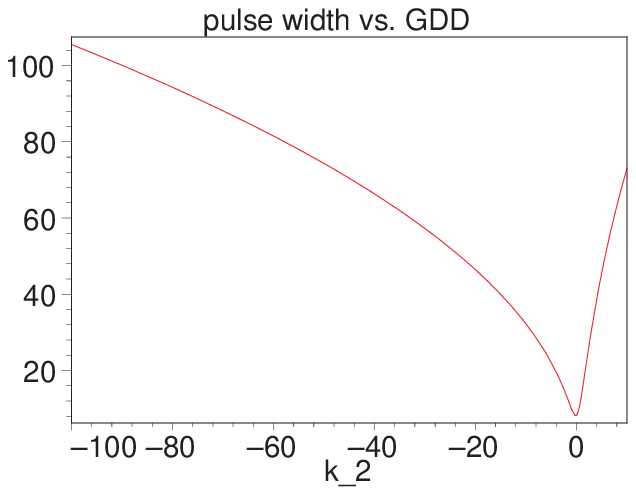}
\end{center}

\emptyline Only solution with $\psi $$>$0 in the region of the
negative GDD and $\psi $$<$0 for the positive GDD (green curve)
has a physical meaning because it corresponds to the positive
square of the pulse width 1/ $\tau $.

\fbox{\parbox{.8\linewidth} { The existence of the pulse duration
minimum in the vicinity of zero GDD pays a very important role in
the pulse shortening technique based on the creation of
appropriate negative GDD in the laser cavity.}}

\emptyline

\indent Now, we shall consider the evolution of ultrashort pulse
parameters on the basis of the obtained system of ODE. We suppose
to solve the system by the standard operator \textit{DEplot. }Let
normalize $\sigma $ and $\beta $ to 1,7* $10^{ - 12}$
$\mathit{cm}^{2}$/W, times to ${t_{f}}$ (2,5 fs for Ti: sapphire
laser), then $\xi $ = 0.0018. The fundamental step is the
assumption about saturation of the Kerr nonlinearity: $\sigma $ =
${\sigma _{0}}$(1 - $\frac {{\rho _{0}}^{2}\,{\sigma _{0}}}{2}$ ),
$\beta $ = ${\beta _{0}}$(1 - $\frac {{\rho _{0}}^{2}\,{\beta
_{0}}}{2}$ ), where ${\sigma _{0}}$ and ${\beta _{0}}$ are the
unsaturated nonlinear parameters.

\emptyline
$>$\#procedure for solution of the obtained system of ODE\\
\indent

ODE\_plot := proc(alphamx,gam,sigma0,beta0,Tr,Pump,xi,disp,tg,n)\\
\indent \indent
 sigma := sigma0*(1 - sigma0*rho0(z)$^{2}$/2 ):\\
\indent \indent \indent
  beta := beta0*(1 - beta0*rho0(z)$^{2}$/2 ):\\
\indent \indent \indent \indent
   sys := [D(alpha)(z) =\\
\indent
alpha(z)*exp(-2*xi*rho0(z)$^{2}$/tau(z)-1/Tr-Pump)+\\
\indent
Pump*alphamx*(1-exp(-1/Tr-Pump))/(Pump+1/Tr)-alpha(z), \\
\indent \indent
D(psi)(z) =\\
\indent
-4*disp*tau(z)$^{2}$-2*tg$^{2}$*psi(z)*tau(z)$^{2}$-4*disp*psi(z)$^{2}$*tau(z)$^{2}$-\\
\indent
2*beta*rho0(z)$^{2}$-2*beta*rho0(z)$^{2}$*psi(z)$^{2}$-2*tg$^{2}$*psi(z)$^{3}$*tau(z)$^{2}$, \\
\indent \indent
D(rho0)(z) =\\
\indent
sigma*rho0(z)$^{3}$-gam*rho0(z)-tg$^{2}$*rho0(z)*tau(z)$^{2}$+alpha(z)*rho0(z)+\\
\indent
disp*rho0(z)*psi(z)*tau(z)$^{2}$, \\
\indent \indent
D(tau)(z) =\\
\indent
-2*tg$^{2}$*tau(z)$^{3}$+sigma*rho0(z)$^{2}$*tau(z)+3*disp*tau(z)$^{3}$*psi(z)+\\
\indent
beta*rho0(z)$^{2}$*psi(z)*tau(z)+tg$^{2}$*psi(z)$^{2}$*tau(z)$^{3}$]:\\
\indent

DEplot(sys,[rho0(z),tau(z),psi(z),alpha(z)],z=0..n,[[rho0(0)=0.001,\\
\indent
tau(0)=0.01,alpha(0)=0,psi(0)=0]],stepsize=1,scene=[z,rho0(z)],\\
\indent
method=classical[foreuler],axes=FRAME,linecolor=BLACK):\\
\indent

end:

\emptyline
\noindent
The next figure demonstrates the autooscillations of pulse amplitude
(quasi-breezer behavior).

\emptyline
$>$display(ODE\_plot(0.5,0.05,10,1,300,0.0004,0.0018,-10,1,15000));

\emptyline
\begin{center}
\mapleplot{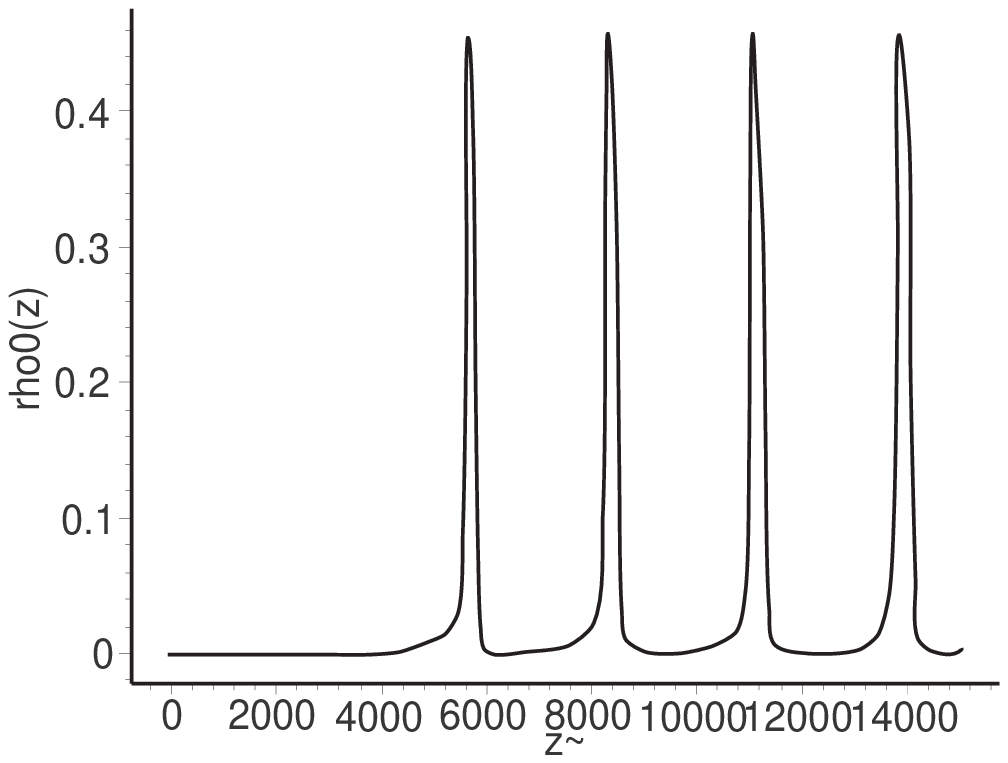}
\end{center}

\emptyline
\noindent
The character of the pulse evolution strongly depends on the
correlations between system's parameters. For example, next figure
demonstrates the pulse stabilization due to negative dispersion
growth.

\emptyline
$>$display(ODE\_plot(0.5,0.05,10,1,300,0.0004,0.0018,-20,1,15000));

\emptyline
\begin{center}
\mapleplot{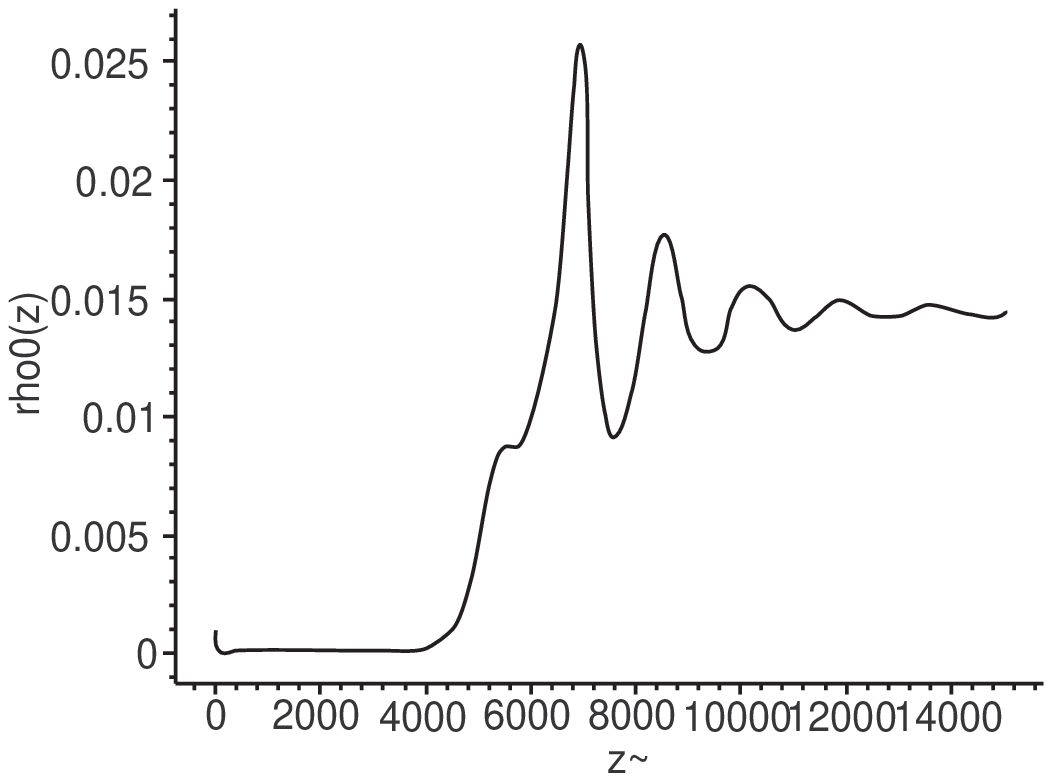}
\end{center}

\emptyline
\noindent
But the pumping growth produces the irregular autooscillations.

\emptyline
$>$ODE\_plot(0.5,0.05,10,1,300,0.00047,0.0018,-20,1,5000));

\emptyline
\begin{center}
\mapleplot{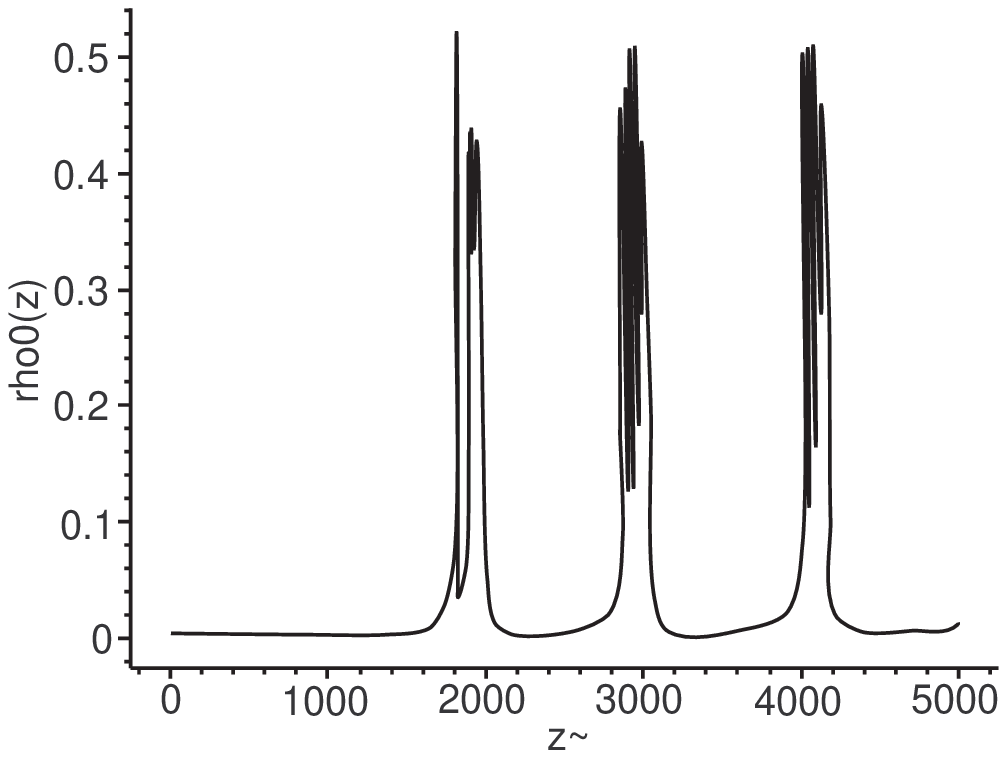}
\end{center}

\emptyline
\noindent
The next parameter's set produces the chaotic oscillations:

\emptyline
$>$display(ODE\_plot(0.5,0.05,1,0,300,0.0004,0.0018,0,1,8000));

\emptyline
\begin{center}
\mapleplot{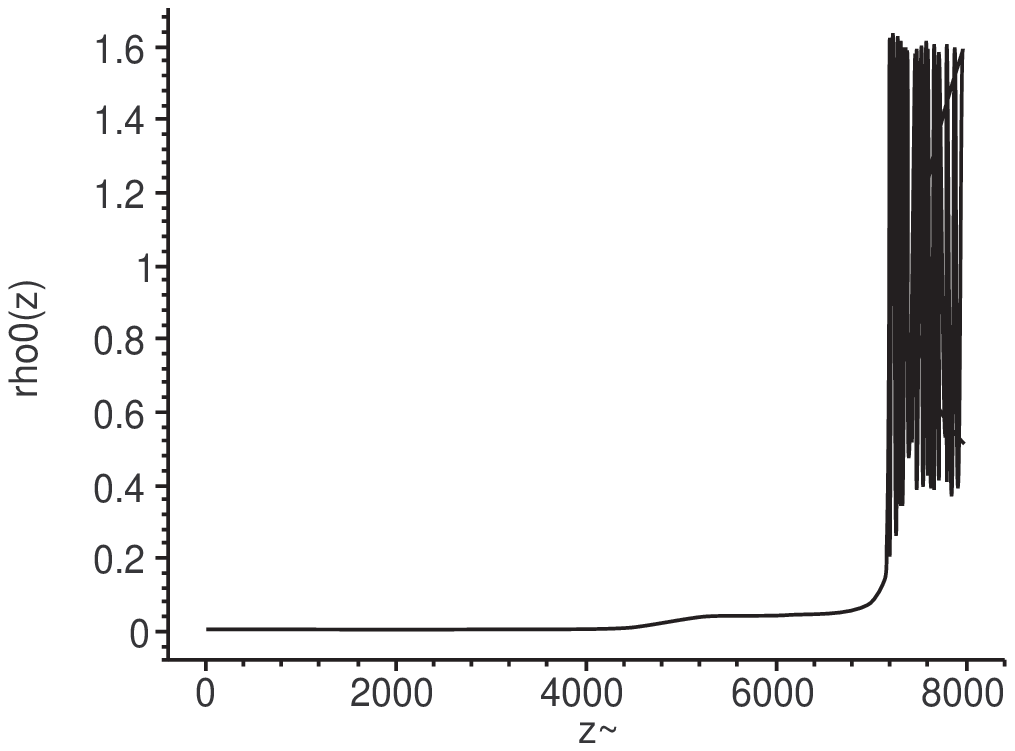}
\end{center}

\emptyline
The pulse parameters (amplitude and inverse pulse duration)
corresponding to irregular oscillations can be shown on the basis of
the iteration procedure, which realizes the direct Euler method for
the solution of the described above ODE system.

\emptyline
$>$\# Attention! This block can take a lot of CPU-time\\
\indent
iterations := proc(iter)\\
\indent
 alphamx := 0.5:\\
\indent \indent
   gam := 0.05:\\
\indent \indent \indent
     sigma0 := 1:\\
\indent \indent \indent \indent
       beta0 := 0:\\
\indent \indent \indent
         Tr := 300:\\
\indent \indent
            Pump := 0.0004:\\
\indent
         xi := 0.0018:\\
\indent \indent
       disp := 0:\\
\indent \indent \indent
     tg := 1:\\
\indent \indent \indent \indent
   rho0n:=0.001;\\
\indent \indent \indent
 taun:=0.01;\\
\indent \indent
    alphan:=0;\\
\indent
      psin:=0;\\
\indent \indent
 for m from 1 to iter do\\
\indent \indent \indent
    rho0old:=rho0n;\\
\indent \indent \indent \indent
       tauold:=taun;\\
\indent \indent \indent
       psiold:=psin;\\
\indent \indent
       alphaold:=alphan;\\
\indent
    sigma := evalhf(sigma0*(1 - sigma0*rho0old$^{2}$/2 )):\\
\indent \indent
       beta := evalhf(beta0*(1 - beta0*rho0old$^{2}$/2 )):\\
\indent \indent \indent
    alphan := evalhf(\\
\indent
alphaold*exp(-2*xi*rho0old$^{2}$/tauold-1/Tr-Pump)+\\
\indent
Pump*alphamx*(1-exp(-1/Tr-Pump))/(Pump+1/Tr));\\
\indent \indent
    psin :=\\
\indent
evalhf(psiold-4*disp*tauold$^{2}$-2*tg$^{2}$*psiold*tauold$^{2}$-\\
\indent
4*disp*psiold$^{2}$*tauold$^{2}$-2*beta*rho0old$^{2}$-\\
\indent
2*beta*rho0old$^{2}$*psiold$^{2}$-2*tg$^{2}$*psiold$^{3}$*tauold$^{2}$);\\
\indent \indent
    rho0n :=\\
\indent
evalhf(rho0old+sigma*rho0old$^{3}$-gam*rho0old-tg$^{2}$*rho0old*tauold$^{2}$\\
\indent
+alphaold*rho0old+disp*rho0old*psiold*tauold$^{2}$);\\
\indent \indent
    taun :=\\
\indent
evalhf(tauold-2*tg$^{2}$*tauold$^{3}$+sigma*rho0old$^{2}$*tauold+3*disp*tauold$^{3}$*\\
\indent
psiold+beta*rho0old$^{2}$*psiold*tauold+tg$^{2}$*psiold$^{2}$*tauold$^{3}$);\\
\indent \indent

if m = iter then pts := [rho0n, taun] fi;\\
\indent \indent \indent

    od;\\
\indent \indent \indent \indent
    pts\\
\indent \indent \indent \indent
 end:\\
\indent

PLOT(seq(POINTS(iterations(i)), i=9800 .. 10000), SYMBOL(POINT));

\emptyline
\begin{center}
\mapleplot{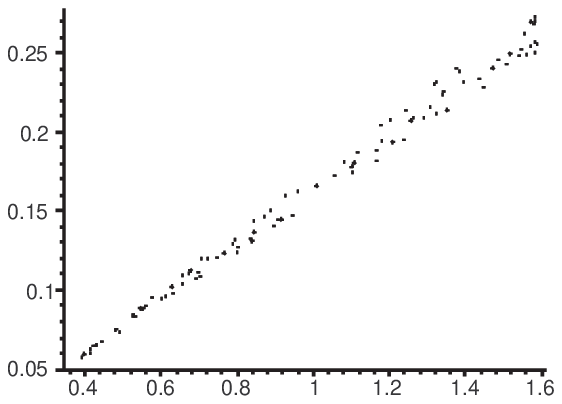}
\end{center}

\emptyline
\noindent
This is the so-called strange attractor, i. e. the chaotic attracting
manifold: the pulse parameters are changed chaotically but within the
limited range.

\emptyline Thus, the pulse oscillations in the Kerr-lens
mode-locked laser were analyzed (see \cite{Jasapara}, and
\docLink{http://xxx.lanl.gov/abs/physics/0009020}{arXiv:
physics/0009020}).

\begin{center}
\fbox{ \parbox{.8\linewidth}{
The oscillations accompany the negative
dispersion decrease and the pump growth and close connect with the
nonlinear factors in the system. The regular oscillation is the
analogue of the high-order soliton propagation and the irregular one
is the analogue of the nonlinear breezer}}
\end{center}

\section{Numerical approaches: ultrashort pulse spectrum, stability
and multipulsing}

\emptyline

\subsection{Kerr-lens mode-locked Cr:LiSGaF-laser with the Raman
self-scattering in active medium}

\emptyline The aberrationless approximation demonstrates the
stability loss in the vicinity of zero dispersions. It is
necessary to interpret the phenomenon. Moreover, we did not
consider a lot of the lasing factors, which affect the ultrashort
pulse dynamics in sub-100 fs region, viz. higher-order dispersion,
stimulated Raman self-scattering (see Part II), strong fast
absorber saturation etc. It is clear, that in order to take into
account these phenomena we need the numerical simulations beyond
the computer algebra abilities. However, Maple can help in the
preparation of the source code and in the interpretation of the
obtained results.

Now we describe the simplest generation model, which is highly
useful for the numerical simulations. This model is based on the
generalized Landau-Ginzburg equation:

\emptyline $>$restart:\\
\indent
 $>$with(codegen,fortran):\\
 \indent \indent
  $>$with('linalg'):\\
\indent \indent \indent
   $>$with(stats):\\
   \indent \indent \indent \indent
    $>$with(plots):\\

\indent \indent $>$master1 := diff(a(z,t),z) = \\
(alpha(z) - rho - gamma/(1+sigma*Phi(z,t)))*a(z,t) +\\
diff(a(z,t),t\$2) +
sum('(-I)$^{k+1}$*D[k]*diff(a(z,t),t\$k)','k'=2..N) -
I*Phi(z,t)*a(z,t);

\# Phi is the field intensity, gamma is the fast absorber
modulation depth, sigma is the inverse saturation intensity, we
used the normalization of the field to the self-phase modulation
coefficient and the time to the inverse gain bandwidth, D[k] are
the dispersion coefficients absorbed the (1/factorial(k)) factors

\emptyline
 \boxed{
\maplemultiline{ \mathit{master1} := {\frac {\partial }{\partial
z}}\,\mathrm{a}(z , \,t)=(\alpha (z) - \rho  - {\displaystyle
\frac {\gamma }{1 + \sigma \,\Phi (z, \,t)}} )\,\mathrm{a}(z, \,t)
+ ({\frac {
\partial ^{2}}{\partial t^{2}}}\,\mathrm{a}(z, \,t)) \\
\mbox{} +  \left(  \! {\displaystyle \sum _{k=2}^{N}} \,( - I)^{(
k + 1)}\,{\mathrm{D}_{k}}\,\mathrm{diff}(\mathrm{a}(z, \,t), \,t
\,\mathrm{\$}\,k) \!  \right)  - I\,\Phi (z, \,t)\,\mathrm{a}(z,
\,t) } }

\emptyline \noindent For the gain evolution we have:

\emptyline $>$master2 := diff(alpha(z),z) = P*(a - alpha(z))
-alpha(z)*Int(Phi(z,t),t=-T[cav]/2..T[cav]/2)/E[s] -
T[cav]*alpha(z)/T[r];

\emptyline
\[ \boxed{
\mathit{master2} := {\frac {\partial }{\partial z}}\,\alpha (z)=P
\,(a - \alpha (z)) - {\displaystyle \frac {\alpha (z)\,
{\displaystyle \int _{ - 1/2\,{T_{\mathit{cav}}}}^{1/2\,{T_{
\mathit{cav}}}}} \Phi (z, \,t)\,dt}{{E_{s}}}}  - {\displaystyle
\frac {{T_{\mathit{cav}}}\,\alpha (z)}{{T_{r}}}} }
\]

\emptyline Here \textit{a} is the maximal gain for the full
population inversion, \textit{P} is the dimensionless pump
(\textit{Pump} from the previous section), ${T_{\mathit{cav}}}$
and ${T_{r}}$ are the cavity period and the gain relaxation time,
respectively, ${E_{s}}$ is the gain saturation energy taking into
account the accepted normalization of time and intensity. The best
methods for the solution of the system (\textit{master1},
\textit{master2}) is the split-step Fourier method. Then in the
Fourier domain we escape to calculate the partial derivatives:

\emptyline $>$op(2,rhs(master1));\\
\indent
 $>$inttrans[fourier](\%, t, omega)*F(omega);\\
 \indent \indent
$>$print(`F(w) takes into account the shape factors for the
gainband and output coupler spectral profiles`);\\
\indent \indent \indent
   $>$subs( N=6,op(3,rhs(master1)) );\# we confine the maximal dispersion
order\\
\indent \indent
    $>$inttrans[fourier](evalc( \% ), t, omega):\\
    \indent
      $>$factor(\%);

\[
{\frac {\partial ^{2}}{\partial t^{2}}}\,\mathrm{a}(z, \,t)
\]

\[
 - \omega ^{2}\,\mathrm{fourier}(\mathrm{a}(z, \,t), \,t, \,
\omega )\,\mathrm{F}(\omega )
\]

\[
{\displaystyle \sum _{k=2}^{6}} \,( - I)^{(k + 1)}\,{\mathrm{D}_{
k}}\,\mathrm{diff}(\mathrm{a}(z, \,t), \,t\,\mathrm{\$}\,k)
\]

\[
 - I\,\omega ^{2}\,\mathrm{fourier}(\mathrm{a}(z, \,t), \,t, \,
\omega )\,({\mathrm{D}_{6}}\,\omega ^{4} + {\mathrm{D}_{5}}\,
\omega ^{3} + {\mathrm{D}_{4}}\,\omega ^{2} + {\mathrm{D}_{3}}\,
\omega  + {\mathrm{D}_{2}})
\]

\emptyline To define the parameters of the simulation we use the
experimental data for Cr:LiSGaF-laser kindly presented by Dr. I.T.
Sorokina and Dr. E. Sorokin and written in the corresponding
*.txt-files. NOTE! ALL DATA FILES HAVE TO BE PLACED IN YOUR
CURRENT DIRECTORY:

\emptyline $>$currentdir();

\[
\mbox{``G:$\backslash$$\backslash$Maple~6''}
\]

\emptyline We used the next experimental setup, which is typical
for the Kerr-lens mode-locked lasers (high reflective plane and
spherical (focal length f=5 cm) mirrors HR, chirped mirrors Ch,
output mirror OC):

\begin{center}
\mapleplot{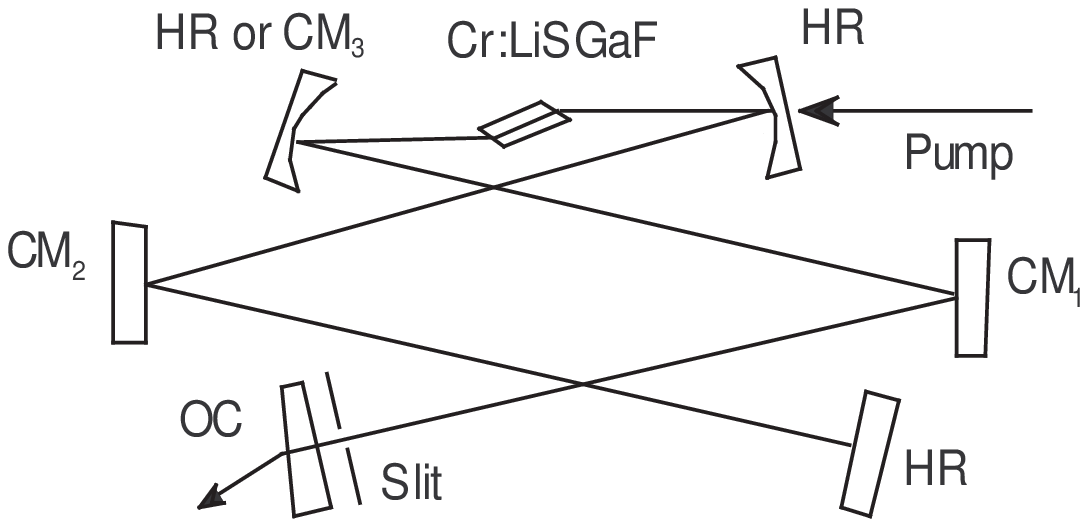}
\end{center}

\emptyline The first step is the analysis of the gain band, which
allows to define the gain bandwidth and the parameters of the time
normalization (\textit{tf} from the previous section) and the
frequency normalization (this is the gain bandwidth).

\emptyline $>$gain\_x := readdata(`gain\_x.dat`,1,float):\#
wavelength\\
\indent
 $>$gain\_y := readdata(`gain\_y.dat`,1,float):\# gain
 cross-section\\
 \indent \indent
     $>$n := vectdim(gain\_y):\\
     \indent \indent \indent
    $>$plot($[[gain\_x[k],gain\_y[k]]$
\$k=1..n],\\
color=red,view=0..3e-20,axes=boxed,title=`gain band profile`);\\
\indent \indent \indent \indent
   $>$max\_cross\_section := max(gain\_y[k] \$k=1..n);\#gain band
   maximum\\
   \indent \indent \indent
  $>$half\_cross\_section := evalf(max\_cross\_section/2);\#half of the gain
band maximum\\
\indent \indent
 $>$P := array(1..n):\\
 \indent
$>$Q := array(1..n):\\

 $>$for i from 2 to n do\\
 \indent
  $>$P[i] := evalf(2*Pi*3*1e10/(gain\_x[i]*1e-7)):\#transition from
wavelength to frequency\\
\indent \indent
   $>$Q[i] := [P[i],gain\_y[i]]:\\
   \indent \indent \indent
    $>$if gain\_y[i]=max\_cross\_section then X\_max := P[i] else
    fi:\\
    \indent \indent
   $>$if gain\_y[i]$>$half\_cross\_section and gain\_y[i-1]$<$half\_cross\_section
then X\_half\_1 :=  evalf((P[i-1]+P[i])/2) else fi:\\
\indent
  $>$if gain\_y[i]$<$half\_cross\_section and gain\_y[i-1]$>$half\_cross\_section
then X\_half\_2 := evalf((P[i-1]+P[i])/2) else fi:\\
 $>$od:
 \indent
$>$print(`position of the gain maximum:`);\\
\indent $>$X\_max;\\
\indent $>$print(`position of the first half of maximum:`);\\
\indent
 $>$X\_half\_1;\\
 \indent
$>$print(`position of the second half of maximum:`);\\
\indent
  $>$X\_half\_2;\\
  \indent
$>$print(`bandwidth:`);\\
\indent
   $>$bandwidth := evalf(abs(X\_half\_2-X\_half\_1));\\
   \indent
$>$print(`inverse bandwidth for Gaussian approximation [s]:`);\\
\indent
    $>$minimal\_pulse\_width:=evalf(4*ln(2)/bandwidth);\\
    \indent
$>$for i from 1 to n do\\
\indent
 $>$P[i] :=
 evalf((2*Pi*3*1e10/(gain\_x[i]*1e-7)-X\_max)/bandwidth):\\
 \indent
  $>$Q[i] := [P[i],gain\_y[i]]:\\
  \indent $>$od:\\
  \indent
   $>$j := 'j':\\
   \indent
     $>$plot([Q[j] \$j=1..n],axes=BOXED, color=red, title=`gain cross
section versus normalized frequency`,view=0..3e-20,color=blue);

\begin{center}
\mapleplot{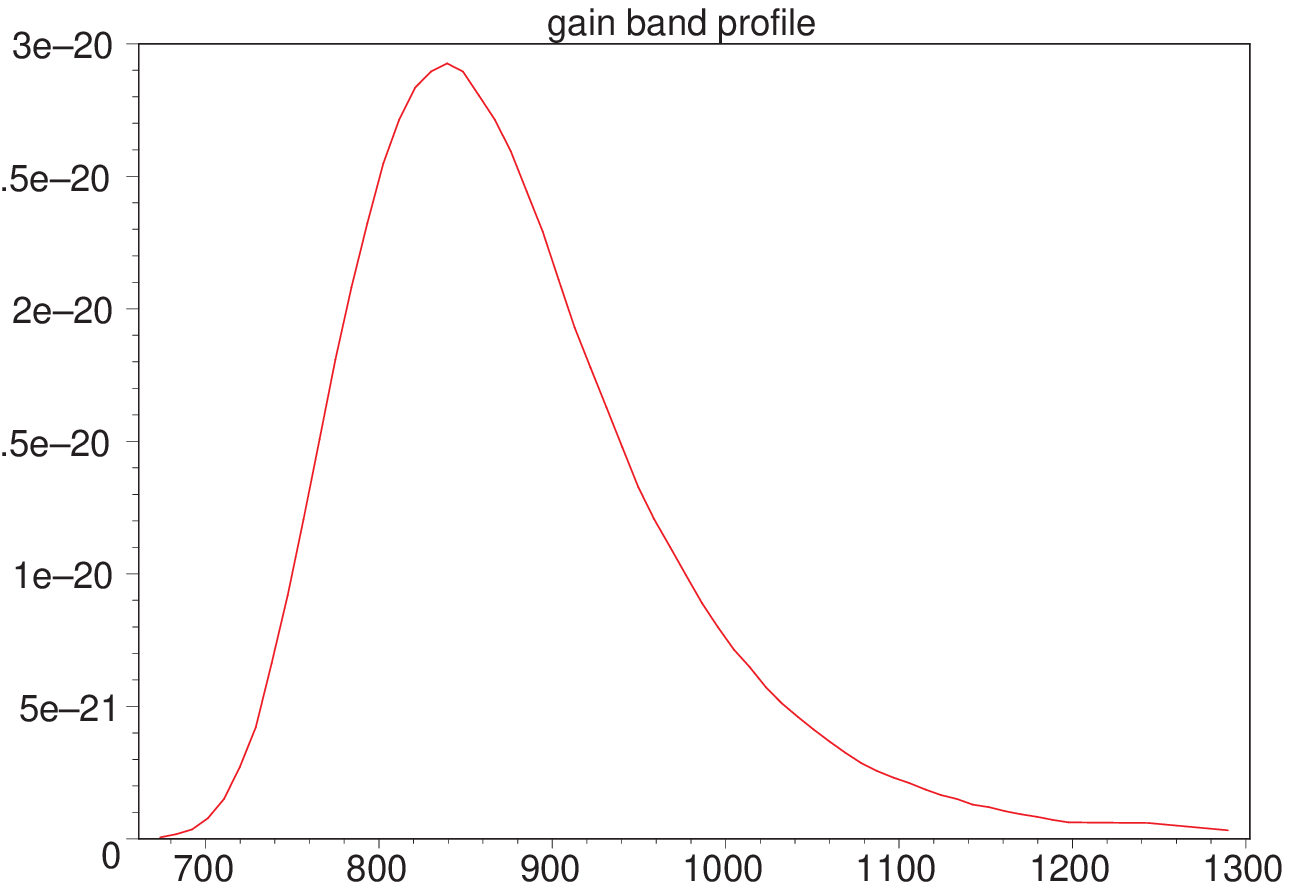}
\end{center}

\[
\mathit{max\_cross\_section} := .2926\,10^{-19}
\]

\[
\mathit{half\_cross\_section} := .1463000000\,10^{-19}
\]

\[
\mathit{position\ of\ the\ gain\ maximum:}
\]

\[
.2245909122\,10^{16}
\]

\[
\mathit{position\ of\ the\ first\ half\ of\ maximum:}
\]

\[
.2476565668\,10^{16}
\]

\[
\mathit{position\ of\ the\ second\ half\ of\ maximum:}
\]

\[
.1994706028\,10^{16}
\]

\[
\mathit{bandwidth:}
\]

\[
\mathit{bandwidth} := .481859640\,10^{15}
\]

\[
\mathit{inverse\ bandwidth\ for\ Gaussian\ approximation\ [s]:}
\]

\[
\mathit{minimal\_pulse\_width} := .5753934325\,10^{-14}
\]

\begin{center}
\mapleplot{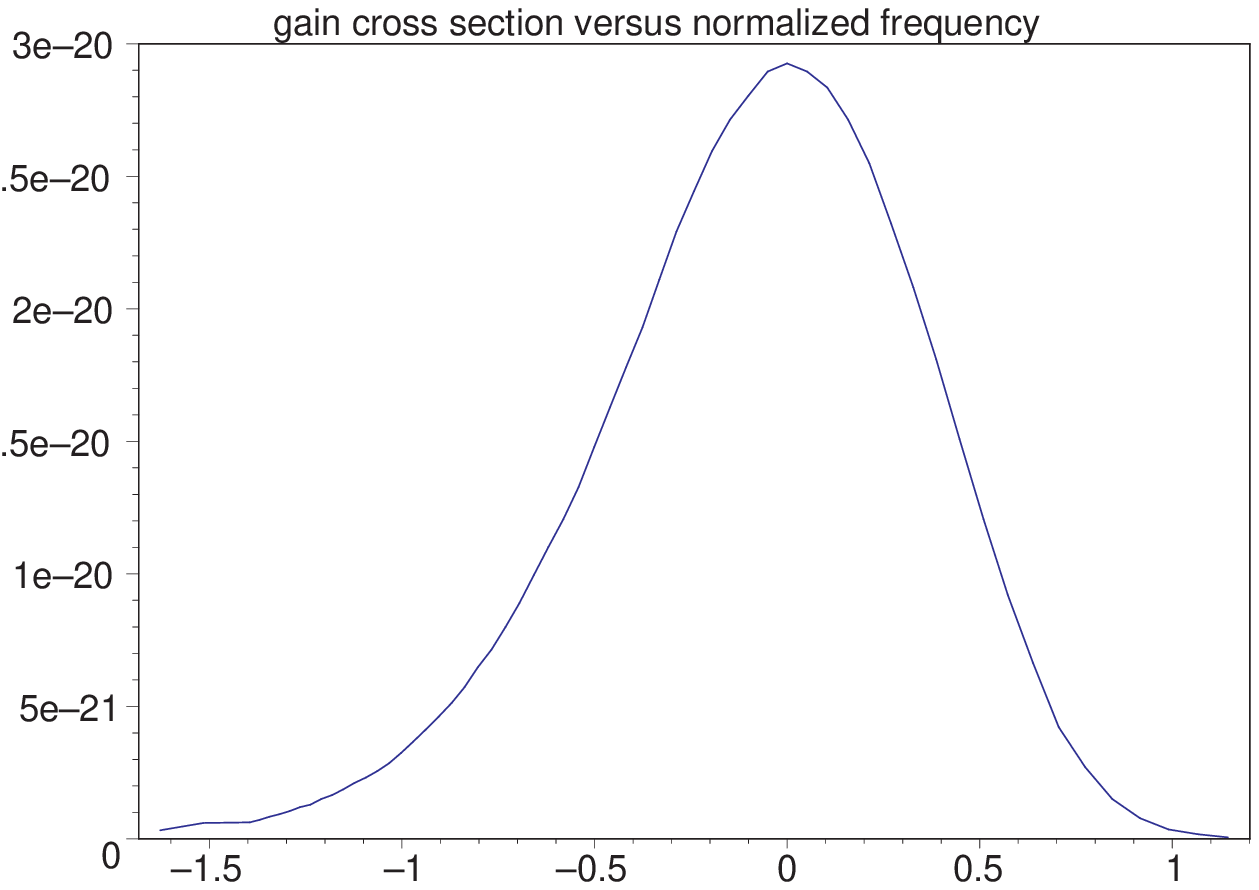}
\end{center}

\emptyline The similar manipulations are performed for the output
coupler. The fit-approximation gives the spectral shape factor for
the output mirror.

\emptyline $>$Out\_x := readdata(`out\_x.dat`,1,float):\\
\indent
 $>$Out\_y := readdata(`out\_y.dat`,1,float):\\
 \indent \indent
  $>$n := vectdim(Out\_y):\\
  \indent \indent \indent
   $>$plot($[[Out\_x[k],Out\_y[k]]$\$k=1..n],color=red,axes=boxed,\\
   title=`experimental reflection profile`);\\
   \indent \indent \indent \indent
    $>$P0 := array(1..n):\\
    \indent \indent \indent
     $>$Q0 := array(1..n):\\
     \indent \indent
      $>$for i from 1 to n do\\
      \indent
     $>$P0[i] :=\\
evalf((2*Pi*3*1e10/(Out\_x[i]*1e-7)-X\_max)/bandwidth):\#transition
from wavelength to frequency\\
\indent
   $>$Q0[i] := evalf(1-Out\_y[i]):\\
   \indent
 $>$od:\\
 \indent
$>$f1 := plot($[[P0[k],Q0[k]]$ \$k=1..n],color=red):\#experimental
output loss profile in the dimensionless frequency domain\\
\indent
 $>$eq0 := fit[leastsquare[[x,y],\\
y=a*x$^6$ + b*x$^5$ + c*x$^4$ + d*x$^3$ + e*x$^2$ + f*x +
g]]($[[P0[k] \$k=1..n], [Q0[k] \$k=1..n]]$);\# fit-approximation\\
\indent
  $>$Q0 := [subs(x=P0[k],rhs(eq0)) \$k=1..n]:\\
  \indent
    $>$f2 := plot($[[P0[k],Q0[k]]$ \$k=1..n],color=blue):\\
    \indent
     $>$display(f1,f2,axes=boxed,title=`comparison of the experimental
and approximated profiles`);

\begin{center}
\mapleplot{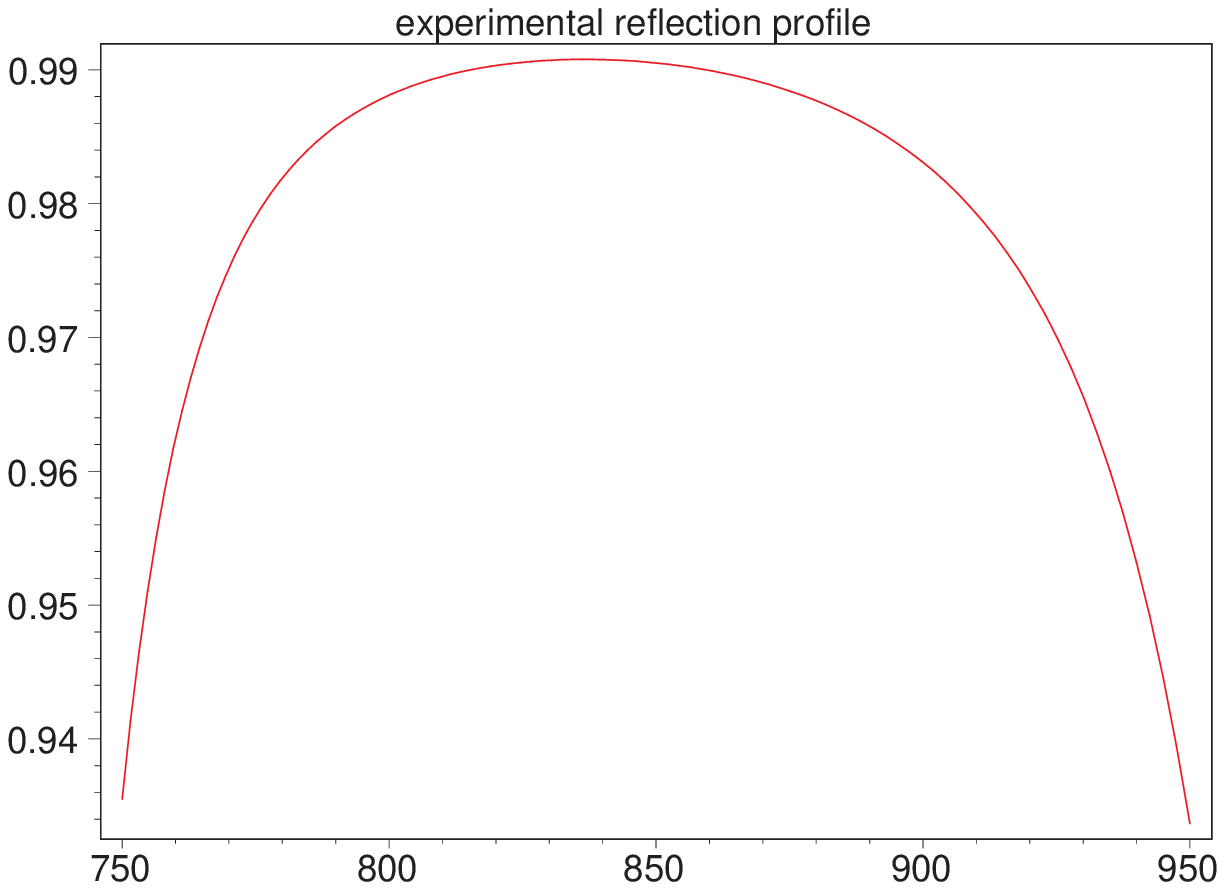}
\end{center}

\maplemultiline{ \mathit{eq0} := y=1.800575680\,x^{6} -
.05353754671\,x^{5} -
.1603422410\,x^{4} - .002641254098\,x^{3} \\
\mbox{} + .06981141739\,x^{2} - .001534877123\,x + .009038433676
 }

\begin{center}
\mapleplot{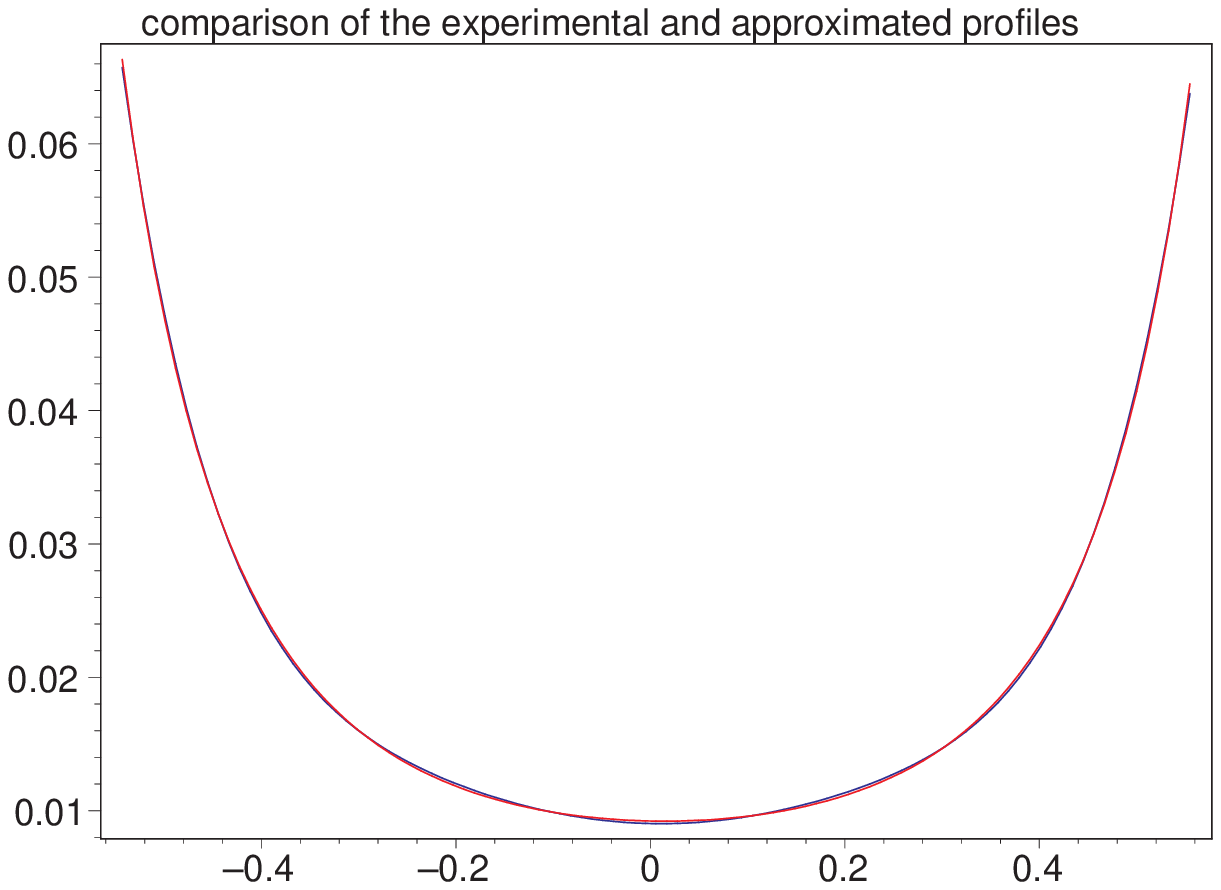}
\end{center}

\emptyline The first picture presents the experimental data, the
second results from the fit-approximation (\textit{eq0}) and the
transition to the dimensionless frequencies .

Now let's find the group-delay dispersion (GDD) in the active
medium from the measured data:

\emptyline $>$\#crystal length is 0.8 cm for double pass\\
\indent $>$GDD\_cry\_x :=
readdata(`gdd\_crystal\_x.dat`,1,float):\\
\indent
 $>$GDD\_cry\_y := readdata(`gdd\_crystal\_y.dat`,1,float):\\
 \indent
  $>$n := vectdim(GDD\_cry\_y):\\
  \indent
   $>$fig1 := plot($[[GDD\_cry\_x[k],0.8*GDD\_cry\_y[k]]$
\$k=1..n],\\ color=red,axes=boxed):\\
\indent \indent
    $>$display(fig1,title=`measured GDD in active crystal`);\\
    \indent \indent \indent
     $>$P1 := array(1..n):\\
     \indent \indent
      $>$Q1 := array(1..n):\\
      \indent
     $>$for i from 1 to n do\\
     \indent
    $>$P1[i] :=\\
evalf((2*Pi*3*1e10/(GDD\_cry\_x[i]*1e-7)-X\_max)/bandwidth):\#
transition\\
 from wavelength to frequency\\
\indent
  $>$od:\\
  \indent
 $>$f1 := plot([[P1[k],0.8*(bandwidth*10$^{-15}$)**2*GDD\_cry\_y[k]]\\
\$k=1..n],color=red):\\
 \indent $>$eq1 := fit[leastsquare[[x,y],
y=a*x$^2$+b*x+c]]([[P1[k] \$k=1..n],\\
evalf(0.8*(bandwidth*10$^{-15}$)**2)*GDD\_cry\_y]);\#
fit-approximation\\
\indent
 $>$Q1 := [subs(x=P1[k],rhs(eq1)) \$k=1..n]:\\
 \indent
  $>$f2 := plot([[P1[k],Q1[k]] \$k=1..n],color=blue,axes=boxed):\\
  \indent
   $>$display(f1,f2,title=`GDD fit-approximation in frequency
domain`);

\begin{center}
\mapleplot{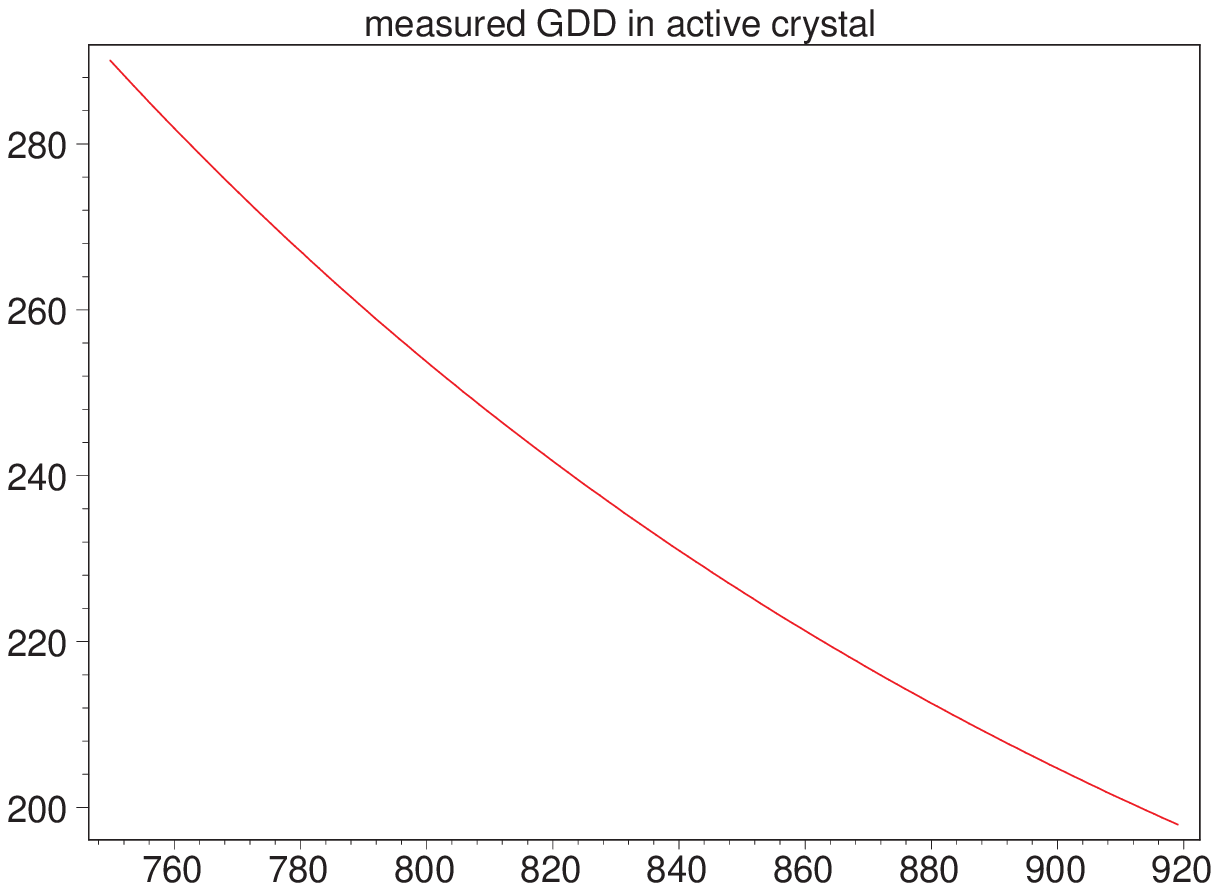}
\end{center}

\[
\mathit{eq1} := y=5.557558914\,x^{2} + 21.42668994\,x +
53.71773838
\]

\begin{center}
\mapleplot{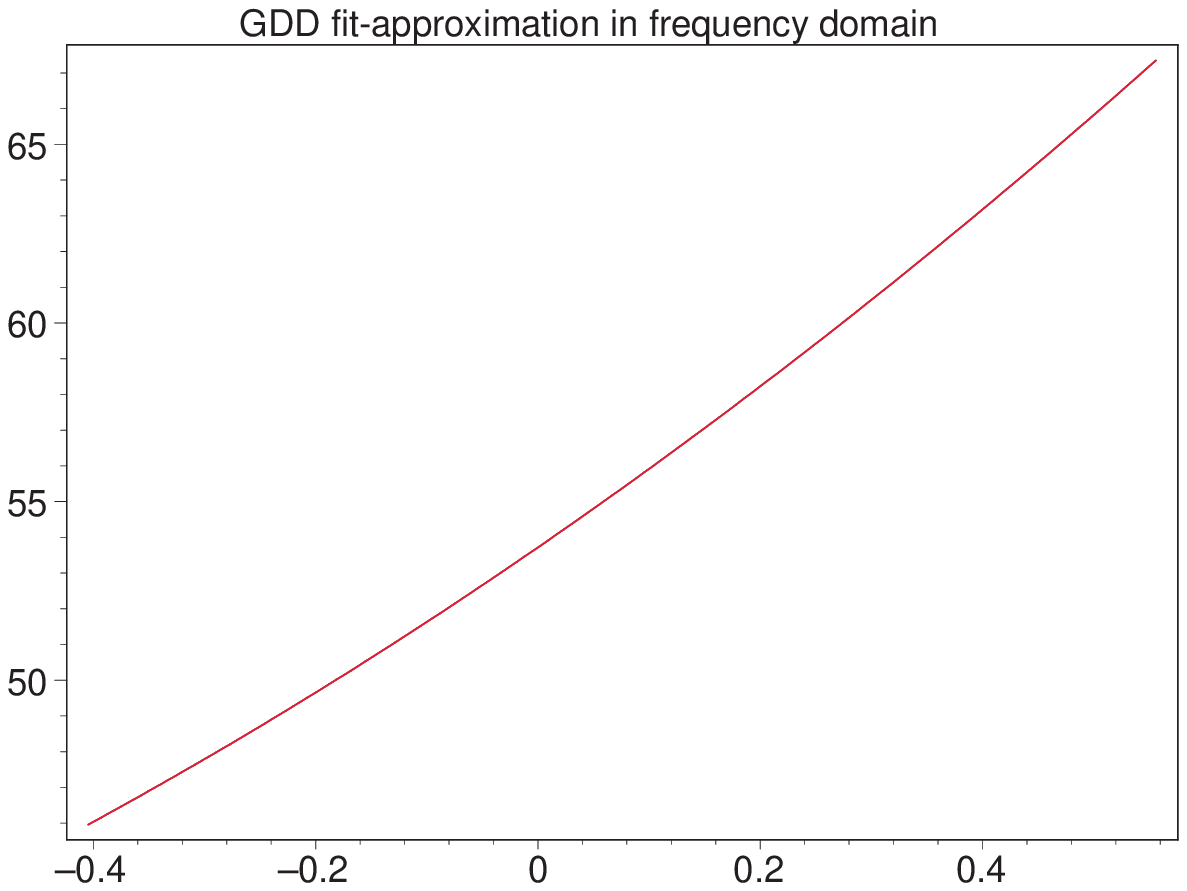}
\end{center}

\emptyline \noindent \textit{eq1} is the result of the
fit-approximation.

The main devices for the GDD manipulation are the chirp mirrors.
Their parameters are:

\emptyline $>$\#Ch1 (double pass)\\
\indent $>$GDD\_Ch1\_x := readdata(`gdd\_ch1\_x.dat`,1,float):\\
\indent
 $>$GDD\_Ch1\_y := readdata(`gdd\_ch1\_y.dat`,1,float):\\
 \indent
  $>$n := vectdim(GDD\_Ch1\_y):\\
  \indent
   $>$plot([[GDD\_Ch1\_x[k],2*GDD\_Ch1\_y[k]] \$k=1..n],\\
   color=red,axes=boxed,title=`measured GDD for Ch1`);\\
   \indent
     $>$P2 := array(1..n-4):\\
     \indent \indent
      $>$Q2 := array(1..n-4):\\
      \indent \indent \indent
     $>$for i from 1 to n-4 do\\
     \indent \indent
    $>$P2[i] :=\\
evalf((2*Pi*3*1e10/(GDD\_Ch1\_x[i]*1e-7)-X\_max)/bandwidth):\#transition\\
from wavelength to frequency\\
\indent
   $>$Q2[i] := 2*(bandwidth*10$^{-15}$)**2*GDD\_Ch1\_y[i] od:\\
   \indent
  $>$f1 := plot([[P2[k],Q2[k]] \$k=1..n-4],color=red):\\
  \indent
 $>$eq2 := fit[leastsquare[[x,y],\\
y=a*x$^6$+b*x$^5$+c*x$^4$+d*x$^3$+e*x$^2$+f*x+g]]([[P2[k]
\$k=1..n-4], [Q2[k] \$k=1..n-4]]);\\
\indent $>$Q2 := [subs(x=P2[k],rhs(eq2)) \$k=1..n-4]:\\
\indent
 $>$f2 := plot([[P2[k],Q2[k]] \$k=1..n-4],color=blue):\\
 \indent
  $>$display(f1,f2,axes=boxed,title=`fit-approximation in frequency
domain`);

\begin{center}
\mapleplot{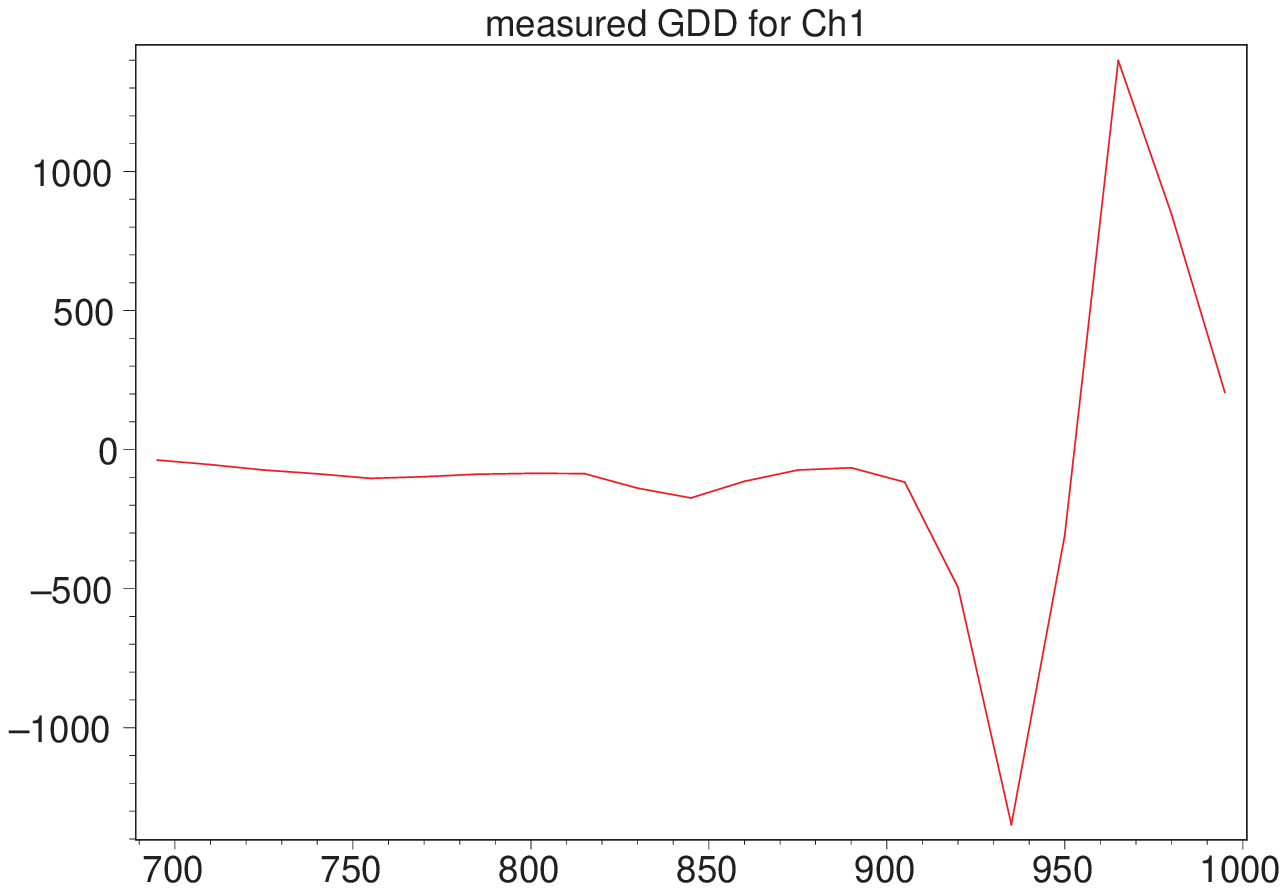}
\end{center}

\maplemultiline{ \mathit{eq2} := y= - 4487.256771\,x^{6} +
8899.324767\,x^{5} -
4466.632623\,x^{4} - 612.3714657\,x^{3} \\
\mbox{} + 692.5781343\,x^{2} - 17.99769657\,x - 37.66103148 }

\begin{center}
\mapleplot{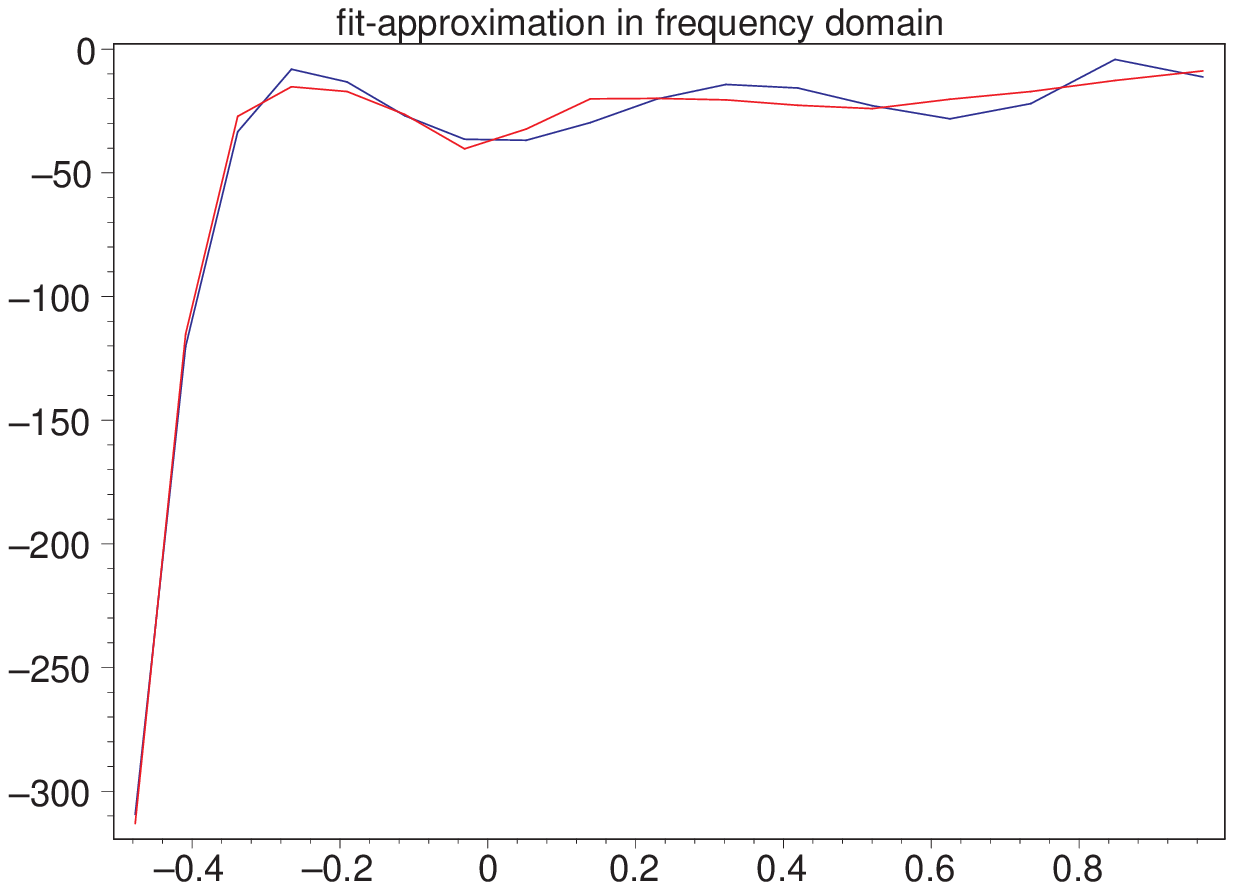}
\end{center}

\emptyline $>$\#Ch2 (four passes)\\
\indent
 $>$GDD\_Ch2\_x := readdata(`gdd\_ch2\_x.dat`,1,float):\\
 \indent
 $>$GDD\_Ch2\_y := readdata(`gdd\_ch2\_y.dat`,1,float):\\
 \indent
  $>$n := vectdim(GDD\_Ch2\_y):\\
  \indent
   $>$plot([[GDD\_Ch2\_x[k],4*GDD\_Ch2\_y[k]] \$k=1..n],\\
color=red,axes=boxed,title=`measured GDD for Ch2`);\\
\indent \indent
     $>$P3 := array(1..n):\\
     \indent \indent \indent
      $>$Q3 := array(1..n):\\
      \indent \indent \indent \indent
       $>$for i from 1 to n do\\
       \indent \indent \indent
      $>$P3[i] :=\\
evalf((2*Pi*3*1e10/(GDD\_Ch2\_x[i]*1e-7)-X\_max)/bandwidth):\#transition\\
from wavelength to frequency\\
\indent \indent
     $>$od:\\
\indent
    $>$f1 :=
    plot([[P3[k],4*(bandwidth*10$^{-15}$)**2*GDD\_Ch2\_y[k]]\\
\$k=1..n],color=red):\\
\indent
   $>$eq3 := fit[leastsquare[[x,y],\\
y=a*x$^6$+b*x$^5$+c*x$^4$+d*x$^3$+e*x$^2$+f*x+g]]([[P3[k] \$k=1..n],\\
evalf(4*(bandwidth*10$^{-15}$)**2)*GDD\_Ch2\_y]);\\
\indent
  $>$Q3 := [subs(x=P3[k],rhs(eq3)) \$k=1..n]:\\
  \indent
 $>$f2 := plot([[P3[k],Q3[k]] \$k=1..n],color=blue):\\
 \indent
$>$display(f1,f2,axes=boxed,title=`fit-approximation in frequency
domain`);

\begin{center}
\mapleplot{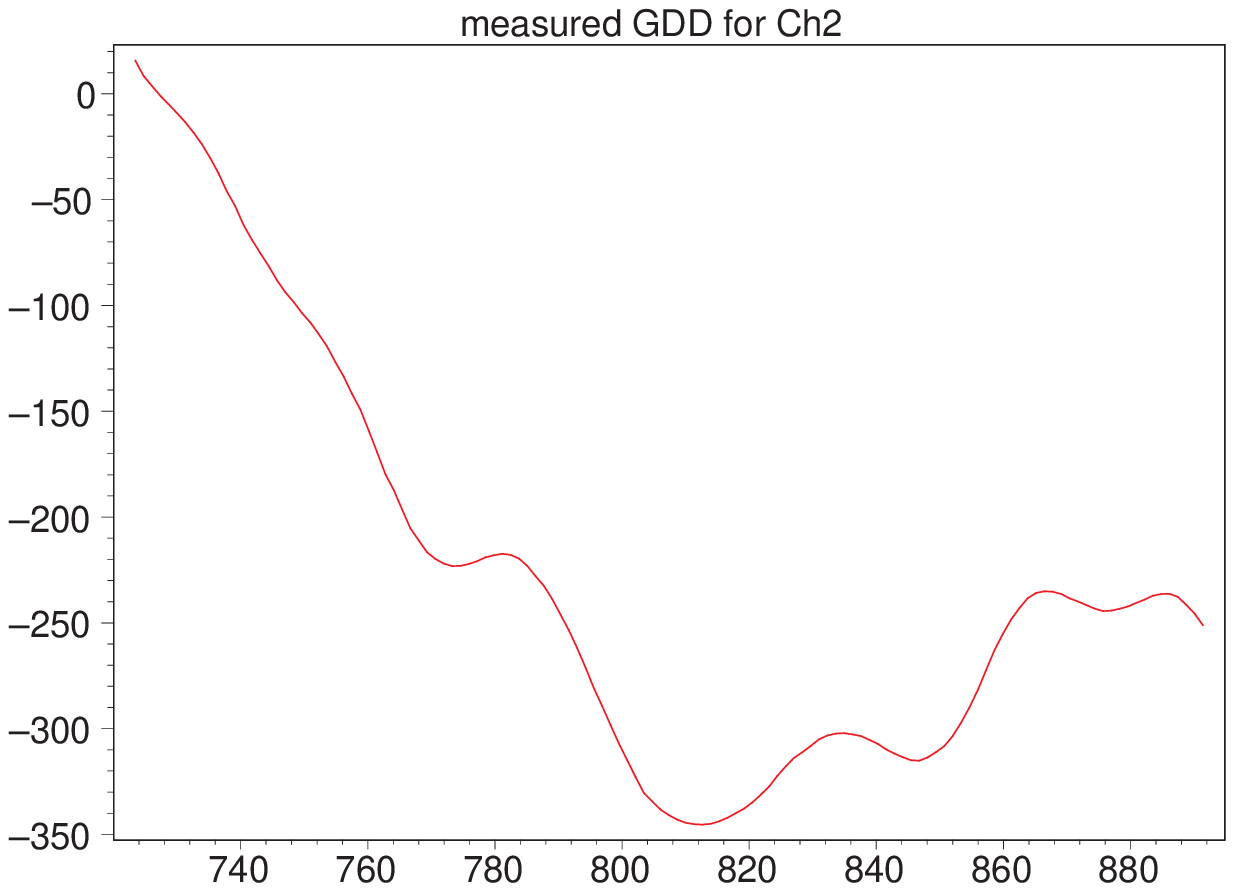}
\end{center}

\maplemultiline{ \mathit{eq3} := y= - 2051.277573\,x^{6} +
4260.310370\,x^{5} -
3057.192323\,x^{4} + 565.9468050\,x^{3} \\
\mbox{} + 385.5186662\,x^{2} - 74.89842760\,x - 73.37050894 }

\begin{center}
\mapleplot{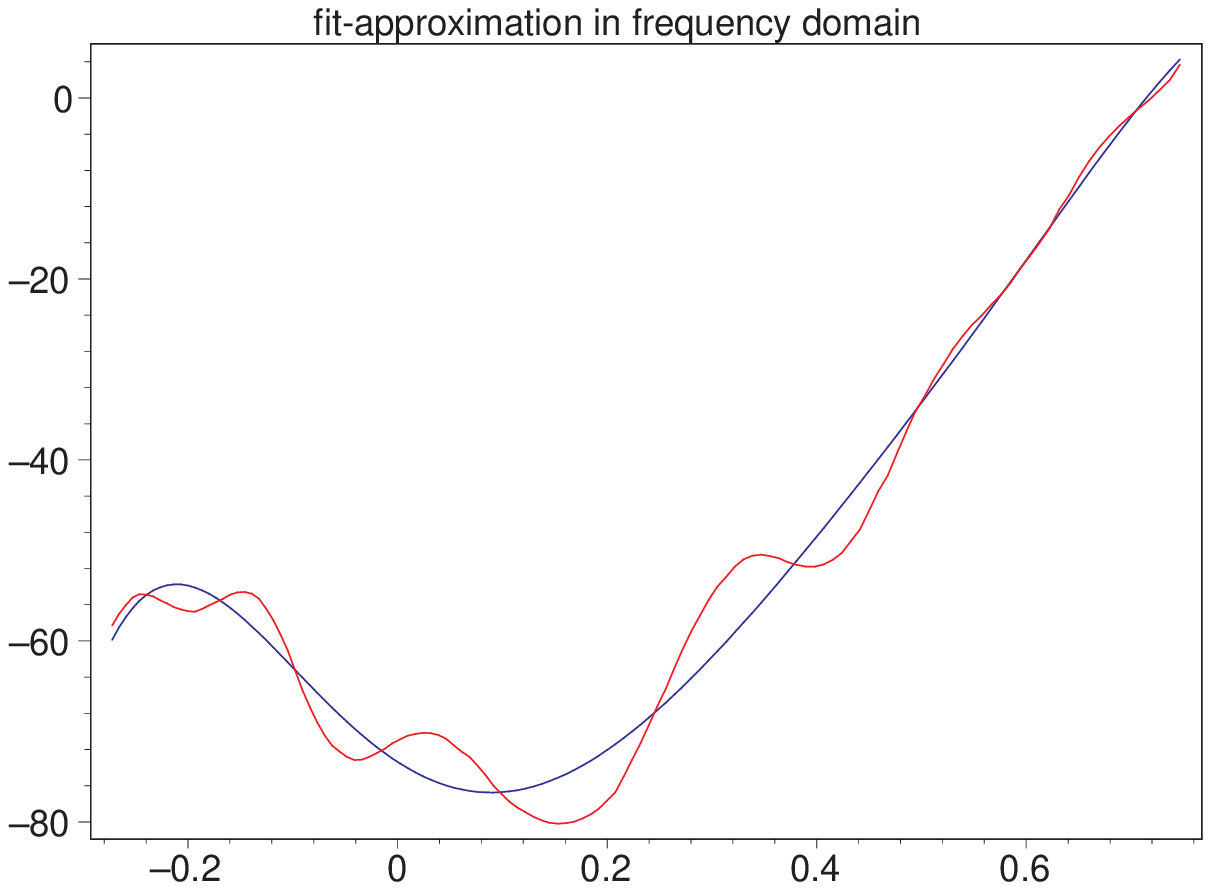}
\end{center}

\emptyline And, at last, the GDD in output coupler and
high-reflective mirrors:

\emptyline $>$\#Oc (single pass)\\
\indent
 $>$GDD\_Oc\_x := readdata(`gdd\_oc\_x.dat`,1,float):\\
 \indent
  $>$GDD\_Oc\_y := readdata(`gdd\_oc\_y.dat`,1,float):\\
  \indent
   $>$n := vectdim(GDD\_Oc\_y):\\
   \indent
    $>$plot([[GDD\_Oc\_x[k],GDD\_Oc\_y[k]]
\$k=1..n],\\
color=red,axes=boxed,title=`measured GDD for OC`);\\
\indent \indent
      $>$P4 := array(1..n):\\
      \indent \indent \indent
       $>$Q4 := array(1..n):\\
       \indent \indent \indent \indent
      $>$for i from 1 to n do\\
      \indent \indent \indent
     $>$P4[i] :=\\
evalf((2*Pi*3*1e10/(GDD\_Oc\_x[i]*1e-7)-X\_max)/bandwidth):\#transition\\
from wavelength to frequency\\
\indent \indent
    $>$od:\\
    \indent
   $>$f1 := plot([[P4[k],GDD\_Oc\_y[k]*(bandwidth*10$^{-15}$)**2]
\$k=1..n],\\
color=red):\\
\indent
  $>$eq4 := fit[leastsquare[[x,y],\\
y=a*x$^6$+b*x$^5$+c*x$^4$+d*x$^3$+e*x$^2$+f*x+g]]\\
([[P4[k] \$k=1..n],
GDD\_Oc\_y*evalf((bandwidth*10$^{-15}$)**2)]);\\
\indent
 $>$Q4 := [subs(x=P4[k],rhs(eq4)) \$k=1..n]:\\
 \indent
$>$f2 := plot([[P4[k],Q4[k]] \$k=1..n],color=blue):\\
\indent $>$display(f1,f2,axes=boxed,title=`fit-approximation in
frequency domain`);

\mapleresult
\begin{center}
\mapleplot{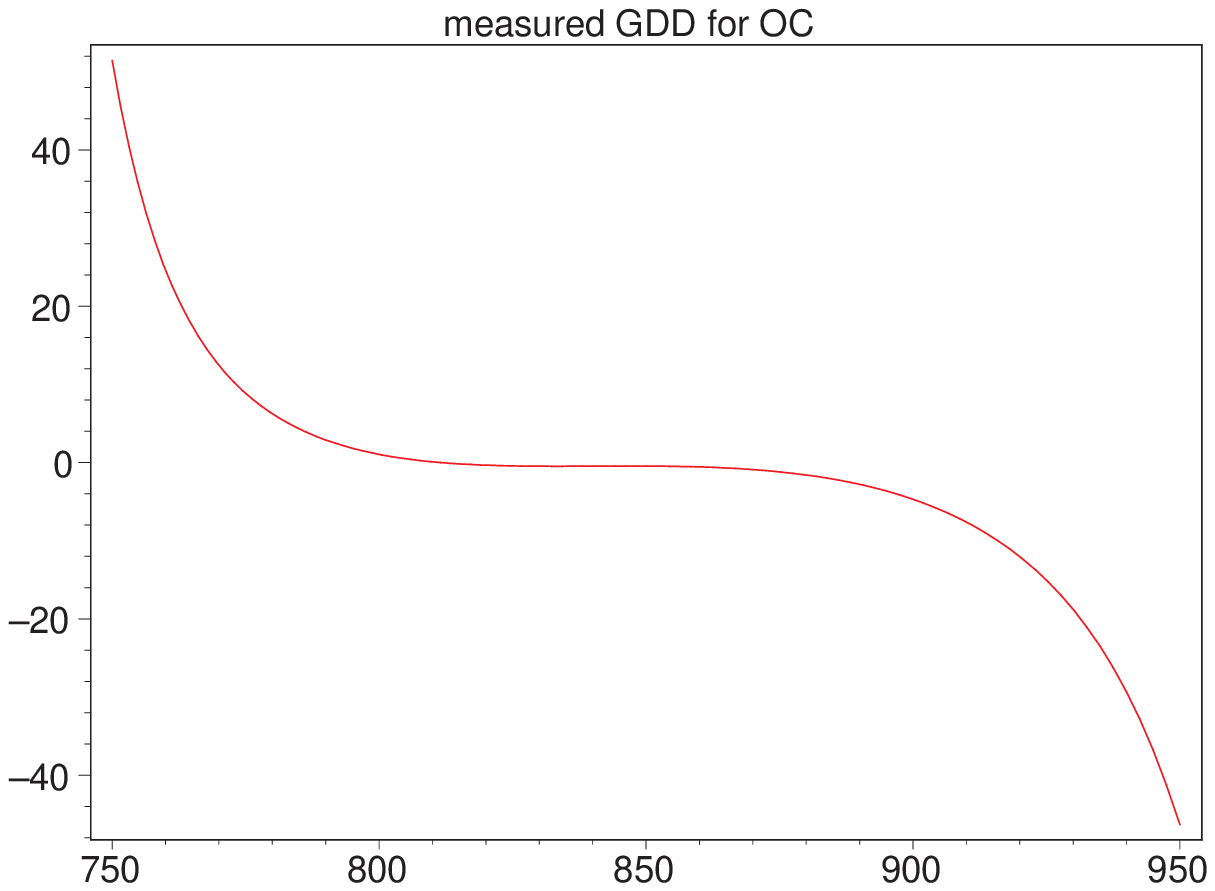}
\end{center}

\maplemultiline{ \mathit{eq4} := y=40.56991627\,x^{6} +
226.5828840\,x^{5} -
15.88361104\,x^{4} - 6.140885675\,x^{3} \\
\mbox{} + 1.492992122\,x^{2} + 1.204668103\,x - .1174906766 }

\begin{center}
\mapleplot{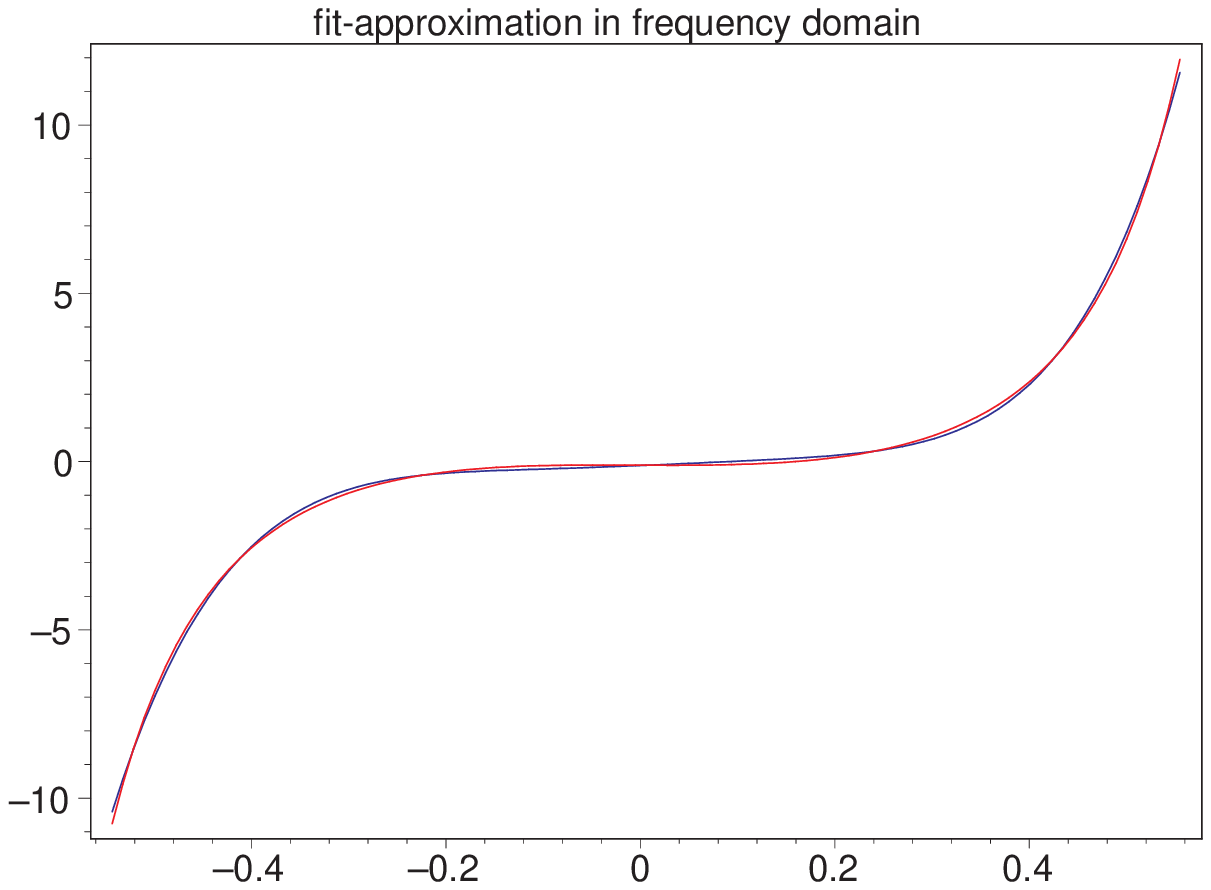}
\end{center}

\emptyline $>$\#HR (four passes)\\
\indent $>$GDD\_Hr\_x := readdata(`gdd\_hr\_x.dat`,1,float):\\
\indent
 $>$GDD\_Hr\_y := readdata(`gdd\_hr\_y.dat`,1,float):\\
 \indent \indent
  $>$n := vectdim(GDD\_Hr\_y):\\
  \indent \indent \indent
   $>$plot([[GDD\_Hr\_x[k],GDD\_Hr\_y[k]]\\
\$k=1..n],color=red,axes=boxed,title=`measured GDD for HR`);\\
\indent \indent
     $>$P5 := array(1..n):\\
     \indent
      $>$Q5 := array(1..n):\\
      \indent \indent
       $>$for i from 1 to n do\\
       \indent \indent \indent
      $>$P5[i] :=\\
evalf((2*Pi*3*1e10/(GDD\_Hr\_x[i]*1e-7)-X\_max)/bandwidth):\#transition\\
from wavelength to frequency\\
\indent \indent
     $>$od:\\
     \indent
    $>$f1 :=
    plot([[P5[k],GDD\_Hr\_y[k]*4*(bandwidth*10$^{-15}$)**2]\\
\$k=1..n],color=red):\\
\indent
   $>$eq5 := fit[leastsquare[[x,y],\\
y=a*x$^6$+b*x$^5$+c*x$^4$+d*x$^3$+e*x$^2$+f*x+g]]([[P5[k]\\
\$k=1..n], GDD\_Hr\_y*evalf(4*(bandwidth*10$^{-15}$)**2)]);\\
\indent
  $>$Q5 := [subs(x=P5[k],rhs(eq5)) \$k=1..n]:\\
  \indent \indent
 $>$f2 := plot([[P5[k],Q5[k]] \$k=1..n],color=blue):\\
 \indent
$>$display(f1,f2,axes=boxed,title=`fit-approximation in frequency
domain`);

\mapleresult
\begin{center}
\mapleplot{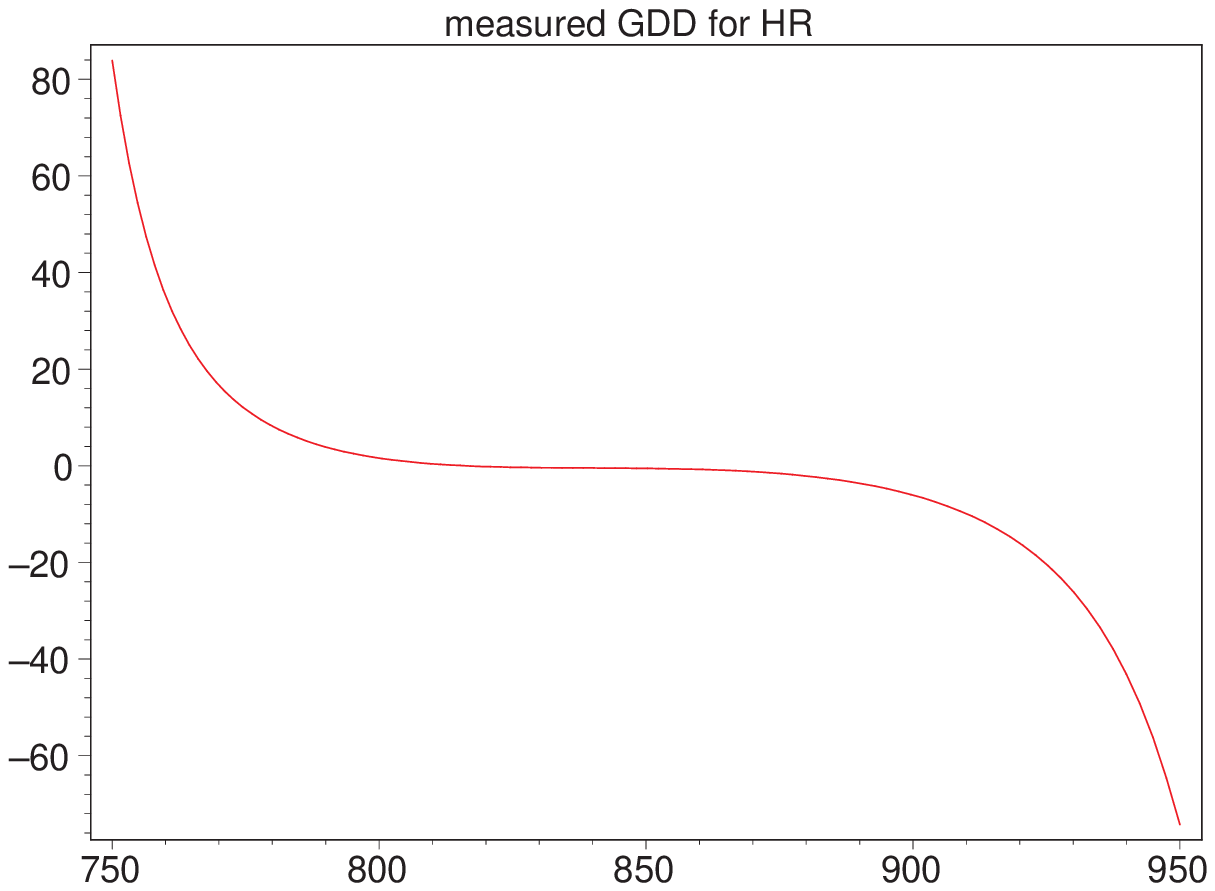}
\end{center}

\maplemultiline{ \mathit{eq5} := y=371.6378105\,x^{6} +
1859.815396\,x^{5} -
146.0722578\,x^{4} - 186.9742695\,x^{3} \\
\mbox{} + 12.76215250\,x^{2} + 13.36697439\,x - .5561160074 }

\begin{center}
\mapleplot{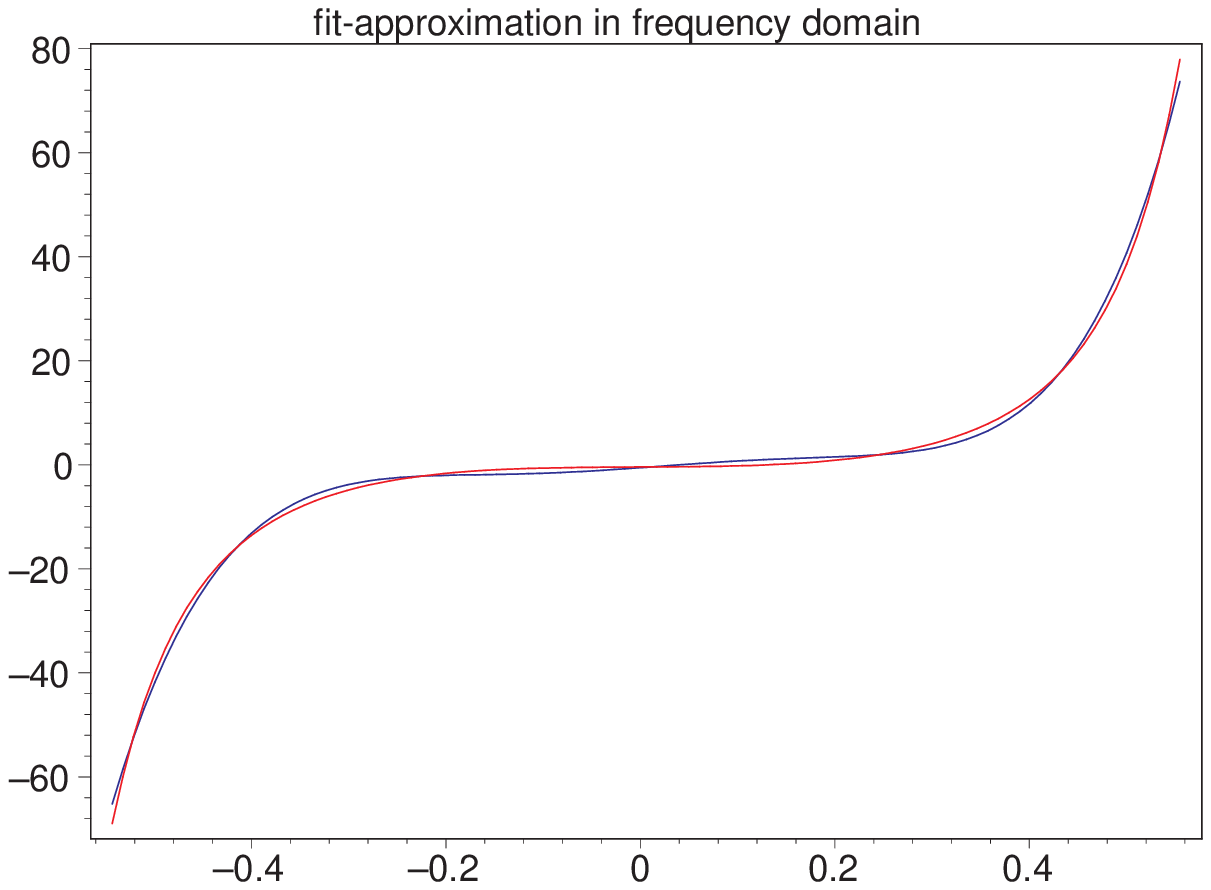}
\end{center}

\emptyline Now, as a result of the obtained fit-approximations, we
have the normalized net-GDD with corresponding FORTRAN-code for
simulation:

\emptyline $>$end\_res :=\\
evalf(rhs(eq1)+rhs(eq2)+rhs(eq3)+rhs(eq4)+rhs(eq5));\\
\indent
 $>$plot(\%,x=-0.45..0.8,axes=boxed,title=`normalized to tf
 net-GDD`);\\
 \indent \indent
  $>$codegen[fortran](end\_res);

\maplemultiline{ \mathit{end\_res} :=  - 57.98740873 -
6126.326617\,x^{6} +
15246.03342\,x^{5} - 7685.780815\,x^{4} \\
\mbox{} - 239.5398159\,x^{3} + 1097.909504\,x^{2} - 56.89779174\,
x }

\begin{center}
\mapleplot{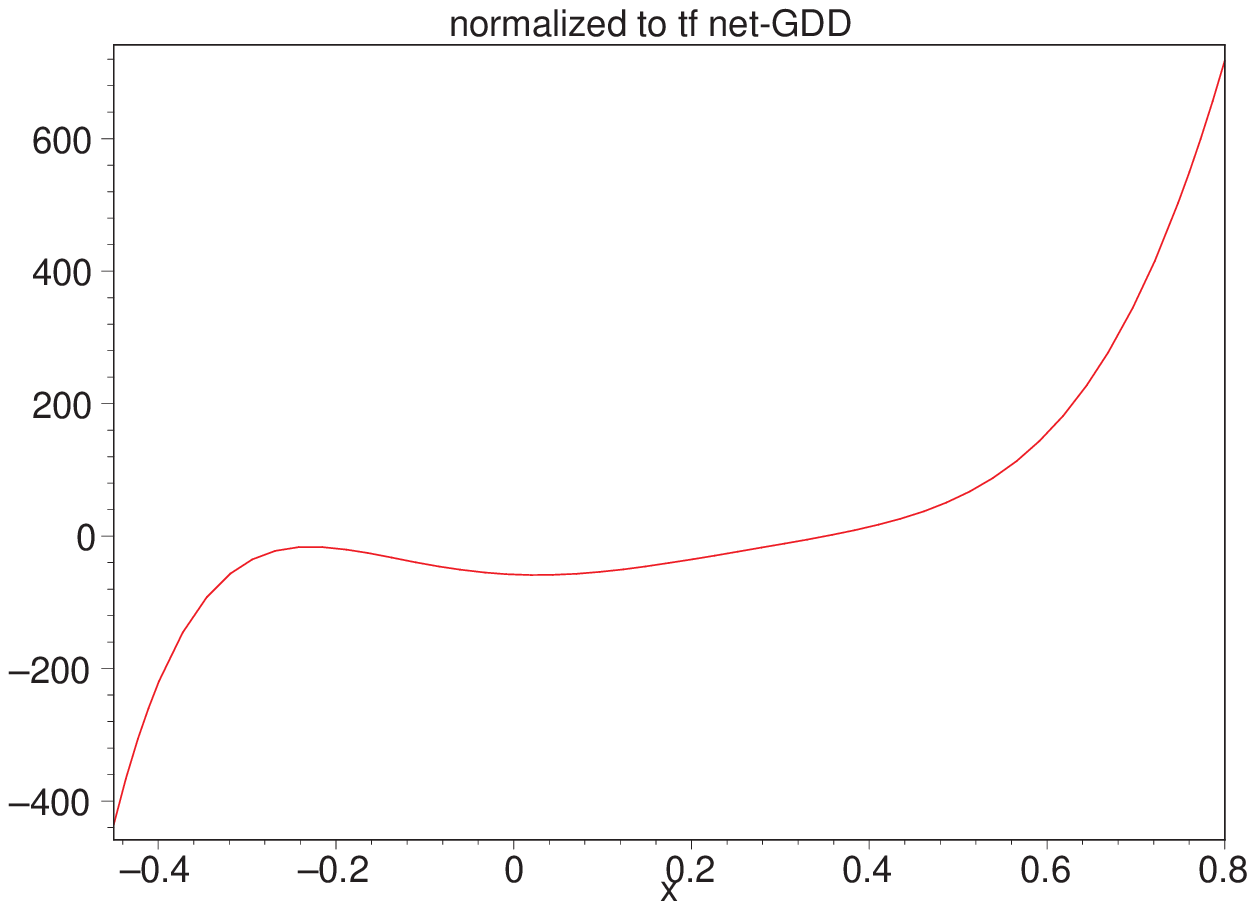}
\end{center}

\begin{maplettyout}
      t0 =
-0.5798741E2-0.6126327E4*x**6+0.1524603E5*x**5-0.7685781E4*x*
\end{maplettyout}

\begin{maplettyout}
     #*4-0.2395398E3*x**3+0.109791E4*x**2-0.5689779E2*x
\end{maplettyout}

\emptyline \noindent Or in the usual co-ordinates:

\emptyline $>$P6 := array(1..100):\\
\indent
 $>$Q6 := array(1..100):\\
 \indent
  $>$for i from 1 to 100 do\\
  \indent
   $>$P6[i] := 2*Pi*3e10*1e7/(bandwidth*(-0.4+0.8*i/100)+X\_max):\#
   from\\
frequency to wavelength [in nm]\\
\indent
    $>$Q6[i] :=\\
evalf(subs(x=-0.4+0.8*i/100,end\_res/(bandwidth*1e-15)$^2$)):\\
\indent
     $>$od:\\
     \indent
      $>$plot([[P6[k],Q6[k]] \$k=1..100],\\
      color=red,axes=BOXED,title=`net-GDD [fs$^2$] vs.
wavelength [nm]`);

\begin{center}
\mapleplot{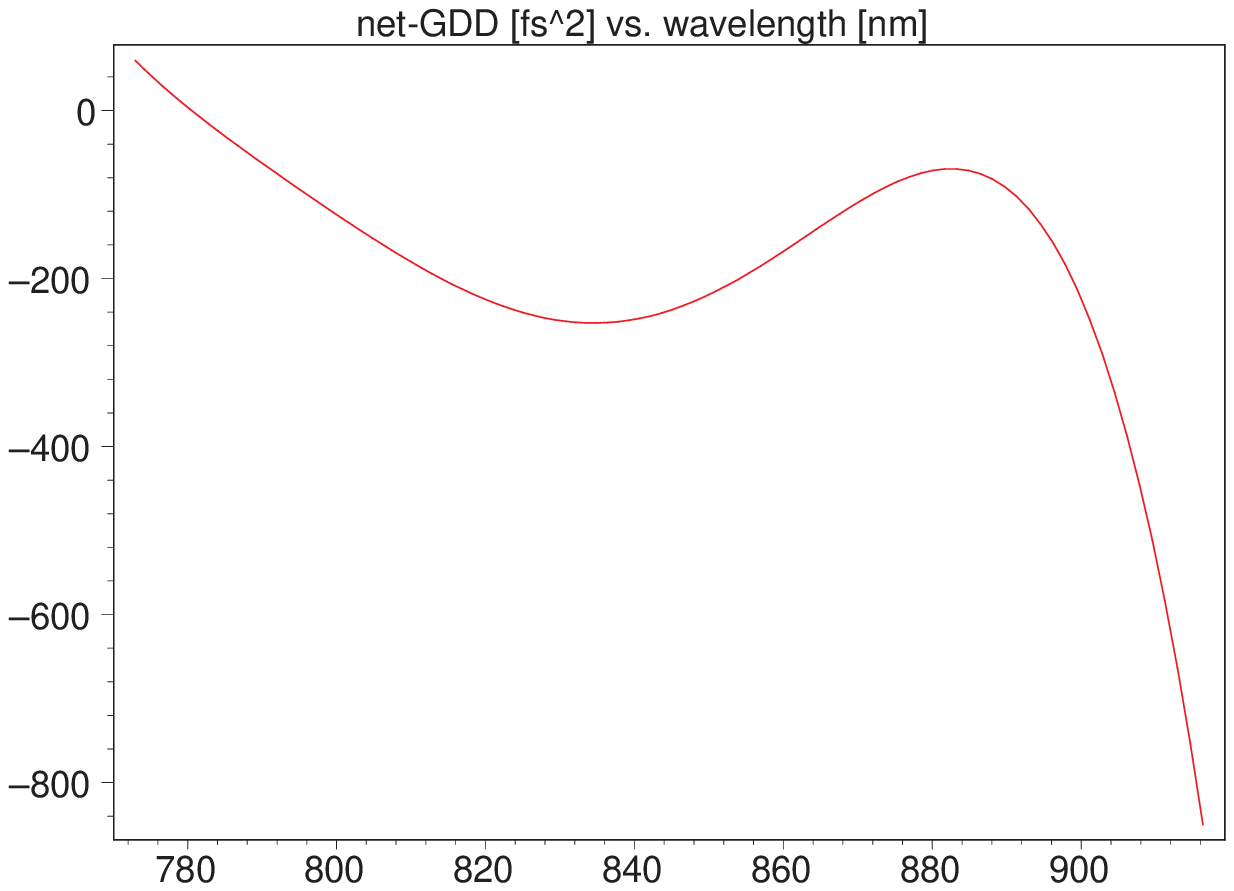}
\end{center}

\emptyline The experimentally observed ultrashort-pulse spectrum
is:

\emptyline $>$Spec\_x := readdata(`LiSGaF\_x.dat`,1,float):\\
\indent
 $>$Spec\_y := readdata(`LiSGaF\_y.dat`,1,float):\\
 \indent \indent
  $>$n := vectdim(Spec\_y):\\
  \indent \indent \indent
   $>$plot([[Spec\_x[k],Spec\_y[k]]\\
\$k=1..n],color=red,axes=boxed,title=`experimental spectrum (a.u.
vs. wavelength)`);\\
\indent \indent \indent \indent
    $>$P5 := array(1..n):\\
    \indent \indent \indent
     $>$Q5 := array(1..n):\\
     \indent \indent
    $>$for i from 1 to n do\\
    \indent
   $>$P5[i] :=\\
evalf((2*Pi*3*1e10/(Spec\_x[i]*1e-7)-X\_max)/bandwidth):\#transition\\
from wavelength to frequency\\
\indent
  $>$Q5[i] := Spec\_y[i]:\\
  \indent
 $>$od:\\
 \indent
$>$plot([[P5[k],Q5[k]]
\$k=1..n],\\
color=green,axes=boxed,title=`experimental spectrum (a.u. vs/
dimensionless frequency)`);

\begin{center}
\mapleplot{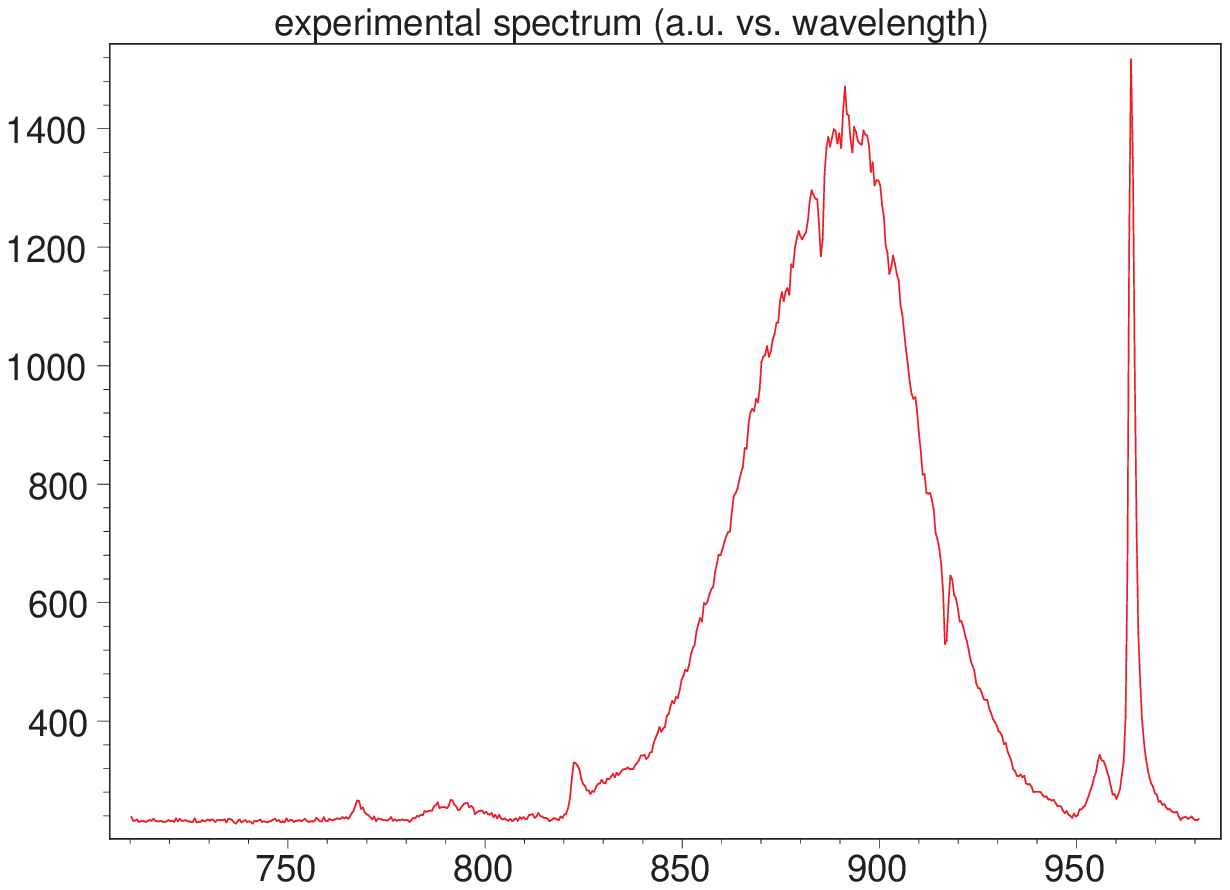}
\end{center}

\begin{center}
\mapleplot{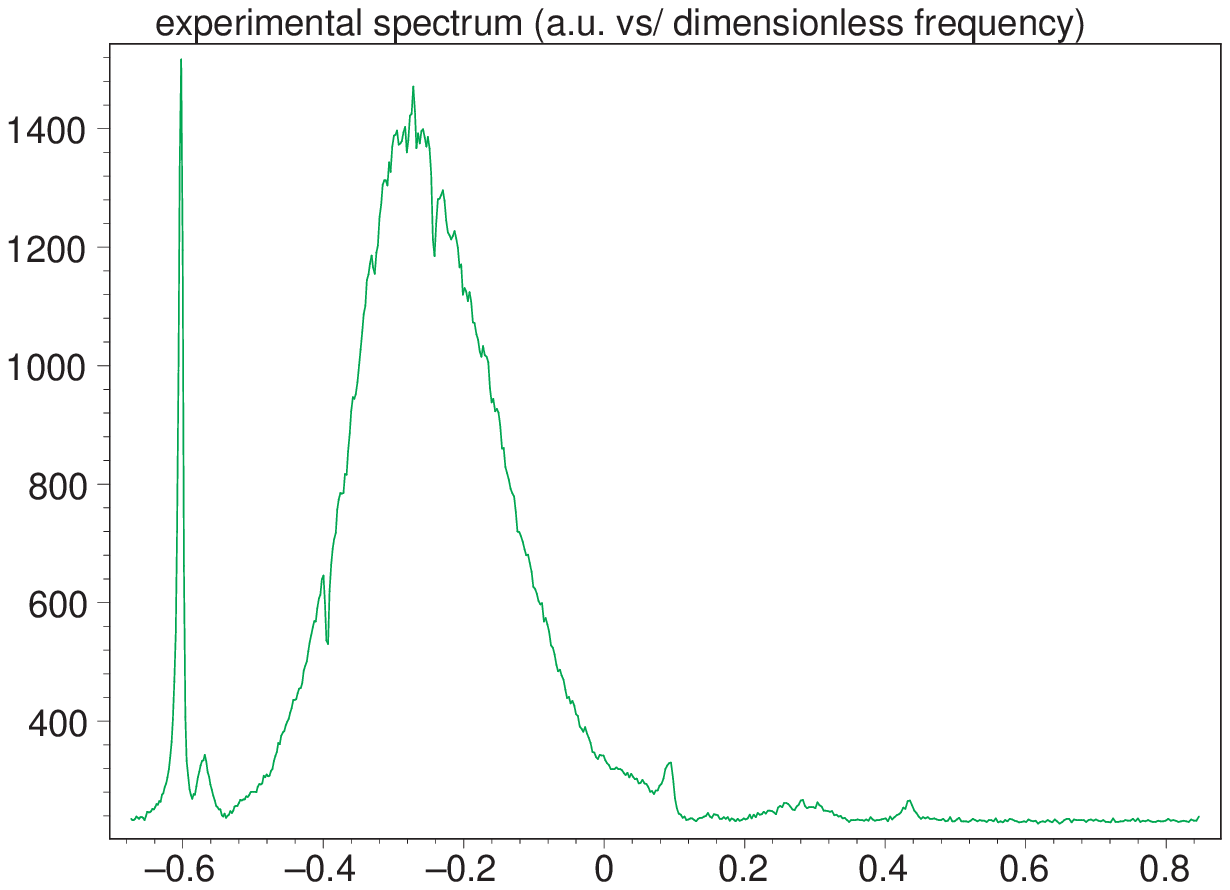}
\end{center}

\emptyline \fbox{\parbox{.8\linewidth} {We can see a pronounced
red shift relatively to the gain band center. The narrow line is
the so-called side-band or dispersive wave.}}

\emptyline The typical results of the numerical simulations based
on the described above model are presented in the next figure
(power is given in the arbitrary units).

\emptyline $>$Spec1\_x := readdata(`num1\_x.dat`,1,float):\\
\indent
 $>$Spec1\_y := readdata(`num1\_y.dat`,1,float):\\
 \indent \indent
  $>$Spec2\_x := readdata(`num2\_x.dat`,1,float):\\
  \indent \indent \indent
   $>$Spec2\_y := readdata(`num2\_y.dat`,1,float):\\
   \indent \indent \indent \indent
    $>$Spec3\_x := readdata(`num3\_x.dat`,1,float):\\
    \indent \indent \indent
     $>$Spec3\_y := readdata(`num3\_y.dat`,1,float):\\
     \indent \indent
      $>$Spec4\_x := readdata(`num4\_x.dat`,1,float):\\
      \indent
       $>$Spec4\_y := readdata(`num4\_y.dat`,1,float):\\
       \indent \indent
        $>$Spec5\_x := readdata(`num5\_x.dat`,1,float):\\
        \indent \indent \indent
         $>$Spec5\_y := readdata(`num5\_y.dat`,1,float):\\
         \indent \indent \indent \indent
          $>$Spec6\_x := readdata(`num6\_x.dat`,1,float):\\
          \indent \indent \indent
           $>$Spec6\_y := readdata(`num6\_y.dat`,1,float):\\
           \indent \indent
          $>$n := vectdim(Spec1\_y);\\
          \indent
         $>$p1 := plot([[Spec1\_x[k],Spec1\_y[k]]
         \$k=1..n/4],color=black):\\
         \indent \indent
        $>$n := vectdim(Spec2\_y);\\
        \indent \indent \indent
       $>$p2 := plot([[Spec2\_x[k],Spec2\_y[k]] \$k=3*(n-1)/4..n],\\
       color=black):\\
       \indent \indent \indent \indent
      $>$n := vectdim(Spec3\_y);\\
      \indent \indent \indent
     $>$p3 := plot([[Spec3\_x[k],2.5*Spec3\_y[k]] \$k=3*(n-1)/4..n],\\
     color=red):\\
     \indent \indent
    $>$n := vectdim(Spec4\_y);\\
    \indent
   $>$p4 := plot([[Spec4\_x[k],2.5*Spec4\_y[k]] \$k=1..n/4],\\
   color=red):\\
   \indent
  $>$n := vectdim(Spec5\_y);\\
  \indent
 $>$p5 := plot([[Spec5\_x[k],25*Spec5\_y[k]] \$k=3*(n-1)/4..n],\\
 color=blue):\\
 \indent
$>$n := vectdim(Spec6\_y);\\
\indent
 $>$p6 := plot([[Spec6\_x[k],25*Spec6\_y[k]] \$k=1..n/4],\\
 color=blue):\\
 \indent
  $>$display(p1,p2,p3,p4,p5,p6,view=0..1,axes=BOXED,title=`numerical
spectra (a.u.)`);

\begin{center}
\mapleplot{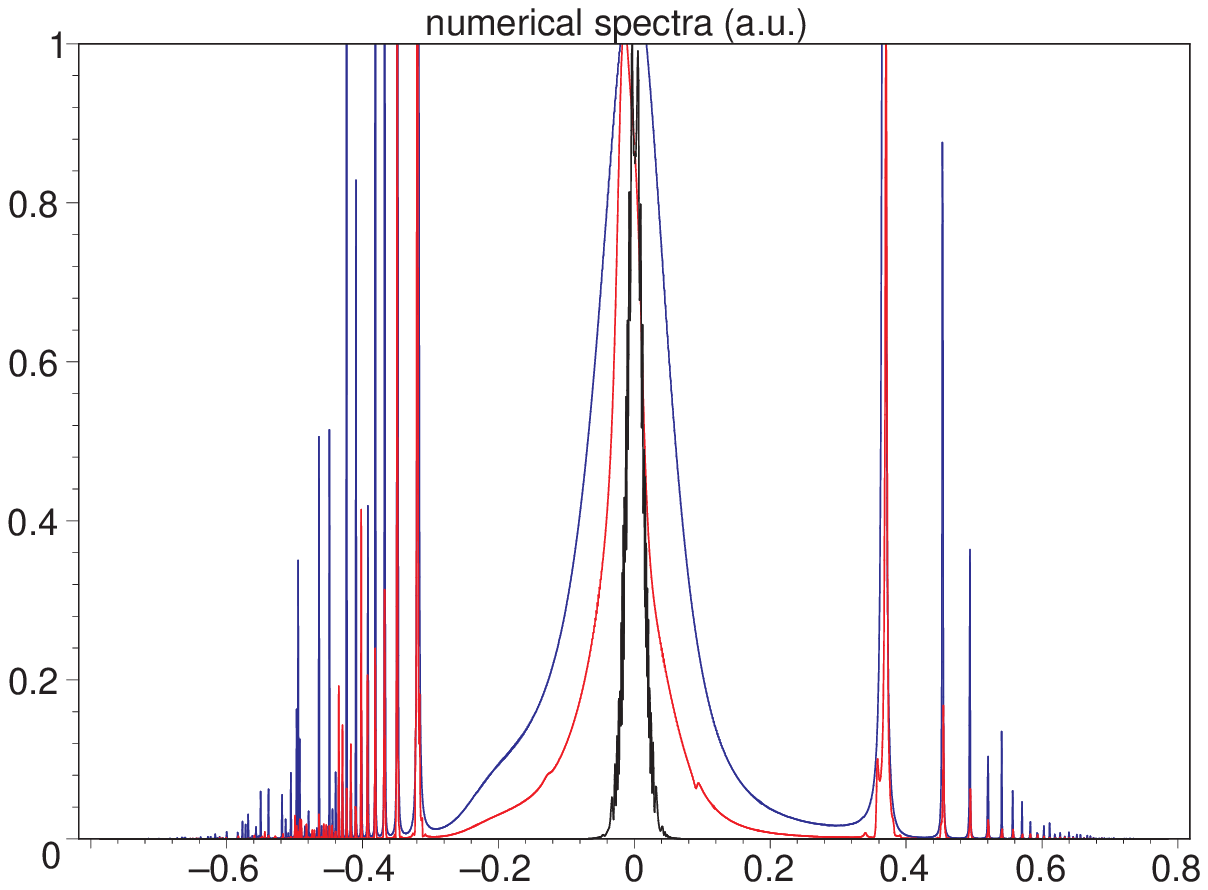}
\end{center}

\emptyline In the "natural" co-ordinates we have:

\emptyline $>$n := vectdim(Spec1\_y);\\
\indent
 $>$p1 :=
 plot([[2*Pi*3e10*1e7/(bandwidth*Spec1\_x[k]+X\_max),Spec1\_y[k]]\\
\$k=1..n/4],color=black):\\
\indent \indent
  $>$n := vectdim(Spec2\_y);\\
  \indent
   $>$p2 :=
   plot([[2*Pi*3e10*1e7/(bandwidth*Spec2\_x[k]+X\_max),Spec2\_y[k]]\\
\$k=3*(n-1)/4..n],color=black):\\
\indent \indent
    $>$n := vectdim(Spec3\_y);\\
    \indent \indent \indent
     $>$p3 :=\\
plot([[2*Pi*3e10*1e7/(bandwidth*Spec3\_x[k]+X\_max),2.5*Spec3\_y[k]]\\
\$k=3*(n-1)/4..n],color=red):\\
\indent \indent \indent \indent
      $>$n := vectdim(Spec4\_y);\\
      \indent \indent \indent
     $>$p4 :=\\
plot([[2*Pi*3e10*1e7/(bandwidth*Spec4\_x[k]+X\_max),2.5*Spec4\_y[k]]\\
\$k=1..n/4],color=red):\\
\indent \indent
    $>$n := vectdim(Spec5\_y);\\
    \indent
   $>$p5 :=\\
plot([[2*Pi*3e10*1e7/(bandwidth*Spec5\_x[k]+X\_max),25*Spec5\_y[k]]\\
\$k=3*(n-1)/4..n],color=blue):\\
\indent
  $>$n := vectdim(Spec6\_y);\\
\indent
 $>$p6 :=\\
plot([[2*Pi*3e10*1e7/(bandwidth*Spec6\_x[k]+X\_max),25*Spec6\_y[k]]\\
\$k=1..n/4],color=blue):\\
\indent
$>$display(p1,p2,p3,p4,p5,p6,view=0..1,axes=BOXED,title=`pulse
spectrum vs. wavelength [nm]`);

\begin{center}
\mapleplot{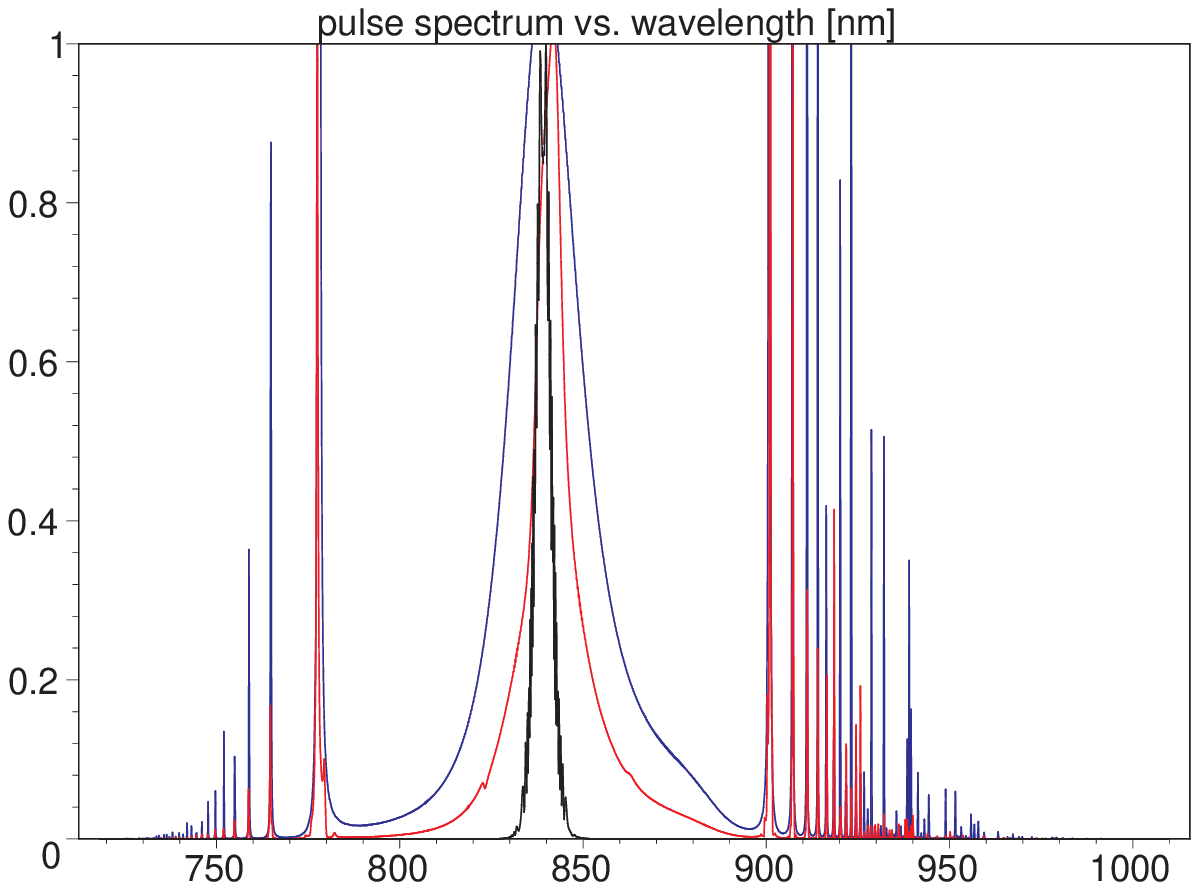}
\end{center}

\emptyline The black curve corresponds to the soliton propagation
without lasing factors, i.e. in the absence of gain, loss
saturation and spectral filtering. The high-order dispersion
slightly transforms the spectrum of the soliton, but there is not
the visible frequency shift. Perhaps, the soliton duration (169
fs) is too large for the high-order dispersions' manifestation.
But we can not change the pulse duration for fixed GDD. Such
possibility is opened by lasing in the presence of the gain and
saturable loss. The red and blue curves correspond to the above
described laser model with 1.2 W absorbed pump power for 20x30
$\mu ^{2}$ pump mode and gaussian mode with 25 $\mu$m diameter,
correspondingly. The pulse durations in these cases are 31 and 27
fs, correspondingly. We can see the appearance of spectral spikes
(dispersion waves), which locate near from zeros and 2 $\pi$m -
values of GDD. The domain of the large gradient of GDD in the red
spectral region is filled by such spikes. But the pulses' spectra
have the very small frequency shifts from gain band maximum,
although there is the spectral "shoulder" in the red region. The
last corresponds to local GDD extremum near from 883 nm (see
above).

\emptyline \fbox{\parbox{.8\linewidth} {We had performed the
various manipulations with the pump and the fast absorber
saturation intensity, but as a result, the self-frequency shift
can not be obtained only due to high-order dispersion.}}

\emptyline So, there is some additional mechanism of the pulse
spectrum shift. As the pulse duration is greater than 10 fs, the
nonlinear dispersion can not cause such shift. Therefore we have
to investigate the stimulated Raman scattering influence on the
pulse spectrum.

Let assume, that the frequency shift results from the stimulated
Raman scattering in the active crystal. The vibrational amplitude
\textit{Q} in the dependence on the pump and signal spectral
amplitudes \textit{A\_m} and \textit{A\_n}, respectively, obeys
the following equation (see \cite{Haus4}):

\emptyline $>$restart:\\
\indent
 $>$eq1 :=\\
diff(Q(t),t\$2)+2*diff(Q(t),t)/T+Omega$^2$*Q(t)=\\
mu*A\_m*exp(I*m*omega*t)*conjugate(A\_n)*exp(-I*n*omega*t);\\
\indent
  $>$Q = subs(\{\_C1=0,\_C2=0\},rhs(dsolve(eq1,Q(t))));\\
  \indent
   $>$amp\_Q =\\
Sum(Sum(expand(numer(rhs(\%))/expand(denom\\
(rhs(\%)))/exp(-I*t*omega*(-m+n))),n=-N/2..N/2),m=-N/2..N/2);\\
\#here we supposed that the pulse width << T (T is the phonon
relaxation time)

\[ \boxed{
\mathit{eq1} := ({\frac {\partial ^{2}}{\partial t^{2}}}\,
\mathrm{Q}(t)) + {\displaystyle \frac {2\,({\frac {\partial }{
\partial t}}\,\mathrm{Q}(t))}{T}}  + \Omega ^{2}\,\mathrm{Q}(t)=
\mu \,\mathit{A\_m}\,e^{(I\,m\,\omega \,t)}\,\overline{(\mathit{
A\_n})}\,e^{( - I\,n\,\omega \,t)} }
\]

\[
Q={\displaystyle \frac {\mu \,\mathit{A\_m}\,e^{(I\,\omega \,t\,(
m - n))}\,\overline{(\mathit{A\_n})}\,T}{T\,\Omega ^{2} + 2\,I\,m
\,\omega  - 2\,I\,n\,\omega  - T\,\omega ^{2}\,m^{2} + 2\,T\,
\omega ^{2}\,m\,n - T\,\omega ^{2}\,n^{2}}}
\]

\begin{gather} \nonumber
\mathit{amp\_Q}=\\
{\displaystyle \sum _{m= - 1/2\,N}^{1/2\,N}} \,
 \left(  \! {\displaystyle \sum _{n= - 1/2\,N}^{1/2\,N}} \,
{\displaystyle \frac {\mu \,\mathit{A\_m}\,\overline{(\mathit{
A\_n})}\,T}{T\,\Omega ^{2} + 2\,I\,m\,\omega  - 2\,I\,n\,\omega
 - T\,\omega ^{2}\,m^{2} + 2\,T\,\omega ^{2}\,m\,n - T\,\omega ^{
2}\,n^{2}}}  \!  \right)\nonumber
\end{gather}

\emptyline Here \textit{m} and\textit{n} are the mode numbers,
$\omega$ is the frequency interval between field spectral
components, $\Omega$ is the Raman frequency, \textit{T} is the
relaxation time, \textit{N} is the number of the frequency
components in the field spectrum (in the case of our numerical
simulations \textit{N}= 2$^{13}$), $\mu$ is the positive real
number expressing photon-phonon coupling. Note, that the last
expression for the amplitude of the vibrational oscillations (the
frequency of these oscillations is \textit{(m-n)}* $\omega$) can
be re-written as

\emptyline
$>$amp\_Q =\\
Sum(Sum(mu*A\_m*conjugate(A\_n)/(Omega$^2$ - \\
2*I*omega*(m-n)/Tr - omega$^2$*(n-m )$^2$),\\
n = -N/2 .. N/2),m = -N/2 .. N/2);\\

\indent
$>$amp\_Q =\\
Sum(Sum(mu*A\_m*conjugate(A\_n)*(Omega$^2$ - omega$^2$*(n-m)$^2$
+\\ 2*I*omega*(m-n)/Tr)/((Omega$^2$ - omega$^2$*(n-m)$^2$)$^2$ +\\
4*omega$^2$*(m-n)$^2$/(Tr$^2$)),n = -N/2 .. N/2),m = -N/2 .. N/2);

\[
\mathit{amp\_Q}={\displaystyle \sum _{m= - 1/2\,N}^{1/2\,N}} \,
 \left(  \! {\displaystyle \sum _{n= - 1/2\,N}^{1/2\,N}} \,
{\displaystyle \frac {\mu \,\mathit{A\_m}\,\overline{(\mathit{
A\_n})}}{\Omega ^{2} + {\displaystyle \frac {-2\,I\,\omega \,(m
 - n)}{\mathit{Tr}}}  - \omega ^{2}\,( - m + n)^{2}}}  \!
 \right)
\]

\[
\mathit{amp\_Q}={\displaystyle \sum _{m= - 1/2\,N}^{1/2\,N}} \,
 \left(  \! {\displaystyle \sum _{n= - 1/2\,N}^{1/2\,N}} \,
{\displaystyle \frac {\mu \,\mathit{A\_m}\,\overline{(\mathit{
A\_n})}\,(\Omega ^{2} - \omega ^{2}\,( - m + n)^{2} +
{\displaystyle \frac {2\,I\,\omega \,(m - n)}{\mathit{Tr}}} )}{(
\Omega ^{2} - \omega ^{2}\,( - m + n)^{2})^{2} + {\displaystyle
\frac {4\,\omega ^{2}\,(m - n)^{2}}{\mathit{Tr}^{2}}} }}  \!
 \right)
\]

\emptyline Now we have to define the values of the used
parameters. There are three Raman lines in LiSGaF with following
characteristics: $\Omega $=551, 349, 230 $\mathit{cm}^{(-1)}$,
\textit{1/T} = 6.2, 7.6, 4.2 $\mathit{cm}^{(-1)}$

\emptyline $>$Omega1 := evalf(2*Pi*3*10$^10$*551):\#[Hz]\\
\indent
 $>$Omega2 := evalf(2*Pi*3*10$^10$*349):\#[Hz]\\
 \indent \indent
  $>$Omega3 := evalf(2*Pi*3*10$^10$*230):\#[Hz]\\
  \indent \indent \indent
   $>$bandwidth := .481859640e15:\#[Hz] the normalization for
   frequencies\\
   \indent \indent
    $>$Omega1 := evalf(Omega1/bandwidth);\#normalized Omega1\\
    \indent
     $>$Omega2 := evalf(Omega2/bandwidth);\#normalized Omega2\\
     \indent \indent
      $>$Omega3 := evalf(Omega3/bandwidth);\#normalized Omega3\\
      \indent \indent \indent
       $>$omega=evalf(2*Pi*bandwidth/N):\#[Hz]\\
       \indent \indent \indent \indent
        $>$print(`normalized w:`);\\
        \indent \indent \indent
         $>$2*Pi/N;\#normalyzed omega\\
         \indent \indent
         $>$T1 := evalf(1/(2*Pi*3*10$^10$*6.2)):\#[s]\\
         \indent
          $>$T2 := evalf(1/(2*Pi*3*10$^10$*7.6)):\#[s]\\
          \indent \indent
         $>$T3 := evalf(1/(2*Pi*3*10$^10$*4.2)):\#[s]\\
         \indent \indent \indent
        $>$T1 := evalf(T1*bandwidth);\#normalized relaxation
        time\\
        \indent \indent \indent \indent
       $>$T2 := evalf(T2*bandwidth);\#normalized relaxation time\\
       \indent \indent \indent
      $>$T3 := evalf(T3*bandwidth);\#normalized relaxation time\\
      \indent \indent
     $>$solve(gain\_s=6*omega\_s*chi/n\_s/c,chi):\#if permittivity chi in
[cm$^2$/W], gain\_s is the Raman signal gain\\
\indent
    $>$mu := simplify(3*omega\_s*\%*2*Omega/n\_s/c/T);\\
    \indent \indent
   $>$mu1 := subs(\{Omega = Omega1,T = T1,gain\_s=evalf(2.5*1.2e-10),\\
   beta=.3379192098e-11\},mu*0.8/beta);\#normalized\\
   \indent
 $>$mu2 := subs(\{Omega = Omega2,T = T2,gain\_s=evalf(.185*1.2e-10),\\
 beta=.3379192098e-11\},mu*0.8/beta);\#normalized\\
\indent
 $>$mu3 := subs(\{Omega = Omega3,T = T3,gain\_s=evalf(.15*1.2e-10),\\
 beta=.3379192098e-11\},mu*0.8/beta);\#normalized

\[
\Omega 1 := .2155421299
\]

\[
\Omega 2 := .1365230551
\]

\[
\Omega 3 := .08997221396
\]

\[
\mathit{normalized\ w:}
\]

\[
2\,{\displaystyle \frac {\pi }{N}}
\]

\[
\mathit{T1} := 412.3136751
\]

\[
\mathit{T2} := 336.3611560
\]

\[
\mathit{T3} := 608.6535205
\]

\[
\mu  := {\displaystyle \frac {\mathit{gain\_s}\,\Omega }{T}}
\]

\[
\mu 1 := .03712810569
\]

\[
\mu 2 := .002133193454
\]

\[
\mu 3 := .0006299236009
\]

So, we can investigate the field evolution on the basis of the
following equation:

\emptyline $>$Diff(A\_n(z),z) =\\
Sum(mu*conjugate(A\_m)*A\_n*(I*(Omega$^2$ - omega$^2$*(n -
m)$^2$)+2*omega*(m - n)/Tr)/((Omega$^2$ - omega$^2$*(n -
m)$^2$)$^2$ + 4*omega$^2$*(m - n)$^2$/(Tr$^2$))*A\_m,m = -N/2 ..
N/2);\\

\indent $>$Diff(A\_n(z),z) =\\
A\_n*Sum(mu*abs(A\_m)$^2$*(I*(Omega$^2$ - omega$^2$*(n - m)$^2$) +
2*omega*(m - n)/Tr)/((Omega$^2$ - omega$^2$*(n - m)$^2$)$^2$ +
4*omega$^2$*(m - n)$^2$/(Tr$^2$)),m = -N/2 .. N/2);

\begin{gather} \nonumber
{\frac {\partial }{\partial z}}\,\mathrm{A\_n}(z)=\\
{\displaystyle \sum _{m= - 1/2\,N}^{1/2\,N}} \,{\displaystyle
\frac {\mathit{ gain\_s}\,\Omega
\,\overline{(\mathit{A\_m})}\,\mathit{A\_n}\,(I \,(\Omega ^{2} -
\omega ^{2}\,( - m + n)^{2}) + {\displaystyle \frac {2\,\omega
\,(m - n)}{\mathit{Tr}}} )\,\mathit{A\_m}}{T\,(( \Omega ^{2} -
\omega ^{2}\,( - m + n)^{2})^{2} + {\displaystyle \frac {4\,\omega
^{2}\,(m - n)^{2}}{\mathit{Tr}^{2}}} )}} \nonumber
\end{gather}

\begin{gather} \nonumber \boxed{
{\frac {\partial }{\partial
z}}\,\mathrm{A\_n}(z)=}\\
\boxed{\mathit{A\_n}\,
 \left(  \! {\displaystyle \sum _{m= - 1/2\,N}^{1/2\,N}} \,
{\displaystyle \frac {\mathit{gain\_s}\,\Omega \, \left|  \! \,
\mathit{A\_m}\, \!  \right| ^{2}\,(I\,(\Omega ^{2} - \omega ^{2}
\,( - m + n)^{2}) + {\displaystyle \frac {2\,\omega \,(m - n)}{
\mathit{Tr}}} )}{T\,((\Omega ^{2} - \omega ^{2}\,( - m + n)^{2})
^{2} + {\displaystyle \frac {4\,\omega ^{2}\,(m - n)^{2}}{
\mathit{Tr}^{2}}} )}}  \!  \right) }\nonumber
\end{gather}

\emptyline \noindent Here \textit{n} is the "signal" mode,
\textit{m} is the "pump" mode, degenerate case
\textit{m}=\textit{n} corresponds to the pure self-phase
modulation, Stokes signal shift \textit{m$>$n} corresponds to an
amplification, anti-Stokes shift \textit{m}$<$\textit{{\large n}}
corresponds to a loss of signal wave. Because of the Raman line is
narrow in the comparison with pulse spectrum (see normalized
\textit{T}), we can re-write the expressions:

\emptyline $>$Diff(A\_n(z),z) =\\
I*A\_n*Sum(mu*abs(A\_m)$^2$/(Omega$^2$ + 2*I*omega*(m -
n)/Tr-omega$^2$*(n - m)$^2$),m = -N/2 .. N/2);\\
\indent $>$Diff(A\_n(z),z) =\\
I*A\_n*Sum(mu*abs(A\_m)$^2$/((Omega - omega*(n - m))*(Omega +
omega*(n -m )) + 2*I*omega*(m - n)/Tr),m = -N/2 .. N/2);\\
\indent $>$Diff(A\_n(z),z) =\\
I*A\_n*Sum(mu*abs(A\_m)$^2$/(2*Omega*(Omega - omega*(n - m)) +\\
2*I*Omega/Tr),m = -N/2 .. N/2);

\[
{\frac {\partial }{\partial z}}\,\mathrm{A\_n}(z)=I\,\mathit{A\_n
}\, \left(  \! {\displaystyle \sum _{m= - 1/2\,N}^{1/2\,N}} \,
{\displaystyle \frac {\mathit{gain\_s}\,\Omega \, \left|  \! \,
\mathit{A\_m}\, \!  \right| ^{2}}{T\,(\Omega ^{2} - \omega ^{2}\,
( - m + n)^{2} + {\displaystyle \frac {2\,I\,\omega \,(m - n)}{
\mathit{Tr}}} )}}  \!  \right)
\]

\begin{gather} \nonumber
{\frac {\partial }{\partial
z}}\,\mathrm{A\_n}(z)=\\
I\,\mathit{A\_n }\, \left(  \!
{\displaystyle \sum _{m= - 1/2\,N}^{1/2\,N}} \, {\displaystyle
\frac {\mathit{gain\_s}\,\Omega \, \left|  \! \, \mathit{A\_m}\,
\!  \right| ^{2}}{T\,((\Omega  - \omega \,( - m
 + n))\,(\Omega  + \omega \,( - m + n)) + {\displaystyle \frac {2
\,I\,\omega \,(m - n)}{\mathit{Tr}}} )}}  \!  \right) \nonumber
\end{gather}

\[
{\frac {\partial }{\partial z}}\,\mathrm{A\_n}(z)=I\,\mathit{A\_n
}\, \left(  \! {\displaystyle \sum _{m= - 1/2\,N}^{1/2\,N}} \,
{\displaystyle \frac {\mathit{gain\_s}\,\Omega \, \left|  \! \,
\mathit{A\_m}\, \!  \right| ^{2}}{T\,(2\,\Omega \,(\Omega  -
\omega \,( - m + n)) + {\displaystyle \frac {2\,I\,\Omega }{
\mathit{Tr}}} )}}  \!  \right)
\]

\emptyline The last expression results from the assumption
\textit{T}$\gg$1. In this case

\emptyline $>$Diff(A\_n(z),z) =\\
I*A\_n*mu*abs(A\_m)$^2$*Int((2*Omega*(Omega - x) -
2*I*Omega/Tr)/\\
(4*Omega$^2$*(Omega - x)$^2$ +
4*Omega$^2$/(Tr$^2$)),x = -infinity .. infinity);

\[
{\frac {\partial }{\partial z}}\,\mathrm{A\_n}(z)={\displaystyle
\frac {I\,\mathit{A\_n}\,\mathit{gain\_s}\,\Omega \, \left|  \!
\,\mathit{A\_m}\, \!  \right| ^{2}\,{\displaystyle \int _{ -
\infty }^{\infty }} {\displaystyle \frac {2\,\Omega \,(\Omega  -
x) + {\displaystyle \frac {-2\,I\,\Omega }{\mathit{Tr}}} }{4\,
\Omega ^{2}\,(\Omega  - x)^{2} + {\displaystyle \frac {4\,\Omega
^{2}}{\mathit{Tr}^{2}}} }} \,dx}{T}}
\]

\emptyline In the last expression \textit{x} is the frequency
difference, \textit{A\_m} corresponds to the pump intensity for -
$\omega $(n-m)= $\Omega $. The contribution of the real part in
the integral (the self-phase modulation from the both sides of the
Raman line) is equal to 0.

\emptyline $>$-(I*Tr/2/Omega)*Int(1/(x$^2$+1),x=0..infinity);\\
\indent
  $>$value(\%);

\[
{\displaystyle \frac {{\displaystyle \frac {-1}{2}} \,I\,\mathit{
Tr}\,{\displaystyle \int _{0}^{\infty }} {\displaystyle \frac {1
}{x^{2} + 1}} \,dx}{\Omega }}
\]

\[
{\displaystyle \frac {{\displaystyle \frac {-1}{4}} \,I\,\mathit{
Tr}\,\pi }{\Omega }}
\]

\emptyline Thus we obtained the simplest expressions for the field
evolution:

\emptyline $>$Diff(A\_n(z),z) = A\_n*gain*Pi*abs(A\_m)$^2$/4;\\
\indent $>$Diff(A\_m(z),z) = -A\_m*gain*Pi*abs(A\_n)$^2$/4;

\emptyline

\[\boxed{
{\frac {\partial }{\partial z}}\,\mathrm{A\_n}(z)={\displaystyle
\frac {1}{4}} \,\mathit{A\_n}\,\mathit{gain}\,\pi \, \left|  \!
\,\mathit{A\_m}\, \!  \right| ^{2}}
\]

\[\boxed{
{\frac {\partial }{\partial z}}\,\mathrm{A\_m}(z)= -
{\displaystyle \frac {1}{4}} \,\mathit{A\_m}\,\mathit{gain}\,\pi
\, \left|  \! \,\mathit{A\_n}\, \!  \right| ^{2}}
\]

\emptyline The stimulated Raman gain parameters for numerical
simulation are:

\emptyline $>$gain\_s1 :=
evalf(0.8*2.5*1.2e-10/.3379192098e-11);\\
\indent
 $>$gain\_s2 := evalf(0.8*.185*1.2e-10/.3379192098e-11);\\
 \indent \indent
  $>$gain\_s3 := evalf(0.8*.15*1.2e-10/.3379192098e-11);

\[
\mathit{gain\_s1} := 71.02289335
\]

\[
\mathit{gain\_s2} := 5.255694108
\]

\[
\mathit{gain\_s3} := 4.261373601
\]

\emptyline Additionally we are to take into consideration the
spontaneous Raman scattering as a source for the stimulated
scattering. The gain coefficients is this case are \cite{Suth}:

\emptyline $>$d\_sigma :=\\
(4*omega\_l/3/omega\_s)*3*n\_s$^2$*h*omega\_s$^3$*\\
gain\_s/(8*Pi$^3$*c$^2$*N)/(1 - exp(-h*(omega\_l -\\
omega\_s)/(2*Pi*kb*Tc)));\#kb and Tc are the Boltzmann's constant
and temperature, correspondingly\\
\indent
 $>$d\_sigma1 :=\\
evalf(sqrt(subs(\\
\{kb=1.38*1e-23,Tc=300,omega\_s=2*Pi*3e10/.9051e-4,\\
omega\_l=2*Pi*3e10/.85e-4,n\_s=1.4,h=6.62e-34,\\
gain\_s=gain\_s1,c=3e10,N=1e20\},d\_sigma*N)));\\
\indent
  $>$d\_sigma2 :=\\
evalf(sqrt(subs(\\
\{kb=1.38*1e-23,Tc=300,omega\_s=2*Pi*3e10/.88e-4,\\
omega\_l=2*Pi*3e10/.85e-4,n\_s=1.4,h=6.62e-34,\\
gain\_s=gain\_s2,c=3e10,N=1e20\},d\_sigma*N)));\\
\indent
   $>$d\_sigma3 :=\\
evalf(sqrt(subs(\\
\{kb=1.38*1e-23,Tc=300,omega\_s=2*Pi*3e10/.87e-4,\\
omega\_l=2*Pi*3e10/.85e-4,n\_s=1.4,h=6.62e-34,\\
gain\_s=gain\_s3,c=3e10,N=1e20\},d\_sigma*N)));

\[
\mathit{d\_sigma} := {\displaystyle \frac {1}{2}} \,
{\displaystyle \frac {\mathit{omega\_l}\,\mathit{omega\_s}^{2}\,
\mathit{n\_s}^{2}\,h\,\mathit{gain\_s}}{\pi ^{3}\,c^{2}\,N\,(1 -
e^{( - 1/2\,\frac {h\,(\mathit{omega\_l} - \mathit{omega\_s})}{
\pi \,\mathit{kb}\,\mathit{Tc}})})}}
\]

\[
\mathit{d\_sigma1} := .0001280998590
\]

\[
\mathit{d\_sigma2} := .00003815471494
\]

\[
\mathit{d\_sigma3} := .00003767032764
\]

\emptyline \noindent \textit{d\_sigma} are the increments of the
spontaneous Stokes components (i.e. Raman spontaneous seeds)
growth .

As a result of the simulations on the basis of this model, we
obtained next spectra:

\emptyline $>$with('linalg'):\\
\indent
 $>$with(plots):\\
 \indent \indent
  $>$R\_Spec1\_x := readdata(`spec1\_x.dat`,1,float):\\
  \indent \indent \indent
   $>$R\_Spec1\_y := readdata(`spec1\_y.dat`,1,float):\\
   \indent \indent \indent \indent
    $>$R\_Spec2\_x := readdata(`spec2\_x.dat`,1,float):\\
    \indent \indent \indent
     $>$R\_Spec2\_y := readdata(`spec2\_y.dat`,1,float):\\
     \indent \indent
      $>$R\_Spec3\_x := readdata(`spec3\_x.dat`,1,float):\\
      \indent
       $>$R\_Spec3\_y := readdata(`spec3\_y.dat`,1,float):\\
       \indent \indent
        $>$R\_Spec4\_x := readdata(`spec4\_x.dat`,1,float):\\
        \indent \indent \indent
         $>$R\_Spec4\_y := readdata(`spec4\_y.dat`,1,float):\\
         \indent \indent \indent \indent
          $>$R\_Spec5\_x := readdata(`spec5\_x.dat`,1,float):\\
          \indent \indent \indent
           $>$R\_Spec5\_y := readdata(`spec5\_y.dat`,1,float):\\
           \indent \indent
            $>$R\_Spec6\_x := readdata(`spec6\_x.dat`,1,float):\\
            \indent
             $>$R\_Spec6\_y := readdata(`spec6\_y.dat`,1,float):\\
             \indent \indent
            $>$n := vectdim(R\_Spec1\_y);\\
            \indent \indent \indent
           $>$p1 := plot([[R\_Spec1\_x[k],R\_Spec1\_y[k]]\\
\$k=1..n/16],color=black):\\
\indent \indent \indent \indent
          $>$n := vectdim(R\_Spec2\_y);\\
          \indent \indent \indent
         $>$p2 := plot([[R\_Spec2\_x[k],R\_Spec2\_y[k]]\\
\$k=7*(n-1)/8..n],color=black):\\
\indent \indent
        $>$n := vectdim(R\_Spec3\_y);\\
        \indent
       $>$p3 := plot([[R\_Spec3\_x[k],R\_Spec3\_y[k]]\\
        \$k=1..n/16],color=red):\\
        \indent \indent
      $>$n := vectdim(R\_Spec4\_y);\\
      \indent \indent \indent
     $>$p4 := plot([[R\_Spec4\_x[k],R\_Spec4\_y[k]]\\
\$k=7*(n-1)/8..n],color=red):\\
\indent \indent \indent \indent
    $>$n := vectdim(R\_Spec5\_y);\\
    \indent \indent \indent
   $>$p5 := plot([[R\_Spec5\_x[2*k],R\_Spec5\_y[2*k]]\\
\$k=1..n/32],color=blue):\\
\indent \indent
  $>$n := vectdim(R\_Spec6\_y);\\
  \indent
 $>$p6 := plot([[R\_Spec6\_x[2*k],R\_Spec6\_y[2*k]]\\
\$k=7*(n-1)/16..n/2],color=blue):\\
\indent $>$display(p1,p2,p3,p4,p5,p6,axes=boxed, title=`pulse
spectrum vs. dimensionless frequency`);

\begin{center}
\mapleplot{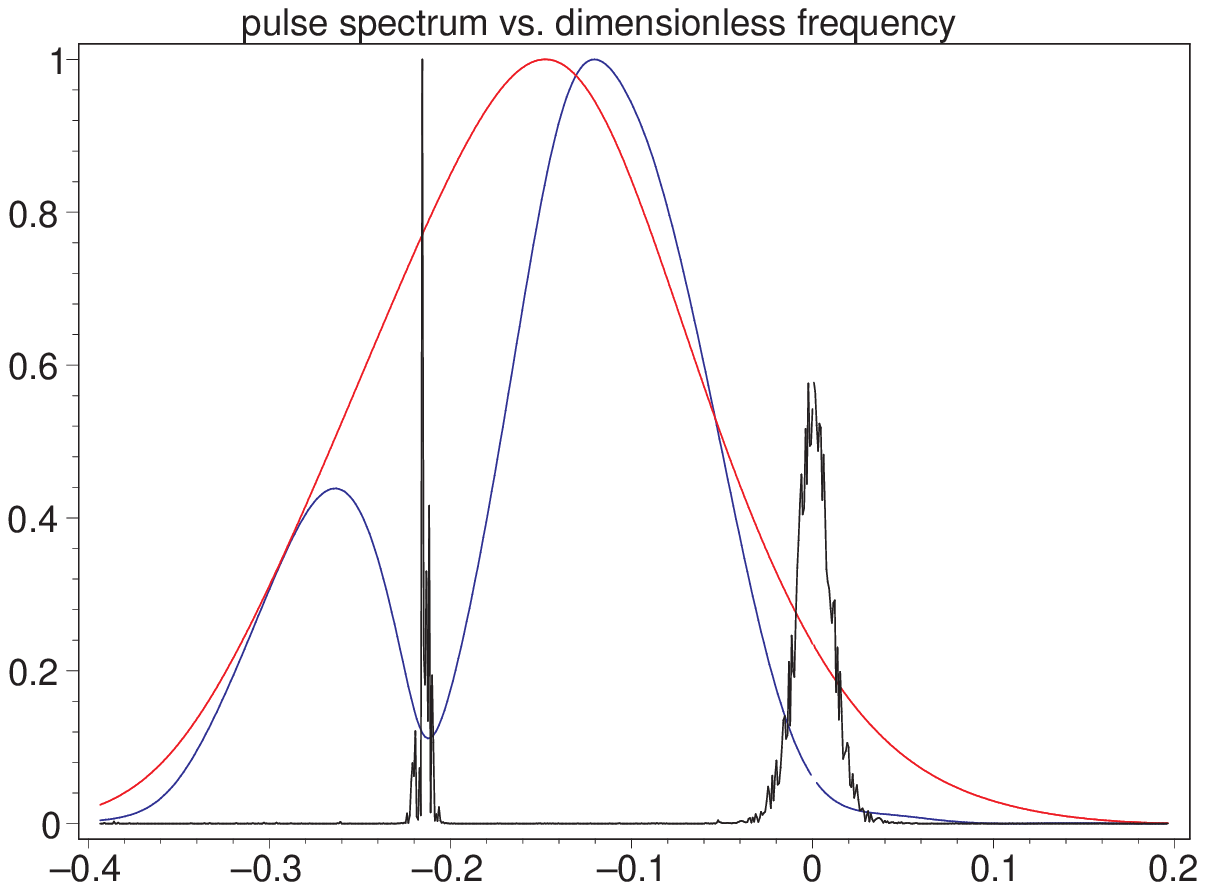}
\end{center}

\emptyline $>$X\_max := .2245909122e16:\\
\indent
 $>$bandwidth := .481859640e15:\\
 \indent \indent
  $>$n := vectdim(R\_Spec1\_y):\\
  \indent \indent \indent
   $>$p1 :=\\
plot([[2*Pi*3e10*1e7/(bandwidth*R\_Spec1\_x[k] +
X\_max),R\_Spec1\_y[k]]\\ \$k=1..n/16],color=black):\\
\indent \indent \indent \indent
    $>$n := vectdim(R\_Spec2\_y):\\
    \indent \indent \indent
     $>$p2 :=\\
plot([[2*Pi*3e10*1e7/(bandwidth*R\_Spec2\_x[k] +
X\_max),R\_Spec2\_y[k]]\\ \$k=7*(n-1)/8..n],color=black):\\
\indent \indent
      $>$n := vectdim(R\_Spec3\_y):\\
      \indent
       $>$p3 :=\\
plot([[2*Pi*3e10*1e7/(bandwidth*R\_Spec3\_x[k] +
X\_max),R\_Spec3\_y[k]]\\ \$k=1..n/16],color=red):\\
\indent \indent
        $>$n := vectdim(R\_Spec4\_y):\\
        \indent \indent \indent
         $>$p4 :=\\
plot([[2*Pi*3e10*1e7/(bandwidth*R\_Spec4\_x[k] +
X\_max),R\_Spec4\_y[k]]\\ \$k=7*(n-1)/8..n],color=red):\\
\indent \indent \indent \indent
          $>$n := vectdim(R\_Spec5\_y):\\
          \indent \indent \indent
           $>$p5 :=\\
plot([[2*Pi*3e10*1e7/(bandwidth*R\_Spec5\_x[2*k] +
X\_max),R\_Spec5\_y[2*k]]\\ \$k=1..n/32],color=blue):\\
\indent \indent
            $>$n := vectdim(R\_Spec6\_y):\\
            \indent
           $>$p6 :=\\
plot([[2*Pi*3e10*1e7/(bandwidth*R\_Spec6\_x[2*k] +
X\_max),R\_Spec6\_y[2*k]]\\ \$k=7*(n-1)/16..n/2],color=blue):\\
\indent \indent
         $>$display(p1,p2,p3,p4,p5,p6,axes=BOXED,\\
         title=`pulse spectrum vs. wavelength [nm]`);

\begin{center}
\mapleplot{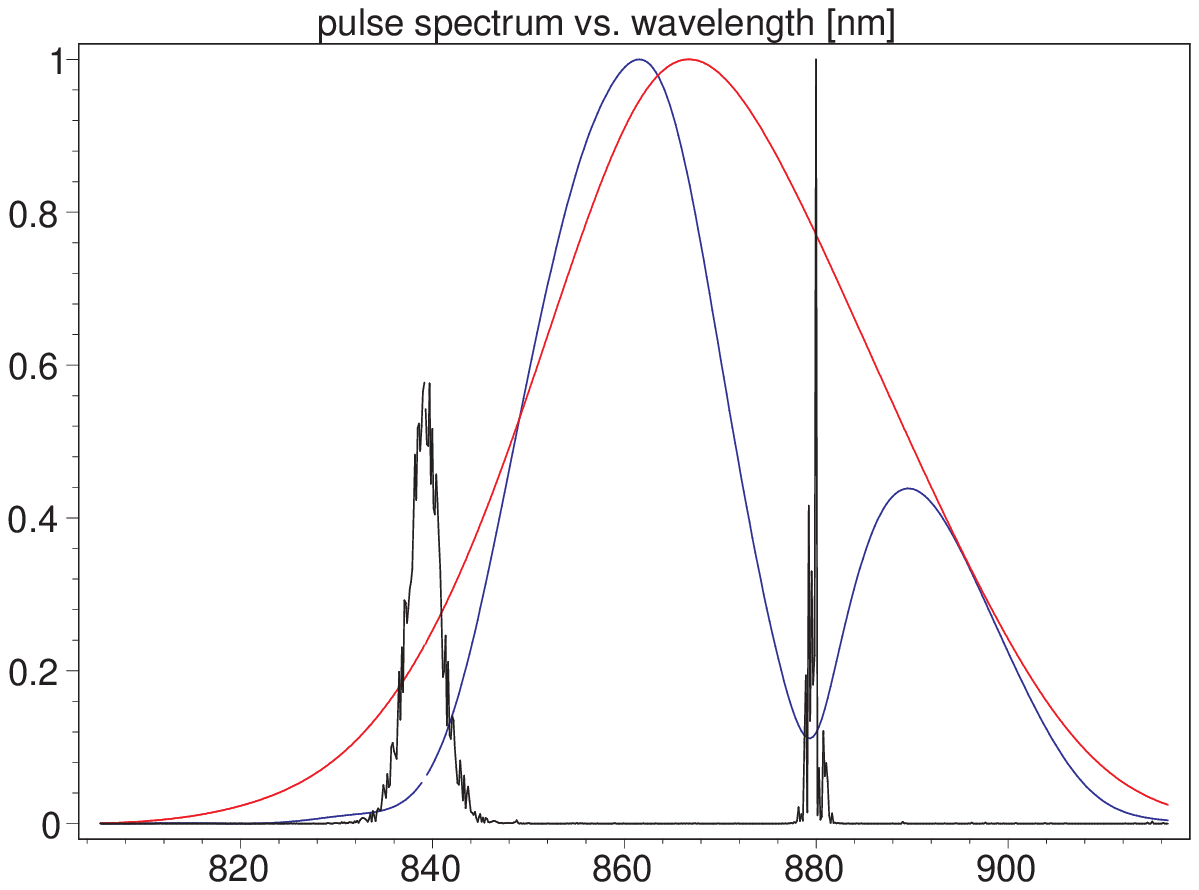}
\end{center}

\emptyline Three different spectra are presented in the figure.
Black curve corresponds to the soliton propagation (200 fs pulse
duration) in the presence of stimulated Raman scattering. As a
result of the propagation, the Stokes component appears at the
main Raman frequency. The strong scattering destroys the soliton
after 10 000 cavity transitions. In the laser, a balance between
all lasing factors stabilizes the ultrashort pulse. But there is a
visible red shift of the field spectrum (red and blue curves) due
to stimulated Raman scattering. The pulse spectrum can be pushed
from the gain band center (red curve, 25 fs pulse duration) as a
result of the Raman self-scattering. There is the possibility of
the generation of the additional Stokes lines (blue curve, 56 fs).

\fbox{\parbox{.8\linewidth} {The frequency shift is comparable
with the experimental one. Hence, the stimulated Raman scattering
is the main source of the Stokes shift of the pulse spectrum in
the Cr:LiSGaF Kerr-lens mode-locked laser (see \cite{LiSGaF}).}}

The collaboration between Maple and external numerical simulators
(based on the FORTRAN-code in our case) proves to be extremely
fruitful:

\fbox{\parbox{.8\linewidth} {1) analytical model building (Maple)
$\Rightarrow$ 2) external code generation (Maple) $\Rightarrow$ 3)
calculation of the simulation parameters (Maple) $\Rightarrow$ 4)
external calculations (FORTRAN-code, the external program can be
started from the Maple directly through the \textit{"system"}
call.) $\Rightarrow$ 5) data processing (Maple) $\Rightarrow$ 6)
analytical interpretation of the results (Maple).}}

As an example of the last step, let's consider the problem of the
pulse stability in the Kerr-lens mode-locked laser.

\emptyline

\subsection{Multipulsing and ultrashort pulse stability}

\emptyline We try to shorten the pulse duration and increase its
energy. For this aim we tend the net-GDD to zero (see Part VIII).
As a result, the ultrashort pulse stability can be lost. Let's
consider this phenomenon in detail (\cite{ZnSe}). In the framework
of the abberationless approximation the stability loss is revealed
as the absence of the soliton-like as well as breezer solution
(the evolutional equations for the pulse parameters diverge). What
is meaning of this divergence?

The answer comes from the numerical simulations based on the above
described model. Let's neglect the stimulated Raman scattering and
the higher-order dispersions. In this case, the typical results of
the simulations demonstrate multipulsing  in the vicinity of zero GDD.

\emptyline \fbox{\parbox{.8\linewidth} {Quasi-soliton
consideration fails due to the appearance of the regular or
irregular multiple pulse generation.}}

\emptyline \noindent The boundary of the soliton-like pulse
nonstability obtained from the numerical simulation is shown here
(the parameters in question can be found in the work of
reference):

\emptyline $>$restart:\\
\indent
 $>$with(plots):\\
\indent
 $>$with(stats):\\

\indent $>$points\_numer\_x :=\\
$[-160,-75,-40,-25,-18,-16,-20,-40,-45,-75,-150]$:\# GDD is
normalized to tf\\
\indent
 $>$points\_numer\_y := $[0.2,0.5,1,2,5,10,20,30,80,200,500]$:\# sigma is
normalized to the self-phase modulation coefficient beta\\
\indent
  $>$statplots[scatterplot](points\_numer\_x, evalf(\\
map(log10,points\_numer\_y) ),axes=boxed,color=red,symbol=box):\\
\indent
    $>$display(\%,color=red,TEXT([-120,1],'`stable single
    pulse`')):\\
    \indent
  $>$fig1 :=\\
display(\\
\%,color=blue,TEXT([-40,2.6],'`multipulsing`'),TEXT([-50,-0.6],\\
'`multipulsing`'),title=`logarithm of boundary sigma vs. GDD`):\\

\indent $>$display(fig1);

\begin{center}
\mapleplot{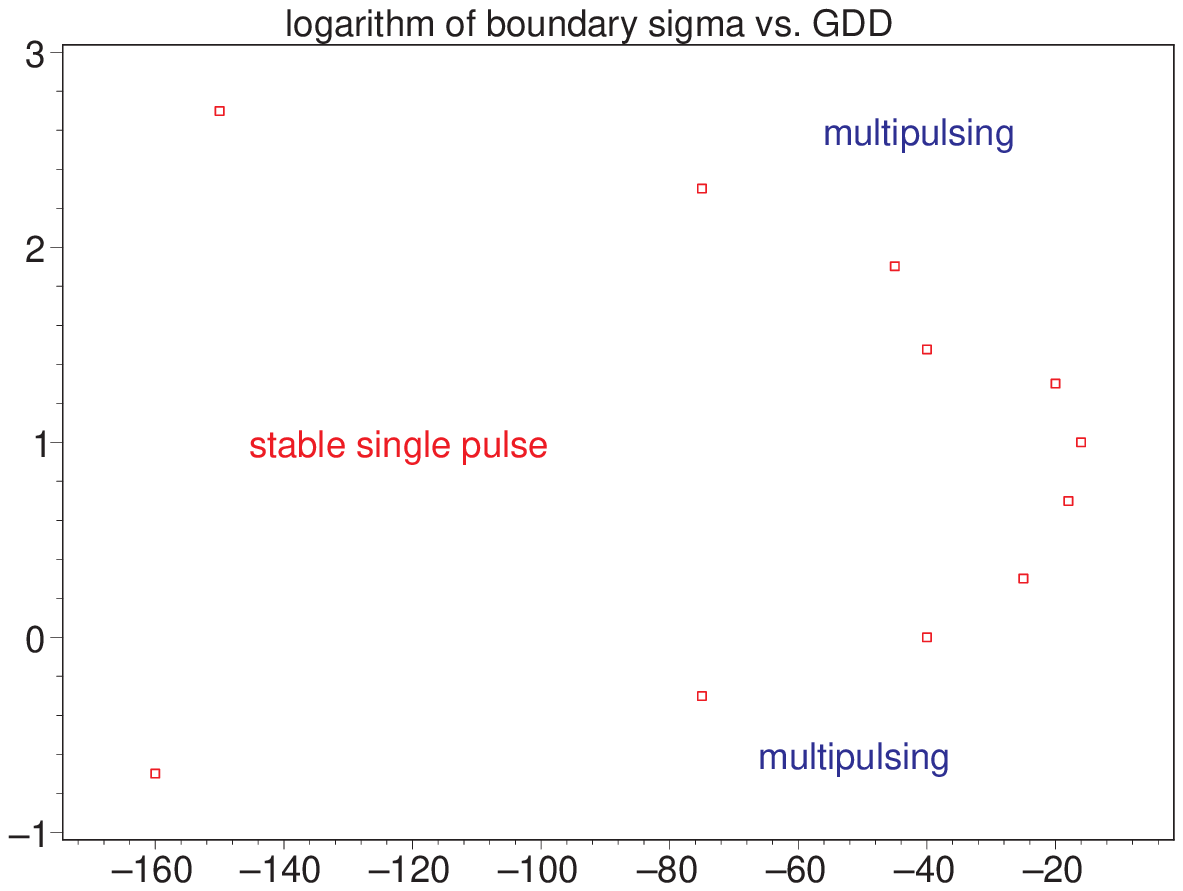}
\end{center}

\emptyline \fbox{\parbox{.8\linewidth} {There are the upper and
lower on $\sigma $ boundaries defining the transition to
multipulsing. There is no single pulse in the vicinity of zero
GDD.}}

\emptyline To interpret this picture let's return to the
generalized Landau-Ginzburg equation in the week-nonlinear limit,
when $\sigma \,\Phi $ $\ll$1 (see Part VII). For the steady-state
ultrashort pulse propagation we have (time and intensity $\Phi $
are normalized):

\emptyline $>$master\_1 := 0 = alpha - gamma + I*phi +
diff(rho(t),`\$`(t,2))/rho(t) +
I*k\_2*diff(rho(t),`\$`(t,2))/rho(t) + gamma*sigma*Phi(t) -
I*Phi(t);\# rho is the filed, Phi is the intensity\\
\indent
 $>$f1 := (t)->rho0*sech(t*tau)$^{1+I*psi}$;\# quasi-soliton
 profile\\
\indent
  $>$f2 := (t)->rho0$^2$*sech(t*tau)$^2$;\# pulse intensity

\[\boxed{
\mathit{master\_1} := 0=\alpha  - \gamma  + I\,\phi  +
{\displaystyle \frac {{\frac {\partial ^{2}}{\partial t^{2}}}\,
\rho (t)}{\rho (t)}}  + {\displaystyle \frac {I\,\mathit{k\_2}\,(
{\frac {\partial ^{2}}{\partial t^{2}}}\,\rho (t))}{\rho (t)}}
 + \gamma \,\sigma \,\Phi (t) - I\,\Phi (t)}
\]

\[\boxed{
\mathit{f1} := t\rightarrow \rho 0\,\mathrm{sech}(t\,\tau )^{(1
 + I\,\psi )}}
\]

\[
\mathit{f2} := t\rightarrow \rho 0^{2}\,\mathrm{sech}(t\,\tau )^{
2}
\]

\emptyline $>$simplify(\\
subs(\{rho(t)=f1(t),Phi(t)=f2(t)\},rhs(master\_1)) ):\\
\indent
 $>$numer(\%):\\
 \indent
  $>$eq1 := collect(\%,cosh(t*tau)$^2$):\\
  \indent
   $>$eq2 := evalc( coeff(eq1,cosh(t*tau),2) );\\
   \indent
    $>$eq3 := evalc( coeff(eq1,cosh(t*tau),0) );

\emptyline
\[
\mathit{eq2} := \alpha  - \gamma  + \tau ^{2} - \tau ^{2}\,\psi
^{2} - 2\,\mathit{k\_2}\,\tau ^{2}\,\psi  + I\,(\phi  + 2\,\tau
^{2}\,\psi  + \mathit{k\_2}\,\tau ^{2} - \mathit{k\_2}\,\tau ^{2}
\,\psi ^{2})
\]

\[
\mathit{eq3} := \gamma \,\sigma \,\rho 0^{2} + \tau ^{2}\,\psi ^{
2} - 2\,\tau ^{2} + 3\,\mathit{k\_2}\,\tau ^{2}\,\psi  + I\,( - 3
\,\tau ^{2}\,\psi  + \mathit{k\_2}\,\tau ^{2}\,\psi ^{2} - \rho 0
^{2} - 2\,\mathit{k\_2}\,\tau ^{2})
\]

\emptyline $>$eq4 := coeff(eq2,I,0);

\[
\mathit{eq4} := \alpha  - \gamma  + \tau ^{2} - \tau ^{2}\,\psi
^{2} - 2\,\mathit{k\_2}\,\tau ^{2}\,\psi
\]

\emptyline As it was shown in Parts VII, VIII, the pulse exists if
$\boxed{\alpha  - \gamma < 0}$. Simultaneously, it is condition of
the cw suppression (the threshold is not exceeded for the noise
out of pulse). As a result, the stability against cw oscillation
is provided with:

\emptyline $>$eq5 := factor( eq4 - (alpha-gamma) )/tau$^2$ $>$ 0;

\emptyline
\[\boxed{
\mathit{eq5} := 0 < 1 - \psi ^{2} - 2\,\mathit{k\_2}\,\psi}
\]

\emptyline The pulse chirp can be found from the equation
\textit{eq3}:

\emptyline $>$eq6 :=
subs(\{rho0$^2$=x,tau$^2$=y\},coeff(eq3,I)=0):\\
\indent
 $>$eq7 := subs(\{rho0$^2$=x,tau$^2$=y\},coeff(eq3,I,0)=0):\\
 \indent \indent
  $>$simplify (subs( x=solve(eq7,x),eq6 ) ):\#we find the
  intensity\\
\indent \indent \indent
   $>$numer(lhs(\%))/y = 0:\\
   \indent \indent \indent \indent
    $>$sol := solve(\%,psi);
\emptyline

\maplemultiline{ \mathit{sol} := {\displaystyle \frac {1}{2}}
\,{\displaystyle \frac {3\,\gamma \,\sigma  - 3\,\mathit{k\_2} +
\sqrt{9\,\gamma ^{2}\,\sigma ^{2} - 2\,\mathit{k\_2}\,\gamma
\,\sigma  + 9\, \mathit{k\_2}^{2} + 8 +
8\,\mathit{k\_2}^{2}\,\gamma ^{2}\,\sigma
 ^{2}}}{1 + \mathit{k\_2}\,\gamma \,\sigma }} ,  \\
{\displaystyle \frac {1}{2}} \,{\displaystyle \frac {3\,\gamma \,
\sigma  - 3\,\mathit{k\_2} - \sqrt{9\,\gamma ^{2}\,\sigma ^{2} -
2\,\mathit{k\_2}\,\gamma \,\sigma  + 9\,\mathit{k\_2}^{2} + 8 + 8
\,\mathit{k\_2}^{2}\,\gamma ^{2}\,\sigma ^{2}}}{1 + \mathit{k\_2}
\,\gamma \,\sigma }}  }

\emptyline In the combination with the condition \textit{eq5} we
have:

\emptyline

$>$eq8 := solve( numer( simplify( subs(psi=sol[1],rhs(eq5)) ) )
=\\
0, sigma ):\\
\indent
 $>$eq9 := solve( numer( simplify( subs(psi=sol[2],rhs(eq5)) ) ) =\\
  0, sigma ):\\
\indent
  $>$plot([log10(subs(gamma=0.01,eq8[1])),log10(subs(gamma=0.01,eq8[2])),\\
  log10(subs(gamma=0.01,eq9[1])),log10(subs(gamma=0.01,eq9[2]))],\\
  k\_2=-160..0,axes=boxed,color=[green,magenta]):\\
\indent $>$display(\%,fig1,view=[-160..0,-1..3]);

\begin{center}
\mapleplot{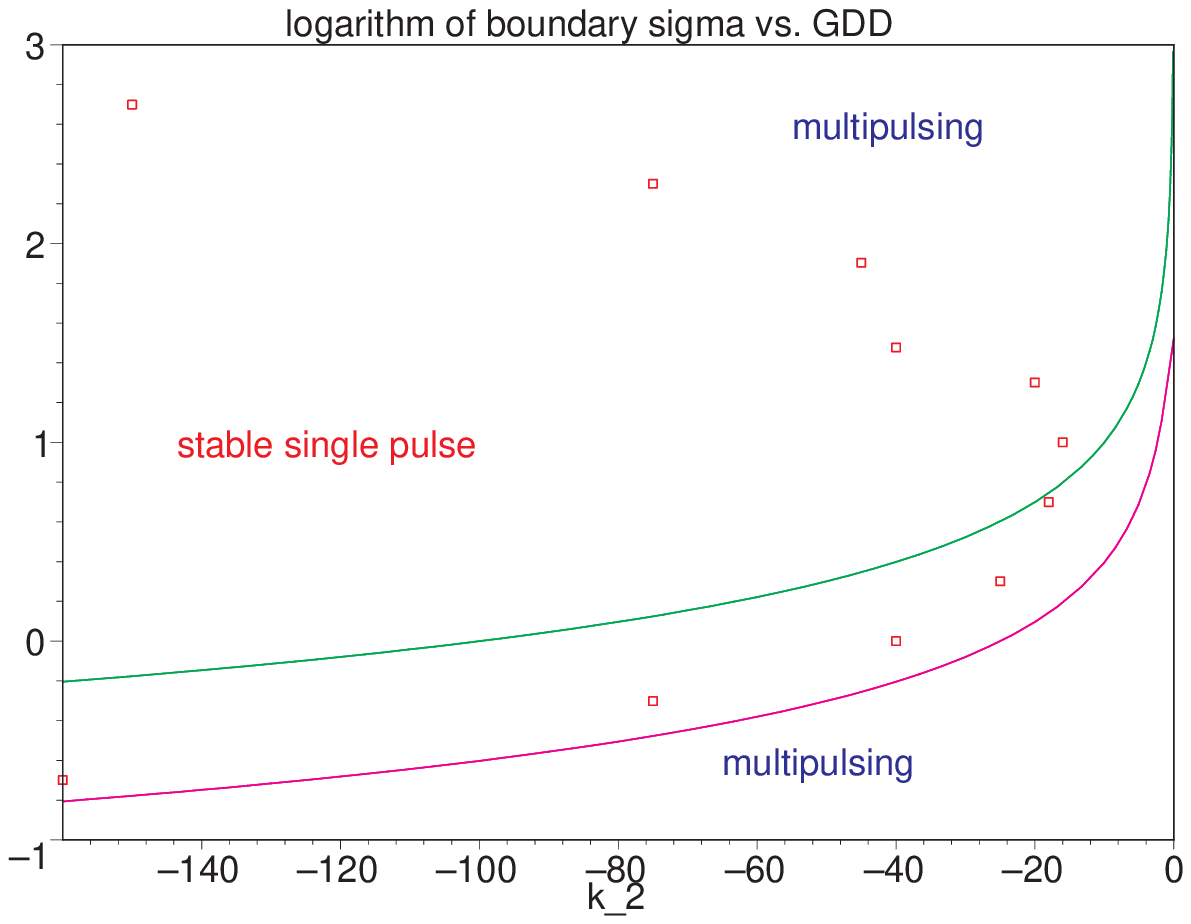}
\end{center}

\emptyline \fbox{\parbox{.8\linewidth} {Hence the lower boundary
of the pulse destabilization is good approximated by the condition
of the cw excitation (magenta curve) for the large negative GDD.
The green curve corresponds to the chirp-free generation in the
soliton model. The intersection of this curve with the stability
boundary (red points) gives the system's parameters corresponding
to the minimal pulse duration.}}

\emptyline \noindent However, there are the differences between
the analytical model and the numerical results: 1) the ultrashort
pulse is chirp-free in the wider region than that predicted from
the soliton model; 2) the $\sigma $ increase providing the pulse
shortening causes the pulse destabilization (upper on the $\sigma
$ parameter boundary of the pulse stability, red points); 3) the
pulse is unstable in the vicinity of zero GDD.

\emptyline \fbox{\parbox{.8\linewidth} {The numerical simulations
demonstrate, that the pulse destabilization on the upper stability
boundary occurs for the negative net-gain coefficient $\alpha  -
\gamma $ $<$0. This prevents the cw excitation.}}

\emptyline Let's consider the toy model of the pulse
destabilization in the absence of the cw excitation. For the sake
of simplicity, we shall consider the totally real Landau-Ginzburg
equation (Part VI).

\emptyline $>$master\_2 := rho(t)*g + diff(rho(t),`\$`(t,2)) +\\
rho(t)$^3$*Sigma;\#g=alpha-gamma, Sigma=gamma*sigma

\emptyline
\[\boxed{
\mathit{master\_2} := \rho (t)\,g + ({\frac {\partial ^{2}}{
\partial t^{2}}}\,\rho (t)) + \rho (t)^{3}\,\Sigma}
\]

\emptyline Let's expand this equation on the small perturbation
$\zeta (t)$:

\emptyline $>$expand( subs( rho(t)=rho(t)+mu*zeta(t),master\_2
)):\\
\indent
 $>$limit( diff(\%,mu),mu=0 ):\# functional Frechet derivative\\
\indent \indent
  $>$master\_3 := master\_2 + \%;

\emptyline
\[\boxed{
\mathit{master\_3} := \rho (t)\,g + ({\frac {\partial ^{2}}{
\partial t^{2}}}\,\rho (t)) + \rho (t)^{3}\,\Sigma  + g\,\zeta (t
) + ({\frac {\partial ^{2}}{\partial t^{2}}}\,\zeta (t)) + 3\,
\Sigma \,\rho (t)^{2}\,\zeta (t)}
\]

\emptyline \noindent and find its steady-state solutions. Thereto
we make the following substitution:

\emptyline $>$f3 := (t)->rho0*sech(t*tau);\\
\indent
 $>$f4 := (t)->epsilon*diff( sech(t*tau) ,t\$2);\#rho0

\emptyline
\[\boxed{
\mathit{f3} := t\rightarrow \rho 0\,\mathrm{sech}(t\,\tau )}
\]

\[\boxed{
\mathit{f4} := t\rightarrow \varepsilon \,({\frac {\partial ^{2}
}{\partial t^{2}}}\,\mathrm{sech}(t\,\tau ))}
\]

\emptyline $>$simplify(\\
subs(\{rho(t)=f3(t),zeta(t)=f4(t)\},master\_3) ):\\
\indent
 $>$expand( numer(\%)/rho0):\\
 \indent
  $>$eq10 := collect( numer(\%),cosh(t*tau) );

\emptyline \maplemultiline{ \mathit{eq10} := (\rho 0\,g +
g\,\varepsilon \,\tau ^{2} + \rho 0 \,\tau ^{2} + \varepsilon
\,\tau ^{4})\,\mathrm{cosh}(t\,\tau )^{
4} \\
\mbox{} + ( - 2\,g\,\varepsilon \,\tau ^{2} + \rho 0^{3}\,\Sigma
 + 3\,\Sigma \,\rho 0^{2}\,\varepsilon \,\tau ^{2} - 2\,\rho 0\,
\tau ^{2} - 20\,\varepsilon \,\tau ^{4})\,\mathrm{cosh}(t\,\tau )
^{2} + 24\,\varepsilon \,\tau ^{4} \\
\mbox{} - 6\,\Sigma \,\rho 0^{2}\,\varepsilon \,\tau ^{2} }

\emptyline We can see, that this substitution obeys the perturbed
steady-state equation ( $\varepsilon$ is the perturbation
amplitude).

\emptyline $>$eq11 := expand(\\
coeff(eq10,cosh(t*tau),0)/epsilon/tau$^2$ );\\
\indent
 $>$eq12 := expand( coeff(eq10,cosh(t*tau),2) );\\
 \indent \indent
  $>$eq13 := simplify( coeff(eq10,cosh(t*tau),4));

\emptyline
\[
\mathit{eq11} := 24\,\tau ^{2} - 6\,\Sigma \,\rho 0^{2}
\]

\[
\mathit{eq12} :=  - 2\,g\,\varepsilon \,\tau ^{2} + \rho 0^{3}\,
\Sigma  + 3\,\Sigma \,\rho 0^{2}\,\varepsilon \,\tau ^{2} - 2\,
\rho 0\,\tau ^{2} - 20\,\varepsilon \,\tau ^{4}
\]

\[
\mathit{eq13} := \rho 0\,g + g\,\varepsilon \,\tau ^{2} + \rho 0
\,\tau ^{2} + \varepsilon \,\tau ^{4}
\]

\emptyline $>$sol :=
solve(\{eq11=0,eq12=0,eq13=0\},\{rho0,tau,epsilon\});

\emptyline \maplemultiline{ \mathit{sol} := \{\tau
=\mathrm{RootOf}(g + \mathit{\_Z}^{2}), \, \rho
0=2\,\mathrm{RootOf}(\Sigma \,\mathit{\_Z}^{2} + g, \,
\mathit{label}=\mathit{\_L1}),  \\
\varepsilon = - {\displaystyle \frac {2}{3}} \,{\displaystyle
\frac {\mathrm{RootOf}(\Sigma \,\mathit{\_Z}^{2} + g, \,\mathit{
label}=\mathit{\_L1})}{g}} \}, \,\{\varepsilon =\varepsilon , \,
\rho 0=0, \,\tau =0\}, \{ \\
\rho 0={\displaystyle \frac {2}{5}} \,g\,\mathrm{RootOf}(\mathit{
\_Z}^{2}\,g\,\Sigma  + 5, \,\mathit{label}=\mathit{\_L2}), \,\\
\varepsilon =2\,\mathrm{RootOf}(\mathit{\_Z}^{2}\,g\,\Sigma  + 5
, \,\mathit{label}=\mathit{\_L2}),  \\
\tau =\mathrm{RootOf}(g + 5\,\mathit{\_Z}^{2})\} }

\emptyline $>$sol1\_tau := allvalues(subs(sol[1],tau));\\
\indent
 $>$sol1\_rho := allvalues(subs(sol[1],rho0));\\
 \indent \indent
  $>$sol1\_e := allvalues(subs(sol[1],epsilon));\\
  \indent \indent \indent
$>$sol2\_tau := allvalues(subs(sol[3],tau));\\
\indent \indent
 $>$sol2\_rho := allvalues(subs(sol[3],rho0));\\
 \indent
  $>$sol2\_e := allvalues(subs(sol[3],epsilon));

\emptyline
\[
\mathit{sol1\_tau} := \sqrt{ - g}, \, - \sqrt{ - g}
\]

\[
\mathit{sol1\_rho} := 2\,\sqrt{ - {\displaystyle \frac {g}{\Sigma
 }} }, \, - 2\,\sqrt{ - {\displaystyle \frac {g}{\Sigma }} }
\]

\[
\mathit{sol1\_e} :=  - {\displaystyle \frac {2}{3}} \,
{\displaystyle \frac {\sqrt{ - {\displaystyle \frac {g}{\Sigma } }
}}{g}} , \,{\displaystyle \frac {2}{3}} \,{\displaystyle \frac
{\sqrt{ - {\displaystyle \frac {g}{\Sigma }} }}{g}}
\]

\[
\mathit{sol2\_tau} := \sqrt{ - {\displaystyle \frac {1}{5}} \,g} ,
\, - \sqrt{ - {\displaystyle \frac {1}{5}} \,g}
\]

\[
\mathit{sol2\_rho} := {\displaystyle \frac {2}{5}} \,g\,\sqrt{ -
5\,{\displaystyle \frac {1}{g\,\Sigma }} }, \, - {\displaystyle
\frac {2}{5}} \,g\,\sqrt{ - 5\,{\displaystyle \frac {1}{g\,\Sigma
 }} }
\]

\[
\mathit{sol2\_e} := 2\,\sqrt{ - 5\,{\displaystyle \frac {1}{g\,
\Sigma }} }, \, - 2\,\sqrt{ - 5\,{\displaystyle \frac {1}{g\,
\Sigma }} }
\]

\emptyline \fbox{\parbox{.6\linewidth} {There exist two types of
the perturbed solutions}}. The first one corresponding to
unperturbed solution for the arbitrary small $\varepsilon $ has a
form

\emptyline $>$sol1 := subs(\\
\{rho0=sol1\_rho[1],tau=sol1\_tau[1],epsilon=sol1\_e[1]\},\\
expand((f3(t) + f4(t))$^2$ ) ):\\
\indent
$>$plot3d(subs(Sigma=1,sol1),t=-10..10,g=-0.05..0,axes=boxed);

\begin{center}
\mapleplot{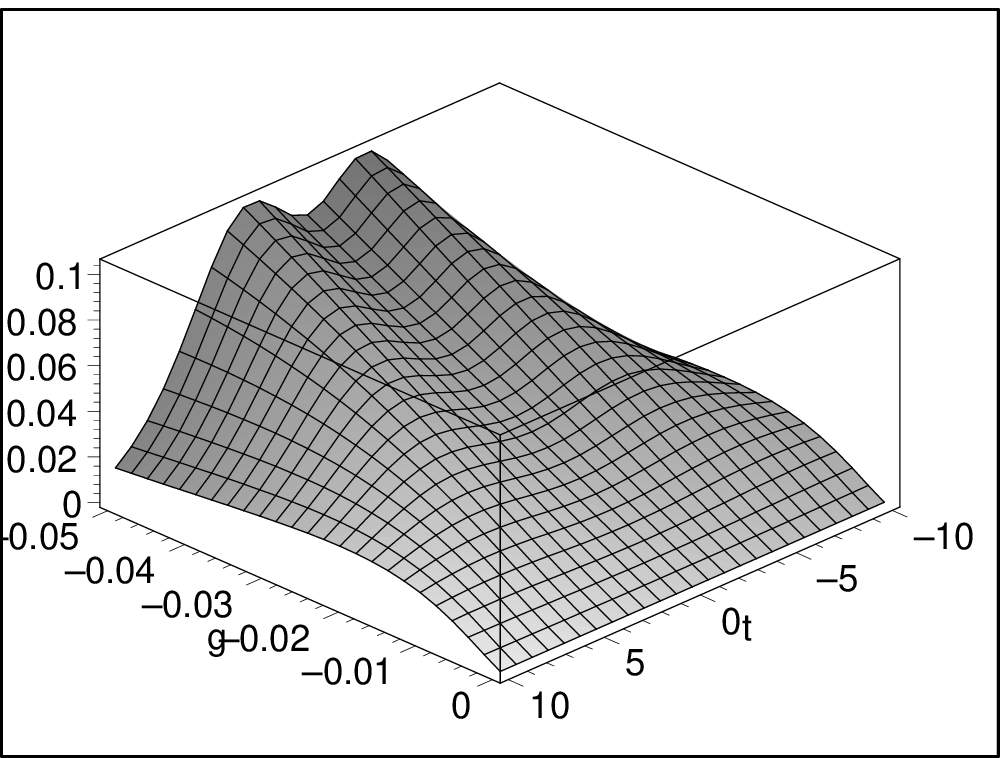}
\end{center}

\emptyline The second solution is:

\emptyline $>$sol2 := subs(\\
\{rho0=sol2\_rho[1],tau=sol2\_tau[1],epsilon=sol2\_e[1]\},expand(\\
(f3(t) + f4(t))$^2$ ) ):\\
\indent
$>$plot3d(subs(Sigma=1,sol2),t=-10..10,g=-0.05..0,axes=boxed);

\begin{center}
\mapleplot{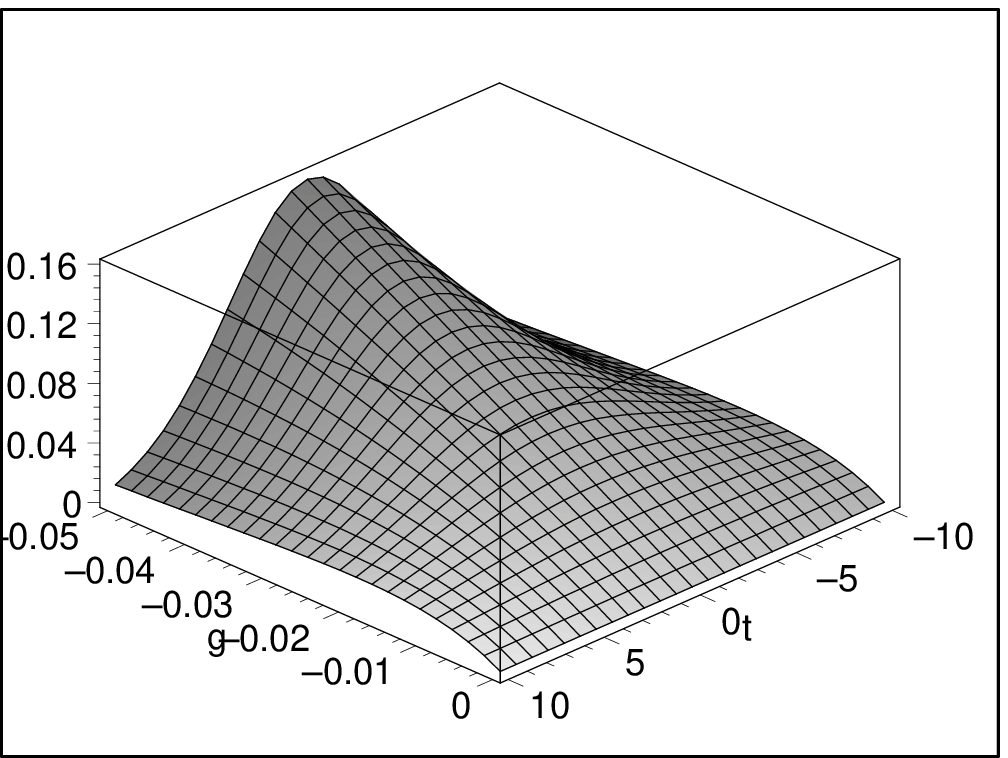}
\end{center}

\emptyline And the unperturbed solution is:

\emptyline $>$sol3 := subs(\\
\{rho0$^2$=-4*g/Sigma,tau=sqrt(-g),epsilon=0\},expand(\\
(f3(t) + f4(t))$^2$ ) ):\\
\indent
$>$plot3d(subs(Sigma=1,sol3),t=-20..20,g=-0.05..0,axes=boxed);

\begin{center}
\mapleplot{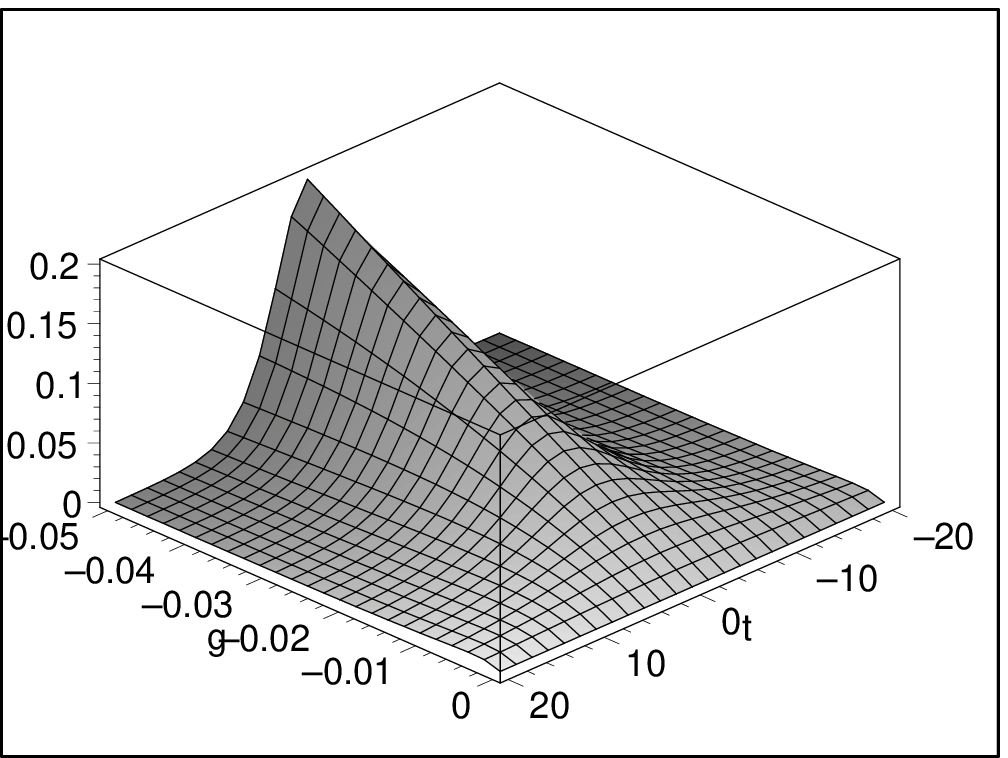}
\end{center}

\emptyline We can see that the perturbations widen the pulse and
reduce its intensity. Moreover, the perturbation of the first type
\underline{splits} the pulse.

\emptyline
\fbox{\parbox{.8\linewidth} {We suppose, that the excitation of
similar perturbations located within the ultrashort pulse
dissociates it for the large $\sigma $, when the contribution of
the higher-order nonlinear terms is essential}}.

\emptyline The important feature of the multiple pulse regimes is
the possibility of a strong correlation of the pulse parameters in
the multipulse complex. This is evidence of the pulse interaction.
An example of such interaction can be illustrated by the following
consideration. Let's take the first momentum of the
Landau-Ginzburg equation by the multiplication to the conjugated
field, adding the conjugated equation and the consequent
integration. The evolution of the energy is described by

\emptyline $>$Diff(E(z),z) = 2*g*E(z) +\\
int(conjugate(rho(z,t))*Diff(rho(z,t),`\$`(t,2)) +\\
rho(z,t)*Diff(conjugate(rho(z,t)),`\$`(t,2)),t=-infinity..infinity)\\
+ 2*Sigma*conjugate(rho(z,t))$^2$*rho(z,t)$^2$;

\emptyline
\begin{gather} \boxed{ \nonumber
{\frac {\partial }{\partial z}}\,\mathrm{E}(z)=2\,g\,\mathrm{E}(z
) +}\\ \nonumber \boxed{{\displaystyle \int _{ - \infty }^{\infty
}} \overline{(\rho (z, \,t))}\,({\frac {\partial ^{2}}{\partial
t^{2}}}\,\rho (z, \, t)) + \rho (z, \,t)\,({\frac {\partial
^{2}}{\partial t^{2}}}\, \overline{(\rho (z, \,t))})\,dt +
2\,\Sigma \,\overline{(\rho (z , \,t))}^{2}\,\rho (z, \,t)^{2}}
\end{gather}

\emptyline \noindent The second term by the virtue of

\emptyline $>$with(student):\\
\indent $>$intparts(int(rho(z,t)*diff(rho(z,t),`\$`(t,2)),t),
rho(z,t));\# the first term vanishes at infinity

\emptyline
\[
\rho (z, \,t)\,({\frac {\partial }{\partial t}}\,\rho (z, \,t))
 - {\displaystyle \int } ({\frac {\partial }{\partial t}}\,\rho (
z, \,t))^{2}\,dt
\]

\emptyline \noindent gives:

\emptyline $>$eq14 := Diff(E(z),z) = 2*g*E(z) -
2*int(Diff(conjugate(rho(z,t)),t)*\\
Diff(rho(z,t),t), t=-infinity..infinity) +
2*Sigma*conjugate(rho(z,t))$^2$*rho(z,t)$^2$;

\emptyline
\begin{gather} \nonumber
\mathit{eq14} :=\\ \nonumber
 {\frac {\partial }{\partial
z}}\,\mathrm{E}(z)=2 \,g\,\mathrm{E}(z) - 2\,{\displaystyle \int
_{ - \infty }^{\infty
 }} ({\frac {\partial }{\partial t}}\,\overline{(\rho (z, \,t))})
\,({\frac {\partial }{\partial t}}\,\rho (z, \,t))\,dt + 2\,
\Sigma \,\overline{(\rho (z, \,t))}^{2}\,\rho (z, \,t)^{2}
\end{gather}

\emptyline Now let's consider the simplest two-pulse complex:

\emptyline $>$fieldr :=\\
rho0*(sech((t-delta)*tau)+sech((t+delta)*tau)*cos(phi));\# real
part\\
\indent
$>$fieldim := rho0*sech((t+delta)*tau)*sin(phi);\#
imaginary part, delta is the distance, phi is the phase
difference\\
 \indent
 $>$print(`the spectral term is defined by:`);\\
 \indent
$>$diff(fieldr,t)$^2$+diff(fieldim,t)$^2$;

\emptyline
\[\boxed{
\mathit{fieldr} := \rho 0\,(\mathrm{sech}((t - \delta )\,\tau )
 + \mathrm{sech}((t + \delta )\,\tau )\,\mathrm{cos}(\phi ))}
\]

\emptyline
\[\boxed{
\mathit{fieldim} := \rho 0\,\mathrm{sech}((t + \delta )\,\tau )\,
\mathrm{sin}(\phi )}
\]

\emptyline
\[
\mathit{the\ spectral\ term\ is\ defined\ by:}
\]

\maplemultiline{ \rho 0^{2}\,( - \mathrm{sech}((t - \delta )\,\tau
)\,\mathrm{tanh }((t - \delta )\,\tau )\,\tau  - \mathrm{sech}((t
+ \delta )\, \tau )\,\mathrm{tanh}((t + \delta )\,\tau )\,\tau
\,\mathrm{cos}(
\phi ))^{2} \\
\mbox{} + \rho 0^{2}\,\mathrm{sech}((t + \delta )\,\tau )^{2}\,
\mathrm{tanh}((t + \delta )\,\tau )^{2}\,\tau ^{2}\,\mathrm{sin}(
\phi )^{2} }

\emptyline \noindent The spectral loss for this complex is (second
term in \textit{eq14}):

\emptyline $>$assume(tau>0):\\

\indent $>$s := 2*rho0$^2$*tau$^2$*(\\
int(sech((t-delta)*tau)$^2$*tanh((t-delta)*tau)$^2$,t=-infinity..infinity)\\
+\\
int(sech((t+delta)*tau)$^2$*tanh((t+delta)*tau)$^2$,t=-infinity..infinity)\\
+\\
2*int(sech((t-delta)*tau)*tanh((t-delta)*tau)*sech((t+delta)*tau)*tanh\\
((t+delta)*tau)*cos(phi),t=-infinity..infinity));

\indent $>$en := int(fieldr$^2$ +
fieldim$^2$,t=-infinity..infinity);

\emptyline
\maplemultiline{
s := 2\rho 0^{2}\,\tau \symbol{126}^{2} \\
 \left(  \! {\displaystyle \frac {4}{3}} \,{\displaystyle \frac {
1}{\tau \symbol{126}}}  - {\displaystyle \frac {4\,(\mathrm{ln}(
\mathrm{\%1})\,e^{(8\,\tau \symbol{126}\,\delta )} - 4\,e^{(8\,
\tau \symbol{126}\,\delta )} + 6\,\mathrm{ln}(\mathrm{\%1})\,
\mathrm{\%1} + 4 + \mathrm{ln}(\mathrm{\%1}))\,e^{(2\,\tau
\symbol{126}\,\delta )}\,\mathrm{cos}(\phi )}{( - 3\,e^{(8\,\tau
\symbol{126}\,\delta )} + 3\,\mathrm{\%1} + e^{(12\,\tau
\symbol{126}\,\delta )} - 1)\,\tau \symbol{126}}}  \!  \right)
 \\
\mathrm{\%1} := e^{(4\,\tau \symbol{126}\,\delta )} }

\emptyline
\[
\mathit{en} := 4\,{\displaystyle \frac {\rho 0^{2}\,(e^{(4\,\tau
\symbol{126}\,\delta )} - 1 + e^{(2\,\tau \symbol{126}\,\delta )}
\,\mathrm{ln}(e^{(4\,\tau \symbol{126}\,\delta )})\,\mathrm{cos}(
\phi ))}{\tau \symbol{126}\,(e^{(4\,\tau \symbol{126}\,\delta )}
 - 1)}}
\]

\emptyline
$>$plot3d(subs(tau=1/10,s/en),delta=0..40,phi=-Pi..Pi,axes=boxed,\\
view=0..0.01,title=`spectral loss vs. phase and distance`

\begin{center}
\mapleplot{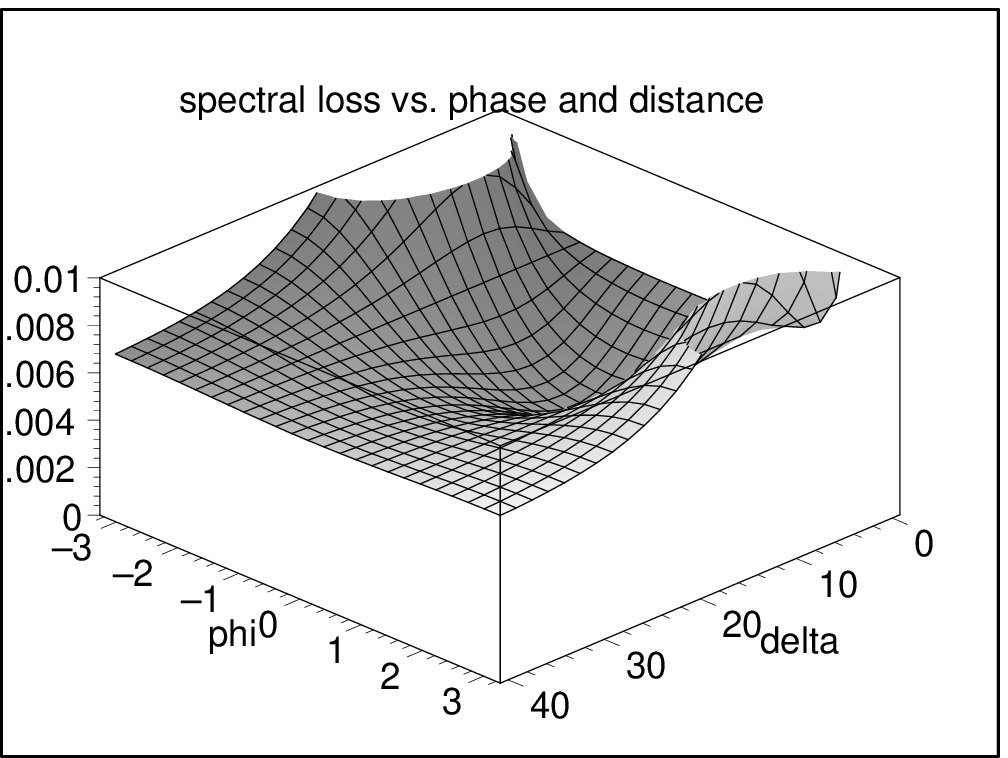}
\end{center}

\emptyline \fbox{\parbox{.8\linewidth} {So, there exists the
potential well, which can absorb the pulses. Also, the last term
in \textit{eq14} contributes to the interpulse attraction due to
the loss saturation enhancement produced by the pulse merging}}

\emptyline We can conclude, that the analytical treatment allowed
the comprehension of the basic features of the ultrashort pulse
dynamics, though they lie out of the quasi-soliton model validity.
Such collaboration between numerical and analytical methods
realized by means of the Maple faculties has not only technical
but also heuristic character.

\section{Mode locking due to a "slow" saturable absorber}

\emptyline

\subsection{Analytical theory and linear stability analysis}

\emptyline
\noindent
In two previous parts we considered the ultrashort pulse formation as
result of the loss saturation by pulse intensity. This supposes the
instant response of the saturable absorber on the signal variation.
The nonresonant (phase) nonlinearity obeys this demand even in
femtosecond domain. But the resonant nonlinearities are more inertial
and the time defining the relaxation of their excitation lies in the
wide region from 100 femtosecond to milliseconds and more. Therefore
the interaction of the ultrashort pulse with such structures differs
essentially from the previously considered.\\
\indent
Let the pulse duration is shorter then the longitudinal relaxation
time. Then the loss saturation is caused by pulse energy flux passed
through absorber. The similar situation had considered for dynamical
gain saturation in part \textit{4}. Here we shall take into
consideration simultaneously the dynamical gain and loss saturation.
These effects can be taken into consideration by the expansion of the
exponential transmission operator
$e^{ \left(  \!  - \frac {g}{1 + \frac {\varepsilon }{{E_{s}}}}
 \!  \right) }$   up to second order on pulse energy
$\varepsilon $ (see \cite{Haus3}). Here
${E_{s}}$ is the gain or loss saturation energy. Then the basic differential
equation is

\emptyline
$>$restart:\\
\indent \indent
 with(plots):\\
\indent \indent \indent
  master := diff(rho(z,t),z) = (alpha - g - l)*rho(z,t) -\\
\indent
chi*alpha*rho(z,t)*int(rho(z,zeta)$^{2}$,zeta=0..t) +\\
\indent
alpha*rho(z,t)*(chi*int(rho(z,zeta)$^{2}$,zeta=0..t))$^{2}$ + \\
\indent
g*rho(z,t)*int(rho(z,zeta)$^{2}$,zeta=0..t) -\\
\indent
g*rho(z,t)*(int(rho(z,zeta)$^{2}$,zeta=0..t))$^{2}$ + diff(rho(z,t),t\$2) +\\
\indent
delta*diff(rho(z,t),t);

\emptyline
\begin{gather*}
\boxed{\mathit{master} := {\frac {\partial }{\partial z}}\,\rho (z, \,t)
=}  \\
\boxed{(\alpha  - g - l)\,\rho (z, \,t) - \chi \,\alpha \,\rho (z, \,t)
\,\mathrm{\%1} + \alpha \,\rho (z, \,t)\,\chi ^{2}\,\mathrm{\%1}
^{2} + g\,\rho (z, \,t)\,\mathrm{\%1}} \\
\boxed{\mbox{} - g\,\rho (z, \,t)\,\mathrm{\%1}^{2} + ({\frac {\partial
^{2}}{\partial t^{2}}}\,\rho (z, \,t)) + \delta \,({\frac {
\partial }{\partial t}}\,\rho (z, \,t))} \\
\boxed{\mathrm{\%1} := {\displaystyle \int _{0}^{t}} \rho (z, \,\zeta )
^{2}\,d\zeta }
\end{gather*}

\emptyline \noindent Here $\chi $ is the ratio of the loss
saturation energy to gain saturation energy (saturation
parameter), \textit{l} is the unsaturable loss coefficient,
\textit{g} and $\alpha $ are the saturable loss and gain
coefficients at pulse peak, respectively, $\delta $ is the pulse
delay on the cavity round-trip. The form of this equation supposes
the normalization of time on ${t_{f}}$ , pulse energy on loss
saturation energy. We shall suppose the soliton-like form of
steady-state solution of \textit{master}:

\emptyline
$>$f1 := (t)$->$rho0*sech(t*tau);\# soliton form\\
\indent \indent
 f2 := (zeta)$->$rho0*sech(zeta*tau);\\
\indent \indent \indent
  ss := rhs(master):\\
\indent \indent
   subs(\{rho(z,t)=f1(t),rho(z,zeta)=f2(zeta)\},ss):\\
\indent
    simplify(\%):\\
\indent \indent
     expand( numer(\%)*2/rho0 ):\\
\indent
      eq := collect( collect( combine(\%,trig),\\
\indent
sinh(2*t*tau)),cosh(2*t*tau) );

\emptyline
\[
\mathit{f1} := t\rightarrow \rho 0\,\mathrm{sech}(t\,\tau )
\]

\[
\mathit{f2} := \zeta \rightarrow \rho 0\,\mathrm{sech}(\zeta \,
\tau )
\]

\begin{eqnarray*}
\mathit{eq} := (\tau ^{2}\,\alpha  - \tau ^{2}\,g + \alpha \,\rho
 0^{4}\,\chi ^{2} - \tau ^{2}\,l - g\,\rho 0^{4} + \tau ^{4})\,
\mathrm{cosh}(2\,t\,\tau )  \\
\mbox{} + ( - \chi \,\alpha \,\rho 0^{2}\,\tau  + g\,\rho 0^{2}\,
\tau  - \delta \,\tau ^{3})\,\mathrm{sinh}(2\,t\,\tau ) - \\
\alpha
\,\rho 0^{4}\,\chi ^{2} + \tau ^{2}\,\alpha  + g\,\rho 0^{4} -
\tau ^{2}\,g - \tau ^{2}\,l - 3\,\tau ^{4}
\end{eqnarray*}

\emptyline
\noindent
Since this equation is valid at any moment, we have the system of the
algebraic equations for the coefficients of hyperbolical functions.

\emptyline
$>$eq1 := coeff(eq,cosh(2*t*tau));\\
\indent \indent
 eq2 := factor(coeff(eq,sinh(2*t*tau)));\\
\indent
  eq3 := expand(eq-eq1*cosh(2*t*tau)-eq2*sinh(2*t*tau));

\emptyline
\[
\mathit{eq1} := \tau ^{2}\,\alpha  - \tau ^{2}\,g + \alpha \,\rho
 0^{4}\,\chi ^{2} - \tau ^{2}\,l - g\,\rho 0^{4} + \tau ^{4}
\]

\[
\mathit{eq2} :=  - \tau \,(\delta \,\tau ^{2} + \chi \,\alpha \,
\rho 0^{2} - g\,\rho 0^{2})
\]

\[
\mathit{eq3} :=  - \alpha \,\rho 0^{4}\,\chi ^{2} + \tau ^{2}\,
\alpha  + g\,\rho 0^{4} - \tau ^{2}\,g - \tau ^{2}\,l - 3\,\tau
^{4}
\]

\emptyline
\noindent
These equations define the inverse pulse duration
$\tau $, pulse intensity
$\rho 0^{2}$, and delay
$\delta $. Let make some manipulations:

\emptyline
$>$eq4 := expand( factor(eq1+eq3)/(2*tau$^{2}$) );\\
\indent \indent
 eq5 := expand((eq1-eq3)/2);

\emptyline
\[
\mathit{eq4} :=  - \tau ^{2} + \alpha  - g - l
\]

\[
\mathit{eq5} := \alpha \,\rho 0^{4}\,\chi ^{2} - g\,\rho 0^{4} +
2\,\tau ^{4}
\]

\emptyline
\noindent
Hence

\emptyline
$>$sol1 := solve(eq4=0,tau$^{2}$);\# solution for tau$^{2}$\\
\indent \indent
solve(subs(tau$^{4}$=sol1$^{2}$,eq5)=0,rho0$^{4}$);\\
\indent \indent \indent
factor(\%);\# solution for rho0$^{4}$

\emptyline
\[ \boxed{
\mathit{sol1} := \alpha  - g - l}
\]
\[
 - 2\,{\displaystyle \frac {\alpha ^{2} - 2\,\alpha \,g - 2\,
\alpha \,l + g^{2} + 2\,g\,l + l^{2}}{\alpha \,\chi ^{2} - g}}
\]
\[
 - 2\,{\displaystyle \frac {(\alpha  - g - l)^{2}}{\alpha \,\chi
^{2} - g}}
\]

\emptyline
\noindent
Note that the last solution needs some consideration. The comparison
with second expression gives for intensity:

\begin{center}
\[ \boxed{
 \frac {\sqrt{2}\,(\alpha  - g - l)}{\sqrt{g - \chi ^{2}\,\alpha
}}}
\]
\end{center}

\emptyline
$>$sol2 :=  sqrt(2)*(alpha-g-l)/sqrt(g-chi$^{2}$*alpha):\# solution for
intensity

\emptyline
\noindent
And at last, for delay we have

\emptyline
$>$subs( \{tau$^{2}$=sol1, rho0$^{2}$=sol2\},expand(eq2/(2*tau)) ):\\
\indent \indent
 simplify(\%):\\
 \indent \indent \indent
  numer(\%):\\
  \indent \indent \indent \indent
   expand(\%/(-alpha+g+l)):\\
   \indent
    sol3 := solve(\%=0,delta);\# solution for delta

\emptyline
\[ \boxed{
\mathit{sol3} :=  - {\displaystyle \frac {\sqrt{2}\,(\chi \,
\alpha  - g)}{\sqrt{g - \alpha \,\chi ^{2}}}} }
\]

\emptyline
\noindent
The gain coefficient at pulse peak is:

\emptyline
$>$assume(tau,positive):\\
\indent \indent
 int(f1(t)$^{2}$,t=-infinity..0):\\
 \indent
  subs(\{tau=sqrt(sol1),rho0$^{2}$=sol2\},\%):\# half-pulse energy\\
  \indent \indent
   numer( simplify( alpha0/(1+\%) - alpha ) ) = 0;\# equation for
saturated gain, alpha0 is nonsaturated gain

\emptyline
\[
\alpha 0\,\sqrt{g - \alpha \,\chi ^{2}} - \alpha \,\sqrt{g -
\alpha \,\chi ^{2}} - \alpha \,\sqrt{2}\,\sqrt{\alpha  - g - l}=0
\]

\emptyline
$>$sol := solve(\%,alpha):\\
\indent \indent
tau := 'tau':

\emptyline
We can plot the pulse intensity and duration versus nonsaturated gain for
different
$\chi $. It should be noted, that
$\chi $ \TEXTsymbol{<} 1 because for the pulse formation the loss saturation
has to leave behind the gain saturation.

\emptyline
$>$plot3d(subs(\{l=0.01,g=0.05\\
\indent
\},subs(alpha=sol[1],sol2)),alpha0=0.061..1,chi=0..1,\\
\indent
axes=boxed,title=`pulse intensity vs nonsaturated gain`);

\emptyline

\emptyline
$>$plot3d(subs(\{l=0.01,g=0.05\\
\indent
\},subs(alpha=sol[1],log(1/sqrt(sol1)))),\\
\indent
alpha0=0.061..1,chi=0..1,axes=boxed,\\
\indent
title=`logarithm of pulse width vs nonsaturated gain`);

\emptyline
\begin{center}
\mapleplot{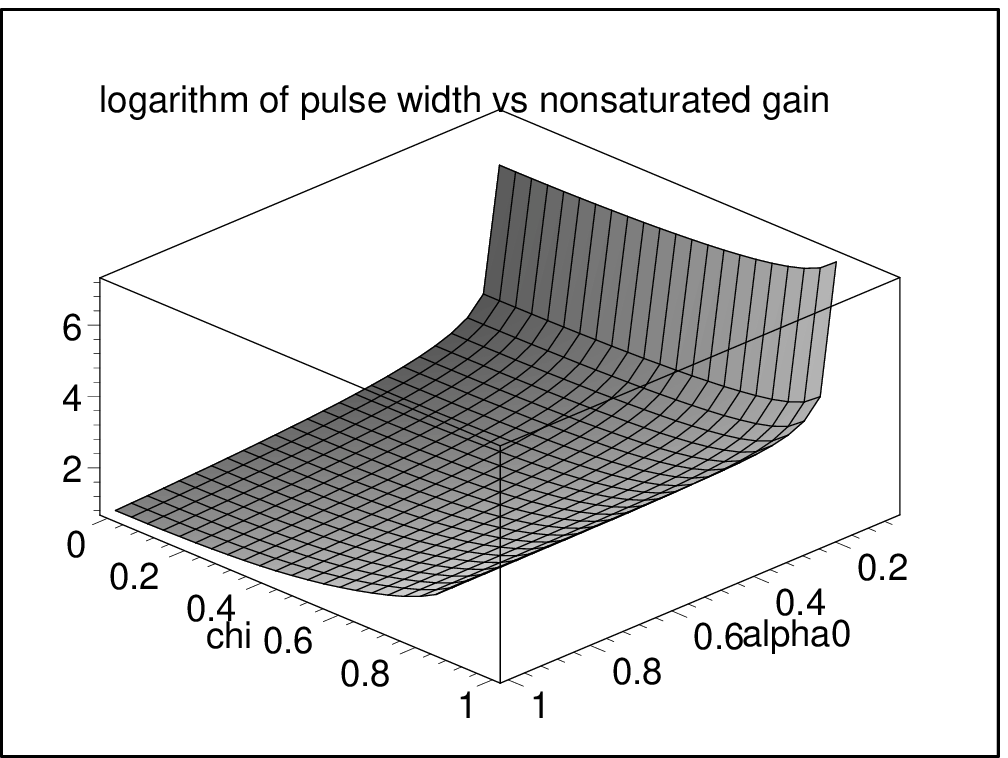}
\end{center}

\emptyline
$>$plot3d(subs(\{l=0.01,g=0.05\\
\indent
\},subs(alpha=sol[1],sol3)),alpha0=0.061..1,chi=0..1,axes=boxed,title=
`logarithm of pulse width vs nonsaturated gain`);

\emptyline
\begin{center}
\mapleplot{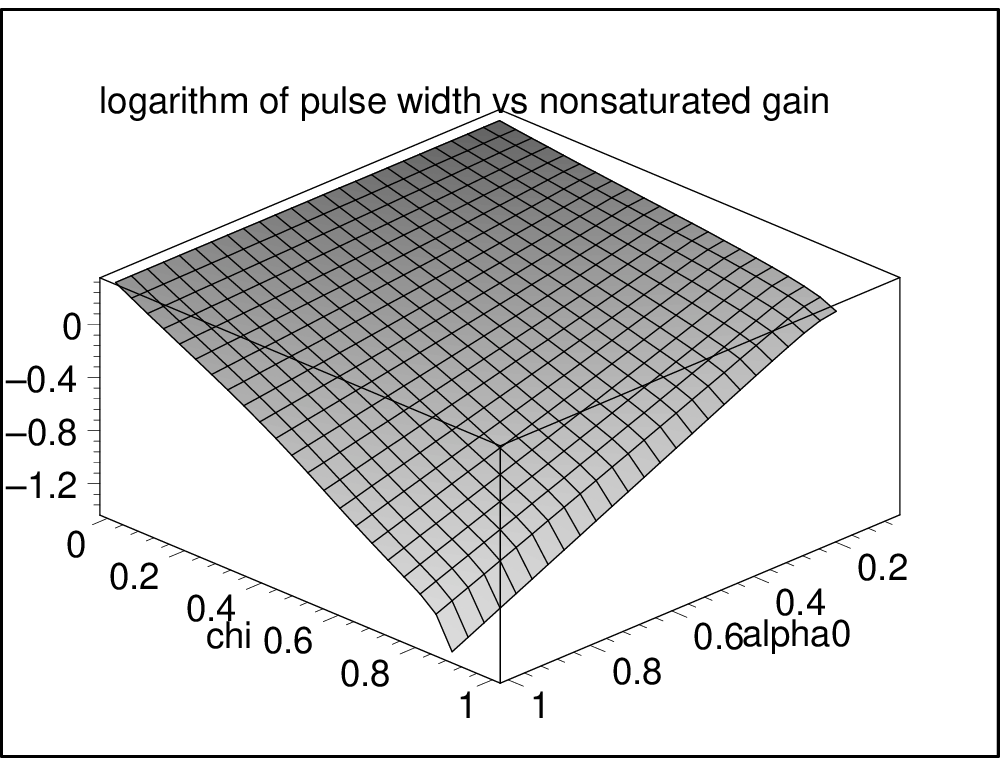}
\end{center}

\emptyline
\noindent
We can see that the pulse width is decreased by gain growth and
decrease of
$\chi $. Additionally there is maximum of the dependence of intensity on
saturation parameter
$\chi $.\\
\indent
Now we shall analyze the ultrashort pulse stability in framework of
linear theory (see part \textit{4}). In this case the equation for
evolution of exponentially growing perturbation
$\xi $(\textit{t}) is

\emptyline
$>$(alpha-g-l-lambda)*xi(t) - chi*alpha*xi(t)*int(rho(zeta)$^{2}$,\\
\indent
zeta = 0.. t) - chi*alpha*rho(t)*int(rho(zeta)*xi(zeta),\\
\indent
zeta = 0 .. t) +\\
\indent
alpha*xi(t)*chi$^{2}$*int(rho(zeta)$^{2}$,zeta = 0 .. t)$^{2}$ +\\
\indent
g*xi(t)*int(rho(zeta)$^{2}$,zeta = 0 .. t) +\\
\indent
g*rho(t)*int(rho(zeta)*xi(t),zeta = 0 .. t) -\\
\indent
g*xi(t)*int(rho(zeta)$^{2}$,zeta = 0 .. t)$^{2}$ + \\
\indent
diff(xi(t),`\$`(t,2)) + delta*diff(xi(t),t);

\emptyline
\begin{gather*}
\boxed{(\alpha  - g - l - \lambda )\,\xi (t) -}  \\
\boxed{\chi \,\alpha \,\xi (t)\,
\mathrm{\%1} - \chi \,\alpha \,\rho (t)\,{\displaystyle \int _{0}
^{t}} \rho (\zeta )\,\xi (\zeta )\,d\zeta  + \alpha \,\xi (t)\,
\chi ^{2}\,\mathrm{\%1}^{2} + g\,\xi (t)\,\mathrm{\%1}} \\
\boxed{\mbox{} + g\,\rho (t)\,{\displaystyle \int _{0}^{t}} \rho (\zeta
)\,\xi (t)\,d\zeta  - g\,\xi (t)\,\mathrm{\%1}^{2} + ({\frac {
\partial ^{2}}{\partial t^{2}}}\,\xi (t)) + \delta \,({\frac {
\partial }{\partial t}}\,\xi (t))} \\
\boxed{\mathrm{\%1} := {\displaystyle \int _{0}^{t}} \rho (\zeta )^{2}\,
d\zeta}
\end{gather*}

\emptyline
\noindent
Here
$\lambda $ is the perturbation's growth increment, its positive value
corresponds to ultrashort pulse destabilization. An
integro-differential character of this equation raises the ODE's order
therefore we shall use some assumptions relatively perturbation's
envelope.\\
\indent
Let consider a long-wave limit for perturbations. In this case the
perturbation's envelope is smooth in compare with pulse envelope. Then
we can to exclude the integration over perturbation.

\emptyline
$>$f1 := (t)$->$rho0*sech(t*tau):\# soliton form\\
\indent \indent
 f2 := (zeta)$->$rho0*sech(zeta*tau):\\
 \indent
  eq1 := (alpha-g-l-lambda)*xi(t)-chi*alpha*xi(t)*int(rho(zeta)$^{2}$,\\
  \indent
  zeta= 0 .. t)-chi*alpha*xi0*rho(t)*int(rho(zeta),\\
  \indent
  zeta = 0 ..t)+alpha*xi(t)*chi$^{2}$*int(rho(zeta)$^{2}$,\\
  \indent
  zeta = 0 ..t)$^{2}$+g*xi(t)*int(rho(zeta)$^{2}$,\\
  \indent
  zeta = 0 ..t)+g*xi0*rho(t)*int(rho(zeta),\\
  \indent
  zeta = 0 ..t)-g*xi(t)*int(rho(zeta)$^{2}$,\\
  \indent
  zeta = 0 ..t)$^{2}$+diff(xi(t),`\$`(t,2))+delta*diff(xi(t),t);\\
  \indent \indent
  \# xi0 is the amplitude of perturbation at t=0

\emptyline
\begin{gather*}
\mathit{eq1} := (\alpha  - g - l - \lambda )\,\xi (t) -  \\
\chi \,\alpha \,\xi (t)\,\mathrm{\%1} - \chi \,\alpha \,\xi 0\,\rho (t)
\,{\displaystyle \int _{0}^{t}} \rho (\zeta )\,d\zeta  + \alpha
\,\xi (t)\,\chi ^{2}\,\mathrm{\%1}^{2} + g\,\xi (t)\,\mathrm{\%1}
 \\
\mbox{} + g\,\xi 0\,\rho (t)\,{\displaystyle \int _{0}^{t}} \rho
(\zeta )\,d\zeta  - g\,\xi (t)\,\mathrm{\%1}^{2} + ({\frac {
\partial ^{2}}{\partial t^{2}}}\,\xi (t)) + \delta \,({\frac {
\partial }{\partial t}}\,\xi (t)) \\
\mathrm{\%1} := {\displaystyle \int _{0}^{t}} \rho (\zeta )^{2}\,
d\zeta
\end{gather*}

\emptyline
\noindent
The long-wave approximation allows to neglect the second-order
derivation in compare with first-order one.

\emptyline
$>$value( subs(\{rho(t)=f1(t),\\
\indent
rho(zeta)=f2(zeta)\},eq1-diff(xi(t),`\$`(t,2))) );

\emptyline
\begin{eqnarray*}
(\alpha  - g - l - \lambda )\,\xi (t) - {\displaystyle \frac {
\chi \,\alpha \,\xi (t)\,\rho 0^{2}\,\mathrm{sinh}(t\,\tau )}{
\tau \,\mathrm{cosh}(t\,\tau )}}  -  \\
{\displaystyle \frac {\chi \,
\alpha \,\xi 0\,\rho 0^{2}\,\mathrm{sech}(t\,\tau )\,\mathrm{
arctan}(\mathrm{sinh}(t\,\tau ))}{\tau }}  \\
\mbox{} + {\displaystyle \frac {\alpha \,\xi (t)\,\chi ^{2}\,\rho
 0^{4}\,\mathrm{sinh}(t\,\tau )^{2}}{\tau ^{2}\,\mathrm{cosh}(t\,
\tau )^{2}}}  + {\displaystyle \frac {g\,\xi (t)\,\rho 0^{2}\,
\mathrm{sinh}(t\,\tau )}{\tau \,\mathrm{cosh}(t\,\tau )}}  + \\
{\displaystyle \frac {g\,\xi 0\,\rho 0^{2}\,\mathrm{sech}(t\,\tau
 )\,\mathrm{arctan}(\mathrm{sinh}(t\,\tau ))}{\tau }}  \\
\mbox{} - {\displaystyle \frac {g\,\xi (t)\,\rho 0^{4}\,\mathrm{
sinh}(t\,\tau )^{2}}{\tau ^{2}\,\mathrm{cosh}(t\,\tau )^{2}}}  +
\delta \,({\frac {\partial }{\partial t}}\,\xi (t))
\end{eqnarray*}

\emptyline
$>$subs(\{op(2,\%)=-chi*alpha*xi(t)*rho0$^{2}$*tanh(t*tau),\\
\indent \indent
op(4,\%)=chi$^{2}$*alpha*xi(t)*rho0$^{4}$*tanh(t*tau)$^{2}$,\\
\indent \indent
op(5,\%)=g*xi(t)*rho0$^{2}$*tanh(t*tau),\\
\indent \indent
op(7,\%)=-g*xi(t)*rho0$^{4}$*tanh(t*tau)$^{2}$\},\%);

\emptyline
\begin{eqnarray*}
(\alpha  - g - l - \lambda )\,\xi (t) - \chi \,\alpha \,\xi (t)\,
\rho 0^{2}\,\mathrm{tanh}(t\,\tau ) -  \\
{\displaystyle \frac {\chi
\,\alpha \,\xi 0\,\rho 0^{2}\,\mathrm{sech}(t\,\tau )\,\mathrm{
arctan}(\mathrm{sinh}(t\,\tau ))}{\tau }}  \\
\mbox{} + \chi ^{2}\,\alpha \,\xi (t)\,\rho 0^{4}\,\mathrm{tanh}(
t\,\tau )^{2} + g\,\xi (t)\,\rho 0^{2}\,\mathrm{tanh}(t\,\tau )
 + \\ \nonumber
 {\displaystyle \frac {g\,\xi 0\,\rho 0^{2}\,\mathrm{sech}(t\,
\tau )\,\mathrm{arctan}(\mathrm{sinh}(t\,\tau ))}{\tau }}  \\
\mbox{} - g\,\xi (t)\,\rho 0^{4}\,\mathrm{tanh}(t\,\tau )^{2} +
\delta \,({\frac {\partial }{\partial t}}\,\xi (t))
\end{eqnarray*}

\emptyline
$>$dsolve(\{\%=0, xi(0)=xi0\} ,xi(t));

\emptyline
\begin{gather*}
\xi (t)= \\
{\displaystyle \int _{0}^{t}} \xi 0\,\rho 0^{2}\,\mathrm{
arctan}(\mathrm{sinh}(u\,\tau ))\,({\displaystyle \frac {\mathrm{
sinh}(u\,\tau ) - \mathrm{cosh}(u\,\tau )}{\mathrm{cosh}(u\,\tau
)}} )^{( - 1/2\,\frac {\rho 0^{2}\,(\mathrm{\%1} + g - \chi \,
\alpha  - g\,\rho 0^{2})}{\delta \,\tau })} \nonumber \\ \nonumber
({\displaystyle \frac {\mathrm{cosh}(u\,\tau ) + \mathrm{sinh}(u
\,\tau )}{\mathrm{cosh}(u\,\tau )}} )^{(1/2\,\frac {\rho 0^{2}\,(
\mathrm{\%1} - g + \chi \,\alpha  - g\,\rho 0^{2})}{\delta \,\tau
 })}\,(\chi \,\alpha  - g) \\
e^{( - \frac { - \alpha \,u\,\tau \,\mathrm{cosh}(u\,\tau ) + g\,
u\,\tau \,\mathrm{cosh}(u\,\tau ) + l\,u\,\tau \,\mathrm{cosh}(u
\,\tau ) + \lambda \,u\,\tau \,\mathrm{cosh}(u\,\tau ) + \alpha
\,\rho 0^{4}\,\chi ^{2}\,\mathrm{sinh}(u\,\tau ) - g\,\rho 0^{4}
\,\mathrm{sinh}(u\,\tau )}{\mathrm{cosh}(u\,\tau )\,\delta \,\tau
 })}
  \\
\left/\right. \!  \! (\delta \,\tau \,\mathrm{cosh}(u\,\tau ))du(\mathrm{tanh}(t\,\tau
) - 1)^{\mathrm{\%2}}\,(1 + \mathrm{tanh}(t\,\tau ))^{( - 1/2\,
\frac {\rho 0^{2}\,(\mathrm{\%1} - g + \chi \,\alpha  - g\,\rho 0
^{2})}{\delta \,\tau })} \\
e^{(\frac { - \alpha \,t\,\tau  + g\,t\,\tau  + l\,t\,\tau  +
\lambda \,t\,\tau  + \alpha \,\rho 0^{4}\,\chi ^{2}\,\mathrm{tanh
}(t\,\tau ) - g\,\rho 0^{4}\,\mathrm{tanh}(t\,\tau )}{\delta \,
\tau })}\mbox{} + (\mathrm{tanh}(t\,\tau ) - 1)^{\mathrm{\%2}}\,
\xi 0 \\
(1 + \mathrm{tanh}(t\,\tau ))^{( - 1/2\,\frac {\rho 0^{2}\,(
\mathrm{\%1} - g + \chi \,\alpha  - g\,\rho 0^{2})}{\delta \,\tau
 })}\, \\
 e^{(\frac { - \alpha \,t\,\tau  + g\,t\,\tau  + l\,t\,\tau
 + \lambda \,t\,\tau  + \alpha \,\rho 0^{4}\,\chi ^{2}\,\mathrm{
tanh}(t\,\tau ) - g\,\rho 0^{4}\,\mathrm{tanh}(t\,\tau )}{\delta
\,\tau })} \\
 \left/ {\vrule height0.44em width0em depth0.44em} \right. \!
 \! (-1)^{\mathrm{\%2}} \\
\mathrm{\%1} := \alpha \,\rho 0^{2}\,\chi ^{2} \\
\mathrm{\%2} := {\displaystyle \frac {1}{2}} \,{\displaystyle
\frac {\rho 0^{2}\,(\mathrm{\%1} + g - \chi \,\alpha  - g\,\rho 0
^{2})}{\delta \,\tau }}
\end{gather*}

\emptyline
\noindent
The last term in this expression has not appropriate asymptotic
behavior at infinity. As consequence, \fbox{there are not long-wave
excitations in our case}.\\
\indent
A short-wave approximation allows to simplify the problem on the basis
of Riemann-Lebesgue theorem:
$\int _{ - \infty }^{\infty }\mathrm{f}(t)\,e^{(I\,\omega \,t)}\,
dt$ =o(1) (
$\omega $--\TEXTsymbol{>}
$\infty $). That is the integral
$\int _{0}^{t}\rho (\zeta )\,\xi (\zeta )\,d\zeta $ is small value if
$\xi (\zeta )$ is quickly oscillating function without steady-state points. In this
case (when
$\tau $\TEXTsymbol{<}\TEXTsymbol{<}1 ) we have (see \cite{Kalashnikov3}):

\emptyline
$>$f1 := (t)$->$rho0*sech(t*tau):\# soliton form\\
\indent \indent
 f2 := (zeta)$->$rho0*sech(zeta*tau):\\
 \indent \indent \indent
  eq1 :=\\
  \indent
(alpha-g-l-lambda)*xi(t) - chi*alpha*xi(t)*int(rho(zeta)$^{2}$,zeta = 0 ..t) +\\
\indent
 alpha*xi(t)*chi$^{2}$*int(rho(zeta)$^{2}$,zeta = 0 .. t)$^{2}$ +\\
 \indent
g*xi(t)*int(rho(zeta)$^{2}$,zeta = 0 .. t) - \\
\indent
g*xi(t)*int(rho(zeta)$^{2}$,zeta= 0 .. t)$^{2}$ + \\
\indent
diff(xi(t),`\$`(t,2)) + delta*diff(xi(t),t);

\emptyline
\begin{eqnarray*}
\mathit{eq1} := (\alpha  - g - l - \lambda )\,\xi (t) - \chi \,
\alpha \,\xi (t)\,\mathrm{\%1} + \alpha \,\xi (t)\,\chi ^{2}\,
\mathrm{\%1}^{2} + \\
g\,\xi (t)\,\mathrm{\%1} - g\,\xi (t)\,
\mathrm{\%1}^{2} + ({\frac {\partial ^{2}}{\partial t^{2}}}\,\xi
(t)) + \delta \,({\frac {\partial }{\partial t}}\,\xi (t)) \\
\mathrm{\%1} := {\displaystyle \int _{0}^{t}} \rho (\zeta )^{2}\,
d\zeta
\end{eqnarray*}

\emptyline
$>$subs(\{rho(t)=f1(t), rho(zeta)=f2(zeta)\},eq1) );

\emptyline
\begin{eqnarray*}
(\alpha  - g - l - \lambda )\,\xi (t) - {\displaystyle \frac {
\chi \,\alpha \,\xi (t)\,\rho 0^{2}\,\mathrm{sinh}(t\,\tau )}{
\tau \,\mathrm{cosh}(t\,\tau )}}  + \\
{\displaystyle \frac {\alpha
\,\xi (t)\,\chi ^{2}\,\rho 0^{4}\,\mathrm{sinh}(t\,\tau )^{2}}{
\tau ^{2}\,\mathrm{cosh}(t\,\tau )^{2}}}  + {\displaystyle
\frac {g\,\xi (t)\,\rho 0^{2}\,\mathrm{sinh}(t\,\tau )}{\tau \,
\mathrm{cosh}(t\,\tau )}} - \\
\mbox{}  {\displaystyle \frac {g\,\xi (t)\,\rho 0^{4}\,\mathrm{
sinh}(t\,\tau )^{2}}{\tau ^{2}\,\mathrm{cosh}(t\,\tau )^{2}}}  +
({\frac {\partial ^{2}}{\partial t^{2}}}\,\xi (t)) + \delta \,(
{\frac {\partial }{\partial t}}\,\xi (t))
\end{eqnarray*}

\emptyline
\noindent
We can rewrite this expression:

\emptyline
$>$(alpha-g-l-lambda)*xi(t)-chi*alpha*xi(t)*rho0$^{2}$*\\
\indent
tanh(t*tau)/tau+alpha*xi(t)*chi$^{2}$*rho0$^{4}$*\\
\indent
tanh(t*tau)$^{2}$/tau$^{2}$+g*xi(t)*rho0$^{2}$*\\
tanh(t*tau)/tau-g*xi(t)*rho0$^{4}$*tanh(t*tau)$^{2}$/tau$^{2}$+\\
diff(xi(t),`\$`(t,2))+delta*diff(xi(t),t):\\
\indent \indent
 eq2 := collect(\%, tanh);\\
 \indent \indent \indent
  A1 = expand( coeff(eq2, tanh(t*tau)$^{2}$)/ xi(t)):\\
  \indent \indent
   A2 = expand( coeff(eq2, tanh(t*tau))/ xi(t)):\\
   \indent
    eq3 := subs( alpha-g-l-lambda=A3,A1*xi(t)*tanh(t*tau)$^{2}$ +\\
    \indent
A2*xi(t)*tanh(tau*t) + coeff(eq2,tanh(t*tau),0) );

\emptyline
\begin{eqnarray*}
\mathit{eq2} := ( - {\displaystyle \frac {g\,\xi (t)\,\rho 0^{4}
}{\tau ^{2}}}  + {\displaystyle \frac {\alpha \,\xi (t)\,\chi ^{2
}\,\rho 0^{4}}{\tau ^{2}}} )\,\mathrm{tanh}(t\,\tau )^{2} + \\
({\displaystyle \frac {g\,\xi (t)\,\rho 0^{2}}{\tau }}  -
{\displaystyle \frac {\chi \,\alpha \,\xi (t)\,\rho 0^{2}}{\tau }
} )\,\mathrm{tanh}(t\,\tau ) \\
\mbox{} + (\alpha  - g - l - \lambda )\,\xi (t) + ({\frac {
\partial ^{2}}{\partial t^{2}}}\,\xi (t)) + \delta \,({\frac {
\partial }{\partial t}}\,\xi (t))
\end{eqnarray*}

\emptyline
\begin{eqnarray*}
\mathit{eq3} := \mathit{A1}\,\xi (t)\,\mathrm{tanh}(t\,\tau )^{2}
 + \mathit{A2}\,\xi (t)\,\mathrm{tanh}(t\,\tau ) + \mathit{A3}\,
\xi (t) + \\
({\frac {\partial ^{2}}{\partial t^{2}}}\,\xi (t)) +
\delta \,({\frac {\partial }{\partial t}}\,\xi (t))
\end{eqnarray*}

\emptyline
$>$dsolve(eq3=0, xi(t));

\emptyline
\begin{gather*}
\xi (t)=\\
\mathit{\_C1}\mathrm{hypergeom}(\\
[\mathrm{\%1} + \mathrm{
\%3} + \mathrm{\%2}, \, - \mathrm{\%3} + 1 + \mathrm{\%2} +
\mathrm{\%1}], \,[ - {\displaystyle \frac {1}{2}} \,
{\displaystyle \frac { - 2\,\tau  - 4\,\mathrm{\%2}\,\tau  +
\delta }{\tau }} ],  \\
 - {\displaystyle \frac {1}{2}} \,\mathrm{tanh}(t\,\tau ) +
{\displaystyle \frac {1}{2}} )(\mathrm{tanh}(t\,\tau ) - 1)^{
\mathrm{\%2}}\,(1 + \mathrm{tanh}(t\,\tau ))^{\mathrm{\%1}}
\mbox{} + \\
\mathit{\_C2}\,( - {\displaystyle \frac {1}{2}} \,
\mathrm{tanh}(t\,\tau ) + {\displaystyle \frac {1}{2}} )^{(1/2\,
\frac { - 4\,\mathrm{\%2}\,\tau  + \delta }{\tau })} \\
\mathrm{hypergeom}(\\
[{\displaystyle \frac {1}{2}} \,
{\displaystyle \frac { - 2\,\mathrm{\%3}\,\tau  + 2\,\tau  +
\delta  + 2\,\mathrm{\%1}\,\tau  - 2\,\mathrm{\%2}\,\tau }{\tau }
} , \,{\displaystyle \frac {1}{2}} \,{\displaystyle \frac {2\,
\mathrm{\%1}\,\tau  + 2\,\mathrm{\%3}\,\tau  + \delta  - 2\,
\mathrm{\%2}\,\tau }{\tau }} ], \\
\mbox{} [{\displaystyle \frac {1}{2}} \,{\displaystyle \frac {2\,\tau  -
4\,\mathrm{\%2}\,\tau  + \delta }{\tau }} ], \, - {\displaystyle
\frac {1}{2}} \,\mathrm{tanh}(t\,\tau ) + {\displaystyle \frac {1
}{2}} )(\mathrm{tanh}(t\,\tau ) - 1)^{\mathrm{\%2}}\,(1 +
\mathrm{tanh}(t\,\tau ))^{\mathrm{\%1}} \\
\mathrm{\%1} := \mathrm{RootOf}(\mathit{A3} - \mathit{A2} +
\mathit{A1} + 2\,\delta \,\mathit{\_Z}\,\tau  + 4\,\mathit{\_Z}^{
2}\,\tau ^{2}) \\
\mathrm{\%2} := \mathrm{RootOf}(4\,\mathit{\_Z}^{2}\,\tau ^{2} -
2\,\delta \,\mathit{\_Z}\,\tau  + \mathit{A2} + \mathit{A3} +
\mathit{A1}) \\
\mathrm{\%3} := \mathrm{RootOf}(\mathit{A1} - \mathit{\_Z}\,\tau
^{2} + \mathit{\_Z}^{2}\,\tau ^{2})
\end{gather*}

\emptyline
$>$coeff1 := allvalues(\\
\indent
-RootOf(A1-\_Z*tau$^{2}$+\_Z$^{2}$*tau$^{2}$)+1+\\
\indent
RootOf(A3-A2+A1+2*delta*\_Z*tau+4*\_Z$^{2}$*tau$^{2}$)+\\
\indent
RootOf(4*\_Z$^{2}$*tau$^{2}$-2*delta*\_Z*tau+A2+A3+A1) ):\# 8 coefficients\\
\indent \indent
coeff2 := allvalues(\\
\indent
-1/2*(2*RootOf(A1-\_Z*tau$^{2}$+\_Z$^{2}$*tau$^{2}$)*tau-2*tau+delta-\\
\indent
2*RootOf(4*\_Z$^{2}$*tau$^{2}$-2*delta*\_Z*tau+A2+A3+A1)*tau+\\
\indent
2*RootOf(A3-A2+A1+2*delta*\_Z*tau+4*\_Z$^{2}$*tau$^{2}$)*tau)/tau ):\# 8 coefficients

\emptyline \noindent We have the set of 16 first coefficients of
hypergeometric functions, which can cause the appropriate
asymptotic behavior at infinity if they are equal to negative
integers (see, in particular, part \textit{4}). Moreover, they are
to be the large negative integers in order to satisfy to our
short-wave approximation. So, we investigate the high-level
excitations in the "potential well" formed by gain and loss
saturation.

\emptyline
$>$\#first coefficients in hypergeometric functions\\
\indent \indent
 c := array(1..16):\\
 \indent \indent \indent
  for i from 1 to 8 do\\
  \indent
   c[i] := subs(\{A1 = -g*rho0$^{4}$/(tau$^{2}$)+alpha*chi$^{2}$*rho0$^{4}$/(tau$^{2}$),\\
   \indent
   A2 = g*rho0$^{2}$/tau-chi*alpha*rho0$^{2}$/tau,\\
   \indent
   A3 = alpha-g-l-lambda\},coeff1[i]):\\
   \indent \indent \indent
    od:\\
    \indent \indent
     for i from 9 to 16 do\\
     \indent
      c[i] := subs(\{A1 =\\
      \indent
-g*rho0$^{4}$/(tau$^{2}$)+alpha*chi$^{2}$*rho0$^{4}$/(tau$^{2}$),\\
\indent
A2 = g*rho0$^{2}$/tau-chi*alpha*rho0$^{2}$/tau,\\
A3 = alpha-g-l-lambda\},coeff2[i-8]):\\
\indent
        od:

\emptyline
$>$\#16 coefficients produce 16 equations for lambda (N is positive
integer)\\
\indent \indent
 s := array(1..16):\\
 \indent \indent \indent
  for j from 1 to 16 do\\
  \indent \indent
   s[j] := solve(c[j] + N = 0, lambda):\\
   \indent
    od:

\emptyline
$>$\# the solutions will be evaluated numerically by variation of chi for
different N\\
\indent
\# Attention! This computational block can take a lot of time!\\
\indent \indent
P := array(1..50,1..16,1..3):\\
\indent \indent \indent
 l := 0.01:\\
 \indent \indent \indent \indent
  g := 0.05:\\
  \indent \indent \indent
   alpha0 := 0.5:\\
   \indent \indent
    Lev := [5, 10, 50]:\\
    \indent
for k from 1 to 3 do\\
\indent
for j from 1 to 16 do\\
\indent
for i from 1 to 50 do\\
\indent
 N := Lev[k]:\\
 \indent \indent
  chi := i/50:\\
  \indent \indent \indent
   alpha := evalf( sol[1] ):\\
   \indent \indent
    tau := evalf( sqrt(sol1) ):\\
    \indent
     rho0 := evalf( sqrt(sol2) ):\\
     \indent \indent
      delta := evalf( sol3 ):\\
      \indent
       P[i,j,k] := evalf( s[j] ):\\
       \indent \indent
od:\\
\indent \indent \indent
od:\\
\indent \indent
print(k); \\
\indent
od:

\emptyline
$>$\# list of plots\\
\indent
macro(usercol = COLOR(RGB, 0.8, 5/Lev[k], 0.5)):\\
\indent \indent
 for k from 1 to 3 do\\
 \indent \indent \indent
  for m from 1 to 16 do\\
  \indent \indent
   p[(k-1)*16+m] := listplot([[n/50,Re(P[n,m,k])]\\
   \indent
\$n=1..50],color=usercol):\\
\indent \indent
    od:\\
    \indent \indent \indent
     od:

\emptyline
$>$display(\{p[ii] \$ii=1..40\},axes=boxed,\\
\indent
title=`stability increment`,view=-2..15);

\emptyline
\begin{center}
\mapleplot{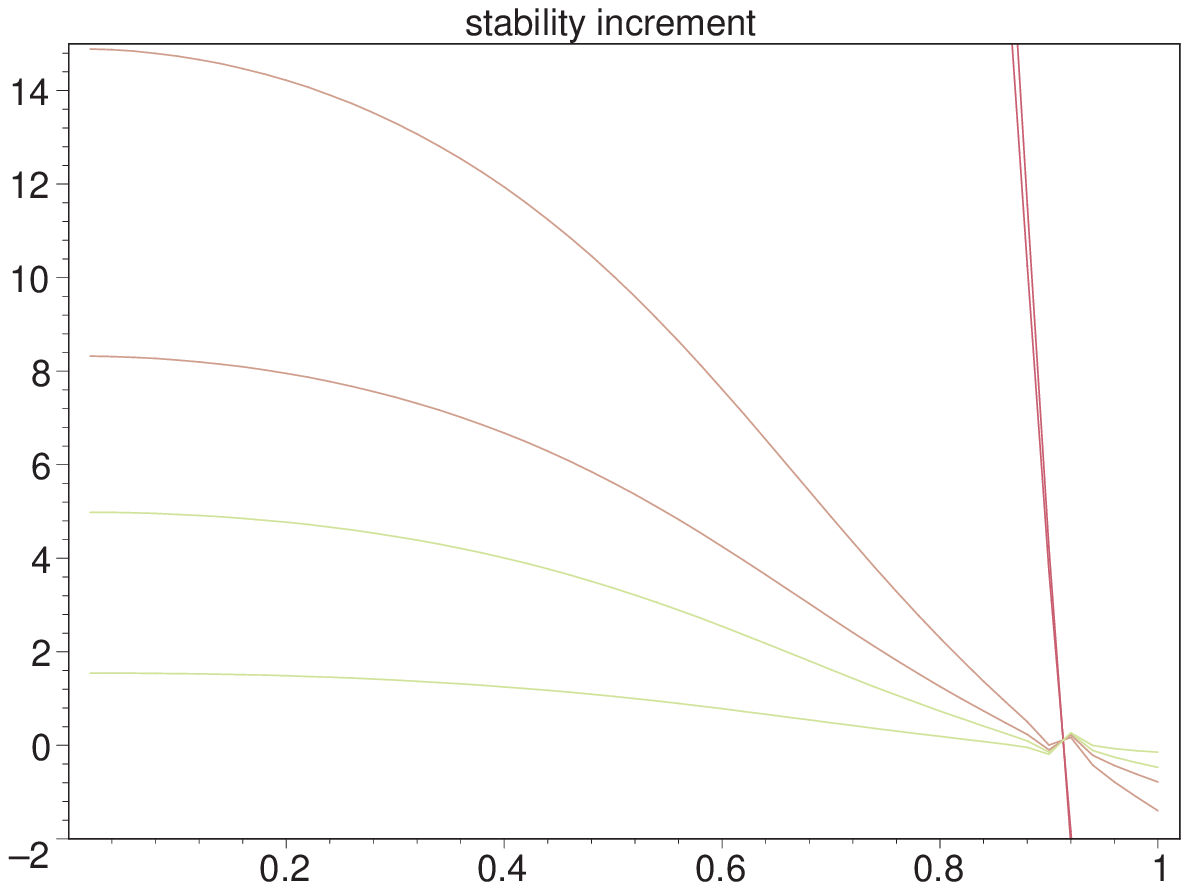}
\end{center}

\emptyline
\noindent
So, we have only two different solutions for real part of
$\lambda $, when "level's number" \textit{N} is fixed. The negative value of
increment corresponds to decaying perturbations, i. e. stable
ultrashort pulse generation.
\begin{center}
\fbox{ \parbox{.8\linewidth}{
Because of the pulse stabilization
results from increase of
$\chi $  there is problem of the shortest pulse generation (note, that
increase of
$\chi $ increases pulse width, see corresponding figure above)}}
\end{center}
\emptyline
\noindent
It is of
interest that the region of the pulse stability can have an
inhomogeneous character (a "kink" on the curves for small \textit{N},
that is "deeper level" in potential well). This fact was confirmed by
numerical simulations (see \cite{Kalashnikov4}) and corresponds to
destabilization due to excitation of "deeper levels", that is the
pulse envelope splitting. Perturbations with large \textit{N} can be
interpreted as continuous-wave generation of noise spikes.

\emptyline

\subsection{Aberrationless approximation}

\emptyline \noindent Now we shall consider one practical
realization of the described here method of ultrashort pulse
generation. The most attractive sort of "slow" absorber is the
semiconductor absorber with comparatively short relaxation time
and small energy of saturation (see next part). But the absorber,
which is based on the impure dielectric crystal, is much more
usable in the picosecond time domain due to its simplicity (as
rule, the semiconductor shutter has a very complicate structure)
and durability. However, the basic disadvantages of the
crystalline absorbers are the large relaxation time (10 ns - 1
$\mu $s) and large saturation energy (up to 1 \textit{J}/
$\mathit{cm}^{2}$). Here we shall analyze the possibility of the
stable ultrashort pulse generation in the forsterite solid-state
laser with YAG: V$^{3+}$ crystalline absorber. As the basic method
the aberrationless approximation will be used. Now we shall
express the loss coefficient in absorber as $g\,e^{( - \frac
{\varepsilon }{{E_{s}}})}$, that differs from phenomenological
expression, which was considered in the beginning of this part,
but is close to it when the expansion on energy is made up to
small orders of $\varepsilon $. Such form of the dependence can be
obtained from Bloch's equations for two-level absorber (see next
part) in noncoherent approximation and in the condition of small
pulse duration in compare with longitudinal relaxation time
\textit{Ta}  (\textit{Ta }= 22 ns for YAG: V$^{3+}$). The
steady-state (there is no dependence on \textit{z}) master
equation is (see above):

\emptyline
$>$restart:\\
\indent \indent
master:=\\
alpha*(rho(t)-rho(t)*chi*epsilon(t)+rho(t)*chi$^{2}$*epsilon(t)$^{2}$/\\
\indent
2)-g*(rho(t)-epsilon(t)*rho(t)+rho(t)*epsilon(t)$^{2}$/2)+\\
\indent
diff(diff(rho(t),t),t)+delta*diff(rho(t),t)-l*rho(t);

\emptyline

\begin{gather*}
\boxed{\mathit{master} := \alpha \,(\rho (t) - \rho (t)\,\chi \,
\varepsilon (t) + {\displaystyle \frac {1}{2}} \,\rho (t)\,\chi
^{2}\,\varepsilon (t)^{2}) -} \\
\boxed{g\,(\rho (t) - \varepsilon (t)\,\rho
 (t) +
 {\displaystyle \frac {1}{2}} \,\rho (t)\,\varepsilon (t)^{
2}) + ({\frac {\partial ^{2}}{\partial t^{2}}}\,\rho (t)) +
\delta \,({\frac {\partial }{\partial t}}\,\rho (t)) - l\,\rho (t)}
\end{gather*}

\emptyline
\noindent
Let suppose the soliton-like shape of the pulse. Then the evolution of
the ultrashort pulse parameters can be found as result of the
expansion at \textit{t}-series:

\emptyline
$>$f1:=(z,t)$->$rho0(z)*sech(t*tau(z));\# pulse amplitude\\
\indent
 f2:=(z,t)$->$rho0(z)$^{2}$*(1+tanh(t*tau(z)))/tau(z);\# pulse energy\\
 \indent

lhs\_master:=subs(\{diff(rho0(z),z)=x,diff(tau(z),z)=y\\
\indent \indent
\},diff(f1(z,t),z)):\# left-hand side of dynamical equation\\
\indent

eq:=collect(series(lhs\_master-subs(\{rho(t)=f1(z,t),epsilon(t)=f2(z,t)\\
\indent
\},master),t=0,3),t):\#dynamical equation\\
\indent \indent
eq1 := coeff(eq,t$^{2}$):\\
\indent \indent \indent
eq2 := coeff(eq,t):\\
\indent \indent
eq3 := coeff(eq,t,0):\\
\indent
sol := factor(solve(\{eq1, eq2, eq3\},\{x, y, delta\}));

\emptyline
\[
\mathit{f1} := (z, \,t)\rightarrow \rho 0(z)\,\mathrm{sech}(t\,
\tau (z))
\]

\[
\mathit{f2} := (z, \,t)\rightarrow {\displaystyle \frac {\rho 0(z
)^{2}\,(1 + \mathrm{tanh}(t\,\tau (z)))}{\tau (z)}}
\]

\begin{gather*}
\mathit{sol} := \{x={\displaystyle \frac {1}{2}} \rho 0(z)\\
(2\,
\alpha \,\tau (z)^{2} - 2\,\alpha \,\rho 0(z)^{2}\,\chi \,\tau (z
) + \alpha \,\rho 0(z)^{4}\,\chi ^{2} - 2\,l\,\tau (z)^{2} - 2\,g
\,\tau (z)^{2} + \\
\mbox{} 2\,g\,\rho 0(z)^{2}\,\tau (z)
 - g\,\rho 0(z)^{4} - 2\,\tau (z)^{4}) \left/ {\vrule
height0.44em width0em depth0.44em} \right. \!  \! \tau (z)^{2},
\,\\
y= - {\displaystyle \frac {1}{2}} \,{\displaystyle \frac {
\alpha \,\rho 0(z)^{4}\,\chi ^{2} - g\,\rho 0(z)^{4} + 4\,\tau (z
)^{4}}{\tau (z)}} ,  \\
\delta ={\displaystyle \frac {\rho 0(z)^{2}\,( - g\,\rho 0(z)^{2}
 - \alpha \,\chi \,\tau (z) + \alpha \,\rho 0(z)^{2}\,\chi ^{2}
 + g\,\tau (z))}{\tau (z)^{3}}} \}
\end{gather*}

\emptyline \noindent The equations for the pulse parameters
evolution have to be supplemented by the equations for the gain (
$\alpha $$\mapsto$ $\alpha $(\textit{z)}) and the saturable loss
(\textit{g}$\mapsto$\textit{g(z)}) evolution at the time intervals
\TEXTsymbol{>}\TEXTsymbol{>}1/ $\tau $ ( $\alpha $0 and
\textit{gmx} are the maximal saturable gain and loss,
respectively, \textit{Ta} and \textit{Tr} are the loss and gain
relaxation times normalized to the cavity period \textit{Tcav},
\textit{Pump} is the dimensionless pump, see part \textit{9}):

\emptyline
$>$eq4:=-4*rho0(z)$^{2}$*g(z)/tau(z)+(gmx-g(z))/Ta;\\
\indent
eq5:=\\
\indent
Pump*(alpha0-alpha(z))-2*chi*alpha(z)*rho0(z)$^{2}$/tau(z)-alpha(z)/Tr;

\emptyline

\[
\mathit{eq4} :=  - 4\,{\displaystyle \frac {\rho 0(z)^{2}\,
\mathrm{g}(z)}{\tau (z)}}  + {\displaystyle \frac {\mathit{gmx}
 - \mathrm{g}(z)}{\mathit{Ta}}}
\]

\[
\mathit{eq5} := \mathit{Pump}\,(\alpha 0 - \alpha (z)) -
{\displaystyle \frac {2\,\chi \,\alpha (z)\,\rho 0(z)^{2}}{\tau (
z)}}  - {\displaystyle \frac {\alpha (z)}{\mathit{Tr}}}
\]

\emptyline
\begin{flushleft}
Then finally:
\end{flushleft}

\emptyline
$>$sys :=\\
\indent
D(g)(z)=eq4,D(a)(z)=eq5,D(rho0)(z)=subs(\{alpha=alpha(z),g=g(z)
\},\\
\indent
subs(sol,x)),D(tau)(z)=subs(\{alpha=alpha(z),g=g(z)\},subs(sol,y));\\
\indent
\# basic systems for evolution of the pulse parameters, gain and loss coefficients

\emptyline
\begin{gather*}
\mathit{sys} := \mathrm{D}(g)(z)= - 4\,{\displaystyle \frac {\rho
 0(z)^{2}\,\mathrm{g}(z)}{\tau (z)}}  + {\displaystyle \frac {
\mathit{gmx} - \mathrm{g}(z)}{\mathit{Ta}}} , \\
\,\mathrm{D}(a)(z)=
\mathit{Pump}\,(\alpha 0 - \alpha (z)) - {\displaystyle \frac {2
\,\chi \,\alpha (z)\,\rho 0(z)^{2}}{\tau (z)}}  - {\displaystyle
\frac {\alpha (z)}{\mathit{Tr}}} ,  \\
\mathrm{D}(\rho 0)(z)={\displaystyle \frac {1}{2}} \rho 0(z)\\
(2\,
\alpha (z)\,\tau (z)^{2} - 2\,\alpha (z)\,\rho 0(z)^{2}\,\chi \,
\tau (z) + \alpha (z)\,\rho 0(z)^{4}\,\chi ^{2} - 2\,l\,\tau (z)
^{2} -\\
\mbox{} 2\,\mathrm{g}(z)\,\tau (z)^{2}
 + 2\,\mathrm{g}(z)\,\rho 0(z)^{2}\,\tau (z) - \mathrm{g}(
z)\,\rho 0(z)^{4} - 2\,\tau (z)^{4}) \left/ {\vrule
height0.44em width0em depth0.44em} \right. \!  \! \tau (z)^{2},
 \\
\mathrm{D}(\tau )(z)= - {\displaystyle \frac {1}{2}} \,
{\displaystyle \frac {\alpha (z)\,\rho 0(z)^{4}\,\chi ^{2} -
\mathrm{g}(z)\,\rho 0(z)^{4} + 4\,\tau (z)^{4}}{\tau (z)}}
\end{gather*}

\emptyline
We shall change the saturation parameter
$\chi $\textit{{\large  }}and to search the stationary points of the pulse
parameter's mapping. These points correspond to the solutions of
\textit{'sys'} with zero left-hand sides

\emptyline
$>$st\_sol1:=solve(\{eq4,eq5\},\{g(z),alpha(z)\}):\\
\indent \indent
 st\_g:=subs(st\_sol1,g(z));\\
 \indent \indent \indent
  st\_a:=subs(st\_sol1,alpha(z));\\
  \indent
   st\_sys1:=[expand(rhs(op(3,$[$sys$]$))*2*tau(z)$^{2}$/rho0(z)),\\
   \indent
    expand(rhs(op(4,[sys]))*2*tau(z))]:
    \indent

st\_sys2:=\{simplify(op(1,st\_sys1)+op(2,st\_sys1))=0,op(2,st\_sys1)=0\};\\
\indent

st\_sys3:=subs(\{rho0(z)=x,tau(z)=y\},subs(\{g(z)=st\_g,\\
\indent
alpha(z)=st\_a\},st\_sys2));

\emptyline
\[
\mathit{st\_g} := {\displaystyle \frac {\tau (z)\,\mathit{gmx}}{4
\,\rho 0(z)^{2}\,\mathit{Ta} + \tau (z)}}
\]

\[
\mathit{st\_a} := {\displaystyle \frac {\mathit{Pump}\,\tau (z)\,
\mathit{Tr}\,\alpha 0}{\mathit{Pump}\,\tau (z)\,\mathit{Tr} + 2\,
\chi \,\rho 0(z)^{2}\,\mathit{Tr} + \tau (z)}}
\]

\begin{gather*}
\mathit{st\_sys2} := \\
\{2\,\alpha (z)\,\tau (z)^{2} - 2\,\alpha (z)\,\rho 0(z)^{2}\,\chi
\,\tau (z) - \\
2\,l\,\tau (z)^{2} - 2\,\mathrm{g}(z)\,\tau (z)^{2}
 + 2\,\mathrm{g}(z)\,\rho 0(z)^{2}\,\tau (z) - 6\,\tau (z)^{4}=0
,  \\
 - \alpha (z)\,\rho 0(z)^{4}\,\chi ^{2} + \mathrm{g}(z)\,\rho 0(z
)^{4} - 4\,\tau (z)^{4}=0\}
\end{gather*}

\begin{gather*}
\mathit{st\_sys3} :=\\
 \{2\,{\displaystyle \frac {\mathit{Pump}\,y
^{3}\,\mathit{Tr}\,\alpha 0}{\mathit{Pump}\,y\,\mathit{Tr} + 2\,
\chi \,x^{2}\,\mathit{Tr} + y}}  - {\displaystyle \frac {2\,
\mathit{Pump}\,y^{2}\,\mathit{Tr}\,\alpha 0\,x^{2}\,\chi }{
\mathit{Pump}\,y\,\mathit{Tr} + 2\,\chi \,x^{2}\,\mathit{Tr} + y}
}  - \\
2\,l\,y^{2} - {\displaystyle \frac {2\,y^{3}\,\mathit{gmx}}{
4\,x^{2}\,\mathit{Ta} + y}}
 + {\displaystyle \frac {2\,y^{2}\,\mathit{gmx}\,x^{2}}{4
\,x^{2}\,\mathit{Ta} + y}}  - 6\,y^{4}=0,\\
\mbox{} \, - {\displaystyle
\frac {\mathit{Pump}\,y\,\mathit{Tr}\,\alpha 0\,x^{4}\,\chi ^{2}
}{\mathit{Pump}\,y\,\mathit{Tr} + 2\,\chi \,x^{2}\,\mathit{Tr} +
y}}  + {\displaystyle \frac {y\,\mathit{gmx}\,x^{4}}{4\,x^{2}\,
\mathit{Ta} + y}}  - 4\,y^{4}=0\}
\end{gather*}

\emptyline
\noindent
The next procedure will be used for numerical solution of
\textit{{\large st\_sys3}}.

\emptyline
$>$num\_sol := proc(alpha0,gmx,chi,l,Tr,Ta,Pump)\\
\indent \indent
    st\_sys :=\\
    \indent
\{-Pump*y*Tr*alpha0*x$^{4}$*chi$^{2}$/\\
\indent
(Pump*y*Tr+2*chi*x$^{2}$*Tr+y)+y*gmx*x$^{4}$/(4*x$^{2}$*Ta+y)-4*y$^{4}$ = 0,\\
\indent
2*Pump*y$^{3}$*Tr*alpha0/(Pump*y*Tr+2*chi*x$^{2}$*Tr+y)-\\
\indent
2*Pump*y$^{2}$*Tr*alpha0*x$^{2}$*chi/(Pump*y*Tr+2*chi*x$^{2}$*Tr+y)\\
\indent
-2*l*y$^{2}$-2*y$^{3}$*gmx/(4*x$^{2}$*Ta+y)+2*y$^{2}$\\
\indent
*gmx*x$^{2}$/(4*x$^{2}$*Ta+y)-6*y$^{4}$ = 0\}:\\
\indent \indent
    fsolve(st\_sys,\{x,y\},\{x=0..1,y=0..1\}):\\
    \indent \indent \indent
end:

\emptyline
$>$v := array(1..100):\\
\indent \indent
 for i from 1 to 100 do  \\
\indent
v[i] := num\_sol(0.5,0.05,1/(1.36+9*i/100),0.01,300,2.2,0.001) od:\\
\indent
\# the normalization of relaxation times to cavity period is supposed (Tcav =
10 ns)

\emptyline
\noindent
Now we can plot the logarithm of the pulse duration versus
\mapleinline{inert}{2d}{chi;}{%
$\chi $%
}.

\emptyline
$>$with(plots):\\
\indent \indent
 ww:=array(1..100):\\
 \indent
  for j from 5 to 100 do ww[j]:=evalf(log10(1/subs(v[j],y))) od:\\
  \indent
   points := \{seq([1/(1.36+9*j/100),ww[j]],j=1..100)\}:\\
   \indent \indent
    f1 := \\
    \indent
    plot(points,x=0.1..0.6,style=point,symbol=circle,color=red):\\
    \indent

display(f1,TEXT([10,3],'`Up=0.0008`'),view=2.4..3.6,\\
\indent
title=`Logarithm of pulse duration versus sigma`,\\
\indent
labels=[`saturation parameter`,``]);

\begin{center}
\mapleplot{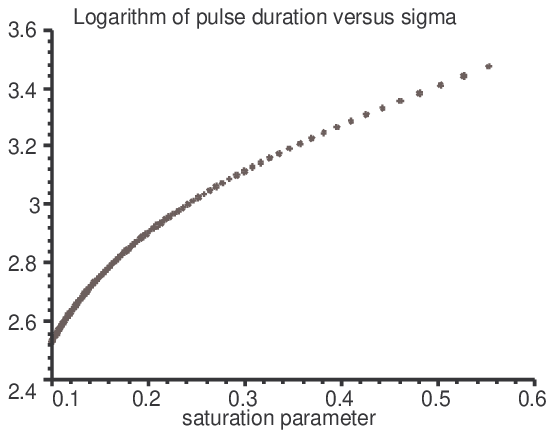}
\end{center}

\emptyline
\noindent
This figure demonstrates the pulse width decrease due to decrease of
\mapleinline{inert}{2d}{chi;}{%
$\chi $%
}\textit{{\large ,}} that corresponds to the predominance of the
loss saturation over gain saturation (see previous subsection).
For the time normalization to the inverse bandwidth of YAG:
V$^{3+}$ absorption line the pulse duration at $\chi $ = 0.3  is about of 50 ps.\\
\indent
Now, we shall consider the ultrashort pulse parameters evolution on
the basis of the obtained system of ODE. This procedure solves the
systems by the standard operator \textit{DEplot}

\emptyline
$>$with(DEtools):\\
\indent
 ODE\_plot := proc(alpha0,gmx,chi,l,Tr,Ta,Pump)\\
 \indent \indent
  sys := [D(g)(z) = \\
  \indent
-4*rho0(z)$^{2}$*g(z)/tau(z)+(gmx-g(z))/Ta, \\
\indent \indent
   D(alpha)(z) =Pump*(alpha0-alpha(z))-\\
   \indent \indent
   2*chi*alpha(z)*rho0(z)$^{2}$/tau(z)-alpha(z)/Tr, \\
   \indent \indent \indent
    D(rho0)(z) =\\
    \indent
1/2*rho0(z)*(-2*l*tau(z)$^{2}$+2*alpha(z)*tau(z)$^{2}$-\\
\indent
2*alpha(z)*rho0(z)$^{2}$*chi*tau(z)+alpha(z)*rho0(z)$^{4}$*chi$^{2}$-\\
\indent
2*tau(z)$^{4}$-2*g(z)*tau(z)$^{2}$+2*g(z)*rho0(z)$^{2}$*tau(z)-\\
\indent
g(z)*rho0(z)$^{4}$)/(tau(z)$^{2}$),\\
\indent \indent
      D(tau)(z) =\\
      \indent
-1/2*(alpha(z)*rho0(z)$^{4}$*chi$^{2}$-g(z)*rho0(z)$^{4}$+\\
4*tau(z)$^{4}$)/tau(z)]:\\
\indent
DEplot(sys,[rho0(z),tau(z),g(z),alpha(z)],z=0..10000,\\
\indent
[[rho0(0)=1e-5,tau(0)=1e-3,g(0)=gmx,alpha(0)=0]],\\
\indent
stepsize=1,scene=[z,rho0(z)],axes=FRAME,linecolor=BLACK):\\
\indent
end:

\emptyline
\noindent
Let vary the saturation parameter
\mapleinline{inert}{2d}{chi;}{%
$\chi $%
} for the fixed pump:

\emptyline
$>$display(ODE\_plot(0.5,0.05,0.3,0.01,300,2.2,0.001));

\begin{center}
\mapleplot{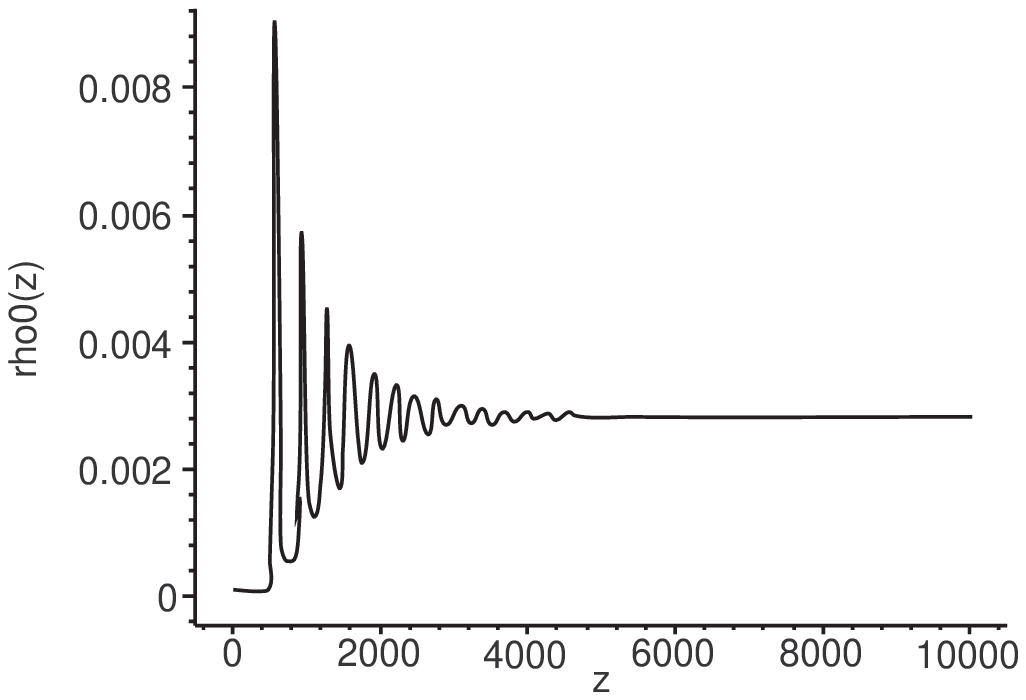}
\end{center}

\emptyline
\noindent
So, we have the decaying oscillations of the ultrashort pulse
amplitude (see part \textit{7}). The decrease of saturation parameter
produces

\emptyline
$>$display(ODE\_plot(0.5,0.05,0.1,0.01,300,2.2,0.001));

\begin{center}
\mapleplot{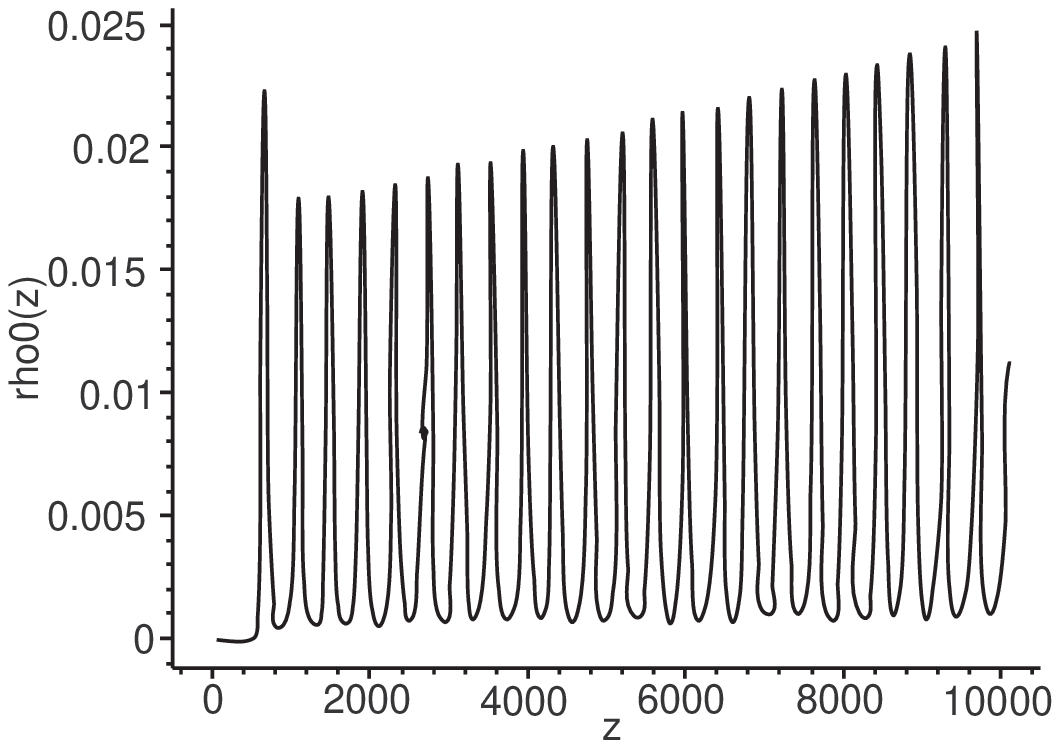}
\end{center}

\begin{center}
\fbox{ \parbox{.8\linewidth}{
One can see the growth of the ultrashort pulse oscillations
(autooscillations or so-called Q-switch mode locking). This is a main
destabilizing factor for the considered type of the mode locking
regime. This result is in agreement with above obtained (note, that
now we are outside of framework of linear perturbation theory)}}
\end{center}
\emptyline
The important obstacle for the pulse generation by slow absorber is
the noise growth. In order to investigate pulse stability against
noise in framework of the considered model we have to add the equation
for the evolution of noise energy \textit{n} to \textit{sys.}

\emptyline
$>$ODE\_noise := proc(alpha0,gmx,chi,l,Tr,Ta,Pump)\\
\indent \indent
sys\_noise := [D(g)(z) =\\
\indent
-4*rho0(z)$^{2}$*g(z)/tau(z)+(gmx-g(z))/Ta, \\
\indent \indent
   D(alpha)(z) =\\
   \indent
Pump*(alpha0-alpha(z))-2*chi*alpha(z)*rho0(z)$^{2}$/tau(z)-alpha(z)/Tr, \\
\indent \indent
    D(rho0)(z) =\\
    \indent
1/2*rho0(z)*(-2*l*tau(z)$^{2}$+2*alpha(z)*tau(z)$^{2}$-\\
\indent
2*alpha(z)*rho0(z)$^{2}$*chi*tau(z)+alpha(z)*rho0(z)$^{4}$*chi$^{2}$-\\
\indent
2*tau(z)$^{4}$-2*g(z)*tau(z)$^{2}$+2*g(z)*rho0(z)$^{2}$*tau(z)-\\
\indent
g(z)*rho0(z)$^{4}$)/(tau(z)$^{2}$),\\
\indent \indent
      D(tau)(z) =\\
      \indent
-1/2*(alpha(z)*rho0(z)$^{4}$*chi$^{2}$-g(z)*rho0(z)$^{4}$+4*tau(z)$^{4}$)/tau(z),\\
        \indent \indent
        D(n)(z) = \\
        \indent
(alpha(z)-l-(gmx+Ta*(g(z)-gmx)*(1-exp(-1/Ta))))*n(z)]:\#see V.L.
Kalashnikov et al. Opt. Commun., v.159, 237 (1999)
\indent

DEplot3d(sys\_noise,[rho0(z),tau(z),g(z),alpha(z),n(z)],\\
\indent
z=0..10000,[[rho0(0)=1e-5,tau(0)=1e-3,g(0)=gmx,alpha(0)=0,n(0)=1e-5]],\\
\indent
stepsize=1,scene=[z,rho0(z),n(z)],axes=FRAME,linecolor=BLACK):\\
\indent \indent
end:
\emptyline
$>$display(ODE\_noise(0.5,0.05,0.3,0.01,300,2.2,0.001),\\
\indent
title=`Noise energy evolution`);

\begin{center}
\mapleplot{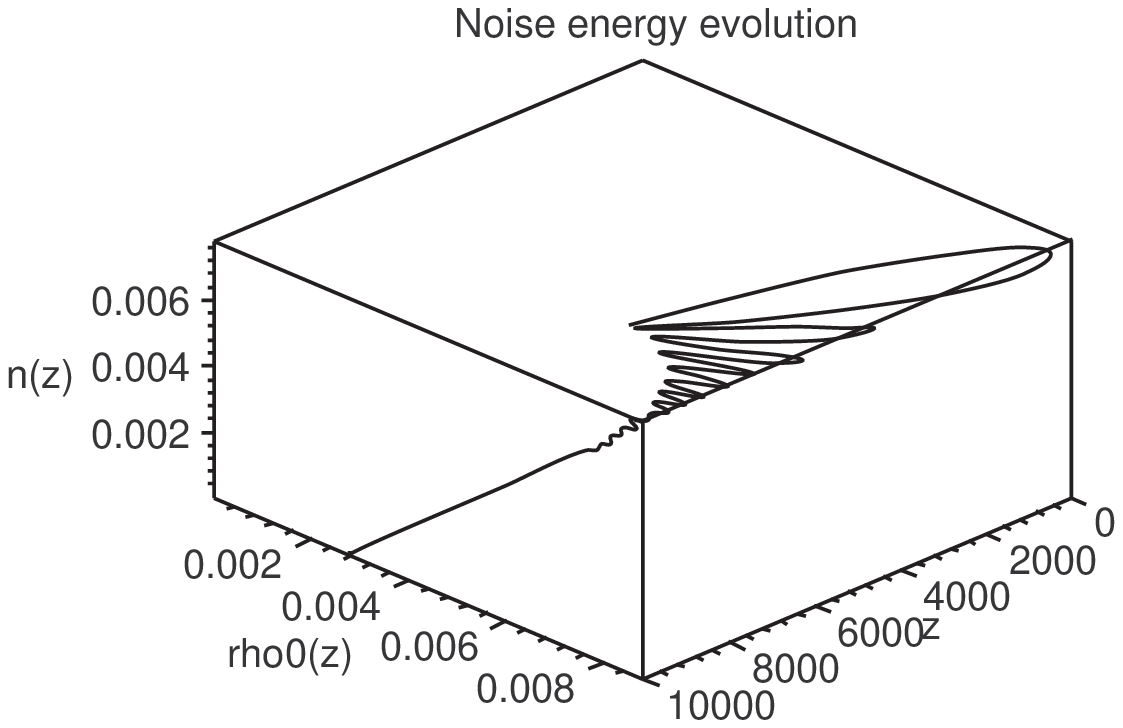}
\end{center}

\emptyline
\begin{center}
\fbox{ \parbox{.8\linewidth}{
As one can see from this picture, the noise energy decays, i. e. 50 ps
"auto-stable" pulse is stable against noise, too}}
\end{center}

\section{Coherent pulses: self-induced transparency in the
laser}

\emptyline
\noindent
Here we shall calculate the main characteristics of ultrashort pulses
in the solid-state laser in the presence of self-focusing in the
active crystal and coherent interaction with a semiconductor absorber.
The good introduction in this problem and the numerical simulations of
the nonlinear dynamics in the laser with coherent absorber can be
found in the article \cite{Kalosha}. \\
\indent
We shall suppose, that as saturable absorber the semiconductor
structure is used (like GaAs/AlGaAs quantum-well structures). The key
characteristics of this absorber, which correspond to the experimental
dates, are the energy flux of the loss saturation
${E_{a}}$ = 5*
$10^{( - 5)}$ J/
$\mathit{cm}^{2}$, and the time of coherency
${t_{\mathit{coh}}}$ = 50 fs. These parameters define the parameter q =
$\sqrt{\frac {\varepsilon 0\,n\,c}{2\,\mathit{Ea}\,\mathit{tcoh}}
}$   (\textit{n} is the refractivity index, \textit{c} is the light
velocity in the vacuum,
$\varepsilon $0 is the dielectric constant).

\emptyline
$>$restart:\\
\indent \indent
 with(plots):\\
 \indent \indent \indent
  with(DEtools):\\
  \indent \indent \indent \indent
parameter\_q := proc()\\
\indent
 local epsilon0,n,c,Ea,q,tcoh:\\
 \indent
  epsilon0 := 8.85e-14:\\
  \indent \indent
   n := 3.32:\\
   \indent \indent \indent
    c := 3e10:\\
    \indent \indent \indent \indent
     Ea := 5e-5:\\
     \indent
      tcoh := 5e-14:\\
      \indent \indent
        sqrt(epsilon0*n*c/(2*Ea*tcoh)): \\
        \indent
end:\\
\indent \indent
q=evalf(parameter\_q());\# The dimension of q is [cm/V/s]

\emptyline
\[
q=.4198714089\,10^{8}
\]

\emptyline
\noindent
The corresponding dipole moment is \textit{d} = \textit{q*h/(2*
$\pi $}):

\emptyline
$>$d=evalf(subs(\{h=6.63e-34,q=parameter\_q()\},q*h/(2*Pi)));\# The
dimension of d is [coulomb*cm]\\
\indent
d/e=evalf(rhs(\%)/1.6e-19*1e7);\#here e is the elementary charge, the
dimension of d/e is [nm]

\emptyline
\[
d=.4430471653\,10^{-26}
\]

\emptyline
\[
{\displaystyle \frac {d}{e}} =.2769044783
\]

\emptyline
\noindent
For a given
${E_{a}}$ and the photon energy 2.5*
$10^{( - 19)}$ J (the wavelength is equal to 800 nm) the absorption cross-section
$\Sigma $ = 5*
$10^{( - 15)}$
$\mathit{cm}^{2}$. The free carrier density \textit{N} =
$10^{19}$
$\mathit{cm}^{( - 3)}$ in the semiconductor produces the loss coefficient \textit{gam\_abs}
= 0.05 for the length of the absorber's layer \textit{z\_abs} = 10 nm.
Other important characteristic is \textit{p} = \textit{2
$\pi $ N d q
$\omega $/c = 4
$\pi ^{2}$ N
$d^{2}$
$\omega $/(c h)} (
$\omega $ is the field carrier frequency). These parameters are connected with
a loss coefficient: \textit{gam\_abs/z\_abs} = \textit{N
$\Sigma $}  = \textit{p*tcoh}, so

\emptyline
$>$p=evalf(subs(\{tcoh=5e-14,N=1e19,Sigma=5e-15\},N*Sigma/tcoh));\\
\indent
\#The dimension of p is [1/cm/s]

\emptyline
\[
p=.1000000000\,10^{19}
\]

\emptyline
\noindent
As the laser we shall consider Ti: sapphire solid-state laser that is
the typical laser for femtosecond generation. Its inverse gain
bandwidth defining the minimal pulse duration
${t_{f}}$ = 2.5 fs. We shall use the next normalizations: 1) the time is
normalized to
${t_{f}}$; 2) the field is normalized to \textit{1/q/
${t_{f}}$} = 0.95*
$10^{7}$ V/cm; 3) the field intensity is normalized to
${\varepsilon _{0}}$\textit{n c/(2q
${t_{f}}$})$^{2}$, that is

\emptyline
$>$intensity\_normalization\_parameter := proc()\\
\indent \indent
 local epsilon0,n,c,q,tg,par:\\
 \indent \indent \indent
  epsilon0 := 8.85e-14:\\
  \indent \indent \indent \indent
   n := 3.32:\\
   \indent \indent \indent
    c := 3e10:\\
    \indent \indent
     tf := 2.5e-15:\\
     \indent
q := parameter\_q():\\
\indent \indent
 epsilon0*n*c/(2*q*tf)$^{2}$: \\
 \indent \indent \indent
end:\\
\indent
evalf(intensity\_normalization\_parameter());\# The dimension is
[W/cm$^{2}$]

\emptyline
\[
.2000000000\,10^{12}
\]

\emptyline \noindent For such normalization the parameter of the
gain saturation for Ti: sapphire (the energy of the gain
saturation is 0.8 J/cm\symbol{94}2) is $\xi $ = 6.25* $10^{( -
4)}$. As it will be seen, the principal factor in our model is the
inverse intensity of the saturation of diffraction loss in the
laser (so-called Kerr - lens mode locking parameter) $\sigma $  =
0.14 (parameter of the diffraction loss saturation is $10^{7}$ W
and the mode cross section of the Gaussian beam is 30 $\mu $m).
The corresponding SPM coefficient in 1 mm active crystal is $\beta
$ = 0.26.

\emptyline
The response of the semiconductor absorber is described on the basis
of Bloch equations \cite{Allen}. In the
absence of the field phase modulation and tuning from the medium
resonance frequency and in the absence of the relaxation over time
interval, which is comparable with the pulse duration, the evolution
of the polarization (components \textit{a(t)} and \textit{b(t)}) and
the population difference between excited and ground states
\textit{w(t)} obey to the following equations:

\emptyline
$>$bloch\_1 := diff(b(t),t)=q*rho(t)*w(t);\\
\indent \indent
 bloch\_2 := diff(a(t),t)=0;\\
 \indent
  bloch\_3 := diff(w(t),t)=-q*rho(t)*b(t);

\emptyline
\begin{center}
\[\boxed{
\mathit{bloch\_1} := {\frac {\partial }{\partial t}}\,\mathrm{b}(
t)=q\,\rho (t)\,\mathrm{w}(t)}
\]

\[\boxed{
\mathit{bloch\_2} := {\frac {\partial }{\partial t}}\,\mathrm{a}(
t)=0}
\]

\[\boxed{
\mathit{bloch\_3} := {\frac {\partial }{\partial t}}\,\mathrm{w}(
t)= - q\,\rho (t)\,\mathrm{b}(t)}
\]
\end{center}

\emptyline
\noindent
The solutions of this system are

\emptyline
$>$sol\_bloch :=\\
\indent \indent
dsolve(\{bloch\_1,bloch\_2,bloch\_3,w(0)=-1,b(0)=0,a(0)=0\},\\
\indent
\{a(t),b(t),w(t)\}):\\
\indent
 sol\_a := subs(sol\_bloch,a(t));\\
 \indent \indent
  sol\_b := subs(sol\_bloch,b(t));\\
  \indent
   sol\_w := subs(sol\_bloch,w(t));

\emptyline
\begin{center}
\[
\mathit{sol\_a} := 0
\]

\[
\mathit{sol\_b} :=  - \mathrm{sin}(q\,{\displaystyle \int _{0}^{t
}} \rho (u)\,du)
\]

\[
\mathit{sol\_w} :=  - \mathrm{cos}(q\,{\displaystyle \int _{0}^{t
}} \rho (u)\,du)
\]
\end{center}

\emptyline
\noindent
The argument of sin/cos is the pulse area
$\psi $:

\emptyline
$>$b(t):= -sin(psi(t));\\
\indent \indent
a(t):= 0;
\emptyline
\begin{center}
\[
\mathrm{b}(t) :=  - \mathrm{sin}(\psi (t))
\]

\[
\mathrm{a}(t) := 0
\]

\end{center}

\emptyline
\noindent
The linear distributed response of the laser with the gain coefficient
$\alpha $, loss coefficient \textit{gam}, frequency filter with the inverse
bandwidth
${t_{f}}$, laser dispersion element with dispersion coefficient
${k_{2}}$ is described by the terms in the right hand side of the wave
equation (see Parts \textit{3}, \textit{5}):

\emptyline
$>$Laser\_linear :=\\
\indent
alpha*rho(z,t)-gam*rho(z,t)+tf$^{2}$*diff(rho(z,t),t,t)+\\
\indent
I*k\_2*diff(rho(z,t),t,t);

\emptyline
\begin{gather*}
\boxed{
\mathit{Laser\_linear} :=}\\
\boxed{ \alpha \,\rho (z, \,t) - \mathit{gam}\,
\rho (z, \,t) + \mathit{tf}^{2}\,({\frac {\partial ^{2}}{
\partial t^{2}}}\,\rho (z, \,t)) + I\,\mathit{k\_2}\,({\frac {
\partial ^{2}}{\partial t^{2}}}\,\rho (z, \,t))}
\end{gather*}

\emptyline \noindent The response of nonlinear laser factors is

\emptyline
$>$Laser\_nonlinear :=\\
\indent
sigma*rho(z,t)*conjugate(rho(z,t))*rho(z,t)-\\
\indent
I*beta*rho(z,t)*conjugate(rho(z,t))*rho(z,t);

\emptyline
\begin{center}
\[\boxed{
\mathit{Laser\_nonlinear} := \sigma \,\rho (z, \,t)^{2}\,
\overline{(\rho (z, \,t))} - I\,\beta \,\rho (z, \,t)^{2}\,
\overline{(\rho (z, \,t))}}
\]
\end{center}

\emptyline
\noindent
Here
$\alpha $ and \textit{gam} are the dimensionless values, that supposes the
normalization of length \textit{z} to the length of the optical medium
(active medium in our case), \textit{gam} includes not only scattering
loss into optical elements, but also the output loss on the laser
mirror.
$\sigma $ and
$\beta $ have dimension of the inverse intensity, i. e.
$\left|  \! \,\rho \, \!  \right| ^{2}$ is the field intensity (we does not write the factor
$\varepsilon $\textit{0*n*c/2}, which corresponds to the transition
$\mathit{field}^{2}$ -\TEXTsymbol{>} intensity).\\
\indent
As the result, the master integro-differential equation for the field
evolution is

\emptyline
$>$master\_1 := diff(rho(z,t),z)+diff(rho(z,t),t)/c =\\
\indent
subs(N=N*z\_abs,-2*Pi*N*d*diff(a(t),t)/c+2*Pi*N*d*b(t)*omega/c)+\\
\indent
Laser\_linear+Laser\_nonlinear;\#see Part 3

\emptyline
\begin{gather*}
\mathit{master\_1} := ({\frac {\partial }{\partial z}}\,\rho (z,
\,t)) + {\displaystyle \frac {{\frac {\partial }{\partial t}}\,
\rho (z, \,t)}{c}} = - 2\,{\displaystyle \frac {\pi \,N\,\mathit{
z\_abs}\,d\,\mathrm{sin}(\psi (t))\,\omega }{c}}  + \\
\alpha \,\rho (z, \,t) - \mathit{gam}\,\rho (z, \,t)
\mbox{} + \mathit{tf}^{2}\,({\frac {\partial ^{2}}{\partial t^{2}
}}\,\rho (z, \,t)) + \\
I\,\mathit{k\_2}\,({\frac {\partial ^{2}}{
\partial t^{2}}}\,\rho (z, \,t)) + \sigma \,\rho (z, \,t)^{2}\,
\overline{(\rho (z, \,t))} - I\,\beta \,\rho (z, \,t)^{2}\,
\overline{(\rho (z, \,t))}
\end{gather*}

\emptyline
\noindent
Let transit to the differential equation:

\emptyline
$>$assume(q,real):\\
\indent
master\_2 := expand(subs(rho(z,t)=diff(psi(t),t)/q,master\_1));

\emptyline
\begin{gather*}
\mathit{master\_2} := {\displaystyle \frac {{\frac {\partial ^{2}
}{\partial t^{2}}}\,\psi (t)}{\mathit{q\symbol{126}}\,c}} = - 2\,
{\displaystyle \frac {\pi \,N\,\mathit{z\_abs}\,d\,\mathrm{sin}(
\psi (t))\,\omega }{c}}  + \\
{\displaystyle \frac {\alpha \,(
{\frac {\partial }{\partial t}}\,\psi (t))}{\mathit{q\symbol{126}
}}}  - {\displaystyle \frac {\mathit{gam}\,({\frac {\partial }{
\partial t}}\,\psi (t))}{\mathit{q\symbol{126}}}}  +
{\displaystyle \frac {\mathit{tf}^{2}\,({\frac {\partial ^{3}}{
\partial t^{3}}}\,\psi (t))}{\mathit{q\symbol{126}}}}  \\
\mbox{} + {\displaystyle \frac {I\,\mathit{k\_2}\,({\frac {
\partial ^{3}}{\partial t^{3}}}\,\psi (t))}{\mathit{q\symbol{126}
}}}  + {\displaystyle \frac {\sigma \,({\frac {\partial }{
\partial t}}\,\psi (t))^{2}\,\overline{({\frac {\partial }{
\partial t}}\,\psi (t))}}{\mathit{q\symbol{126}}^{3}}}  +
{\displaystyle \frac { - I\,\beta \,({\frac {\partial }{\partial
t}}\,\psi (t))^{2}\,\overline{({\frac {\partial }{\partial t}}\,
\psi (t))}}{\mathit{q\symbol{126}}^{3}}}
\end{gather*}

\emptyline
We shall consider only steady-state field states, i. e.
${\frac {\partial }{\partial z}}\,\rho $(z, t) = 0, and to introduce the time delay on the cavity round-trip
$\delta $

\emptyline
$>$master\_3 := rhs(master\_2)+delta*diff(psi(t)/q,`\$`(t,2));

\emptyline
\begin{gather*}
\mathit{master\_3} :=  - 2\,{\displaystyle \frac {\pi \,N\,
\mathit{z\_abs}\,d\,\mathrm{sin}(\psi (t))\,\omega }{c}}  + \\
{\displaystyle \frac {\alpha \,({\frac {\partial }{\partial t}}\,
\psi (t))}{\mathit{q\symbol{126}}}}  - {\displaystyle \frac {
\mathit{gam}\,({\frac {\partial }{\partial t}}\,\psi (t))}{
\mathit{q\symbol{126}}}}  + {\displaystyle \frac {\mathit{tf}^{2}
\,({\frac {\partial ^{3}}{\partial t^{3}}}\,\psi (t))}{\mathit{q
\symbol{126}}}}  \\
\mbox{} + {\displaystyle \frac {I\,\mathit{k\_2}\,({\frac {
\partial ^{3}}{\partial t^{3}}}\,\psi (t))}{\mathit{q\symbol{126}
}}}  + {\displaystyle \frac {\sigma \,({\frac {\partial }{
\partial t}}\,\psi (t))^{2}\,\overline{({\frac {\partial }{
\partial t}}\,\psi (t))}}{\mathit{q\symbol{126}}^{3}}}  + \\
{\displaystyle \frac { - I\,\beta \,({\frac {\partial }{\partial
t}}\,\psi (t))^{2}\,\overline{({\frac {\partial }{\partial t}}\,
\psi (t))}}{\mathit{q\symbol{126}}^{3}}}  + {\displaystyle
\frac {\delta \,({\frac {\partial ^{2}}{\partial t^{2}}}\,\psi (t
))}{\mathit{q\symbol{126}}}}
\end{gather*}

\emptyline
\noindent
Let introduce the parameter
$\lambda $, which can be 1) ratio of mode cross-section on active medium to one
on semiconductor absorber, or 2) coefficient of refractivity (for
field amplitude) of multilayer mirror on the semiconductor absorber
(so-called semiconductor saturable absorber mirror - SESAM). Then

\emptyline
$>$master\_4 :=\\
\indent
subs(\{sigma=sigma/lambda$^{2}$,beta=beta/lambda$^{2}$\},master\_3);

\emptyline
\begin{gather*}
\mathit{master\_4} :=  - 2\,{\displaystyle \frac {\pi \,N\,
\mathit{z\_abs}\,d\,\mathrm{sin}(\psi (t))\,\omega }{c}}  + \\
{\displaystyle \frac {\alpha \,({\frac {\partial }{\partial t}}\,
\psi (t))}{\mathit{q\symbol{126}}}}  - {\displaystyle \frac {
\mathit{gam}\,({\frac {\partial }{\partial t}}\,\psi (t))}{
\mathit{q\symbol{126}}}}  + {\displaystyle \frac {\mathit{tf}^{2}
\,({\frac {\partial ^{3}}{\partial t^{3}}}\,\psi (t))}{\mathit{q
\symbol{126}}}}  \\
\mbox{} + {\displaystyle \frac {I\,\mathit{k\_2}\,({\frac {
\partial ^{3}}{\partial t^{3}}}\,\psi (t))}{\mathit{q\symbol{126}
}}}  + {\displaystyle \frac {\sigma \,({\frac {\partial }{
\partial t}}\,\psi (t))^{2}\,\overline{({\frac {\partial }{
\partial t}}\,\psi (t))}}{\lambda ^{2}\,\mathit{q\symbol{126}}^{3
}}}  + \\
{\displaystyle \frac { - I\,\beta \,({\frac {\partial }{
\partial t}}\,\psi (t))^{2}\,\overline{({\frac {\partial }{
\partial t}}\,\psi (t))}}{\lambda ^{2}\,\mathit{q\symbol{126}}^{3
}}}  + {\displaystyle \frac {\delta \,({\frac {\partial ^{2}}{
\partial t^{2}}}\,\psi (t))}{\mathit{q\symbol{126}}}}
\end{gather*}

\emptyline
\noindent
Now let transit to coordinates 'pulse amplitude - pulse area' and
eliminate fast saturable absorber, GDD and SPM. These suppositions
lead to the second-order ODE. That is

\emptyline
$>$master\_5 :=\\
\indent
collect(expand(subs(\{k\_2=0,beta=0,sigma=0\\
\indent
\},master\_4)/(-2*Pi*N*d*omega*tf*z\_abs/c)),\\
\indent
diff(psi(t),t));\# reduced master equation coefficients\\
\indent
 sub1 := a1=-1/2*c/(Pi*N*d*omega*tf*q*z\_abs):\\
 \indent \indent
  sub2 := a2=-1/2*c*delta/(Pi*N*d*omega*tf*q*z\_abs):\\
  \indent \indent \
   sub3 :=
a3=-1/2*c*alpha/(Pi*N*d*omega*tf*q*z\_abs)+1/2*c*gam/\\
\indent \indent \indent
(Pi*N*d*omega*tf*q*z\_abs):
\indent

master\_6 := a1*diff(psi(t),`\$`(t,3))+a2*diff(psi(t),`\$`(t,2))+\\
\indent \indent
a3*diff(psi(t),t)+sin(psi(t));\\
\indent

dsolve(master\_6=0,psi(t));\# try to solve master\_6 and find a very
useful change of the variables\\
\indent \indent

master\_7 :=\\
\indent
expand(numer(simplify(subs(\{a1=rhs(sub1),a2=rhs(sub2),a3=rhs(sub3)
\},\\
\indent
a1*(diff(rho(psi),psi,psi)*rho(psi)$^{2}$+diff(rho(psi),psi)$^{2}$*\\
\indent
rho(psi))+a2*diff(rho(psi),psi)*rho(psi)+a3*rho(psi)+sin(psi))))/(-c));\\
\indent
\# use the founded change to reduce the order of ODE

\emptyline
\begin{gather*}
\mathit{master\_5} :=  \left(  \!  - {\displaystyle \frac {1}{2}
} \,{\displaystyle \frac {c\,\alpha }{\pi \,N\,d\,\omega \,
\mathit{tf}\,\mathit{z\_abs}\,\mathit{q\symbol{126}}}}  +
{\displaystyle \frac {{\displaystyle \frac {1}{2}} \,c\,\mathit{
gam}}{\pi \,N\,d\,\omega \,\mathit{tf}\,\mathit{z\_abs}\,\mathit{
q\symbol{126}}}}  \!  \right) \,({\frac {\partial }{\partial t}}
\,\psi (t)) +\\
\mbox{} {\displaystyle \frac {\mathrm{sin}(\psi (t))}{
\mathit{tf}}}
 - {\displaystyle \frac {1}{2}} \,{\displaystyle \frac {
\mathit{tf}\,c\,({\frac {\partial ^{3}}{\partial t^{3}}}\,\psi (t
))}{\pi \,N\,d\,\omega \,\mathit{z\_abs}\,\mathit{q\symbol{126}}}
}  - {\displaystyle \frac {1}{2}} \,{\displaystyle \frac {c\,
\delta \,({\frac {\partial ^{2}}{\partial t^{2}}}\,\psi (t))}{\pi
 \,N\,d\,\omega \,\mathit{tf}\,\mathit{z\_abs}\,\mathit{q
\symbol{126}}}}
\end{gather*}

\[
\mathit{master\_6} := \mathit{a1}\,({\frac {\partial ^{3}}{
\partial t^{3}}}\,\psi (t)) + \mathit{a2}\,({\frac {\partial ^{2}
}{\partial t^{2}}}\,\psi (t)) + \mathit{a3}\,({\frac {\partial }{
\partial t}}\,\psi (t)) + \mathrm{sin}(\psi (t))
\]

\begin{gather*}
\psi (t)=\mathit{\_a}\,\mathrm{\&where} \left[ {\vrule
height1.50em width0em depth1.50em} \right. \!  \!  \left\{
{\vrule height1.31em width0em depth1.31em} \right. \!  \! (
{\frac {\partial ^{2}}{\partial \mathit{\_a}^{2}}}\,\mathrm{\_b}(
\mathit{\_a}))\,\mathrm{\_b}(\mathit{\_a})^{2} + {\frac {1}{\mathit{a1}}}\\
\mbox{}  (({\frac {\partial }{\partial
\mathit{\_a}}}\,\mathrm{\_b}(\mathit{\_a}))^{2}\,\mathrm{\_b}(
\mathit{\_a})\,\mathit{a1} + \mathit{a2}\,({\frac {\partial }{
\partial \mathit{\_a}}}\,\mathrm{\_b}(\mathit{\_a}))\,\mathrm{\_b
}(\mathit{\_a}) + \mathit{a3}\,\mathrm{\_b}(\mathit{\_a}) +\\
\mathrm{sin}(\mathit{\_a})) =0 \! \! \left.
{\vrule height1.31em width0em depth1.31em} \right\} ,  \\
\{\mathrm{\_b}(\mathit{\_a})={\frac {\partial }{\partial t}}\,
\psi (t), \,\mathit{\_a}=\psi (t)\}, \, \left\{  \! t=
{\displaystyle \int } {\displaystyle \frac {1}{\mathrm{\_b}(
\mathit{\_a})}} \,d\mathit{\_a} + \mathit{\_C1}, \,\psi (t)=
\mathit{\_a} \!  \right\}  \! \! \left. {\vrule
height1.50em width0em depth1.50em} \right]
\end{gather*}

\begin{gather*}
\boxed{\mathit{master\_7} := ({\frac {\partial ^{2}}{\partial \psi ^{2}
}}\,\rho (\psi ))\,\rho (\psi )^{2} + ({\frac {\partial }{
\partial \psi }}\,\rho (\psi ))^{2}\,\rho (\psi ) + \delta \,(
{\frac {\partial }{\partial \psi }}\,\rho (\psi ))\,\rho (\psi )
 +} \\
\boxed{ \rho (\psi )\,\alpha  - \rho (\psi )\,\mathit{gam}
\mbox{} - {\displaystyle \frac {2\,\mathrm{sin}(\psi )\,\pi \,N\,
d\,\omega \,\mathit{tf}\,\mathit{q\symbol{126}}\,\mathit{z\_abs}
}{c}}  }
\end{gather*}
\emptyline
\noindent
As it is known (see, for example, \cite{Allen}) the
propagation of the extremely short laser pulse in the coherent
absorber causes the effect of the self-induced transparency, when the
pulse does not suffer the decay and there is not a transformation of
the pulse shape in the absorber. This effect is the result of the
coherent interaction of the pulse with the atoms and is described on
the basis of Bloch equations. In the beginning we shall consider the
steady-state ultra-short pulse in the presence of the coherent
interaction with absorber, but without any lasing factors. In this
case, the modified master equation \textit{master\_7} contains only
two terms: the term describing the pulse time delay and the term
corresponding to the coherent polarization response. Note, that
$\rho $ is the field amplitude multiplied by parameter \textit{q},\textit{
$\psi $} is the pulse area.

\emptyline
$>$ode1 := delta*rho(psi)*diff(rho(psi),psi) - p*sin(psi);

\emptyline
\begin{center}
\[ \boxed{
\mathit{ode1} := \delta \,({\frac {\partial }{\partial \psi }}\,
\rho (\psi ))\,\rho (\psi ) - p\,\mathrm{sin}(\psi )}
\]
\end{center}
\emptyline
\noindent
The natural initial condition is
$\rho $\textit{(0) }= 0, that is before pulse front its amplitude and area
are 0. Then solutions of the master equation are

\emptyline
$>$sol := dsolve(\{ode1,rho(0)=0\},rho(psi));

\emptyline
\[
\mathit{sol} :=\boxed{ \rho (\psi )={\displaystyle \frac {\sqrt{\delta
\,( - 2\,p\,\mathrm{cos}(\psi ) + 2\,p)}}{\delta }}} , \,\rho (
\psi )= - {\displaystyle \frac {\sqrt{\delta \,( - 2\,p\,\mathrm{
cos}(\psi ) + 2\,p)}}{\delta }}
\]

\emptyline
\noindent
As one can see, only one solution (
$\delta $ \TEXTsymbol{>} 0) is physical, because of the amplitude is to be
real.\\
\indent
The distinct pulse velocity defines the pulse amplitude. This fact is
illustrated by the next figure:

\emptyline
$>$animate(evalf(abs(subs(p=5e-4,subs(sol[1],rho(psi)))/10)),\\
\indent
psi=0..2*Pi,delta=0.3e-5..1e-4,\\
\indent
frames=100,axes=boxed,color=red,\\
\indent
labels=[`pulse area`,`y, MV/cm`],title=`Pulse amplitude versus its area`);

\begin{center}
\mapleplot{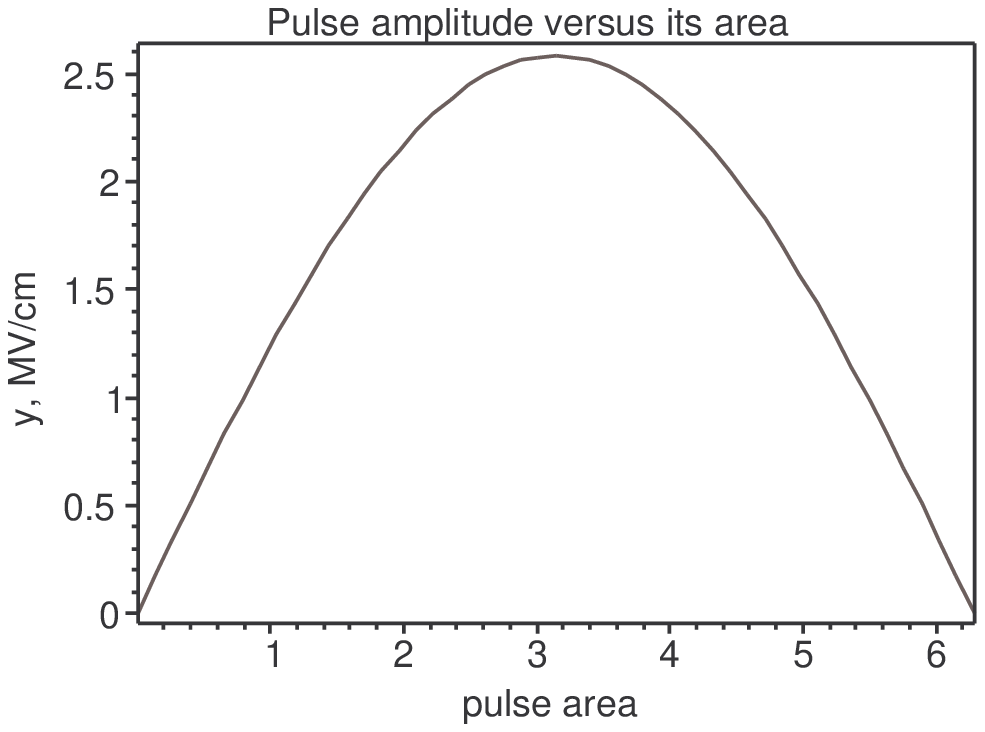}
\end{center}

\emptyline
\begin{center}
\fbox{ \parbox{.8\linewidth}{
The figure demonstrates, that the full pulse area is equal to 2
$\pi $. Such pulse is named as 2
$\pi $ - pulse or 2
$\pi $ - soliton and the process of its formation is the so-called
phenomenon of the self-induced transparency}}
\end{center}
\emptyline
More natural representation of the coherent solitons in our case is
produced by the transition to the coordinates "pulse area - time".
Then the master equation (see \textit{master\_4}) can be written as
follows:

\emptyline
$>$ode2 := diff(psi(t),t\$2)=a*sin(psi(t));\# analog of nonlinear
Klein-Gordon equation

\emptyline
\begin{center}
\[ \boxed{
\mathit{ode2} := {\frac {\partial ^{2}}{\partial t^{2}}}\,\psi (t
)=a\,\mathrm{sin}(\psi (t))}
\]
\end{center}

\emptyline
\noindent
Here \textit{a=p/
$\delta $}. This is the equation, which is analog of the equation of the
pendulum rotation (the angle variable \textit{x} in our case is the
ultra-short pulse area) and the angle is measured beginning from the
upper equilibrium point if
$\delta $\textit{ }\TEXTsymbol{>} 0 (see above).

\emptyline
$>$ode2 := diff(x(t),t\$2)=a*sin(x(t));

\begin{center}
\[
\mathit{ode2} := {\frac {\partial ^{2}}{\partial t^{2}}}\,
\mathrm{x}(t)=a\,\mathrm{sin}(\mathrm{x}(t))
\]
\end{center}

\emptyline
\noindent
The solution of this equation is well known.

\emptyline
$>$sol1 := dsolve(ode2,x(t));

\emptyline
\begin{gather*}
\mathit{sol1} := {\displaystyle \int _{\ }^{\mathrm{x}(t)}}  -
{\displaystyle \frac {1}{\sqrt{ - 2\,a\,\mathrm{cos}(\mathit{\_a}
) + \mathit{\_C1}}}} \,d\mathit{\_a} - t - \mathit{\_C2}=0,  \\
{\displaystyle \int _{\ }^{\mathrm{x}(t)}} {\displaystyle \frac {
1}{\sqrt{ - 2\,a\,\mathrm{cos}(\mathit{\_a}) + \mathit{\_C1}}}}
\,d\mathit{\_a} - t - \mathit{\_C2}=0
\end{gather*}

\emptyline
\noindent
Make an explicit integration:

\emptyline
$>$sol2\_a := value(lhs(sol1[1]));\\
\indent \indent
sol2\_b := value(lhs(sol1[2]));

\emptyline
\begin{gather*}
\mathit{sol2\_a} := 2\sqrt{ - ( - 4\,a\,\mathrm{\%1} + 2\,a +
\mathit{\_C1})\,( - 1 + \mathrm{\%1})}\,\sqrt{1 - \mathrm{\%1}}\,\\
\sqrt{{\displaystyle \frac { - 4\,a\,\mathrm{\%1} + 2\,a +
\mathit{\_C1}}{2\,a + \mathit{\_C1}}} }
\mathrm{EllipticF}(\mathrm{cos}({\displaystyle \frac {1}{2}} \,
\mathrm{x}(t)), \,2\,\sqrt{{\displaystyle \frac {a}{2\,a +
\mathit{\_C1}}} }) \left/ {\vrule
height0.84em width0em depth0.84em} \right. \!  \!  \\
\sqrt{(4\,a\,\mathrm{cos}({\displaystyle \frac {1}{2}} \,\mathrm{x
}(t))^{4} - 6\,a\,\mathrm{\%1} + 2\,a - \mathit{\_C1}\,\mathrm{
\%1} + \mathit{\_C1}}\,\\
\mathrm{sin}({\displaystyle \frac {1}{2}}
\,\mathrm{x}(t))\,\sqrt{ - 4\,a\,\mathrm{\%1} + 2\,a + \mathit{
\_C1}})\mbox{} - t
 - \mathit{\_C2} \\
\mathrm{\%1} := \mathrm{cos}({\displaystyle \frac {1}{2}} \,
\mathrm{x}(t))^{2}
\end{gather*}

\begin{gather*}
\mathit{sol2\_b} :=  - 2\,{\displaystyle \frac {\sqrt{1 -
\mathrm{cos}({\displaystyle \frac {1}{2}} \,\mathrm{x}(t))^{2}}\,
\sqrt{{\displaystyle \frac { - 4\,a\,\mathrm{cos}({\displaystyle
\frac {1}{2}} \,\mathrm{x}(t))^{2} + 2\,a + \mathit{\_C1}}{2\,a
 + \mathit{\_C1}}} }\,\%1)}{\mathrm{sin}(
{\displaystyle \frac {1}{2}} \,\mathrm{x}(t))\,\sqrt{ - 4\,a\,
\mathrm{cos}({\displaystyle \frac {1}{2}} \,\mathrm{x}(t))^{2} +
2\,a + \mathit{\_C1}}}}  \\
\mbox{} - t - \mathit{\_C2} \\
\mathrm{\%1} := \mathrm{EllipticF}(\mathrm{cos}(
{\displaystyle \frac {1}{2}} \,\mathrm{x}(t)), \,2\,\sqrt{
{\displaystyle \frac {a}{2\,a + \mathit{\_C1}}}}
\end{gather*}

\emptyline
\noindent
The result is expressed through elliptic integrals. The simplification
of the radicals produces:

\emptyline
$>$sol3\_a := simplify(sol2\_a+t+\_C2,radical,symbolic)=t+\_C2;\\
\indent
sol3\_b := simplify(sol2\_b+t+\_C2,radical,symbolic)=t+\_C2;

\emptyline
\begin{gather*}
\mathit{sol3\_a} := \\
2\,{\displaystyle \frac {\sqrt{1 - \mathrm{
cos}({\displaystyle \frac {1}{2}} \,\mathrm{x}(t))^{2}}\,\mathrm{
EllipticF}(\mathrm{cos}({\displaystyle \frac {1}{2}} \,\mathrm{x}
(t)), \,2\,{\displaystyle \frac {\sqrt{(2\,a + \mathit{\_C1})\,a}
}{2\,a + \mathit{\_C1}}} )}{\sqrt{2\,a + \mathit{\_C1}}\,\mathrm{
sin}({\displaystyle \frac {1}{2}} \,\mathrm{x}(t))}} \\
= t + \mathit{\_C2}
\end{gather*}

\begin{gather*}
\mathit{sol3\_b} :=\\
  - 2\,{\displaystyle \frac {\sqrt{1 -
\mathrm{cos}({\displaystyle \frac {1}{2}} \,\mathrm{x}(t))^{2}}\,
\mathrm{EllipticF}(\mathrm{cos}({\displaystyle \frac {1}{2}} \,
\mathrm{x}(t)), \,2\,{\displaystyle \frac {\sqrt{(2\,a + \mathit{
\_C1})\,a}}{2\,a + \mathit{\_C1}}} )}{\sqrt{2\,a + \mathit{\_C1}}
\,\mathrm{sin}({\displaystyle \frac {1}{2}} \,\mathrm{x}(t))}} \\
= t + \mathit{\_C2}
\end{gather*}

\emptyline
\noindent
So, we have:

\emptyline
$>$sol3\_a :=\\
\indent
2*EllipticF(cos(1/2*x(t)),2*sqrt(a*(2*a-\_C1))/(2*a-\_C1))/\\
\indent
sqrt(-2*a+\_C1) = t+\_C2;\\
\indent \indent
sol3\_b :=\\
\indent
-2*EllipticF(cos(1/2*x(t)),2*sqrt(a*(2*a-\_C1))/(2*a-\_C1))/\\
\indent
sqrt(-2*a+\_C1) = t+\_C2;

\emptyline
\[
\mathit{sol3\_a} := 2\,{\displaystyle \frac {\mathrm{EllipticF}(
\mathrm{cos}({\displaystyle \frac {1}{2}} \,\mathrm{x}(t)), \,2\,
{\displaystyle \frac {\sqrt{a\,(2\,a - \mathit{\_C1})}}{2\,a -
\mathit{\_C1}}} )}{\sqrt{ - 2\,a + \mathit{\_C1}}}} =t + \mathit{
\_C2}
\]

\[
\mathit{sol3\_b} :=  - 2\,{\displaystyle \frac {\mathrm{EllipticF
}(\mathrm{cos}({\displaystyle \frac {1}{2}} \,\mathrm{x}(t)), \,2
\,{\displaystyle \frac {\sqrt{a\,(2\,a - \mathit{\_C1})}}{2\,a -
\mathit{\_C1}}} )}{\sqrt{ - 2\,a + \mathit{\_C1}}}} =t + \mathit{
\_C2}
\]

\emptyline
Now we define the initial conditions. Let suppose, that  \textit{x(0)
}=
$\pi $, \textit{dx(0)/dt }=
$\alpha $, where
$\alpha $ is the some positive value (we measure the time from lower
equilibrium point). Then

\emptyline
$>$i\_C1 :=\\
\indent
solve(simplify(subs(\{diff(x(t),t)=alpha,x(t)=Pi\},\\
\indent
expand(simplify(diff(lhs(sol3\_a),t)))))=1,\_C1);\\
\indent \indent
i\_C1 :=\\
\indent
solve(simplify(subs(\{diff(x(t),t)=alpha,x(t)=Pi\},\\
expand(simplify(diff(lhs(sol3\_b),t)))))=1,\_C1);

\begin{center}
\[
\mathit{i\_C1} := 2\,a + \alpha ^{2}
\]

\[
\mathit{i\_C1} := 2\,a + \alpha ^{2}
\]
\end{center}
\emptyline
\noindent
The second constant of integration can be found as follows:

\emptyline
$>$i\_C2 := simplify(subs(\{\_C1=i\_C1,x(t)=Pi\},lhs(sol3\_a)))=C2;\\
\indent
i\_C2 := simplify(subs(\{\_C1=i\_C1,x(t)=Pi\},lhs(sol3\_b)))=C2;

\begin{center}
\[
\mathit{i\_C2} := 0=\mathit{C2}
\]

\[
\mathit{i\_C2} := 0=\mathit{C2}
\]
\end{center}
\emptyline
\noindent
The implicit solution is

\emptyline
$>$sol4\_a :=\\
\indent
simplify(subs(\_C1=i\_C1,lhs(sol3\_a)),radical,symbolic)=t+lhs(i\_C2);\\
\indent
sol4\_b :=
simplify(subs(\_C1=i\_C1,lhs(sol3\_b)),radical,symbolic)\\
\indent \indent
=t+lhs(i\_C2);

\begin{center}
\[
\mathit{sol4\_a} := 2\,{\displaystyle \frac {\mathrm{EllipticF}(
\mathrm{cos}({\displaystyle \frac {1}{2}} \,\mathrm{x}(t)), \,2\,
{\displaystyle \frac {\sqrt{ - a}}{\alpha }} )}{\alpha }} =t
\]

\[
\mathit{sol4\_b} :=  - 2\,{\displaystyle \frac {\mathrm{EllipticF
}(\mathrm{cos}({\displaystyle \frac {1}{2}} \,\mathrm{x}(t)), \,2
\,{\displaystyle \frac {\sqrt{ - a}}{\alpha }} )}{\alpha }} =t
\]
\end{center}
\emptyline
\noindent
Let consider a special situation, when \textit{2
$\frac {\sqrt{\frac {p}{\delta }}}{\alpha }$}= 1:

\emptyline
$>$sol5\_a :=\\
\indent
solve(simplify(lhs(subs(2*sqrt(-a)/alpha=1,sol4\_a)))=rhs(sol4\_a),x(t));\\
\indent
sol5\_b :=
solve(simplify(lhs(subs(2*sqrt(-a)/alpha=1,sol4\_b)))\\
\indent \indent
=rhs(sol4\_b),x(t));

\begin{center}
\[
\mathit{sol5\_a} := 2\,\mathrm{arccos}(\mathrm{tanh}(
{\displaystyle \frac {1}{2}} \,t\,\alpha ))
\]

\[
\mathit{sol5\_b} := 2\,\pi  - 2\,\mathrm{arccos}(\mathrm{tanh}(
{\displaystyle \frac {1}{2}} \,t\,\alpha ))
\]
\end{center}

\emptyline
$>$if sign(sol5\_a) $<$ 0 then     \# we choose only the root
corresponding to growing area\\
\indent \indent
   sol := sol5\_a\\
   \indent \indent \indent
                    else\\
                    \indent \indent
   sol := sol5\_b\\
   \indent
fi:

\emptyline
\noindent
The dependence of the pulse area on the time is:

\emptyline
$>$pulse\_area := \\
\indent
subs(alpha=solve(2*sqrt(p/delta)/alpha=1,alpha),sol);\\
\indent

animate(evalf(subs(p=5e-4,pulse\_area)),\\
\indent
t=-10..10,delta=0.3e-5..1e-1,\\
\indent
frames=100,axes=boxed,color=red,labels=[`time, t/tf`,`psi`],\\
\indent
title=`Pulse area versus time`);\\
\indent \indent
\#time is normalized to tf

\begin{center}
\[
\mathit{pulse\_area} := 2\,\pi  - 2\,\mathrm{arccos}(\mathrm{tanh
}(t\,\sqrt{{\displaystyle \frac {p}{\delta }} }))
\]
\end{center}

\begin{center}
\mapleplot{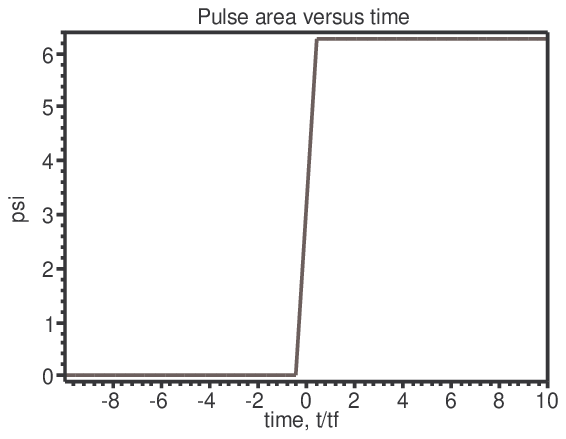}
\end{center}

\emptyline
\noindent
The pulse profile is:

\emptyline
$>$field := value(simplify(convert(diff(pulse\_area,t),sincos)));

\begin{center}
\[ \boxed{
\mathit{field} := 2\,{\displaystyle \frac {\mathrm{csgn} \left(
 \! \mathrm{cosh} \left(  \! {\displaystyle \frac { \left|  \! \,
{\displaystyle \frac {t^{2}\,p}{\delta }} \, \!  \right| }{t\,
\sqrt{{\displaystyle \frac {p}{\delta }} }}}  \!  \right)  \!
 \right) \,\sqrt{{\displaystyle \frac {p}{\delta }} }}{\mathrm{
cosh}(t\,\sqrt{{\displaystyle \frac {p}{\delta }} })}} }
\]
\end{center}

\emptyline
\noindent
This is the coherent soliton with duration
\fbox{${t_{p}}$ =
$\sqrt{\frac {\delta }{p}}$} and amplitude
\fbox{$\frac {2}{q\,{t_{p}}}$}

\emptyline
$>$animate(evalf(subs(p=5e-4,field)/10),t=-10..10,\\
\indent
delta=0.3e-3..1e-1,frames=100,axes=boxed,color=red,\\
\indent
labels=[`time, t/tf`,`rho,MV/cm`],title=`Pulse envelope`);

\begin{center}
\mapleplot{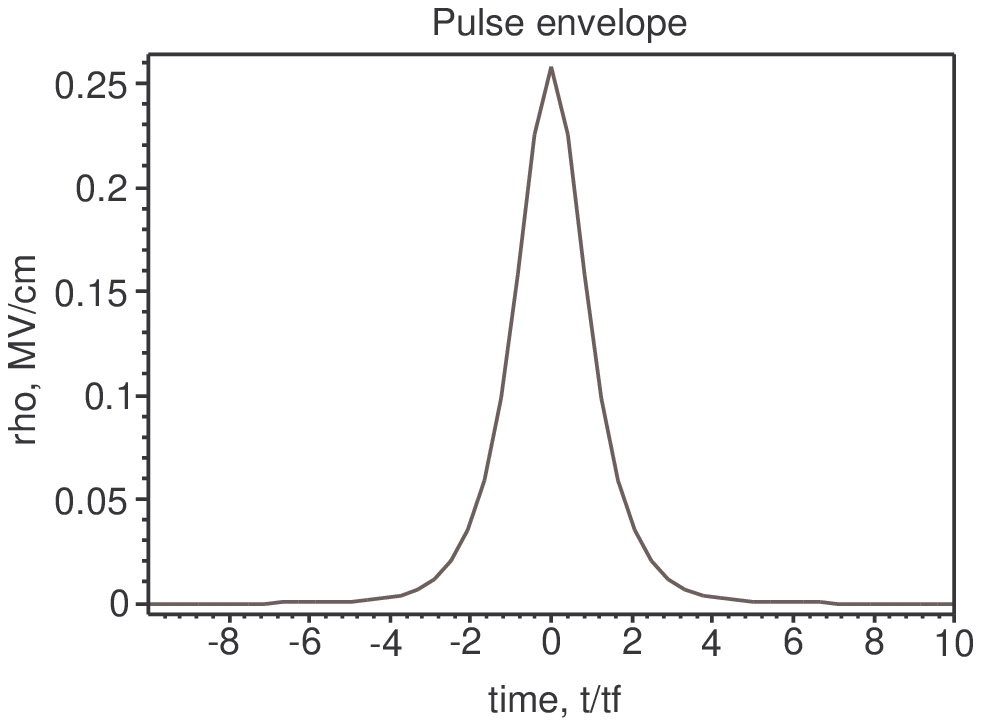}
\end{center}

\begin{center}
\fbox{ \parbox{.8\linewidth}{
We have shown that there is the coherent soliton with \textit{sech} -
shape profile in the condition of the coherent propagation in the
media described on the basis of Bloch equations}}
\end{center}
\emptyline
\noindent
This pulse may be
described in the coordinates 'field - area' or 'field - time'. The
first is formally very simple, but the second representation is
physically more obvious and corresponds to the model of nonlinear
pendulum.\\
\indent
Now we return to the laser model (\textit{master\_7}). This
second-order nonlinear nonautonomous ODE can be solved numerically
(\textit{pz} = \textit{2
$\pi $ N
$d^{2}$(
$\omega $*tf)*z\_abs/(c h)} = \textit{gam\_abs*
$\frac {\mathit{tf}}{\mathit{tcoh}}$}, we supposed \textit{gam\_abs} = 0.01)

\emptyline
$>$de := subs(\{\\
\indent \indent
alpha=0.1,\\
\indent \indent \indent
gam=0.04,\\
\indent \indent \indent \indent
delta=0.0042,\\
\indent \indent \indent
pz=5e-4\},\\
\indent
subs(\{op(6,master\_7)=-pz*sin(psi),rho(psi)=rho(psi)/10\},master\_7)):\\
\indent
\#here 10 results from the time normalization to tf (i.e. the field is
measured in 1/(q*tf) [MV/cm] - units)\\
\indent \indent
fig :=\\
\indent
DEplot([de=0],rho(psi),psi=0.01..1.985*Pi,[[rho(Pi)=0.76*10,\\
\indent
D(rho)(Pi)=1e-15]],rho=0..0.76*10,stepsize=0.001,linecolor=green):\\
\indent
display(fig,labels=[`pulse area`,`rho, MV/cm`],\\
\indent
title=`Pulse amplitude versus its area`,view=0..7.6);

\begin{center}
\mapleplot{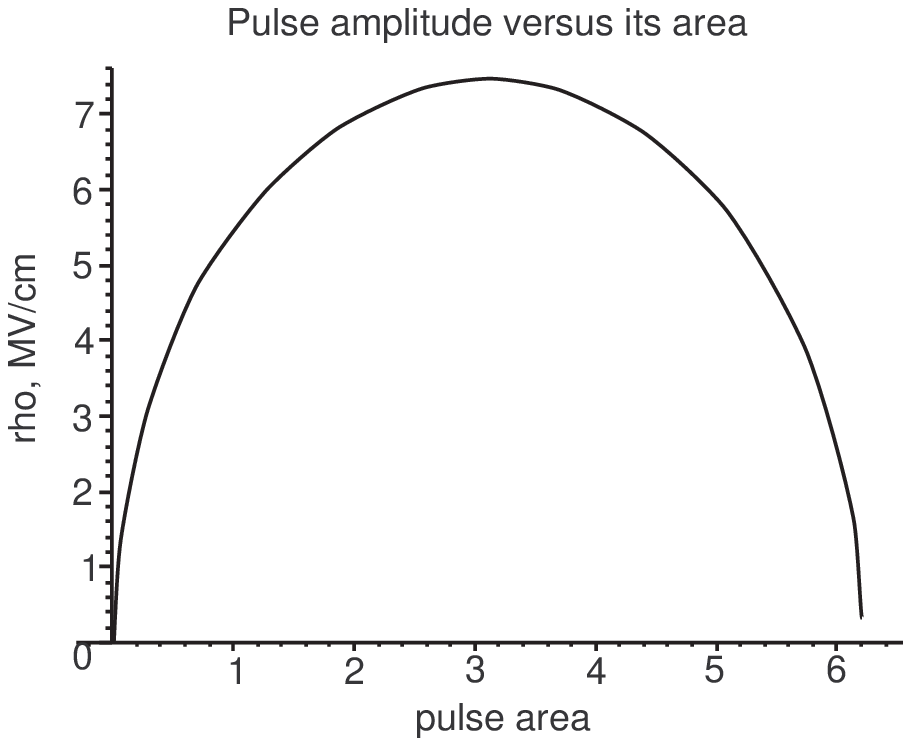}
\end{center}

\emptyline
\noindent
Let compare this result with the \textit{sech}-shape profile (blue
color) corresponding to the coherent soliton:

\emptyline
$>$plot(subs(am1=0.76,am1*sin(psi/2)*10),psi=0..2*Pi,color=blue):\\
\indent \indent
display(fig,\%,labels=[`pulse area`,`rho, MV/cm`],\\
\indent \indent
title=`Pulse amplitude versus its area`);

\begin{center}
\mapleplot{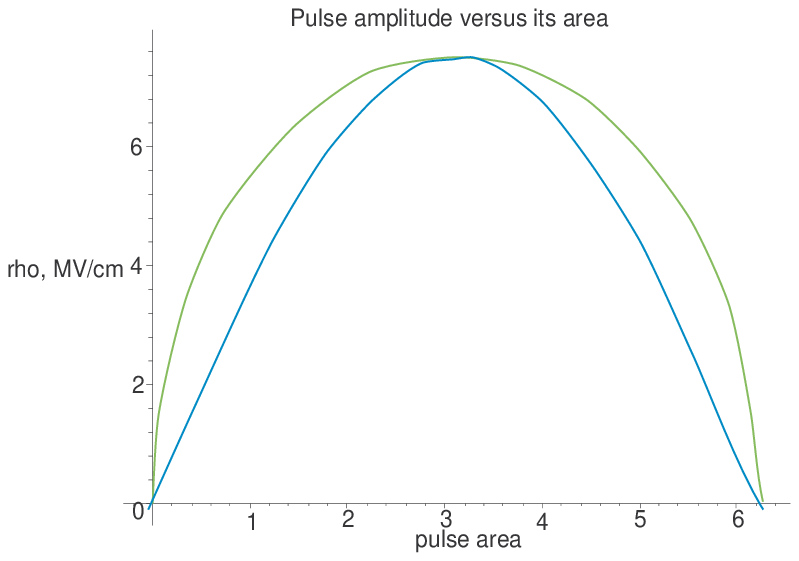}
\end{center}

\begin{center}
\fbox{ \parbox{.8\linewidth}{
We obtained 2
$\pi $ - pulse, but, as it can see from previous figure, such pulse is not
quasi-soliton with \textit{sech}-shape profile (which is shown by the
blue color, see below about connection between coordinates 'field -
area' and 'field - time' in this case)}}
\end{center}
\emptyline \noindent The field $10^{6}$ V/cm corresponds to the
intensity in vacuum 1.3 GW/ $\mathit{cm}^{2}$, that is the typical
intracavity intensity for the mode-locked solid-state laser. Now
make a try for obtaining of an approximate solution of the master
equation. We shall use the solution form, which is typical for the
analysis of the equations describing the autooscillations
(harmonic approximation):

\emptyline
$>$approx\_sol := \\
\indent
am1*sin(psi/2)+am2*sin(psi);\# approximation\\
\indent \indent
f7 :=\
\indent
numer(combine(expand(subs(rho(psi)=\\
\indent
approx\_sol,master\_7)),trig));\\
\indent
\# substitution into initial equation

\begin{center}
\[
\mathit{approx\_sol} := \mathit{am1}\,\mathrm{sin}(
{\displaystyle \frac {1}{2}} \,\psi ) + \mathit{am2}\,\mathrm{sin
}(\psi )
\]
\end{center}
\begin{gather*}
\mathit{f7} := 4\,\delta \,\mathit{am1}^{2}\,\mathrm{sin}(\psi )
\,c + 8\,\delta \,\mathit{am2}^{2}\,\mathrm{sin}(2\,\psi )\,c +
12\,\delta \,\mathit{am1}\,\mathit{am2}\,\mathrm{sin}(
{\displaystyle \frac {3}{2}} \,\psi )\,c \\
\mbox{} - 4\,\delta \,\mathit{am1}\,\mathit{am2}\,\mathrm{sin}(
{\displaystyle \frac {1}{2}} \,\psi )\,c + 11\,\mathit{am1}^{2}\,
\mathit{am2}\,\mathrm{sin}(2\,\psi )\,c - 16\,\mathit{gam}\,
\mathit{am2}\,\mathrm{sin}(\psi )\,c \\
\mbox{} - 16\,\mathit{gam}\,\mathit{am1}\,\mathrm{sin}(
{\displaystyle \frac {1}{2}} \,\psi )\,c + 16\,\alpha \,\mathit{
am2}\,\mathrm{sin}(\psi )\,c + 16\,\alpha \,\mathit{am1}\,
\mathrm{sin}({\displaystyle \frac {1}{2}} \,\psi )\,c \\
\mbox{} - 9\,\mathit{am1}\,\mathit{am2}^{2}\,\mathrm{sin}(
{\displaystyle \frac {3}{2}} \,\psi )\,c + 17\,\mathit{am1}\,
\mathit{am2}^{2}\,\mathrm{sin}({\displaystyle \frac {5}{2}} \,
\psi )\,c - \\
32\,\mathrm{sin}(\psi )\,\pi \,N\,d\,\omega \,
\mathit{tf}\,\mathit{q\symbol{126}}\,\mathit{z\_abs}
\mbox{} - 8\,\mathit{am2}^{3}\,\mathrm{sin}(\psi )\,c - 10\,
\mathit{am1}^{2}\,\mathit{am2}\,\mathrm{sin}(\psi )\,c + \\
2\,
\mathit{am1}^{3}\,\mathrm{sin}({\displaystyle \frac {3}{2}} \,
\psi )\,c - 2\,\mathit{am1}^{3}\,\mathrm{sin}({\displaystyle
\frac {1}{2}} \,\psi )\,c
\mbox{} - 10\,\mathit{am1}\,\mathrm{sin}({\displaystyle \frac {1
}{2}} \,\psi )\,\mathit{am2}^{2}\,c + \\
8\,\mathit{am2}^{3}\,
\mathrm{sin}(3\,\psi )\,c
\end{gather*}

\emptyline
\noindent
We have to collect the coefficients of sin(
$\psi $/2) and sin(
$\psi $):

\emptyline
$>$f8 := coeff(f7,sin(psi/2));\\
\indent \indent
f9 := coeff(f7,sin(psi));
\begin{gather*}
\mathit{f8} :=\\
  - 4\,\delta \,\mathit{am1}\,\mathit{am2}\,c - 16
\,\mathit{gam}\,\mathit{am1}\,c + 16\,\alpha \,\mathit{am1}\,c -
2\,\mathit{am1}^{3}\,c - 10\,\mathit{am1}\,\mathit{am2}^{2}\,c
\end{gather*}
\begin{gather*}
\mathit{f9} := 4\,\delta \,\mathit{am1}^{2}\,c - 16\,\mathit{gam}
\,\mathit{am2}\,c + 16\,\alpha \,\mathit{am2}\,c - 32\,\pi \,N\,d
\,\omega \,\mathit{tf}\,\mathit{q\symbol{126}}\,\mathit{z\_abs}\\
 - 8\,\mathit{am2}^{3}\,c - 10\,\mathit{am1}^{2}\,\mathit{am2}\,c
\end{gather*}

\emptyline
\noindent
Note, that in the absence of the lasing factors \textit{approx\_sol}
is the exact solution with the parameters \textit{am2} = 0,
\fbox{\textit{am1} = 2
$\sqrt{\frac {\mathit{pz}}{\delta }}$}. Below we will suppose, that the approximate solution is close to
the symmetrical shape, i. e. \textit{am2} = 0. Then

\emptyline
$>$f10 := expand(factor(subs(am2=0,f8))/(-2*c*am1));\\
\indent \indent
 f11 := subs(am2=0,f9);\\
 \indent \indent \indent
  symmetrical\_sol\_1 :=\\
  \indent
allvalues(solve(\{f10=0,f11=0\},\{am1,delta\}));

\begin{center}
\[
\mathit{f10} := \mathit{am1}^{2} + 8\,\mathit{gam} - 8\,\alpha
\]

\[
\mathit{f11} := 4\,\delta \,\mathit{am1}^{2}\,c - 32\,\pi \,N\,d
\,\omega \,\mathit{tf}\,\mathit{q\symbol{126}}\,\mathit{z\_abs}
\]
\end{center}

\begin{gather*}
\mathit{symmetrical\_sol\_1} := \{\boxed{ \delta ={\displaystyle \frac {
\pi \,N\,d\,\omega \,\mathit{tf}\,\mathit{q\symbol{126}}\,
\mathit{z\_abs}}{c\,( - \mathit{gam} + \alpha )}} , \,\mathit{am1
}=\sqrt{ - 8\,\mathit{gam} + 8\,\alpha }}\},  \\
\{\delta ={\displaystyle \frac {\pi \,N\,d\,\omega \,\mathit{tf}
\,\mathit{q\symbol{126}}\,\mathit{z\_abs}}{c\,( - \mathit{gam} +
\alpha )}} , \,\mathit{am1}= - \sqrt{ - 8\,\mathit{gam} + 8\,
\alpha }\}
\end{gather*}

\emptyline
\begin{center}
\fbox{ \parbox{.8\linewidth}{
So, the pulse amplitude is defined by the laser parameters, although
the relation between the pulse amplitude, duration and delay
corresponds to coherent soliton (see above) }}
\end{center}
\emptyline \noindent Now return to the coordinates 'amplitude -
time' for the harmonic approximation.

\emptyline
$>$symmetrical\_sol\_2 :=\\
\indent
dsolve(diff(psi(t),t)-subs(\{psi=psi(t),am2=0\},\\
\indent
approx\_sol),psi(t));\\

\begin{gather*}
\mathit{symmetrical\_sol\_2} :=\\
\psi (t)=2\,\mathrm{arctan}(2\,
{\displaystyle \frac {e^{(1/2\,t\,\mathit{am1} + 1/2\,\mathit{
\_C1}\,\mathit{am1})}}{1 + e^{(t\,\mathit{am1} + \mathit{\_C1}\,
\mathit{am1})}}} , \,{\displaystyle \frac { - e^{(t\,\mathit{am1}
 + \mathit{\_C1}\,\mathit{am1})} + 1}{1 + e^{(t\,\mathit{am1} +
\mathit{\_C1}\,\mathit{am1})}}} )
\end{gather*}

\emptyline
\noindent
The normalized field envelope is:

\emptyline
$>$symmetrical\_sol\_3:=\\
\indent \indent
simplify(diff(subs(symmetrical\_sol\_2,psi(t)),t));

\begin{center}
\[
\mathit{symmetrical\_sol\_3} := 2\,{\displaystyle \frac {\mathit{
am1}\,e^{(1/2\,\mathit{am1}\,(t + \mathit{\_C1}))}}{e^{(\mathit{
am1}\,(t + \mathit{\_C1}))} + 1}}
\]
\end{center}

\emptyline
\noindent
The initial condition is
$\rho $(0) = \textit{am1}. Then

\emptyline
$>$in\_C := solve(subs(t=0,symmetrical\_sol\_3)=am1,\_C1);

\begin{center}
\[
\mathit{in\_C} := 0, \,0
\]
\end{center}

\emptyline
\noindent
And finally we have:

\emptyline
$>$symmetrical\_sol := subs(\_C1=0,symmetrical\_sol\_3);

\begin{center}
\[ \boxed{
\mathit{symmetrical\_sol} := 2\,{\displaystyle \frac {\mathit{am1
}\,e^{(1/2\,t\,\mathit{am1})}}{e^{(t\,\mathit{am1})} + 1}} }
\]
\end{center}

\emptyline
\noindent
Naturally, this is a \textit{sec}\textit{h} - shape pulse with the
amplitude \fbox{\textit{am1} = \[
\frac{{2\sqrt {\alpha  - \gamma } }}{{qt_f }}
\]} and the duration \fbox{\textit{tp} = \[
\frac{2}{{am_1 q}} = \frac{{t_f }}{{\sqrt {2(\alpha  - \gamma )} }}
\]}. The pulse profile in the dependence on the gain coefficient can be
shown by the next function:

\emptyline
$>$plot3d(subs(gam=0.04,2*sqrt(2*(alpha-gam))*\\
\indent
sech(t*sqrt(2*(alpha-gam))/2.5e-15)*10),t=-1e-13..1e-13,\\
\indent
alpha=0.04..0.1,axes=boxed,orientation=[290,70],labels=\\
\indent
[`t, s`,`alpha`,`rho`],title=`Pulse amplitude (MV/cm) versus time`);


\emptyline
\noindent
Now we can found the dependence of the pulse parameters on the
critical laser parameter, that is the pump. For this aim, we have to
express the gain coefficient from the pump intensity. We shall
suppose, that active medium operates as a four level scheme. In this
case the steady-state saturated gain is described as follows
\cite{Herrmann}:

\emptyline
$>$alpha=Pump*alphamx/(Pump+tau*Energy+1/Tr);

\begin{center}
\[
\alpha ={\displaystyle \frac {\mathit{Pump}\,\mathit{alphamx}}{
\mathit{Pump} + \tau \,\mathit{Energy} + {\displaystyle \frac {1
}{\mathit{Tr}}} }}
\]
\end{center}

\emptyline
\noindent
Here Pump =
$\sigma $\_\textit{ab*Tc*Ip/h}*
$\nu $ is the normalized pump intensity,
$\sigma $ \_\textit{ab} is the absorption cross-section at the pump
wavelength, \textit{Tc} is the cavity period, \textit{Ip} is the pump
intensity, \textit{h}*
$\nu $ is the pump photon energy, \textit{alphamx} is the maximal gain,
\textit{Energy} is the normalized pulse energy, \textit{Tr} is the
gain recovery time normalized to \textit{Tc} (dimensionless\textit{
Tr} = 300 for Ti: sapphire laser with cavity period 10 ns).\\
\indent
For harmonical approximation:

\emptyline
$>$Energy=2*am1$^{2}$*tp:\#pulse energy\\
\indent \indent
f12 := Pump*alphamx/(Pump+tau*Energy+1/Tr)-alpha:\\
\indent \indent \indent
f13 :=\\
\indent
numer(simplify(subs(\\
\indent
\{am1=2*sqrt(2*(alpha-gam)),tp=1/sqrt(2*(alpha-gam))\\
\indent
\},subs(Energy=2*am1$^{2}$*tp,f12)))):\\
\indent
alpha\_sol := solve(f13=0,alpha): \#solution for the saturated gain

\emptyline
\noindent
The dependence of the pulse duration (two physical solutions
correspond to two different pulse energy) versus dimensionless pump
coefficient is

\emptyline
$>$fig := plot(\{\\
\indent
Re(evalf(subs(lambda=1,subs(\\
\indent
\{alphamx=0.1,Tr=300,tau=6.25e-4/lambda$^{2}$,gam=0.01\\
\indent
\},subs(alpha=alpha\_sol[1],2.5/sqrt(2*(alpha-gam))))))),\\
\indent
Re(evalf(subs(lambda=1,subs(\\
\indent
\{alphamx=0.1,Tr=300,tau=6.25e-4/lambda$^{2}$,gam=0.01\\
\indent
\},subs(alpha=alpha\_sol[3],2.5/sqrt(2*(alpha-gam)))))))\},\\
\indent
Pump=0.0005..0.005,axes=boxed,labels=[`Pump, a.u.`, `tp,\\
\indent
fs`],title=`Pulse duration versus pump`,color=red):\\
\indent \indent
display(fig,view=5..40);

\begin{center}
\mapleplot{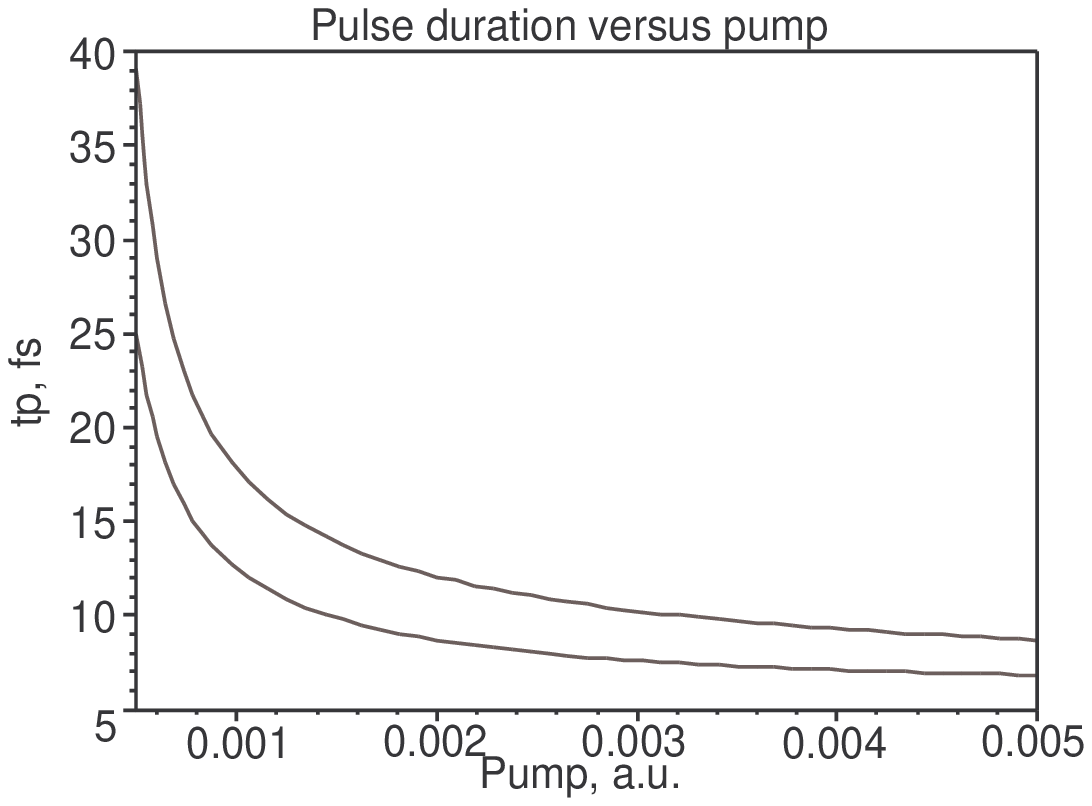}
\end{center}

\begin{center}
\fbox{ \parbox{.8\linewidth}{
As we can see, the pump growth decreases the pulse duration up to
sub-10 fs region}}
\end{center}

\emptyline \noindent The quasi-soliton in the absence of the
coherent absorber in the laser with self-focusing is described in
parts \textit{VII}, \textit{VIII}. Now we shall demonstrate, that
the essential features of the lasing in the presence of both
factors is the possibility of the quasi-soliton generation
(compare with above discussed situation). We shall search such
solutions.

\emptyline
$>$assume(t,real):\\
\indent \indent
 assume(tp,real):\\
 \indent \indent \indent
  a0 := 2/tp: \# This is the pulse amplitude\\
  \indent \indent
   sol := int(a0/cosh(t/tp),t): \# This is the pulse area psi\\
   \indent
     subs(\{disp=0,beta=0,psi(t)=sol,k\_2=0,beta=0\},master\_4):\\
     \indent \indent
       expand(\%):\\
       \indent \indent \indent
         numer(\%):\\
         \indent
           eq1 := expand(\%/(4*exp(t/tp)));

\emptyline
\begin{gather*}
\mathit{eq1} :=
  - 2\,\pi \,N\,\mathit{z\_abs}\,d\,\omega \,
\mathit{tp\symbol{126}}^{3}\,\mathit{q\symbol{126}}^{3}\,\lambda
^{2} + \\
2\,\pi \,N\,\mathit{z\_abs}\,d\,\omega \,\mathit{tp
\symbol{126}}^{3}\,\mathit{q\symbol{126}}^{3}\,\lambda ^{2}\,(e^{
(\frac {\mathit{t\symbol{126}}}{\mathit{tp\symbol{126}}})})^{4}
 + \alpha \,c\,\mathit{tp\symbol{126}}^{2}\,\mathit{q\symbol{126}
}^{2}\,\lambda ^{2} \\
\mbox{} + 2\,\alpha \,c\,\mathit{tp\symbol{126}}^{2}\,\mathit{q
\symbol{126}}^{2}\,\lambda ^{2}\,(e^{(\frac {\mathit{t
\symbol{126}}}{\mathit{tp\symbol{126}}})})^{2} + \alpha \,c\,
\mathit{tp\symbol{126}}^{2}\,\mathit{q\symbol{126}}^{2}\,\lambda
^{2}\,(e^{(\frac {\mathit{t\symbol{126}}}{\mathit{tp\symbol{126}}
})})^{4} - \mathit{gam}\,c\,\mathit{tp\symbol{126}}^{2}\,\mathit{
q\symbol{126}}^{2}\,\lambda ^{2} \\
\mbox{} - 2\,\mathit{gam}\,c\,\mathit{tp\symbol{126}}^{2}\,
\mathit{q\symbol{126}}^{2}\,\lambda ^{2}\,(e^{(\frac {\mathit{t
\symbol{126}}}{\mathit{tp\symbol{126}}})})^{2} - \mathit{gam}\,c
\,\mathit{tp\symbol{126}}^{2}\,\mathit{q\symbol{126}}^{2}\,
\lambda ^{2}\,(e^{(\frac {\mathit{t\symbol{126}}}{\mathit{tp
\symbol{126}}})})^{4} + \mathit{tf}^{2}\,c\,\mathit{q\symbol{126}
}^{2}\,\lambda ^{2} \\
\mbox{} - 6\,\mathit{tf}^{2}\,c\,\mathit{q\symbol{126}}^{2}\,
\lambda ^{2}\,(e^{(\frac {\mathit{t\symbol{126}}}{\mathit{tp
\symbol{126}}})})^{2} + \mathit{tf}^{2}\,(e^{(\frac {\mathit{t
\symbol{126}}}{\mathit{tp\symbol{126}}})})^{4}\,c\,\mathit{q
\symbol{126}}^{2}\,\lambda ^{2} + 16\,\sigma \,(e^{(\frac {
\mathit{t\symbol{126}}}{\mathit{tp\symbol{126}}})})^{2}\,c +\\
\delta \,c\,\mathit{tp\symbol{126}}\,\mathit{q\symbol{126}}^{2}\,
\lambda ^{2}
\mbox{} - \delta \,c\,\mathit{tp\symbol{126}}\,\mathit{q
\symbol{126}}^{2}\,\lambda ^{2}\,(e^{(\frac {\mathit{t
\symbol{126}}}{\mathit{tp\symbol{126}}})})^{4}
\end{gather*}

\emptyline
\noindent
Collect the terms with equal degrees of \textit{exp(t/tp)}. As result
we obtain the equations for the pulse and system parameters.

\emptyline
$>$e1 := coeff(eq1,exp(t/tp)$^{4}$);\\
\indent \indent
e2 := coeff(eq1,exp(t/tp)$^{2}$);\\
\indent
e3 := expand(eq1-e1*exp(t/tp)$^{4}$-e2*exp(t/tp)$^{2}$);

\begin{gather*}
\mathit{e1} := \\
2\,\pi \,N\,\mathit{z\_abs}\,d\,\omega \,\mathit{
tp\symbol{126}}^{3}\,\mathit{q\symbol{126}}^{3}\,\lambda ^{2} +
\alpha \,c\,\mathit{tp\symbol{126}}^{2}\,\mathit{q\symbol{126}}^{
2}\,\lambda ^{2} - \mathit{gam}\,c\,\mathit{tp\symbol{126}}^{2}\,
\mathit{q\symbol{126}}^{2}\,\lambda ^{2} + \mathit{tf}^{2}\,c\,
\mathit{q\symbol{126}}^{2}\,\lambda ^{2} \\
\mbox{} - \delta \,c\,\mathit{tp\symbol{126}}\,\mathit{q
\symbol{126}}^{2}\,\lambda ^{2}
\end{gather*}
\emptyline
\begin{center}
\[
\mathit{e2} := 2\,\alpha \,c\,\mathit{tp\symbol{126}}^{2}\,
\mathit{q\symbol{126}}^{2}\,\lambda ^{2} - 2\,\mathit{gam}\,c\,
\mathit{tp\symbol{126}}^{2}\,\mathit{q\symbol{126}}^{2}\,\lambda
^{2} - 6\,\mathit{tf}^{2}\,c\,\mathit{q\symbol{126}}^{2}\,\lambda
 ^{2} + 16\,\sigma \,c
\]
\end{center}

\begin{gather*}
\mathit{e3} :=\\
  - 2\,\pi \,N\,\mathit{z\_abs}\,d\,\omega \,
\mathit{tp\symbol{126}}^{3}\,\mathit{q\symbol{126}}^{3}\,\lambda
^{2} + \alpha \,c\,\mathit{tp\symbol{126}}^{2}\,\mathit{q
\symbol{126}}^{2}\,\lambda ^{2} - \mathit{gam}\,c\,\mathit{tp
\symbol{126}}^{2}\,\mathit{q\symbol{126}}^{2}\,\lambda ^{2} +
\mathit{tf}^{2}\,c\,\mathit{q\symbol{126}}^{2}\,\lambda ^{2} \\
\mbox{} + \delta \,c\,\mathit{tp\symbol{126}}\,\mathit{q
\symbol{126}}^{2}\,\lambda ^{2}
\end{gather*}

\emptyline
$>$e4 := simplify(e1-e3);\\
\indent \indent
e5 := simplify(e1+e3);\\
\indent \indent \indent
e6 := simplify(e2-e5);

\begin{center}
\[
\mathit{e4} := 4\,\pi \,N\,\mathit{z\_abs}\,d\,\omega \,\mathit{
tp\symbol{126}}^{3}\,\mathit{q\symbol{126}}^{3}\,\lambda ^{2} - 2
\,\delta \,c\,\mathit{tp\symbol{126}}\,\mathit{q\symbol{126}}^{2}
\,\lambda ^{2}
\]

\[
\mathit{e5} := 2\,\alpha \,c\,\mathit{tp\symbol{126}}^{2}\,
\mathit{q\symbol{126}}^{2}\,\lambda ^{2} - 2\,\mathit{gam}\,c\,
\mathit{tp\symbol{126}}^{2}\,\mathit{q\symbol{126}}^{2}\,\lambda
^{2} + 2\,\mathit{tf}^{2}\,c\,\mathit{q\symbol{126}}^{2}\,\lambda
 ^{2}
\]

\[
\mathit{e6} :=  - 8\,\mathit{tf}^{2}\,c\,\mathit{q\symbol{126}}^{
2}\,\lambda ^{2} + 16\,\sigma \,c
\]

\end{center}

\emptyline
$>$allvalues(solve(\{e4=0,e5=0,e6=0\},\{tp,delta,sigma\}));

\emptyline
\begin{gather*}
\{ \boxed{\mathit{tp\symbol{126}}=\sqrt{ - {\displaystyle \frac {1}{ -
\mathit{gam} + \alpha }} }\,\mathit{tf}, \,\delta = - 2\,
{\displaystyle \frac {\mathit{tf}^{2}\,\pi \,N\,\mathit{z\_abs}\,
d\,\omega \,\mathit{q\symbol{126}}}{c\,( - \mathit{gam} + \alpha
)}} , \,\sigma ={\displaystyle \frac {1}{2}} \,\mathit{tf}^{2}\,
\mathit{q\symbol{126}}^{2}\,\lambda ^{2}}\},  \\
\{\mathit{tp\symbol{126}}= - \sqrt{ - {\displaystyle \frac {1}{
 - \mathit{gam} + \alpha }} }\,\mathit{tf}, \,\delta = - 2\,
{\displaystyle \frac {\mathit{tf}^{2}\,\pi \,N\,\mathit{z\_abs}\,
d\,\omega \,\mathit{q\symbol{126}}}{c\,( - \mathit{gam} + \alpha
)}} , \,\sigma ={\displaystyle \frac {1}{2}} \,\mathit{tf}^{2}\,
\mathit{q\symbol{126}}^{2}\,\lambda ^{2}\}
\end{gather*}

\emptyline
\fbox{ \parbox{.8\linewidth}{
We see the essential differences from the previous situation: 1) there
is the pulse with \textit{sech}-shape (quasi-soliton); 2) the pulse
exists, when
$\alpha $ \TEXTsymbol{<}
$\gamma $, i. e. the linear loss exceeds the saturated gain. This is an
essential demand to the pulse stabilization and breaks the limitations
for the loss coefficient in the semiconductor absorber; 3) the
quasi-soliton exists only for the defined value of
$\sigma $, which can be changed for the fixed absorber properties by the
variation of
$\lambda $, i. e. by variation of mode cross-section in active medium and in
absorber or by variation of the absorber mirror reflectivity}}\\

\emptyline \noindent Note,that the pulse duration is defined by
the formula, which is similar to one for Kerr-lens mode locking
(see part \textit{VII}). Let find the pulse duration as the
function of pump.

\emptyline
$>$Energy=2*a0$^{2}$*tp:\#energy of quasi-soliton\\
\indent \indent
f14 :=\\
\indent
numer(simplify(subs(tp=1/sqrt(gam-alpha),\\
\indent \indent
subs(Energy=2*a0$^{2}$*tp,f12)))):\\
\indent
alpha\_sol2 := solve(f14=0,alpha):\#saturated gain\\
\indent \indent
  fig2 := plot(\\
  \indent
Re(evalf(subs(lambda=0.5,subs(\\
\indent
\{alphamx=0.1,Tr=300,tau=6.25e-4/lambda$^{2}$,gam=0.01\\
\indent
\},subs(alpha=alpha\_sol2[2],2.5/sqrt(gam-alpha)))))),\\
\indent
Pump=0.0005..0.005,axes=boxed,labels=[`Pump, a.u.`, `tp,fs`],\\
\indent
title=`Pulse duration versus pump`, color=blue):\\
\indent \indent
     display(fig,fig2,view=5..20);

\begin{center}
\mapleplot{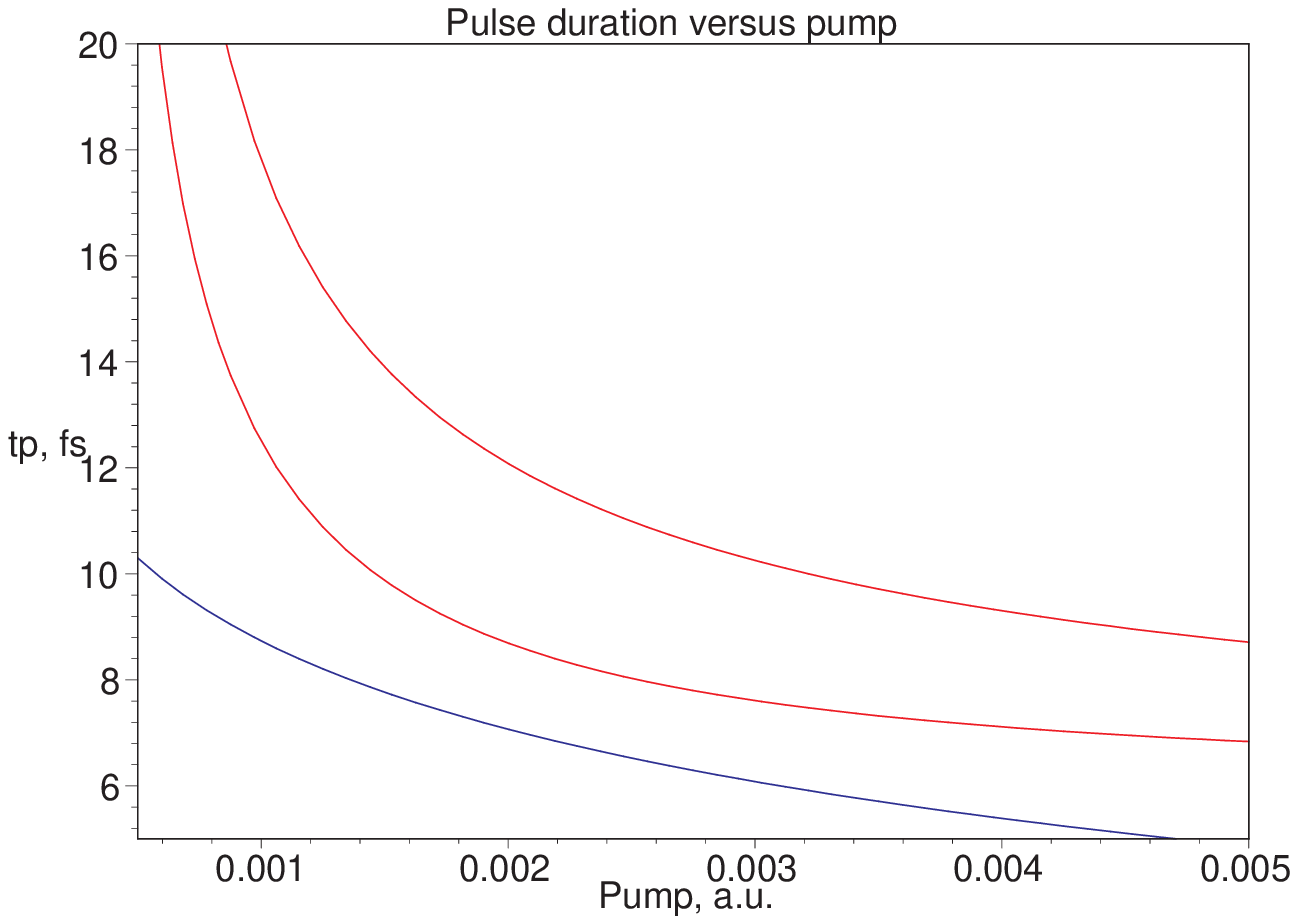}
\end{center}

\emptyline
\noindent
Thus, the Kerr-lensing (lower curve) allows to reduce the pulse
duration and to generate the sub-10 fs quasi-soliton.

\emptyline

\fbox{ \parbox{.8\linewidth}{So, we had demonstrated, that an joint action
of the lasing factors and coherent absorber objectives to the
generation of the coherent soliton. The ultrashort pulse in this case
has 2
$\pi $ area, but it is not \textit{sech}-shape pulse (quasi-soliton). The
obtained values of the pulse duration are placed within interval of 8
- 30 fs. The contribution of the self-focusing changes the pulse shape
essentially. In this case, there exists the stable \textit{sech}-shape
quasi-soliton with duration, which depends on the absorber mirror
reflectivity (or ratio of the laser mode cross-section in the active
medium and absorber). The obtained result is very attractive for the
elaboration of compact, all-solid-state, "hand-free" femtosecond
lasers}}

\section{Conclusion}

\emptyline
\noindent
The powerful computation abilities of Maple 6 allowed to demonstrate
the basic conceptions of the modern femtosecond technology. The
application of these conceptions to the Kerr-lens mode locking and
mode locking due to coherent semiconductor absorber leaded to the new
scientific results (see, for example, \cite{Jasapara}), which are very useful
for elaboration of the high-stable generators of sub-10 fs laser
pulses. We can see, that the combination of the symbolical, numerical
approaches and programming opens a door for a new opinion on the
ultrashort pulse generation. This opinion is based on the search of
the soliton and quasi-soliton states of nonlinear dynamical equations
and on the analysis of their evolution as the evolution of the
breezer's type.

\end{document}